\documentclass[12pt,a4paper]{article}
\usepackage{longtable}
\usepackage{afterpage} %
\usepackage[draft, textwidth=1.75cm]{todonotes} %
\usepackage{ifthen} %
\usepackage{bm} %
\newboolean{pdflatex}
\setboolean{pdflatex}{true} %

\newboolean{articletitles}
\setboolean{articletitles}{true} %

\newboolean{uprightparticles}
\setboolean{uprightparticles}{false} %

\usepackage{microtype}
\usepackage{lineno}  %
\usepackage{xspace} %
\usepackage{caption} %

\usepackage{graphicx}  %
\usepackage{color}
\usepackage{colortbl}
\graphicspath{{./figs/*}} %

\usepackage{amsmath} %
\usepackage{amssymb}
\usepackage{amsfonts}
\usepackage{upgreek} %

\newcommand*\patchAmsMathEnvironmentForLineno[1]{%
\expandafter\let\csname old#1\expandafter\endcsname\csname #1\endcsname
\expandafter\let\csname oldend#1\expandafter\endcsname\csname
end#1\endcsname
 \renewenvironment{#1}%
   {\linenomath\csname old#1\endcsname}%
   {\csname oldend#1\endcsname\endlinenomath}%
}
\newcommand*\patchBothAmsMathEnvironmentsForLineno[1]{%
  \patchAmsMathEnvironmentForLineno{#1}%
  \patchAmsMathEnvironmentForLineno{#1*}%
}
\AtBeginDocument{%
\patchBothAmsMathEnvironmentsForLineno{equation}%
\patchBothAmsMathEnvironmentsForLineno{align}%
\patchBothAmsMathEnvironmentsForLineno{flalign}%
\patchBothAmsMathEnvironmentsForLineno{alignat}%
\patchBothAmsMathEnvironmentsForLineno{gather}%
\patchBothAmsMathEnvironmentsForLineno{multline}%
\patchBothAmsMathEnvironmentsForLineno{eqnarray}%
}

\usepackage{hyperref}    %
\usepackage[all]{hypcap} %

\usepackage{xspace} 
\usepackage{upgreek}

\def\cleo   {\mbox{CLEO}\xspace}

\def\MagUp {\mbox{\em Mag\kern -0.05em Up}\xspace}

\ifthenelse{\boolean{uprightparticles}}%
{

 \def\Ppi         {\ensuremath{\uppi}\xspace}

 \def\Ppsi        {\ensuremath{\uppsi}\xspace}

 \def\PDelta      {\ensuremath{\Delta}\xspace}                 
 \def\PXi      {\ensuremath{\Xi}\xspace}                 
 \def\PLambda      {\ensuremath{\Lambda}\xspace}                 
 \def\PSigma      {\ensuremath{\Sigma}\xspace}                 
 \def\POmega      {\ensuremath{\Omega}\xspace}                 
 \def\PUpsilon      {\ensuremath{\Upsilon}\xspace}

 \def\PB      {\ensuremath{\mathrm{B}}\xspace}                 
                  
 \def\PD      {\ensuremath{\mathrm{D}}\xspace}

 \def\PJ      {\ensuremath{\mathrm{J}}\xspace}                 
 \def\PK      {\ensuremath{\mathrm{K}}\xspace}

 \def\Pi      {\ensuremath{\mathrm{i}}\xspace}

}
{

 \def\Ppi         {\ensuremath{\pi}\xspace}

 \def\Ppsi        {\ensuremath{\psi}\xspace}                 
                  
 \mathchardef\PDelta="7101
 \mathchardef\PXi="7104
 \mathchardef\PLambda="7103
 \mathchardef\PSigma="7106
 \mathchardef\POmega="710A
 \mathchardef\PUpsilon="7107
                  
 \def\PB      {\ensuremath{B}\xspace}                 
                  
 \def\PD      {\ensuremath{D}\xspace}

 \def\PJ      {\ensuremath{J}\xspace}                 
 \def\PK      {\ensuremath{K}\xspace}

 \def\Pi      {\ensuremath{i}\xspace}

}

\makeatletter
\ifcase \@ptsize \relax%
  \newcommand{\miniscule}{\@setfontsize\miniscule{4}{5}}%
\or%
  \newcommand{\miniscule}{\@setfontsize\miniscule{5}{6}}%
\or%
  \newcommand{\miniscule}{\@setfontsize\miniscule{5}{6}}%
\fi
\makeatother

\DeclareRobustCommand{\optbar}[1]{\shortstack{{\miniscule (\rule[.5ex]{1.25em}{.18mm})}
  \\ [-.7ex] $#1$}}

\def\pion   {{\ensuremath{\Ppi}}\xspace}

\def\pip    {{\ensuremath{\pion^+}}\xspace}
\def\pim    {{\ensuremath{\pion^-}}\xspace}

\def\kaon    {{\ensuremath{\PK}}\xspace}
  \def\Kbar    {{\kern 0.2em\overline{\kern -0.2em \PK}{}}\xspace}

\def\KorKbar    {\kern 0.18em\optbar{\kern -0.18em K}{}\xspace}

\def\Kp      {{\ensuremath{\kaon^+}}\xspace}
\def\Km      {{\ensuremath{\kaon^-}}\xspace}

  \def\Dbar    {{\kern 0.2em\overline{\kern -0.2em \PD}{}}\xspace}
\def\D       {{\ensuremath{\PD}}\xspace}
\def\Db      {{\ensuremath{\Dbar}}\xspace}
\def\DorDbar    {\kern 0.18em\optbar{\kern -0.18em D}{}\xspace}
\def\Dz      {{\ensuremath{\D^0}}\xspace}
\def\Dzb     {{\ensuremath{\Dbar{}^0}}\xspace}
\def\Dp      {{\ensuremath{\D^+}}\xspace}

\def\B       {{\ensuremath{\PB}}\xspace}
\def\Bbar    {{\ensuremath{\kern 0.18em\overline{\kern -0.18em \PB}{}}}\xspace}

\def\BorBbar    {\kern 0.18em\optbar{\kern -0.18em B}{}\xspace}

\def\Bub     {{\ensuremath{\B^-}}\xspace}

\def\Bm      {{\ensuremath{\Bub}}\xspace}

\def\jpsi     {{\ensuremath{{\PJ\mskip -3mu/\mskip -2mu\Ppsi\mskip 2mu}}}\xspace}

  \def\Y#1S{\ensuremath{\PUpsilon{(#1S)}}\xspace}%

\def\Lbar        {{\ensuremath{\kern 0.1em\overline{\kern -0.1em\PLambda}}}\xspace}
\def\LorLbar    {\kern 0.18em\optbar{\kern -0.18em \PLambda}{}\xspace}

\newcommand{\decay}[2]{\ensuremath{#1\!\to #2}\xspace}         %

\def\to                 {\ensuremath{\rightarrow}\xspace}

\def\eps   {{\ensuremath{\varepsilon}}\xspace}

\def\CP                {{\ensuremath{C\!P}}\xspace}

\def\AT#1     {\ensuremath{A_{\mathrm{T}}^{#1}}\xspace}           %

\def\C#1      {\ensuremath{\mathcal{C}_{#1}}\xspace}                       %
\def\Cp#1     {\ensuremath{\mathcal{C}_{#1}^{'}}\xspace}                    %
\def\Ceff#1   {\ensuremath{\mathcal{C}_{#1}^{\mathrm{(eff)}}}\xspace}        %
\def\Cpeff#1  {\ensuremath{\mathcal{C}_{#1}^{'\mathrm{(eff)}}}\xspace}       %
\def\Ope#1    {\ensuremath{\mathcal{O}_{#1}}\xspace}                       %
\def\Opep#1   {\ensuremath{\mathcal{O}_{#1}^{'}}\xspace}                    %

\newcommand{\tev}{\ifthenelse{\boolean{inbibliography}}{\ensuremath{~T\kern -0.05em eV}\xspace}{\ensuremath{\mathrm{\,Te\kern -0.1em V}}}\xspace}
\newcommand{\gev}{\ensuremath{\mathrm{\,Ge\kern -0.1em V}}\xspace}
\newcommand{\mev}{\ensuremath{\mathrm{\,Me\kern -0.1em V}}\xspace}
\newcommand{\kev}{\ensuremath{\mathrm{\,ke\kern -0.1em V}}\xspace}
\newcommand{\ev}{\ensuremath{\mathrm{\,e\kern -0.1em V}}\xspace}
\newcommand{\gevc}{\ensuremath{{\mathrm{\,Ge\kern -0.1em V\!/}c}}\xspace}
\newcommand{\mevc}{\ensuremath{{\mathrm{\,Me\kern -0.1em V\!/}c}}\xspace}
\newcommand{\gevcc}{\ensuremath{{\mathrm{\,Ge\kern -0.1em V\!/}c^2}}\xspace}
\newcommand{\gevgevcccc}{\ensuremath{{\mathrm{\,Ge\kern -0.1em V^2\!/}c^4}}\xspace}
\newcommand{\mevcc}{\ensuremath{{\mathrm{\,Me\kern -0.1em V\!/}c^2}}\xspace}

\newcommand{\stat}{\ensuremath{\mathrm{\,(stat)}}\xspace}
\newcommand{\syst}{\ensuremath{\mathrm{\,(syst)}}\xspace}

\newcommand{\chisq}{\ensuremath{\chi^2}\xspace}

\def\gsim{{~\raise.15em\hbox{$>$}\kern-.85em
          \lower.35em\hbox{$\sim$}~}\xspace}
\def\lsim{{~\raise.15em\hbox{$<$}\kern-.85em
          \lower.35em\hbox{$\sim$}~}\xspace}

\def\tell1  {TELL1\xspace}
\def\ukl1   {UKL1\xspace}

\newcommand{\eg}{\mbox{\itshape e.g.}\xspace}
\newcommand{\ie}{\mbox{\itshape i.e.}\xspace}

\newcommand{\cf}{\mbox{\itshape cf.}\xspace}

\usepackage{cite} %
\usepackage{mciteplus}
\newcommand{\phs}{\ensuremath{\Phi_4}}  %
\newcommand{\phsd}{\ensuremath{\phi_4}} %
\newcommand{\dphs}{\ensuremath{\mathrm{d}\phs}}  

\newcommand{\phsPoint}{\ensuremath{\mathbf{x}}}
\newcommand{\dphsPoint}{\ensuremath{\mathrm{d}^5x}}
\newcommand{\phsPointCP}{\ensuremath{\overline{\mathbf{x}}}}

\newcommand{\cleoc}{{C}{L}{E}{O}{-}{c}}

\newcommand{\prt}[1]{\ensuremath{#1}} %

\newcommand{\fourpi}{\prt{\pi^+\pi^-\pi^+\pi^-}}
\newcommand{\KKpipi}{\prt{K^+K^-\pi^+\pi^-}}

\newcommand{\eqnPRDref}[1]{Eq.~(\ref{#1})}

\newcommand{\EqnPRDref}[1]{Equation~(\ref{#1})}

\newcommand{\secref}[1]{Sec.~\ref{#1}}

\reversemarginpar %

\usepackage{pifont}
\definecolor{darkgreen}{rgb}{0.0,0.5,0.0}
\definecolor{darkred}{rgb}{0.5,0.0,0.0}

\renewcommand{\thefigure}{\arabic{section}.\arabic{figure}}
\makeatletter \@addtoreset{figure}{section} \makeatother

\makeatletter \@addtoreset{equation}{section} \makeatother

\renewcommand{\thetable}{\arabic{section}.\arabic{table}}
\makeatletter \@addtoreset{table}{section} \makeatother

\usepackage[labelfont=bf]{caption}

\usepackage{booktabs}
\usepackage[nottoc]{tocbibind}

\usepackage{subfig}
\usepackage{subfloat}
\usepackage{rotating}
\usepackage{enumitem}   

\begin{document}

\renewcommand{\thefootnote}{\fnsymbol{footnote}}
\setcounter{footnote}{1}

%
%
%
% start input /Users/pnaik/Documents/CLEO/Papers/latexpand/latex/title.tex
%

%
%
%
%
%
%

%
%
%
\begin{titlepage}

\vspace*{-1.5cm}

\hspace*{-0.5cm}
\flushright\today

\centering\hrulefill

\vspace{\fill}
{\bf\boldmath\huge
\begin{center}
  Amplitude Analyses of $\Dz \to \fourpi \, $ and $\Dz \to \KKpipi \, $ Decays
\end{center}
}
\vspace{\fill}
\begin{center}
P.~d'Argent$^1$,
N.~Skidmore$^2$,
J.~Benton$^2$,
J.~Dalseno$^2$,
E.~Gersabeck$^1$, \\
S.T.~Harnew$^2$,
P.~Naik$^2$, 
C.~Prouve$^2$, 
J.~Rademacker$^2$
\bigskip\\
{\it\footnotesize
$ ^1$Physikalisches Institut, Ruprecht-Karls-Universit\"{a}t Heidelberg, Heidelberg, Germany\\
$ ^2$H.H. Wills Physics Laboratory, University of Bristol, Bristol, United Kingdom\\
}
\end{center}
\vspace{\fill}
\begin{abstract}
  \noindent
The resonant substructure of $\Dz \to \fourpi \, $ decays is studied using data collected by the CLEO-c detector.
An amplitude analysis is performed in order to disentangle the various intermediate state contributions.
To limit the model complexity a data driven 
regularization procedure is applied.
The prominent contributions are the decay modes $\Dz \to a_{1}(1260)^{+} \, \pi^{-}$, $\Dz \to \sigma \, f_0(1370)$ 
and $\Dz \to \rho(770)^{0} \, \rho(770)^{0}$. %
The broad resonances $a_1(1260)^+$, 
$\pi(1300)^+$ 
and 
$a_1(1640)^+$ 
are studied in detail, 
including quasi-model-independent parametrizations of their lineshapes.
The mass and width of the $a_{1}(1260)^{+}$ meson are determined to be 
$m_{a_{1}(1260)^{+}} = [1225 \pm 9  \textrm{ (stat)}  \pm 17 \textrm{ (syst)} \pm 10 \textrm{ (model)}] \, \mev/c^{2}$ and
 $\Gamma_{a_{1}(1260)^{+}} = [430 \pm 24 \textrm{ (stat)} \pm 25 \textrm{ (syst)} \pm 18 \textrm{ (model)}] \, \mev$.

The amplitude model of $\Dz \to \KKpipi \, $ decays obtained from CLEO
II.V, CLEO III, and CLEO-c data is revisited with improved lineshape
parametrizations. The largest components are the decay modes $\Dz \to \phi(1020) \rho(770)^{0}$, $\Dz
\to K_1(1270)^+ K^-$ and $\Dz \to K(1400)^+ K^-$. %

The fractional \CP-even content of the decay $\Dz \to \fourpi \,$ is calculated 
from the amplitude model to be 
$F_{+}^{4\pi} = [72.9 \pm 0.9 \textrm{ (stat)} \pm 1.5 \textrm{ (syst)} \pm 1.0 \textrm{ (model)}] \, \%$, 
consistent with that obtained from a previous model-independent measurement.
For  $\Dz \to \KKpipi \,$ decays, the \CP-even fraction is measured for the first time and found to be $F_{+}^{KK\pi\pi} = [75.3 \pm 1.8
\textrm{ (stat)} \pm 3.3 \textrm{ (syst)} \pm 3.5 \textrm{ (model)}]
\, \%$.

The global decay rate asymmetries between \Dz and \Dzb decays are measured to be 
$\mathcal A^{4\pi}_{CP} = [+0.54 \pm 1.04 \stat \pm 0.51 \syst] \%$
and 
$\mathcal A^{KK\pi\pi}_{CP} = [+1.84 \pm 1.74 \stat \pm 0.30 \syst] \%$.
A search for \CP asymmetries in the amplitude components %
yields no evidence for \CP violation in either decay mode.
\end{abstract}
\vspace{\fill}
Published in JHEP 05 (2017) 143
\end{titlepage}

\pagestyle{empty}  %

\newpage
\setcounter{page}{2}
\mbox{~}

\cleardoublepage
 % end input /Users/pnaik/Documents/CLEO/Papers/latexpand/latex/title.tex
 %
%

\renewcommand{\thefootnote}{\arabic{footnote}}
\setcounter{footnote}{0}

\pagenumbering{gobble}
\tableofcontents
\cleardoublepage

\pagestyle{plain} %
\setcounter{page}{1}
\pagenumbering{arabic}

%
%
%\linenumbers

\newpage
% start input /Users/pnaik/Documents/CLEO/Papers/latexpand/latex/intro.tex
%

\clearpage\section{Introduction}
\label{sec:introduction}

We present amplitude analyses for $\Dz \to h^+ h^- \pip \pim$ decays,
where $h^{\pm}$ is either a pion or a kaon.
These decay modes
have the potential to make an important contribution to the determination of
the \CP-violating phase $\gamma~(\phi_3) \equiv -\arg(V_{ud}V^*_{ub}/V_{cd}V^*_{cb})$ in \prt{\Bm \to \D \Km} 
and related decays~\cite{GLW1, GLW2, ADS, DalitzGamma1, DalitzGamma2, Rademacker}. 
The all-charged
final states (impossible in three-body decays of \Dz) particularly
suit the environment of hadron collider experiments, such as LHCb. 
The sensitivity to the weak phase can be significantly improved with 
a measured  
$D$-decay %
amplitude model, either to be used directly in the $\gamma$ extraction, or in
order to optimize model-independent
measurements~\cite{DalitzGamma1,Atwood:coherenceFactor,Bondar:2005ki,Bondar:2007ir,coherenceFromMixing2}. 

\begin{figure}[b]
  \centering
  \includegraphics[width=0.49\textwidth,height=!]{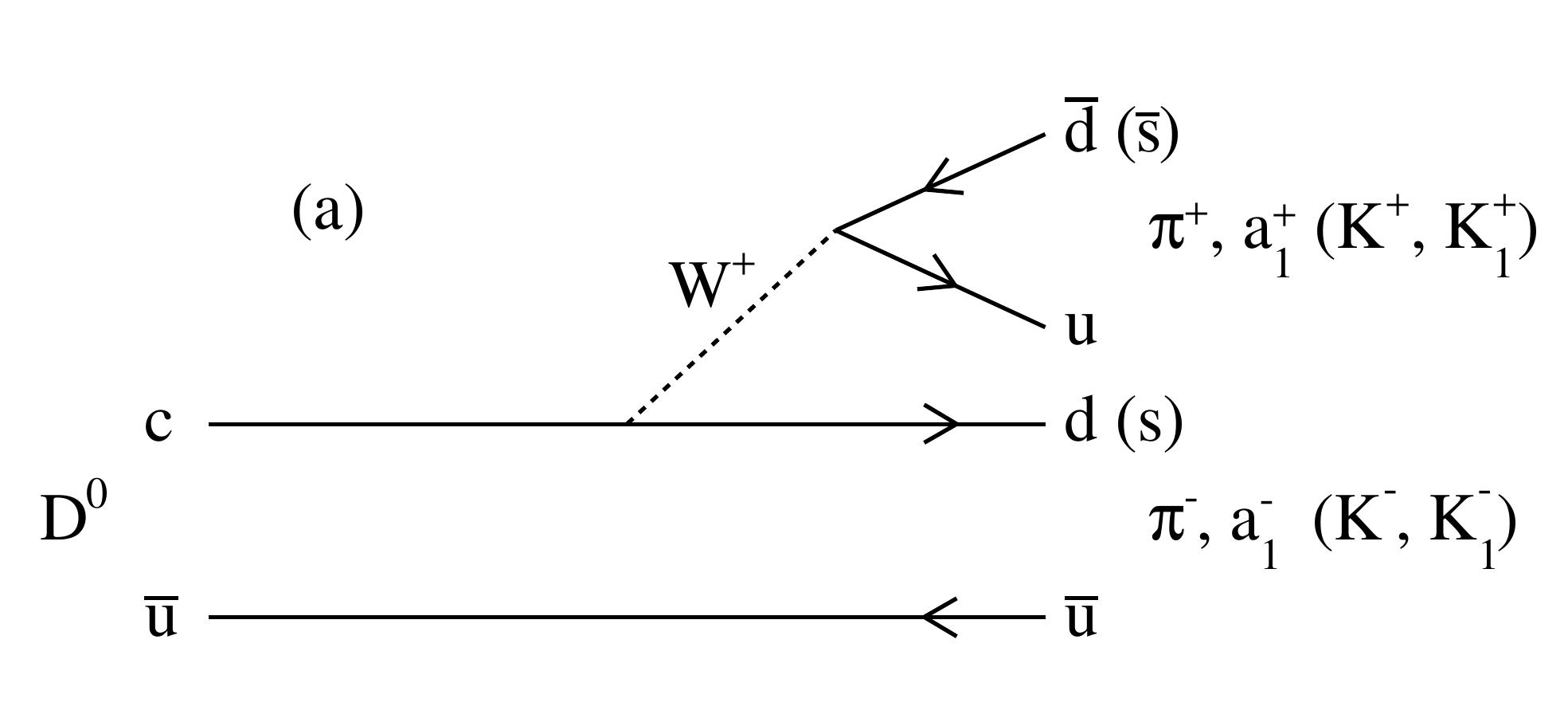}
  \includegraphics[width=0.49\textwidth,height=!]{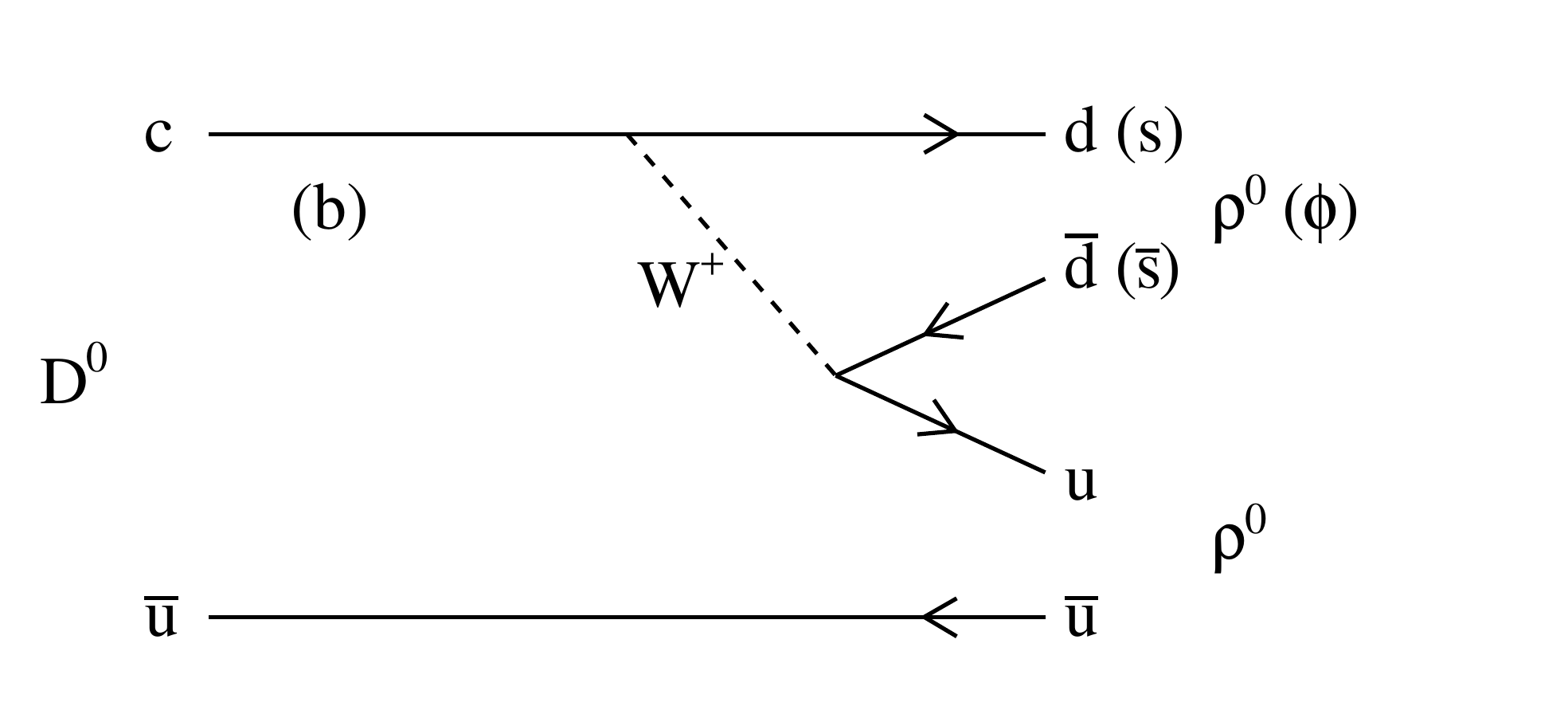}
  \includegraphics[width=0.49\textwidth,height=!]{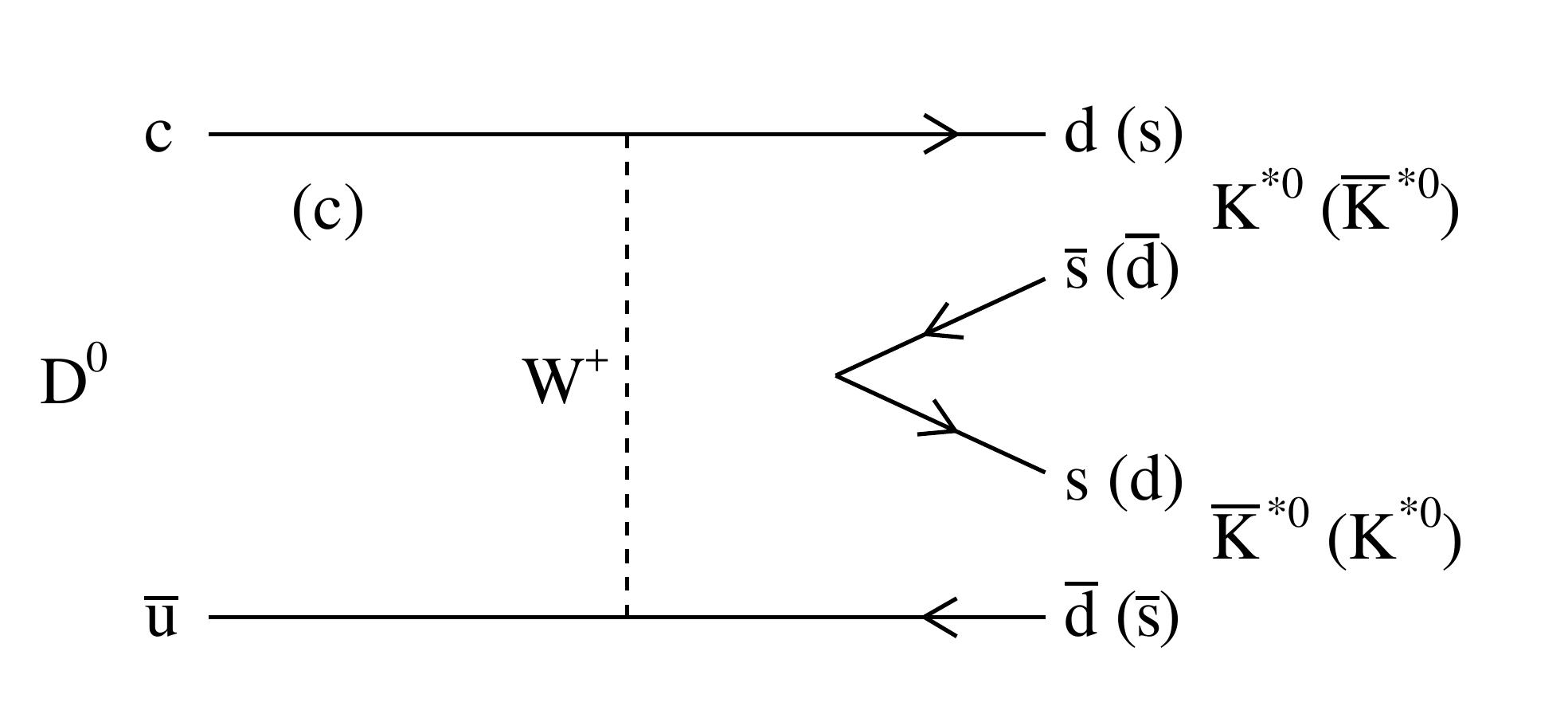}
  \caption{Examples of the color-favored (a), color-suppressed (b) and $W$-exchange (c) diagrams that contribute towards the resonant structure in \prt{\Dz\ \to \fourpi} and \prt{\Dz \to \KKpipi} decays.}
  \label{fig:feynman}
\end{figure}

A study of the rich resonance structure of these four-body decays is also of considerable
interest in its own right. Figure \ref{fig:feynman} shows the dominant processes that contribute to the visible structure in the phase space. 
The color-favored tree diagram manifests as a cascade whereby a resonance decays into 
another resonance before decaying into the final state. Due to the identical quark content produced in the 
weak and spectator interactions, a given process and its \CP-conjugate may arise even from the same initial state. 
Such processes, which we refer to as non-self-conjugate, are also known as flavor-non-specific 
decays as flavor-tagging is required to distinguish between the source of these two partners despite not being \CP\ eigenstates. 
The color-suppressed tree diagram and the $W$-exchange diagram result in self-conjugate intermediate states such as $\rho(770)^{0} \rho(770)^{0}$ or $\rho(770)^{0} \phi(1020)$ whose partial waves are eigenstates of \CP.
Certain intermediate states in $\Dz \to \KKpipi$ decays, for instance $K^{*}(892)^{0} \, \bar{K}^{*}(892)^{0}$, are only accessible via the $W$-exchange diagram.

The decay \prt{\Dz \to \fourpi} provides an
excellent environment to study the properties of the $a_1(1260)^+$ meson, whose
width is an unresolved question, currently given as $250-600\mev$ in
the Particle Data Group's {\it Review of Particle Physics} (PDG)~\cite{PDG2016}.
The only previous analysis of the \prt{\Dz \to \fourpi} amplitude
structure was published by the FOCUS collaboration based on approximately $6000$
\prt{\Dz,\Dzb \to \fourpi} signal events~\cite{FOCUS4pi}. 
The analysis presented here benefits from the ability to distinguish
\Dz\ from \Dzb\ decays and a larger data sample of approximately $7000$ signal events.

Based on the four-body amplitude formalism and analysis software
used in
the \prt{\Dz \to \KKpipi} amplitude analysis 
performed by the \cleo collaboration~\cite{KKpipi}, we 
introduce significant improvements especially in the
parametrization of three-body resonances. Using a state-of-the-art
parametrization of the $a_1(1260)^+$ lineshape, we present new
measurements of the $a_1(1260)^+$ mass and width. By utilizing different
parametrizations, we confirm a significant dependence of the
measured width on the lineshape itself. We
also observe contributions from the decay modes \prt{\Dz\ \to a_1(1640)^+ \, \pi^-} and
\prt{\Dz\ \to \pi(1300)^+ \, \pi^-}, not seen in previous analyses and provide
model-independent complex lineshapes for the $a_1(1260)^+$, 
$a_1(1640)^+$ and $\pi(1300)^+$ mesons. %

In addition to our new \prt{\Dz \to \fourpi}
analysis, we also revisit the \cleo 
\prt{\Dz \to \KKpipi} 
data using the improved formalism and analysis procedures presented in this paper. 
Prior to the \cleo analysis, an amplitude analysis of the decay \prt{\Dz \to \KKpipi} was also performed by FOCUS~\cite{FOCUSKKpipi}.

This article is structured as follows: After an introduction to
the CLEO~II.V, CLEO~III, and \cleoc\ experiments in \secref{sec:data} and a description of the event selection in \secref{sec:event}, the amplitude
formalism and its implementation is described in
\secref{sec:analysis} and \secref{sec:likelihood}, respectively. The results of the fit
to data, including a
model-dependent measurement of the fractional \CP-even content and search for direct \CP\ violation, 
are presented in \secref{sec:results4pi} and \secref{sec:resultsKKpi}. Systematic uncertainties are outlined in \secref{sec:systematics}, 
and
our
conclusions are given in \secref{sec:conclusions}.
Additional technical details of the analyses can be found in the appendices and supplementary material.
 % end input /Users/pnaik/Documents/CLEO/Papers/latexpand/latex/intro.tex
 %
% start input /Users/pnaik/Documents/CLEO/Papers/latexpand/latex/selection.tex
%

\clearpage\section{Data Set and CLEO Detector}
\label{sec:data}

The data analyzed in this paper were produced in symmetric $e^+e^-$ collisions at CESR between 1995 and 2008, and %
collected with three different configurations of the CLEO detector:
CLEO II.V, CLEO III, and CLEO-c.

In CLEO II.V~\cite{Kubota:1991ww,Hill:1998ea} tracking was provided by a three-layer double-sided silicon vertex detector, and two %
drift chambers. Charged particle identification came from $dE/dx$ information in the drift chambers, and %
time-of-flight (TOF) counters inserted before the calorimeter.  For CLEO III~\cite{Viehhauser:2001ue} a new silicon  vertex %
detector was installed, and a ring imaging Cherenkov (RICH) detector was deployed to enhance the particle %
identification abilities~\cite{Artuso:2005dc}.  In CLEO-c, the vertex detector was replaced with a low-mass wire drift %
chamber \cite{ZD}.  A superconducting solenoid supplied a 1.5 T magnetic field for CLEO II.V and III, and 1~T for %
CLEO-c operation, where the average particle momentum was lower.
In all detector configurations, neutral pion and photon identification was provided by a 7800-crystal  CsI %
electromagnetic calorimeter.

Four distinct data sets are analyzed in the present study:
\begin{enumerate}[label=(\arabic*)]
\item{
approximately 9~$\rm fb^{-1}$  accumulated at $\sqrt{s} \approx 10$~GeV by the CLEO II.V detector;
}
\item{
a total of 15.3~$\rm fb^{-1}$  accumulated by the CLEO III detector  in an energy range $\sqrt{s} = 7.0-11.2$~GeV, %
with over 90\% of this sample taken at  $\sqrt{s} = 9.5-10.6$~GeV;
}
\item{
 818~$\rm pb^{-1}$ collected at the $\psi(3770)$ resonance by the CLEO-c detector;
}
\item{
a further 600~$\rm pb ^{-1}$  taken by CLEO-c at $\sqrt{s} = 4170$~MeV,}
\end{enumerate}
where $\sqrt{s}$ is the total energy delivered by the beam in the center-of-mass system (CMS).
These samples are referred to as the CLEO II.V, CLEO III, CLEO-c 3770 and CLEO-c 4170 data %
sets, respectively.

Detector response is studied with GEANT-based~\cite{GEANT3} Monte Carlo (MC) simulations of each detector configuration, %
in which the MC events are processed with the same reconstruction algorithm as used for data.

\clearpage\section{Event Selection}
\label{sec:event}

We select events where one neutral \D meson decays either into a \fourpi\ or \KKpipi\ final state. The analysis considers two classes of signal decays, for both of which information on the quantum numbers of the %
meson decaying to the signal mode is provided by an event tag.
\begin{enumerate}[label=(\roman*)]
\item{
  Flavor-tagged decays are selected from the CLEO II.V and CLEO III data sets,  in which the flavor of the %
decaying meson is determined by the charge of the `slow pion', $\pi_s$, in the $D^{*+} \to \Dz \pi^+_s$ decay %
chain. Flavor-tagged decays are also selected from the two CLEO-c data sets, where here the tag is obtained through %
the charge of a kaon associated with the decay of the other $D$ meson in the event. The wrong tag fractions for each data set are represented by the parameter $w$, given in Ref.~\cite{KKpipi}. 
}
\item{
 \CP-tagged  decays are selected in the CLEO-c 3770 data set alone.  In $\psi(3770)$ decays the %
  $D-\overline{D}$ pair is produced coherently.  Therefore, the \CP\ of the signal $D$ can be determined if the %
other $D$ meson is reconstructed in a decay to a \CP-eigenstate.  Useful information is also obtained if the %
tagging meson is reconstructed decaying into the modes
$K^0_S \pi^+\pi^-$ or $K^0_L \pi^+\pi^-$, for which the relative contribution of \CP-even and \CP-odd states is %
known~\cite{Libby:2010nu}.
}
\end{enumerate}

The $\Dz\ \to \fourpi$ analysis uses only the flavor-tagged subset of the CLEO-c 3770 data sample, while $\Dz\ \to \KKpipi$ makes use of all the data sets described. The selection criteria for producing the data sets of each of these classes is discussed in detail in Ref.~\cite{KKpipi} and is identical to that used in our analysis, except for a few improvements that will be highlighted where applicable.

\subsection{\texorpdfstring{$\Dz\ \to \fourpi$ Selection}{D0 to pi+ pi- pi+ pi- Selection}}
\label{sec:event4pi}

Apart from other backgrounds, there is a source of peaking background arising from
$\Dz \to K_{S}^{0} (\to \pip \pim) \, \pip \, \pim$ decays.
Although this has the same final state as signal, it is an incoherent process since
the $K_{S}^{0}$ lifetime is much longer than those of any other possible intermediate resonance.
Therefore, $K_{S}^{0}$ decays are rejected if the invariant mass of any $\pip \, \pim$ combination is within $7.5\,\mev/c^2$ 
of the world-average $K_{S}^{0}$ mass~\cite{PDG2016}.

Two nearly uncorrelated kinematic variables are used to define a signal and two sideband background regions. 
These variables are defined as the beam-constrained mass,
\begin{equation}
  m_{bc} \equiv \sqrt{ {\left(\frac{\sqrt{s}}{2}\right)}^2 -  {\vec{p}_{D}}{}^2 }, %
\end{equation}
where 
$\vec{p}_{D}$ is the reconstructed three-momentum of the candidate \D\ in the CMS;
and the missing energy $\Delta E$,
\begin{equation}
	\Delta E \equiv E_{D} - \frac{\sqrt{s}}{2},
\end{equation}
where $E_{D}$ is the total reconstructed energy of candidate \D\ in the CMS. Signal events should have missing energy close to zero and beam-constrained mass close to that of the nominal $\Dz$ mass, $m_{D}$~\cite{PDG2016}. By construction, the $m_{bc}$ width is a measure of the beam-energy spread while the $\Delta E$ width is dominated by the detector resolution. 
Candidates that satisfy $m_{bc} > 1.83$ $\gev/c^2$ 
and $|\Delta E| < 0.1 \; \gev$ are retained for further analysis.

\begin{figure}[t]
	\centering
	\includegraphics[width=0.8\textwidth, height = !]{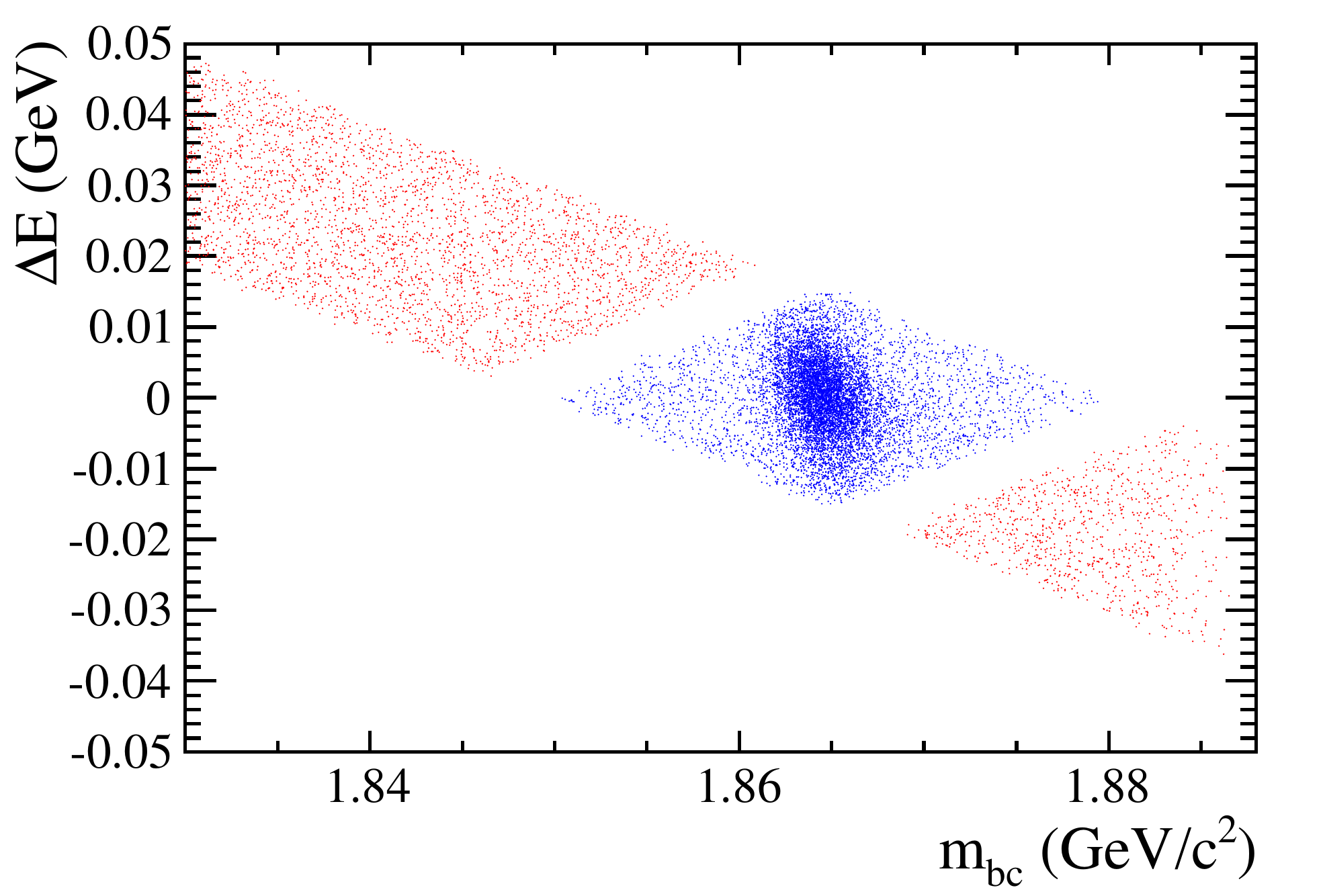} 
	\caption{Distribution of $\Dz\ \to \fourpi$ candidate events in missing energy $\Delta E$ and beam constrained mass within the
	selection regions, which are bounded by the annulus of constant invariant mass and lines normal to it. The central region (blue) is defined as the signal region,
	with sideband regions (red) providing background samples.
        }
	\label{fig:signalRegion}
\end{figure}

As the sideband events are used to study the background contribution within the signal region, it is crucial 
to select signal and background regions with a mutual and constant invariant mass, \ie that of the \D meson. 
First, a 
region of constant invariant mass is obtained by selecting events with 
\begin{equation}
  \left \vert \sqrt{ {{\Delta E}}^2 + {\Delta E} {\sqrt{s}} \, + {m_{bc}^{2}} } - {m_{D}} \right \vert < 15 \mev/c^2 .
  \label{eq:bentonbox}
\end{equation}
This relation describes an 
annulus
in $m_{bc}$ and $\Delta E$ space.
Lines normal to this 
annulus of constant invariant mass
have
an angle 
of inclination %
\begin{equation}
	\theta = \arctan \left( \frac{{\sqrt{s}}+2\,\Delta E }{2\,m_{bc}} \right) 
\end{equation}
about
the center of the annulus. %
A signal region around the \D mass peak is then defined by requiring 
$\vert \theta - \theta_{D} \vert < 0.004$, 
where $m_{bc} = m_{D}$ and $\Delta E = 0$ \gev\ at $\theta_{D}$,
as shown in Fig.~\ref{fig:signalRegion}. 
Similarly, sideband regions are defined with $ \vert \theta - \theta_{D} \vert > 0.006$.
These criteria preserve the range of invariant mass selected throughout the kinematic variables $m_{bc}$ and $\Delta E$, 
ensuring the distribution of events in phase space are consistent between regions. 
The signal region contains $9247$ $D \to \fourpi$ candidates.

\begin{figure}[t]
	\centering
	\includegraphics[width=0.45\textwidth, height=!]{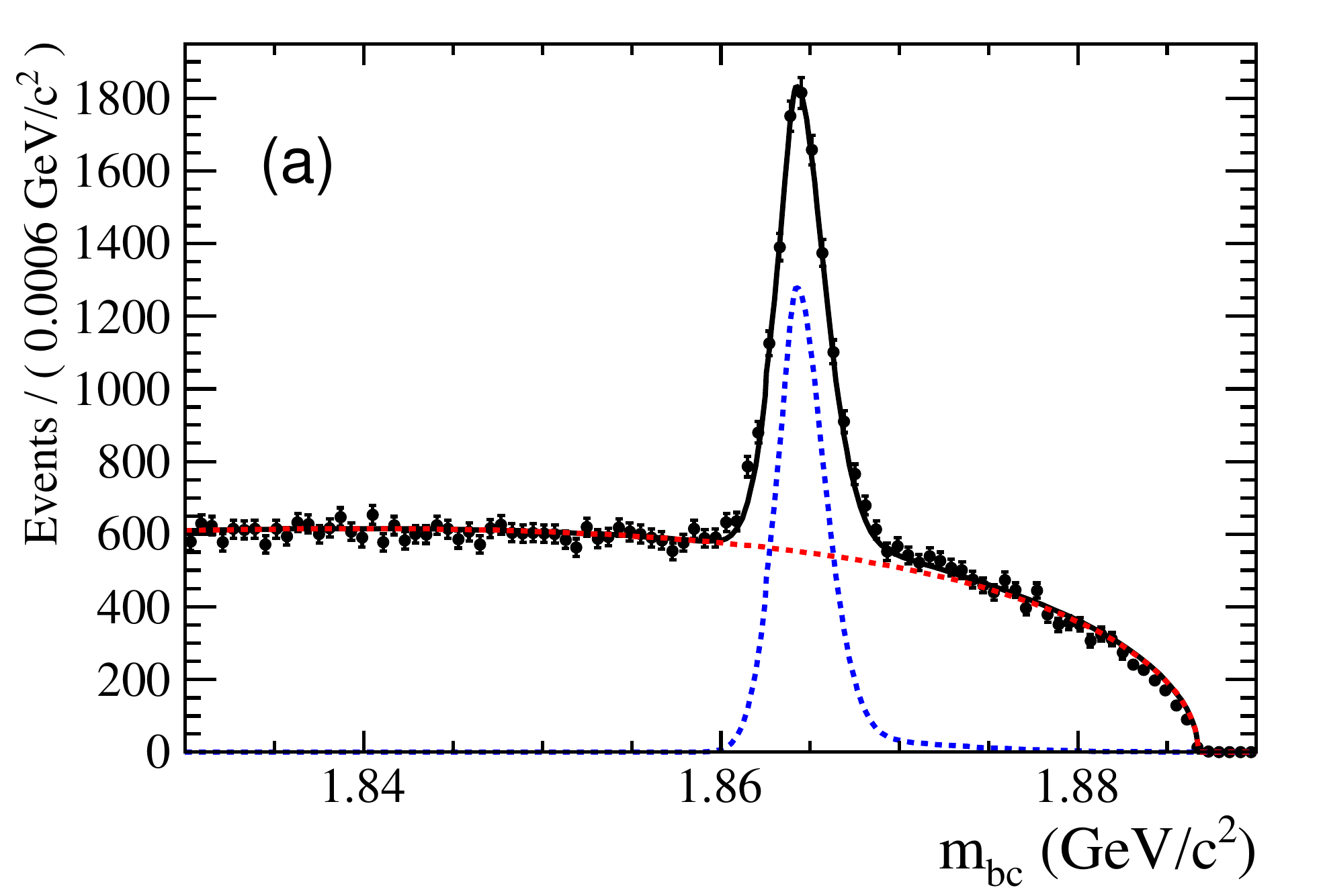} 
	\includegraphics[width=0.45\textwidth, height=!]{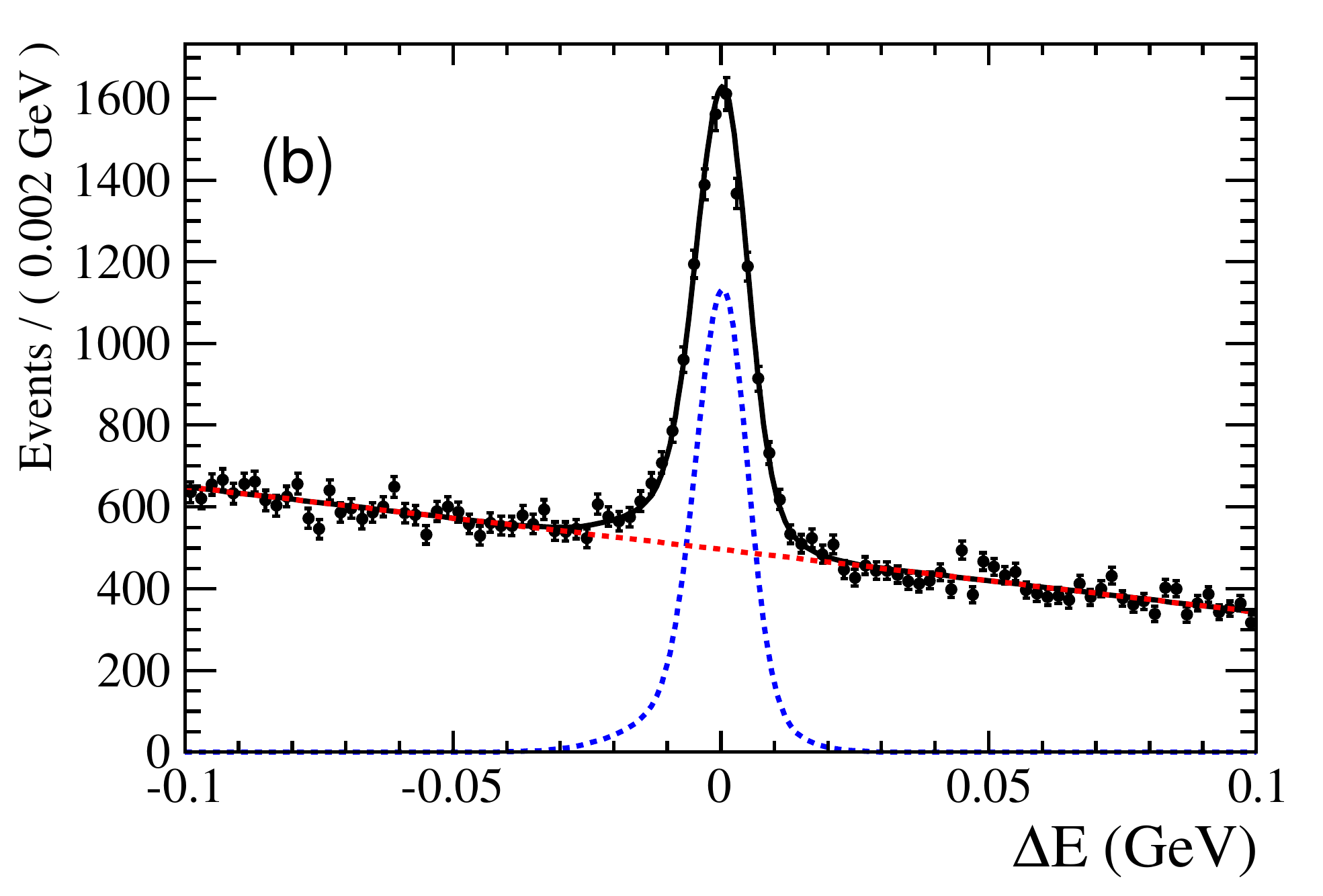} 
	\caption{Beam constrained mass (a) and missing energy (b) distribution of $\Dz\ \to \fourpi$ candidates, 
	overlaid with the projections of the fitted PDF (solid black line).
	The signal component is shown in blue (dashed) and the background component in red (dashed).}
	\label{fig:signalRegionFit}
\end{figure}

To estimate the signal purity of the sample, a two-dimensional unbinned maximum likelihood fit to $m_{bc}$ and $\Delta E$ is performed in the whole range.
While the signal peak is modeled with a sum of three (two) Gaussian functions,
the combinatorial background is described by an ARGUS~\cite{Albrecht:1990am} (linear) function in $m_{bc}$ ($\Delta E$).
The number of signal events within the signal region is estimated from the fit result
displayed in Fig.~\ref{fig:signalRegionFit}, to be $7250 \pm 56 \textrm{ (stat)} \pm 46 \textrm{ (syst)}$  events, where the first uncertainty is statistical and second is systematic.
The signal fraction $f_{\rm Sig}$, in this region is 
$f_{\rm Sig} = (78.4 \pm 0.6 \textrm{ (stat)} \pm 0.5 \textrm{ (syst)}) \%$. These systematic uncertainties are estimated by repeating the fit with different appropriate probability density function (PDF) hypotheses.
As we observed a negligible impact of the background on our analysis, further improvements of the signal purity were not studied.

\subsection{\texorpdfstring{$\Dz\ \to \KKpipi$ Selection}{D0 to K+ K- pi+ pi- Selection}}
\label{sec:eventkkpipi}

With respect to Ref.~\cite{KKpipi},
we veto the \pip \pim invariant mass region around the $K_{S}^{0}$ mass, which removes essentially all 
peaking background from $\Dz \to K_{S}^{0} (\to \pip \pim) \, \Kp \, \Km$, greatly 
simplifying our analysis.
The $K_{S}^{0}$ veto depends on the CLEO configuration, as the mass resolution is better for data collected with the CLEO-c configurations. For data collected with CLEO (CLEO-c), the $\pip \, \pim$ invariant mass combination does not fall within $16.5\,(12) \mev/c^2$ of the world-average $K_{S}^{0}$ mass. 

In addition, for the flavor-tagged data, 
several
changes have been applied with respect to Ref.~\cite{KKpipi}. 
The CLEO II.V minimum track momenta cut for the $D$ daughters is raised to $275 \mev/c$ as the MC was found not to represent the data 
sufficiently well below this value. 
As in Ref.~\cite{KKpipi}, the kinematic variables that describe signal in the CLEO II.V and CLEO III samples are the 
reconstructed $D$ mass $m_{KK\pi\pi}$, and the mass difference between the $D^*$ and $D$ candidates, 
$\Delta m$. 
We take advantage of the possibility to ensure a constant $D$-candidate invariant mass range across different kinematic regions. 
For CLEO II.V, we choose $|m_{KK\pi\pi} - m_D| < 5 \; \mev/c^2$; 
for CLEO III, we require that $m_{KK\pi\pi}$ is between ($m_D - 11.2$) and ($m_D + 8.3) \, \mev/c^2$. 
For CLEO-c 3770, %
we utilize the criteria given in \eqnPRDref{eq:bentonbox}; 
for CLEO-c 4170 \eqnPRDref{eq:bentonbox} is also used, but the 
tolerance of 
the annulus of constant invariant mass, %
with respect to $m_D$, 
is 
reduced from $15$ to $10 \, \mev/c^2$ in order to boost the signal purity in this sample. 
The signal and sidebands definitions in the respective accompanying 
kinematic variables ($\Delta m$ or $m_{bc}$) are defined accordingly in Table~\ref{tab:sideband}.
  In the CLEO II.V and CLEO III (CLEO-c 3770) samples, signal candidates are chosen to have $\Delta m$ ($m_{bc}$) 
  near the expected value for signal $D$ decays. 
  In the CLEO-c 4170 sample, we isolate our signal $D$ candidates from 
  $\D^* \Db{}^*$ 
  events, 
  which have the highest rate and intrinsic purity~\cite{PhysRevD.80.072001}.

The procedure to measure the purity in the signal region of each sample is identical to 
that of the previous analysis~\cite{KKpipi}. The events retained for the amplitude analysis and 
signal fractions for the improved selection criteria are given in Table~\ref{tab:kkpipiyield}.

\begin{table}[h]
  \footnotesize
  \centering
  \caption{Signal region and sideband definitions in the $\Delta m$ or $m_{bc}$ kinematic variable, 
  for flavor-tagged $\Dz\ \to \KKpipi$ data in the different CLEO configurations.}
  \begin{tabular}
    {@{\hspace{0.5cm}}c@{\hspace{0.25cm}}  @{\hspace{0.25cm}}c@{\hspace{0.25cm}}  @{\hspace{0.25cm}}c@{\hspace{0.5cm}}}
    \hline \hline
    Sample & Signal region & Sideband\\
    \hline
    CLEO II.V & $144.6 \; \mev/c^2 < \Delta m < 146.2 \; \mev/c^2$ & $148.5 \; \mev/c^2 < \Delta m < 160.0 \; \mev/c^2$\\
    CLEO III & $144.6 \; \mev/c^2 < \Delta m < 146.1 \; \mev/c^2$ & $148.5 \; \mev/c^2 < \Delta m < 160.0 \; \mev/c^2$\\
    CLEO-c 3770 & $|m_{bc}-m_D| < 0.005 \; \gev/c^2$ & $1.834 \; \gev/c^2 < m_{bc} < 1.854 \; \gev/c^2$\\
    & & $1.876 \; \gev/c^2 < m_{bc} < 1.890 \; \gev/c^2$\\
    CLEO-c 4170 & $2.005 \; \gev/c^2 < m_{bc} < 2.030 \; \gev/c^2$ & $1.880 \; \gev/c^2 < m_{bc} < 1.920 \; \gev/c^2$\\
    \hline \hline
  \end{tabular}
  \label{tab:sideband}
\end{table}

\begin{table}[h]
  \footnotesize
  \centering
  \caption{Updated number of signal candidates and fractions in the signal region, for flavor-tagged $\Dz\ \to \KKpipi$ data in the different CLEO configurations.}
  \begin{tabular}
    {@{\hspace{0.5cm}}c@{\hspace{0.25cm}}  @{\hspace{0.25cm}}c@{\hspace{0.25cm}}  @{\hspace{0.25cm}}c@{\hspace{0.5cm}}}
    \hline \hline
    Sample & Signal candidates & $f_{\rm Sig}$\\
    \hline
    CLEO II.V & 237 & $0.759 \pm 0.019$\\
    CLEO III & 1163 & $0.898 \pm 0.004$\\
    CLEO-c 3770 & 1300 & $0.871 \pm 0.005$\\
    CLEO-c 4170 & 598 & $0.694 \pm 0.010$\\
    \hline \hline
  \end{tabular}
  \label{tab:kkpipiyield}
\end{table}
 % end input /Users/pnaik/Documents/CLEO/Papers/latexpand/latex/selection.tex
 %
% start input /Users/pnaik/Documents/CLEO/Papers/latexpand/latex/amplitude.tex
%

\clearpage\section{Amplitude Analysis Formalism}
\label{sec:analysis}

Previous four-body amplitude analyses of \D\ decays have been
performed by the Mark~III collaboration for \prt{\D \to
  \overline{K}\pi\pi\pi} comprising a total of four Cabibbo-favored
decay modes modes of \Dz\ and \Dp~\cite{MarkIIIK3pi}, FOCUS for
\prt{\Dz \to \pi^+\pi^-\pi^+\pi^-}, \prt{K^+K^-\pi^+\pi^-},
\prt{K^-K^-K^+\pi^+}~\cite{FOCUS4pi,FOCUSKKpipi,FOCUS3Kpi} and most
recently, for the decay \prt{\Dz \to K^+K^-\pi^+\pi^-}, by
\cleo~\cite{KKpipi}. Here, we further develop the formalism and
analysis software used in Ref.~\cite{KKpipi}.  Key differences are in
the formalism used for the spin factors, where we now use a more
consistent and intuitive implementation of the Zemach
formalism~\cite{Zemach,Rarita,helicity3}, and an
improved 
description of the lineshapes of resonances
decaying to three-body final states.

The differential decay rate of a $\Dz$ meson 
with mass, $m_{\Dz}$,
decaying into four pseudoscalar particles with four-momenta $p_{i}= (E_{i},\, \vec p_{i}) \, (i=1,2,3,4)$
is given by
\begin{equation}
	\text{d}\Gamma = \frac{1}{2 \, m_{\Dz}} \,  \vert A_{\Dz}(\phsPoint) \vert^{2} \, \dphs   \, ,
	\label{eq:decayRate}
\end{equation}
where the transition amplitude $A_{\Dz}(\phsPoint)$, describes the dynamics of the interaction, 
\dphs\ is the four-body phase space element \cite{Peskin}, and 
$\phsPoint$ 
represents a unique set of kinematic conditions within the phase space of the decay.
Each final state particle contributes three observables,
manifesting in their three-momentum,
summing up to twelve observables in total.
Four of them are redundant 
due to four-momentum conservation and
the overall orientation of the system can be integrated out.
The remaining  five independent degrees of freedom unambiguously determine the kinematics of the decay.
Convenient choices for the kinematic observables
include the invariant mass combinations of the final state particles
\begin{align}
	\nonumber
	m^{2}_{ij} &= (p_{i}+p_{j})^{2} , \\
	m^{2}_{ijk} &= (p_{i}+p_{j}+p_{k})^{2} \, 
\end{align}
or acoplanarity and helicity angles \cite{Beneke:2006hg,Aaij:2015kba}.
It is however important
to take into account that, while $m^2_{12}, m^2_{23}$ are sufficient
to fully describe a three-body decay, the obvious extension to
four-body decays with $m^{2}_{ij}, m^{2}_{ijk}$ requires additional
care, as these variables alone are insufficient to describe the parity-odd
moments possible in four-body kinematics.
In practice, we do not need to choose a particular five-dimensional
basis, but use the full four-vectors of the decay in our
analysis. 
The dimensionality is handled by the phase space element which can be written in terms of any set of five independent kinematic observables, $\phsPoint = (x_1, \ldots, x_5)$, as
\begin{equation}
	\dphs = \phsd(\phsPoint) \, \dphsPoint ,
\end{equation}
where $\phi_{4}(\phsPoint ) = \left\vert  \frac{\partial \phs}{\partial(x_1, \ldots x_5)} \right\vert$ is the phase space density.
In contrast to three-body decays, the four-body phase space density
function is not flat in the usual kinematic variables.  
Therefore, an analytic expression for \phsd\ is
taken from Ref.~\cite{kinematics}.

The total amplitude for the 
$\Dz \to h_{1}\,h_{2}\,h_{3} \, h_{4}$
decay is given by the coherent sum over all 
intermediate state amplitudes $A_{i}(\phsPoint )$, each weighted by a complex coefficient $a_{i} = \vert a_{i} \vert \, e^{i \, \phi_i}$
to be measured from data,
\begin{equation}
	A_{\Dz}(\phsPoint ) = \sum_{i}  a_{i} \, A_{i}(\phsPoint )   \, .
\end{equation}

To construct $A_{i}(\phsPoint )$,
the isobar approach is used, which 
assumes that
the decay process can be factorized into subsequent two-body decay amplitudes \cite{isobar1,isobar,isobar2}.
This gives rise to two different decay topologies;
quasi two-body decays
$\Dz \to (R_{1} \to h_{1}\,h_{2}) \, (R_{2} \to h_{3}\,h_{4})$ 
or cascade decays
$\Dz \to h_{1} \, \left[R_{1} \to h_{2} \,  (R_{2} \to h_{3} \, h_{4}) \right]$.
In either case, the intermediate state amplitude is parameterized as a product of
form factors $B_{L}$, included for each vertex of the decay tree, 
Breit-Wigner propagators $T_{R}$,  included for each resonance $R$,
and an overall angular distribution represented by a spin factor $S$,
\begin{equation}
	A_{i}(\phsPoint ) =  B_{L_{D}}(\bold x) \, [B_{L_{R_{1}}}(\phsPoint )  \, T_{R_{1}}(\phsPoint )] \, [B_{L_{R_{2}}}(\phsPoint ) \, T_{R_{2}}(\phsPoint )]  \,  S_{i}(\phsPoint )  \, .
	\label{eq:amp4}
\end{equation}
As the $\pip \pim \pim \pip$ final state 
involves two pairs of indistinguishable pions,
the amplitudes are Bose-symmetrized
and therefore symmetric under exchange of like-sign pions.

We define the \CP-conjugate phase space point $\phsPointCP$ such that it is mapped onto $\phsPoint$ by the
interchange of final state charges, and the reversal of three-momenta. If
$\phsPoint$, $\phsPointCP$ are expressed as a function of the
four-momenta $(E_i, \vec{p}_i)$ (where $i$ labels the particle), this
implies for \prt{\Dz \to K^+ K^- \pi^+ \pi^-} that
\begin{align}
\lefteqn{\phsPointCP\left[ (E_{K^+}, \vec{p}_{K^+}), (E_{K^-}, \vec{p}_{K^-}), (E_{\pi^+}, \vec{p}_{\pi^+}), (E_{\pi^-}, \vec{p}_{\pi^-})\right]} & \nonumber \\
 &\equiv
\phsPoint\left[ (E_{K^-}, -\vec{p}_{K^-}), (E_{K^+}, -\vec{p}_{K^+}), (E_{\pi^-}, -\vec{p}_{\pi^-}), (E_{\pi^+}, -\vec{p}_{\pi^+})\right],
\end{align} 
and equivalently for \prt{\Dz \to \pi^+\pi^-\pi^+\pi^-}. 
The \CP-conjugate of a given intermediate state amplitude, $A_i(x)$, is defined as
\begin{equation}
\label{eq:AiAibar}
\overline{A}_i (\phsPoint)  \equiv A_i(\phsPointCP),
\end{equation}
and the total \Dzb\ decay amplitude is defined as
\begin{align}
\label{eq:AdAdbar}
A_{\Dzb}(\phsPoint ) &\equiv \sum_{i}  \bar{a}_{i} \, \overline{A}_{i}(\phsPoint ) = \sum_{i} \bar{a}_i A_{i}(\phsPointCP ).
\end{align}
Unless stated otherwise, we assume \CP\ conservation in the \Dz\ decay, implying $\bar{a}_i = a_i$.
Moreover, \CP\ conservation in the strong interaction is implemented in the
cascade topology by the sharing of couplings between related
quasi-two-body final states. For example, given the two $a_i$ parameters required for
\prt{\Dz\ \to \pi^- a_1(1260)^+} with \prt{a_1(1260)^+ \to \rho(770)^0 \, \pi^+}
and \prt{a_1(1260)^+ \to \sigma \, \pi^+}, the amplitude \prt{\Dz\ \to
  \pi^+ \, a_1(1260)^-} with \prt{a_1(1260)^- \to \rho(770)^0 \, \pi^-} and
\prt{a_1(1260)^- \to \sigma \, \pi^-} only requires one additional global
complex parameter to represent the different weak processes of
\prt{D^0 \to a_1(1260)^+ \, \pi^-} and \prt{D^0 \to a_1(1260)^- \, \pi^+},
while the relative magnitude and phase of \prt{a_1(1260)^- \to \rho(770)^0 \,
  \pi^-} and \prt{a_1(1260)^- \to \sigma \, \pi^-} are the same as for
\prt{a_1(1260)^+ \to \rho(770)^0 \, \pi^+} and \prt{a_1(1260)^+ \to \sigma \,
  \pi^+}. For historical reasons, this constraint is only applied to
the \fourpi\ final state, but, as discussed in \secref{sec:resultsKKpipi}, the
results we obtain for the \KKpipi\ final state are also compatible with \CP\
conservation in the strong interaction.

\subsection{Form Factors and Resonance Lineshapes}
\label{ssec:lineshapes}

To account for the finite size of the decaying resonances,
the Blatt-Weisskopf penetration factors, 
derived in Ref.~\cite{Bl2}
by assuming a square well interaction potential with radius $r_{\rm BW}$,
are used as form factors, $B_L$.
They depend on
the breakup momentum $q$,
and the orbital angular momentum $L$, between the resonance daughters.
Their explicit expressions are
\begin{align}
         \nonumber
	B_{0}(q)  &= 1 ,  \\ \nonumber
	B_{1}(q)  &= 1 / \sqrt{{1+ (q \, r_{\rm BW})^{2}}} ,  \\
	B_{2}(q)  &= 1 / \sqrt{9+3\,(q \, r_{\rm BW})^{2}+(q \, r_{\rm BW})^{4}} . 
\end{align}
Resonance lineshapes
are described as function of the energy-squared, $s$, by Breit-Wigner propagators
\begin{equation}
	T(s) = \frac{1}
	{M^{2}(s) - s - i\,m_{0}\,\Gamma(s)}   \, ,
	\label{eq:BW}
\end{equation}
featuring the energy-dependent mass $M(s)$ (defined below), and total width, $\Gamma(s)$.
The latter is normalized to give the nominal width, $\Gamma_{0}$, when evaluated at the nominal mass $m_{0}$, 
\ie $\Gamma_{0} = \Gamma(s = m_{0}^{2})$.

For a decay into two stable particles $R \to AB$, the energy dependence of the decay width can be described by 
\begin{equation}
	\Gamma_{R \to AB}^{(2)}(s) = \Gamma_{0} \, \frac{m_{0}}{\sqrt s} \, \left(\frac{q}{q_{0}}\right)^{2L+1} \, \frac{B_{L}(q)^{2}}{B_{L}(q_{0})^{2}}  \, ,
	\label{eq:gamma2}
\end{equation}
where $q_{0}$ is the value of the breakup momentum at the resonance pole \cite{BW}.

The energy-dependent width for a three-body decay $R \to ABC$, on the other hand, is considerably more complicated and has no
analytic expression in general. However, it 
can be obtained numerically by integrating the transition amplitude-squared over the phase space,
\begin{equation}
	\Gamma_{R \to ABC}^{(3)}(s) =  \frac{1}{2 \, \sqrt s} \, \int \vert A_{R \to ABC} \vert^{2} \, \text{d}\Phi_{3}   ,
	\label{eq:gamma3}
\end{equation}
and therefore requires knowledge of the resonant substructure. 
The three-body amplitude $A_{R \to ABC}$ can be parameterized 
similarly to
the four-body amplitude in \eqnPRDref{eq:amp4}.
In particular, it includes form factors and propagators of intermediate two-body resonances.

Both \eqnPRDref{eq:gamma2} and \eqnPRDref{eq:gamma3} give only the partial width for the decay into a specific channel.
To obtain the total width, a sum over all possible decay channels has to be performed,
\begin{equation}
	\Gamma(s) = \sum_{i} g_{i} \, \Gamma_{i}(s) ,
\end{equation}
where the coupling strength to channel $i$, is given by $g_{i}$.
Branching fractions ${\cal B}_{i}$ are related to the couplings $g_{i}$ via the equation \cite{PDG2016}
\begin{equation}
	{\cal B}_{i} = \int_{s_{min}}^{\infty} \frac{g_{i} \, m_{0} \, \Gamma_{i}(s)}{ \vert M^{2}(s) - s - i \, m_{0} \, \sum_{j} g_{j} \, \Gamma_{j}(s) \vert^{2}} \, \text{d}s  .
	\label{eq:BF}
\end{equation}
As experimental values are usually only available for the branching fractions, \eqnPRDref{eq:BF} needs to be inverted to obtain values for the couplings.
In practice, this is solved by minimizing the quantity $\chi^{2}(g) = \sum_{i}  \left[ \mathcal B_{i} - \mathcal I_{i}(g) \right]^{2} / \Delta\mathcal B_{i}^{2}$, 
where $\mathcal I_{i}(g)$ denotes the right-hand side of \eqnPRDref{eq:BF}.

The energy-dependent mass follows from the decay width via the Kramers-Kronig dispersion relation \cite{PhysRevD.39.1357,Vojik:2010ua}:
\begin{equation}
	M^{2}(s) %
	= m_{0}^{2} + \frac{m_{0}}{\pi} \,
	 \int_{s_{min}}^{\infty}   \left( \frac{\Gamma(s^{\prime})}{s - s^{\prime}} 
	 - 	\frac{\Gamma(s^{\prime})}{m_{0}^{2} - s^{\prime}}\right) \, \text{d}s^{\prime}.
	 \label{eq:dispersion}
\end{equation}
Here, the energy-dependent mass is renormalized such that $M^{2}(s=m_{0}^{2}) = m_{0}^{2}$.
In practice, the energy-dependent mass is often approximated as being constant, \ie $M^{2}(s) = m_{0}^{2}$, since its calculation requires a detailed understanding of
the decay width for arbitrarily large energies and is computationally expensive.
This is usually justified as the energy-dependent mass needs to satisfy the condition,
\begin{equation}
	\frac{\text{d}M^{2}(s)}{\text{d}s} \bigg \rvert_{s = m_{0}^{2}} = 0,
	\label{eq:M-deriv}
\end{equation}
such that $M^{2}(s)$ is indeed, approximately constant near the on-shell mass \cite{Lichard:2006ky}.
Larger dispersive effects are thus only expected for very broad resonances. 

The treatment of the lineshape for various resonances considered in this analysis is described in what follows.
The nominal masses and widths of the resonances are taken from the PDG \cite{PDG2016} with the exceptions described below.
We assume an energy-independent mass unless otherwise stated.

 For the broad scalar resonance $\sigma$,
     		the model from Bugg is used \cite{BuggSigma}.
		Besides $\sigma \to \pi \pi$ decays, it includes contributions from the decay modes $\sigma \to K K$, $\sigma \to \eta \eta$ and $\sigma \to \pi \pi \pi \pi$ 
		as well as dispersive effects 
		due to the channel opening of the latter.
	We use the Gournaris-Sakurai parametrization for the $\rho(770)^{0} \to \pi \pi$ propagator which provides an analytical description of the dispersive term, $M^{2}(s)$  \cite{GS}.
	The energy-dependent width of the $f_{0}(980)$ resonance is given by the sum of the partial widths into the $\pi\pi$ and $KK$ channels \cite{Flatte},
		\begin{equation}
			\Gamma_{f_{0}(980)}(s) = g_{\pi\pi} \, \Gamma^{(2)}_{f_{0}(980) \to \pi \pi}(s) + g_{KK} \, \Gamma^{(2)}_{f_{0}(980) \to KK}(s),
		\end{equation}
		where the coupling constants $g_{\pi\pi}$ and $g_{KK}$, as well as the mass and width are taken from a measurement performed by the BES Collaboration~\cite{Flatte2}.
	        The total decay widths for both the $f_{2}(1270)$ and the $f_{0}(1370)$ meson take the channels $\pi  \pi, K  K, \eta  \eta$ and $\pi \pi \pi \pi$ into account. 
		While the two-body partial widths are described by \eqnPRDref{eq:gamma2}, a model for the partial width for a decay into four pions is taken from Ref.~\cite{Buggf0}.
		The corresponding branching fractions are taken from the PDG \cite{PDG2016}.
		The nominal mass and width of the $f_{0}(1370)$ resonance are taken from an LHCb measurement \cite{LHCb:2012ae}.
		\EqnPRDref{eq:gamma2} is used for all other resonances decaying into a two-body final state.

\begin{figure}[h]
  \includegraphics[width=0.49\textwidth, height = !]{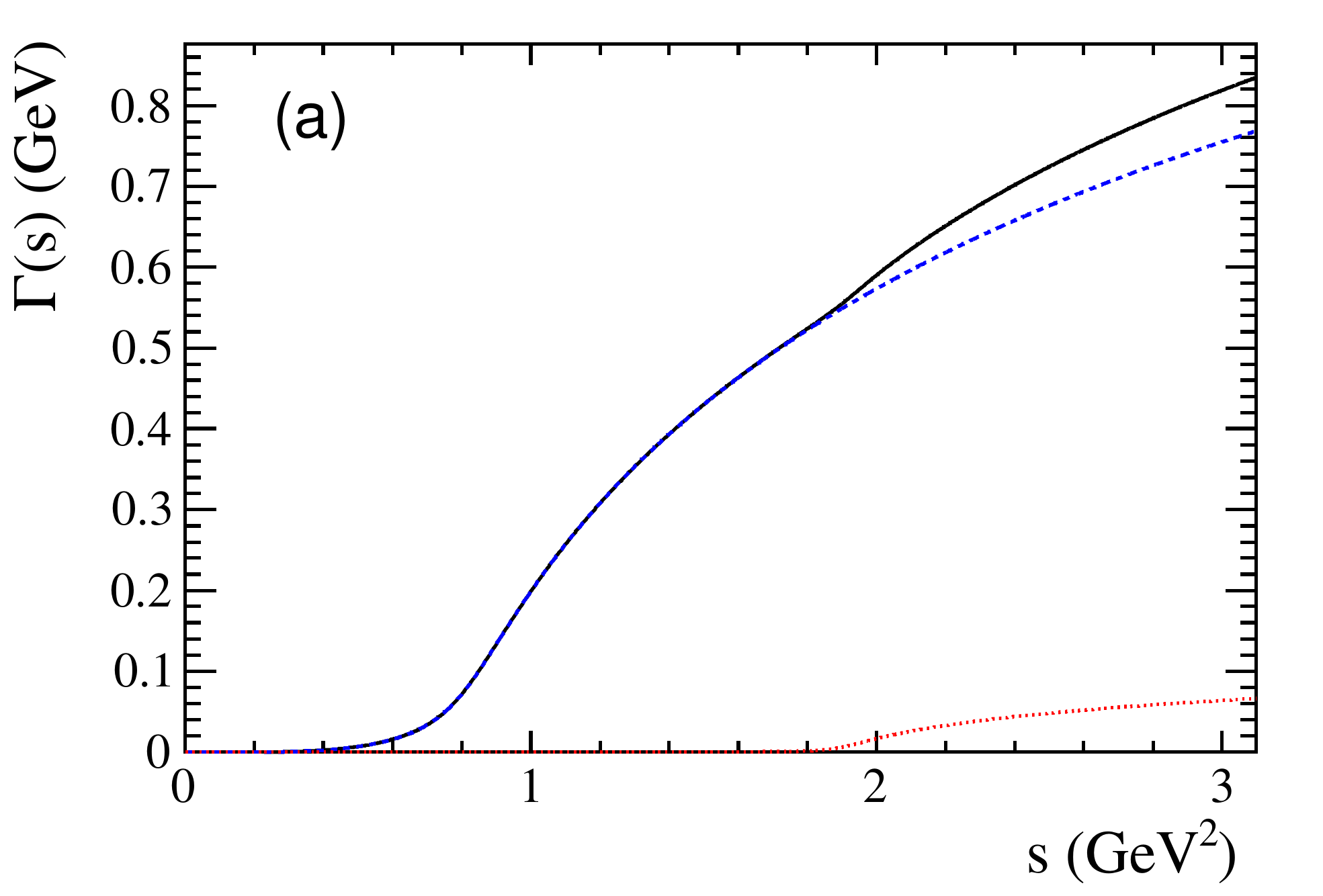} 
  \includegraphics[width=0.49\textwidth, height = !]{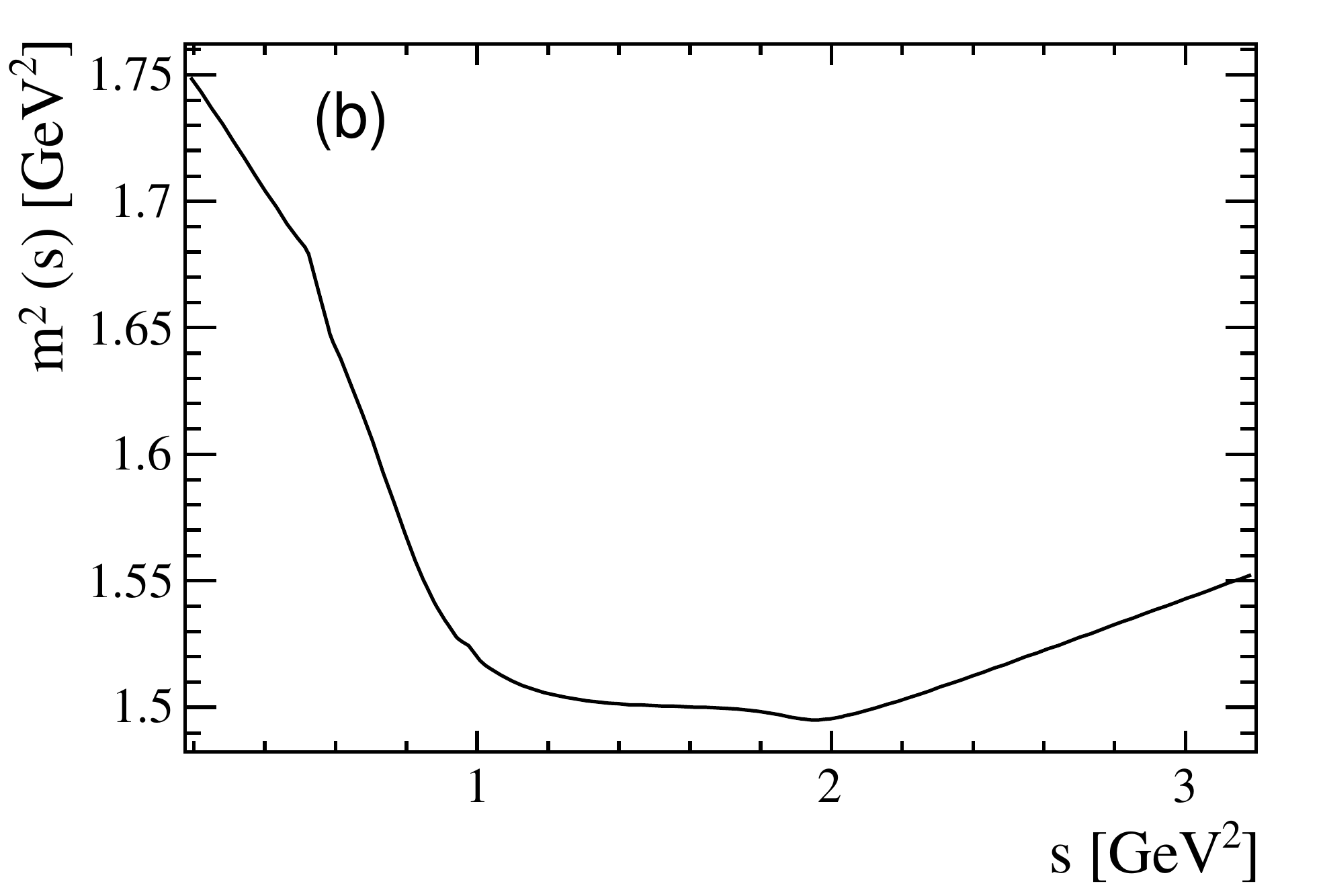} 

  \caption{Energy-dependent width (a) and energy-dependent mass (b) for the $a_{1}(1260)$ resonance. 
    The total width is shown in black (solid), while the partial widths $\Gamma^{(3)}_{a_{1}(1260) \to \pi \pi \pi}(s)$ and $ \Gamma^{(3)}_{a_{1}(1260) \to K \bar K \pi}(s)$
    are shown in blue (dashed) and red (dotted), respectively. }
  \label{fig:gamma_a1}
\end{figure}

To describe the decay width of the axial vector resonance $a_{1}(1260)$, the decay channels $\pi \pi \pi$ and $K \bar K \pi$ are considered,
		\begin{equation}
			\Gamma_{a_{1}(1260)}(s) = g_{\pi \pi \pi} \, \Gamma^{(3)}_{a_{1}(1260) \to \pi \pi \pi}(s) +  g_{K \bar K \pi} \, \Gamma^{(3)}_{a_{1}(1260) \to K \bar K \pi}(s) ,
		\end{equation}
		where isospin symmetry is assumed, \ie $ \Gamma^{(3)}_{a_{1}(1260)^{+} \to \pip \pim \pip}(s)  =  \Gamma^{(3)}_{a_{1}(1260)^{+} \to \pi^{0} \pi^{0} \pip}(s) $.
		The partial width $\Gamma^{(3)}_{a_{1}(1260) \to K \bar K \pi}(s)$ is calculated from \eqnPRDref{eq:gamma3} assuming the decay proceeds entirely 
		via $a_{1}(1260) \to K^{*}(892) \, K$. The corresponding branching fraction is taken from a CLEO analysis of hadronic $\tau$ decays \cite{Asner:1999kj}.
		The calculation of the partial width $\Gamma^{(3)}_{a_{1}(1260) \to \pi \pi \pi}(s)$ is more complicated 
		due to the fact that it requires information about the three pion Dalitz plot structure of the $a_{1}(1260)$ resonance
		whose determination in turn, needs the propagator as input.
		For this reason, we follow an iterative approach. 
		The initial amplitude fit, described in Sec.~\ref{sec:results4pi}, is performed using
		an energy-dependent width distribution 
		derived from an uniform phase space population. 
		Afterwards, the energy-dependent width is recalculated with the results of the substructure analysis and the amplitude fit is subsequently repeated with the new propagator.
		It is found that the energy-dependent width is not highly sensitive to the details of the Dalitz plot as this procedure converges after a few iterations.
		As the $a_{1}(1260)$ resonance is very broad, the dispersive term is calculated as well.
		Figure \ref{fig:gamma_a1} shows the final iteration of the energy-dependent width and mass.
		The energy-dependent width varies strongly around $s \approx 0.8 \; \gev^{2}$ where the energy of the $\pip \, \pim$ subsystem is equal to the $\rho(770)^{0}$ on-shell mass. 
		Around $s = 2 \; \gev^{2}$, a small hump develops due to the opening of the $K \bar K \pi$ channel.
		The energy-dependent mass indeed shows a plateau around the nominal mass as expected.
		Note that as the condition of \eqnPRDref{eq:M-deriv} is not explicitly enforced by \eqnPRDref{eq:dispersion}, 
		it serves as an independent check of whether the main thresholds have been included ~\cite{PhysRevD.39.1357, Asner:1999kj}.

	      For the resonances $\pi(1300)$, 
                $a_{1}(1640)$ and $\pi_{2}(1670)$, 
			the energy-dependent width is obtained via the same 
			 iterative procedure as for the $a_{1}(1260)$ resonance.
			In case of the $\pi_{2}(1670)$ meson, the $K \bar K \pi$ and $\omega \rho(770)^{0}$ thresholds are included with the PDG branching fractions taken from 						Ref.~\cite{PDG2016}, 
			otherwise only decays to three pions are considered.                      
                In the $\Dz \to \KKpipi$ analysis, resonant decays of the $K_1(1270)$ and $K_1(1400)$ mesons into the $K \rho(770)^{0}$, $K^{*}(892) \pi$, $K_0^{*}(1430) \pi$, 
                $K f_0(1370)$ and $K \omega$ decay channels are taken into account assuming the lowest possible angular momentum state.
                		For the purpose of evaluating the energy-dependent widths of the excited kaons, these decay channels are assumed to be incoherent and the branching fractions from the PDG are used \cite{PDG2016}.
			The same procedure is applied to obtain the energy-dependent width for the $K^{*}(1410)$ and $K^{*}(1680)$ resonances.
			In their case, the decay channels  $K \rho(770)^{0}$, $K^{*}(892) \pi$ and $K \pi$ are considered.
			For the $K^{*}(1410)$ meson 
			there are only upper limits for the branching fractions into the $K \rho(770)^{0}$ and $K^{*}(892) \pi$ channels available.
			We assume no $K^{*}(1410) \to K \rho(770)^{0}$ contribution                        
                        and $\mathcal{B}[K^{*}(1410) \to K^{*}(892) \pi] = 1- \mathcal{B}[K^{*}(1410) \to K \pi] = (93.4 \pm 1.3)\, \%$~\cite{PDG2016}. All energy-dependent widths 
not shown in this Chapter 
are shown in Appendix~\ref{a:rw}.

Some particles may not originate from a resonance but are in a state of relative orbital angular momentum.
We denote such non-resonant states by surrounding the particle system with brackets  and indicate the partial wave state with an subscript;
for example $(\pi \pi)_S$ refers to a non-resonant di-pion $S$-wave.
The lineshape for non-resonant states is set to unity.

\subsection{Spin Densities}

The spin amplitudes are phenomenological descriptions
of decay processes that 
are required to be Lorentz invariant,
compatible with angular momentum conservation and,
where appropriate, parity conservation.
They are constructed in the covariant Zemach (Rarita-Schwinger) tensor formalism
\cite{Zemach,Rarita,helicity3}.
At this point, we briefly introduce 
the fundamental objects of the covariant tensor formalism 
which connect the particle's four-momenta to the spin dynamics of the reaction
and give a general recipe to calculate the spin factors for arbitrary decay trees.
Further details can be found in Refs.~\cite{Zou, Filippini}.

A spin-$S$ particle with four-momentum $p$, and spin projection $\lambda$, is represented 
by the polarization tensor $\epsilon_{(S)}(p,\lambda)$, which is symmetric, traceless and orthogonal to $p$.
These so-called Rarita-Schwinger conditions reduce the a priori $4^{S}$  elements of the rank-$S$ tensor to 
$2S +1$ independent  elements in accordance with the number of degrees of freedom of a spin-$S$ state\cite{Rarita,Zhu}.

The spin projection operator $P^{\mu_{1} \dots \mu_{S} \nu_{1} \dots \nu_{S}}_{(S)}(p_{R})$,  
for a resonance $R$, with spin $S = \{0,1,2\}$, and four-momentum $p_{R}$,
is given by \cite{Filippini}:
\begin{align}
	\nonumber
	P_{(0)}^{\mu \nu}(p_{R}) &= 1 \\
	\nonumber
	P_{(1)}^{\mu \nu}(p_{R}) &= %
	- \, g^{\mu \nu} + \frac{p_{R}^{\mu} \, p_{R}^{\nu}}{p_{R}^{2}}    \\
	P_{(2)}^{\mu \nu \alpha \beta}(p_{R})  &=
	 \frac{1}{2} \,  \left[ P_{(1)}^{\mu \alpha}(p_{R})  \, P_{(1)}^{\nu \beta}(p_{R})  + P_{(1)}^{\mu \beta}(p_{R})  \, P_{(1)}^{\nu \alpha}(p_{R}) \right] - \frac{1}{3} \, P_{(1)}^{\mu \nu}(p_{R}) 
	 \, P_{(1)}^{\alpha \beta}(p_{R})    \,,
	\label{eq:pol1}
\end{align}
where $ g^{\mu \nu}$ %
is the Minkowski metric.
Contracted with an arbitrary tensor, the projection operator selects 
the part of the tensor which satisfies the Rarita-Schwinger conditions.

For a decay process $R \to A B$, with relative orbital angular momentum $L$, between particle $A$ and $B$,
the angular momentum tensor is obtained by 
projecting 
the rank-$L$ tensor $q_R^{\nu_{1}}   \,  q_R^{\nu_{2}}  \dots  \,  q_R^{\nu_{L}} $, constructed from the relative momenta
$q_{R} = p_{A} - p_{B}$,
onto the spin-$L$ subspace,
\begin{equation}
	L_{(L)\mu_{1}  \dots \mu_{L}}(p_{R},q_{R}) = (-1)^{L}  \, P_{(L)\mu_{1}  \dots \mu_{L} \nu_{1}  \dots \nu_{L}}(p_R)  \, 
	 q_R^{\nu_{1}}     \dots  \,  q_R^{\nu_{L}}  .
\end{equation}
Their $\vert \vec q_{R} \vert^{L} $ dependence accounts for the influence of the centrifugal barrier on the transition amplitudes.
For the sake of brevity, the following 
 notation is introduced,
\begin{align}
\nonumber
\eps_{(S)}(R) & \equiv \eps_{(S)}(p_{R},\lambda_{R}) , \\ \nonumber
P_{(S)}(R) & \equiv P_{(S)}(p_{R}), \\
L_{(L)}(R) & \equiv L_{(L)}(p_{R},q_{R})  .
\end{align}

Following the isobar approach, a four-body decay amplitude is described as a product of two-body decay amplitudes.
Each sequential two-body decay $R \to A \, B$, 
with relative orbital angular momentum $L_{AB}$, and total intrinsic spin $S_{AB}$,
contributes a term to the overall spin factor given by
\begin{align}
	\nonumber
	S_{R \to A B}(\bold x \vert L_{AB}, S_{AB} ; \lambda_{R}, \lambda_{A}, \lambda_{B})  &=
	 \eps_{(S_{R})}(R) \, X(S_{R},L_{AB},S_{AB}) \,  L_{(L_{AB})}(R) \,    \\
          &\times \Phi(\bold x \vert S_{AB} ; \lambda_{A} , \lambda_{B}) ,
          \label{eq:spin1}
\end{align}
where
\begin{align}
	 \Phi(\bold x \vert S_{AB} ; \lambda_{A} , \lambda_{B})  &=  P_{(S_{AB})}(R) \, X(S_{AB},S_{A},S_{B})  \, \eps^{*}_{(S_{A})}(A)  \, \eps^{*}_{(S_{B})}(B)  \,   .
          \label{eq:spin2}
\end{align}
Here, a polarization vector is assigned to the decaying particle 
and the complex conjugate vectors for each decay product.
The spin and orbital angular momentum couplings are described by the tensors $P_{(S_{AB})}(R)$
and $L_{(L_{AB})}(R)$, respectively.
Firstly, the two spins $S_{A}$ and $S_{B}$, are coupled to a total spin-$S_{AB}$ state, $\Phi(\bold x \vert S_{AB})$,
by projecting the corresponding polarization vectors  onto the spin-$S_{AB}$
subspace transverse to the momentum of the decaying particle.
Afterwards, the spin and orbital angular momentum tensors are properly contracted with the
polarization vector of the decaying particle to give a Lorentz scalar.
This requires in some cases to include the tensor $\eps_{\alpha\beta\gamma\delta} \, p_{R}^{\delta}$ via
\begin{equation}
	  X(j_{a},j_{b},j_{c}) = 		
	  \begin{cases}
			  1 & \mbox{if } j_{a} + j_{b} + j_{c} \; {\rm even} \\
			   \eps_{\alpha \beta \gamma \delta} \, p_{R}^{\delta} & \mbox{if } j_{a} + j_{b} + j_{c} \; {\rm odd}
	 \end{cases} \, ,
\end{equation}
where $\eps_{\alpha\beta\gamma\delta}$ is the Levi-Civita symbol and $j$ refers to the arguments of $X$ defined in Eqs.~\ref{eq:spin1}~and~\ref{eq:spin2}.
Its antisymmetric nature ensures the correct parity 
transformation behavior of the amplitude. 
The spin factor for a whole decay chain, for example $R \to (R_{1} \to AB) \, (R_{2} \to CD)$,
 is obtained by combining the two-body terms and performing a sum over all unobservable, intermediary spin projections
\begin{equation}
	\sum_{\lambda_{R_{1}},\lambda_{R_{2}}}   S_{R \to R_{1} R_{2}}(\bold x \vert L_{R_{1}R_{2}} ; \lambda_{R_{1}} , \lambda_{R_{2}}) \, 
	S_{R_{1} \to AB}(\bold x \vert L_{AB} ; \lambda_{R_{1}}) 
	\, S_{R_{2} \to CD}(\bold x \vert L_{CD} ; \lambda_{R_{2}}) ,
\end{equation}
where $\lambda_{R} = \lambda_{A} = \lambda_{B}  = \lambda_{C}  = \lambda_{D} = 0$,  $S_{AB} = S_{CD} = 0$ and $S_{R_{1}R_{2}} =  L_{R_{1}R_{2}}$, as only pseudoscalar initial/final states are involved.

The main difference to the formalism used in Ref.~\cite{KKpipi} is the inclusion of additional projection operators,
\ie $P_{(S_{AB})}(R)$ and the one intrinsic to $L_{(L_{AB})}(R)$,
which ensure pure spin and angular momentum tensors.
The spin factors for all decay topologies considered in this analysis are explicitly given in Appendix~\ref{a:sf}.

\subsection{Measurement Quantities}
\label{subsec:MQ}
Here, we define all quantities derived from the amplitude model that are of physical importance. In order to provide implementation-independent measurements in addition to the complex coefficients $a_i$, we define two quantities. Firstly, the fit fractions
\begin{equation}
\label{eq:DefineFitFractions}
	F_{i} \equiv \frac{\int \left\vert   a_{i} \, A_{i}(\phsPoint) \right\vert^{2} \, \text{d}\Phi_{4} }
	{\int \left\vert  A_{\Dz}(\phsPoint) \right\vert^{2} \, \text{d}\Phi_{4}}, 
\end{equation}
which are a measure of the relative strength between the different transitions. Secondly, the interference fractions are given by
\begin{equation}
\label{eq:DefineInterferenceFractions}
	I_{ij} \equiv \frac{\int  2\,\Re[a_{i}a^*_{j} \, A_{i}(\phsPoint) A^*_{j}(\phsPoint) ] \, \text{d}\Phi_{4} }
	{\int \left\vert  A_{\Dz}(\phsPoint) \right\vert^{2} \, \text{d}\Phi_{4}} ,
\end{equation}
which measures the interference effects between amplitude pairs. Constructive interference leads to $I_{ij} > 0$, while destructive interference leads to $I_{ij} < 0$. Note that $\sum_i F_{i} + \sum_{j<k} I_{j,k} = 1$.

The global fractional \CP-even content is defined as,
\begin{align}
  F_{+} \equiv \frac{\int \vert A_{+} \vert^{2}  \, \text{d}\Phi_{4}   }
  {\int \vert  A_{+}  \vert^{2}  + \vert  A_{-}  \vert^{2}  \, \text{d}\Phi_{4} } \label{eqn:CPcont}%
\end{align}
where $A_{\pm} \equiv A_{\Dz}(\phsPoint) \pm A_{\Dzb}(\phsPoint)$ is the decay amplitude for a \D meson in a \CP-even / \CP-odd state.  
The parameter $F_{+}$, can be determined from an amplitude model (\eqnPRDref{eqn:CPcont}) or by using model-independent methods~\cite{Malde:2015mha}; 
the consistency of the two techniques provides a useful cross-check of the amplitude model.
The fractional \CP-even content also provides useful input to the determination of the CKM phase $\gamma$~($\phi_3$) in $B^\pm \to D K^\pm$ and related decays. 
Additionally, knowledge of $F_{+}$ for all $D$ decay final states can be used to determine the net \CP-content of the \D meson system, which is related to the charm-mixing parameter $y_D$~\cite{Gershon:2015xra}.

Finally, measurements of direct \CP violation will also be reported. For this purpose, the amplitude coefficients are expressed in terms of 
a \CP-conserving ($c_{i}$) and a \CP-violating ($\Delta c_{i}$) parameter,
\begin{equation}
\label{eq:aicidefinition}
	a_{i} \equiv c_{i} \, (1 + \Delta c_{i}), \, \, \bar{a}_{i} \equiv c_{i} \, (1 - \Delta c_{i}) .
\end{equation}
For $\Delta c_{i} = 0$ there is no \CP violation between the corresponding \Dz and \Dzb intermediate state amplitudes.
Note that the \CP-violating parameters are included only for distinct weak decay processes as the strong interaction is assumed to be \CP-conserving 
such that \eg the amplitudes for the processes 
$\Dz \to \pim \, \left[ a_{1}(1260)^{+}\to \pip \, \rho(770)^{0} \right] $ and 
$\Dz \to \pim \, \left[ a_{1}(1260)^{+} \to \pip \, \sigma \right]$ share a common 
$\Delta c_{i}$, while having different \CP-conserving parameters. 
As we do not measure the time distribution, we have no sensitivity to the overall phase difference between $\Dz$ and $\Dzb$ and thus, the phase difference between $A_\Dz(\phsPoint)$ and $A_\Dzb(\phsPoint)$ is fixed to null. From these separate amplitudes, the direct \CP violation in each amplitude is simply calculated from the fit coefficients as
\begin{equation}
  \mathcal A_{CP}^{i} \equiv\frac{|a_{i}|^2 - |\bar{a}_{i}|^2}{|a_{i}|^2 + |\bar{a}_{i}|^2}.
\end{equation}
In principle, the global direct \CP asymmetry can be calculated from
\begin{equation}
	\mathcal A_{CP} \equiv  \frac{\int \vert A_{\Dz}(\phsPoint) \vert^{2}  \, \text{d}\Phi_{4}  -  \int \vert A_{\Dzb}(\phsPoint) \vert^{2}  \, \text{d}\Phi_{4} }
	         {\int \vert A_{\Dz}(\phsPoint) \vert^{2}  \, \text{d}\Phi_{4}  +  \int \vert A_{\Dzb}(\phsPoint) \vert^{2}  \, \text{d}\Phi_{4} },
                 \label{eq:AcpModel}
\end{equation}
however to avoid an unnecessary systematic uncertainty arising from the amplitude model, this will instead be determined from an asymmetry in the integrated decay rates, 
\begin{equation}
	\mathcal A_{CP} \equiv \frac{\Gamma(\Dz \to h^+ h^- \pip \pim)-{\Gamma}(\Dzb \to h^+ h^- \pip \pim)}{\Gamma(\Dz \to h^+ h^- \pip \pim)+{\Gamma}(\Dzb \to h^+ h^- \pip \pim)} = \frac{\bar{\eps}_{\rm Tag} N_{\Dz} - \eps_{\rm Tag} N_{\Dzb}}{\bar{\eps}_{\rm Tag} N_{\Dz} + \eps_{\rm Tag} N_{\Dzb}},
\end{equation}
composed of the number of signal candidates tagged as \Dz(\Dzb) mesons, $N_{\Dz}$ ($N_{\Dzb}$). 
For the CLEO-c data, the signal tagging efficiency ratio,  
\begin{equation}
  \frac{\bar{\eps}_{\rm Tag}}{{\eps}_{\rm Tag}}  = 0.9899 \pm 0.0015,
\end{equation}
has been determined from an average over the $D \to K\pi$, $K\pi\pi^0$ and $K\pi\pi\pi$ efficiencies given in Ref.~\cite{Bonvicini:2013vxi}. 
No asymmetry in pion identification is found in the preceding CLEO data samples and thus the tagging efficiency ratio is set to unity with an uncertainty of 1.5\%~\cite{Coan:1999kh}.

 % end input /Users/pnaik/Documents/CLEO/Papers/latexpand/latex/amplitude.tex
 %
% start input /Users/pnaik/Documents/CLEO/Papers/latexpand/latex/model.tex
%

\clearpage
\section{Likelihood Fit}
\label{sec:likelihood}

Due to flavor tagging, there are two independent data sets available;
$\Dz \to h^+ h^- \pip \pim$  and $\Dzb \to h^- h^+ \pim \pip$ events
which can be described by the amplitudes
$	A_{\Dz}(\phsPoint)$
and
$	A_{\Dzb}(\phsPoint)$, respectively. 
In general, the signal PDF for events tagged as $\Dz \to h^+ h^- \pip \pim$ is given by
\begin{equation}
	\mathcal P_{\rm Sig}(\phsPoint) %
	=  
	\frac{ [(1-w)\left\vert   A_{\Dz}(\phsPoint) \right\vert^{2} + w \, \left\vert    A_{\Dzb}(\phsPoint) \right\vert^{2} ]\,\epsilon_{\rm Sig}(\phsPoint) \, \phsd(\phsPoint) }
	{\int [\left\vert  A_{\Dz}(\phsPoint) \right\vert^{2} + \left\vert    A_{\Dzb}(\phsPoint) \right\vert^{2} ]\, \epsilon_{\rm Sig}(\phsPoint) \, \text{d}\Phi_{4} } , 
\label{eq:sigPDF}
\end{equation}
where 
$\epsilon_{\rm Sig}(\phsPoint)$ is the phase-space efficiency and $w$ is the wrong tag fraction as defined in Sec.~\ref{sec:event}. 
In the case of no $\CP$ violation, the integrals over the \Dz\ and \Dzb\ amplitudes will be equal. For the $\CP$-tagged data sets used in the $\Dz \to K^+ K^- \pip \pim$ analysis, the signal PDFs are given in Ref.~\cite{KKpipi}.
We do not account for effects of neutral charm meson oscillations, as we expect these to be negligible in these analyses.

Note that the efficiency in the numerator appears as an additive constant in the $\log {\cal L}$ that does not depend on any fit parameters such that it can be ignored.
However, the efficiency function still enters via the normalization integrals. 
These normalization terms are determined numerically by a MC integration technique.
For this purpose, we use simulated events generated according to a preliminary model, pass them 
through the full detector simulation and apply the same selection criteria as for data 
in order to perform the MC integrals.
For example, the first integral in \eqnPRDref{eq:sigPDF} can be approximated as 
\begin{equation}
	\int \left\vert   A_{\Dz}(\phsPoint) \right\vert^{2} \, \epsilon_{\rm Sig}(\phsPoint) \, \text{d}\Phi_{4}   \approx 
	\frac{1}{N_{\rm MC}} \, \sum_{k}^{N_{\rm MC}}    \frac{\left\vert   A_{\Dz}(\bold{x_{k}}) \right\vert^{2}}
	{\left\vert A_{\Dz}^{\prime}(\bold{x_{k}}) \right\vert^{2}}
\end{equation}
where $A_{\Dz}^{\prime}$ labels the preliminary amplitude model and
$x_{k}$ is the $k$-th MC event. As a result, the efficiency can be included in the amplitude fit without explicitly modeling it.
For $\Dz \to \pi^+ \pi^- \pip \pim$, we use a sample of $N_{\rm MC}  = 600 000$  MC events to 
ensure that the uncertainty on the integral is less than $0.5 \%$.
For $\Dz \to K^+ K^- \pip \pim$, we use samples of $N_{\rm MC} \approx 900 000$ events each, produced under each of the CLEO III and CLEO-c 
detector conditions. MC representing the CLEO II.V detector conditions is 
simulated from CLEO III MC via the reweighting process discussed in Ref.~\cite{KKpipi}. 
The uncertainty on the integral for each $\Dz \to K^+ K^- \pip \pim$ MC sample is less than $0.5 \%$.

The background PDF,
\begin{equation}
	\mathcal P_{\rm Bkg}(\phsPoint) %
	= \frac{ \epsilon_{\rm Sig}(\bold x)  \,B(\phsPoint) \, \phsd(\phsPoint) }{\int  \epsilon_{\rm Sig}(\phsPoint) \, B(\phsPoint) \, \text{d}\Phi_{4} } ,
\label{eq:bkgPDF}
\end{equation} is determined in \secref{sec:bkgPDF} from sideband data. Note that because of the integration method, the background parameters only have meaning relative to the signal efficiency.
The event likelihood is constructed from the signal 
PDF and the background PDF,
\begin{equation}
	\mathcal L = f_{\rm Sig} \, \mathcal P_{\rm Sig}(\phsPoint \vert \theta) + (1-f_{\rm Sig}) \, \mathcal P_{\rm Bkg}(\phsPoint \vert \theta),
\end{equation}
where $f_{\rm Sig}$ is the signal fraction as determined in Sec.~\ref{sec:event4pi} and $\theta$ is the set of fit parameters.
\subsection{Background Model}
\label{sec:bkgPDF}

Background events arise from randomly combined particles from various processes such as other $D$ decays or continuum which, by chance, fulfill all required selection criteria.
Some of them may even contain resonances that do not arise from the signal $\Dz$ decay. 
The chosen background PDF for the $\Dz \to \fourpi$ mode
includes Breit-Wigner (BW) contributions from the resonances $\sigma, \rho(770)^{0}, f_{0}(980)$ and two ad-hoc scalar resonances ($S^{0}_{1}, S^{-}_{2}$) with free masses and widths.
They are added incoherently on top of two non-resonant components.
In addition, several exponential and polynomial functions are included to allow for more flexibility.
The background function is explicitly given by
\begin{equation}
	B(\phsPoint) = \sum_{i=1}^{7} b_{i} \,  \vert B_{i}(\phsPoint)  \vert^{2},
\end{equation}
where,
\begin{align}
	B_{1}(\phsPoint) &=  {\rm BW}_{\sigma}(s_{12})  \cdot {\rm BW}_{\sigma}(s_{34}),  \nonumber \\
	B_{2}(\phsPoint) &=  {\rm BW}_{\rho(770)^{0}}(s_{12}) \cdot \exp(- \, \alpha_{1} \cdot s_{34}) , \nonumber\\ 
	B_{3}(\phsPoint) &=  {\rm BW}_{f_{0}(980)}(s_{12})  \cdot {\rm BW}_{f_{0}(980)}(s_{34}), \nonumber \\ 
	B_{4}(\phsPoint) &=  {\rm BW}_{S^{0}_{1}}(s_{12}) \cdot \left(  \sum_{i=0}^{5}  \, c_{i} \cdot s_{34}^{i} \right), \nonumber \\ 
	B_{5}(\phsPoint) &=  {\rm BW}_{S_{2}^{-}}(s_{124}), \nonumber  \\ 
	B_{6}(\phsPoint) &= \exp(- \, \alpha_{2} \cdot s_{14}) \cdot \exp(- \, \alpha_{3} \cdot s_{23}), \nonumber\\ 
	B_{7}(\phsPoint) &=  \left(  \sum_{i=0}^{4}  \, d_{i} \cdot s_{124}^{i} \right) \cdot \left(  \sum_{i=0}^{5}  \, e_{i} \cdot s_{12}^{i} \right),
\end{align}
with $s_{ij} = m^{2}(\pi_{i} \,\pi_{j})$, $s_{ijk} = m^{2}(\pi_{i} \,\pi_{j}\, \pi_{k})$ and $\Dz \to \pi^{+}_1 \pi^{-}_2 \pi^{+}_3 \pi^{-}_4$.
The real parameters $b_{i}, \alpha_{i}, c_{i}, d_{i}$ and $ e_{i}$ are extracted from a fit to the sideband samples defined in Sec.~\ref{sec:event4pi}.%

For $\Dz \to K^+ K^- \pip \pim$ decays, the background shape is determined for each data set and is simply modeled by an incoherent sum of the $K_1(1400)^+ \to K^*(892)^0 \pip$, $\phi(1020)$, $K^*(892)^0$, $\bar K^*(892)^0$, $\rho(770)^{0}$ resonances and a constant term with relative couplings determined from the relevant sidebands.

\subsection{Signal Model Construction}
\label{sec:LASSO}

The light meson spectrum comprises multiple resonances which are expected to contribute to $\Dz \to h^+ h^- \pip \pim$  decays as intermediate states. 
Apart from clear contributions coming from resonances such as $a_{1}(1260) \to \rho(770)^{0} \pi$, $\phi(1020)$ and $K^*(892)^0$, 
the remaining structure is impossible to infer due to
the cornucopia of broad, overlapping and interfering resonances 
within the phase space boundary.
The complete list of considered amplitudes can be found in Appendix \ref{a:decays}.

To build the amplitude model, one could successively add amplitudes on top of one another until a reasonable agreement between data and fit was achieved.
However, this step-wise approach is not particularly suitable for amplitude analyses as discussed in Ref.~\cite{Guegan:2015mea}.
Instead, we include the whole pool of amplitudes in the first instance and use the 
Least Absolute Shrinkage and Selection Operator~\cite{Tibshirani94regressionshrinkage,Guegan:2015mea} (LASSO) approach to limit the model complexity.
In this method, the event likelihood is extended by a penalty term
\begin{equation}
	-2 \, \log \mathcal L \to -2 \, \log \mathcal L + \lambda \, \sum_{i} \sqrt{ \int \vert a_{i} \, A_{i}(\phsPoint) \vert^{2} \, \text{d}\Phi_{4}  },
\end{equation}
which 
shrinks the amplitude coefficients
towards zero.
The amount of shrinkage is controlled by the parameter $\lambda$, to be tuned on data.
Higher values for $\lambda$ encourage sparse models, \ie models with only a few non-zero amplitude coefficients.
The optimal value for $\lambda$ is found by minimizing the Bayesian information criteria~\cite{BIC} (BIC),
\begin{equation}
	\text{BIC}(\lambda) = - 2 \, \log \mathcal L + r  \, \log N_{\rm Sig},
\end{equation}
where $N_{\rm Sig}$ is the number of signal events and $r$ is the number of amplitudes with a decay fraction above 
a certain threshold.
In this way, the optimal $\lambda$ balances
the fit quality ($- 2 \, \log  \mathcal L$) against the model complexity.
The LASSO penalty term is only used to select the model. 
Afterwards, this term must be discarded in the final amplitude fit with the selected model, otherwise the parameter uncertainties would be biased. 

\begin{figure}[b]
  \centering
  \includegraphics[width=0.49\textwidth, height=!]{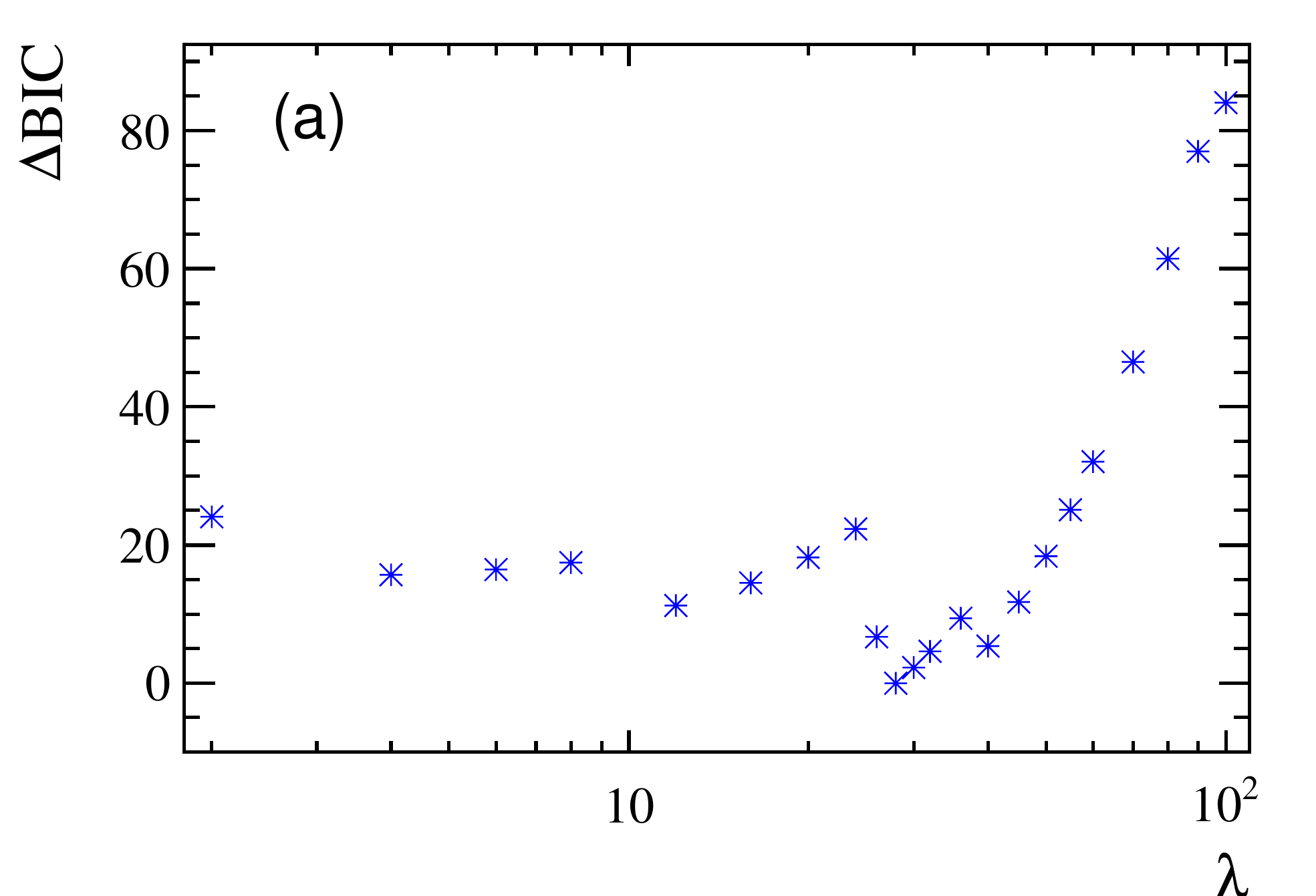} 
  \includegraphics[width=0.49\linewidth, height=!]{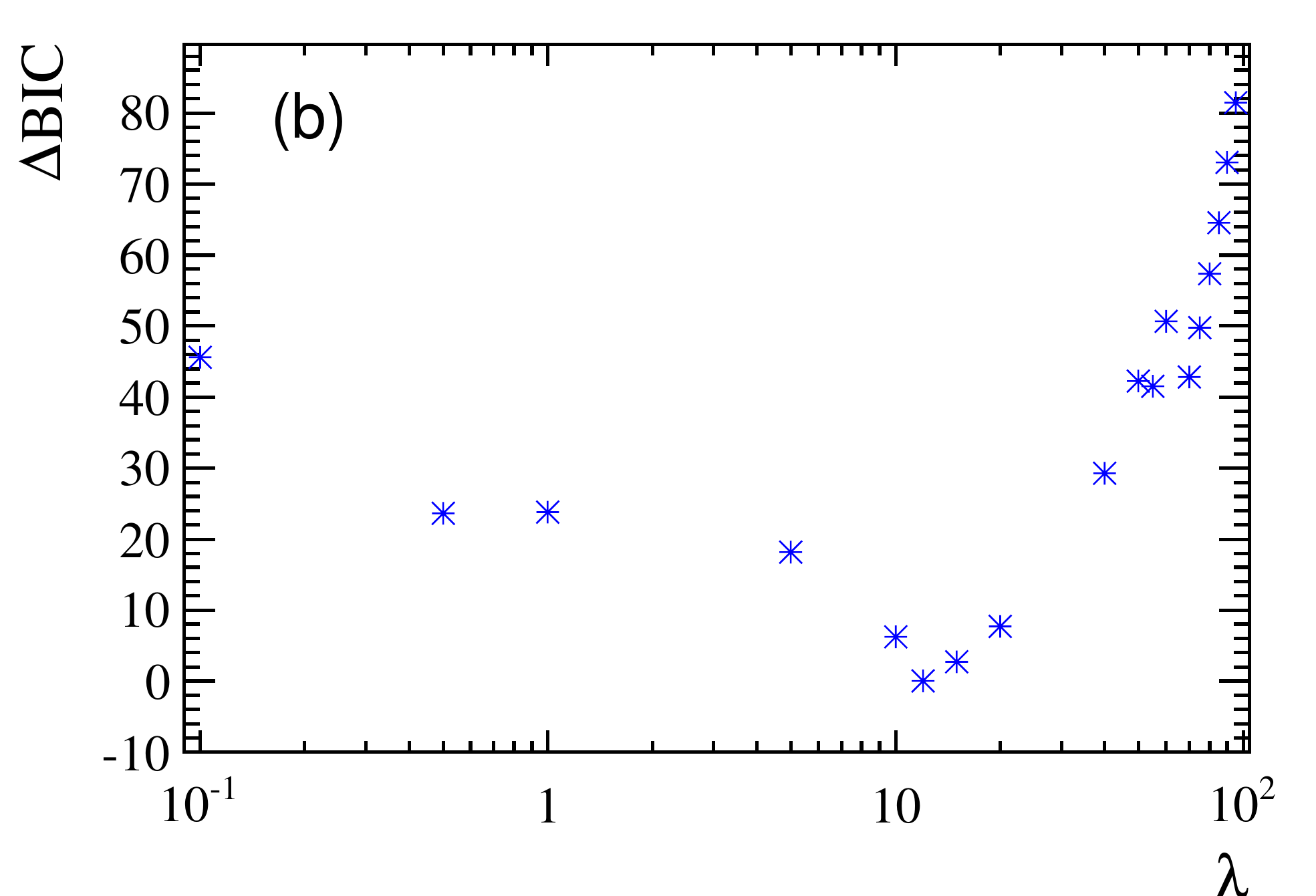}
  \caption{Difference in the BIC value from its minimum as function of the LASSO parameter $\lambda$ for $\Dz \to \fourpi$ (a) and Stage 1 $\Dz \to \KKpipi$ (b).}
  \label{fig:BIC}
\end{figure}

The implementation of the LASSO procedure differs between the $\Dz \to h^+ h^- \pip \pim$ analyses. For $\Dz \to \fourpi$ decays, the set of amplitudes is selected using the optimal value of $\lambda=28$, and is henceforth called the LASSO model; 
Figure \ref{fig:BIC}(a) shows the distribution of BIC values obtained by scanning over $\lambda$
where we choose the decay fraction threshold to be $0.5 \%$.
It is important to note that there are certain groups of amplitudes with the same angular distribution 
that are prone to produce artificially high interference effects.
Amongst them are the di-scalar amplitudes: 
$D \to (\pi \, \pi)_{S} \, (\pi \, \pi)_{S}$, $D \to (\pi \, \pi)_{S} \, \sigma$, $D \to \sigma \, \sigma$, $D \to \sigma \, f_{0}(1370)$
and $D \to f_{0}(1370) \, f_{0}(1370)$
as well as the di-vector amplitudes:  $D \to (\pi \, \pi)_{P} \, (\pi \, \pi)_{P}$, $D \to (\pi \, \pi)_{P} \, \rho(1450)^{0}$ and $D \to \rho(1450)^{0} \, \rho(1450)^{0}$.
In these cases, only one amplitude of the group is included at a time and the model selection is performed for each choice.
It was further observed that the inclusion of the $D \to \pi [\pi(1300) \to \pi (\pi \, \pi)_{P}]$ amplitude leads to a  $D \to \rho(770)^{0} \, \rho(770)^{0}$ D-wave fraction much larger than 
the S-wave fraction with a large destructive interference. 
As we consider this as unphysical we do not include it in our default approach but in an alternative model presented in Appendix~\ref{a:alternative}.
In addition, we repeated the model selection procedure under multiple different conditions:
\begin{enumerate}[label=(\alph*)]
	\item The fit fraction threshold for inclusion in the final model was varied within the interval $[0.05, 5] \%$.
		The set of selected amplitudes is stable for thresholds between $0.1\%$ and $1\%$. 
		Other choices result in marginally different models containing one component more or less.
	\item Instead of BIC, the Akaike information criteria ($\text{AIC}(\lambda) = -2 \, \log  \mathcal L + 2 \, r$ \cite{AIC}) was used to optimize $\lambda$.
		For a given threshold, the AIC method tends to prefer %
		lower $\lambda$ values.
		However, the set of models obtained varying the threshold within the interval $[0.05, 5] \%$
		is identical to the BIC method. 
	\item The amplitudes selected under nominal conditions were excluded one-by-one from the set of all amplitudes considered.  
\end{enumerate}
From that we obtained a set of alternative models shown in Appendix~\ref{a:alternative}.

Due to the vast number of potential amplitude components and computational limits imposed by the consideration of multiple data samples in the $\Dz \to \KKpipi$ analysis, a staged LASSO method using only the flavor-tagged data, 
representing over 90\% of the available statistics, is employed. The approach taken is based on the assumption that the signal decay proceeds primarily by doubly resonant decays, \ie cascade and quasi-two-body decays, rather than decay amplitudes with non-resonant components. In Stage 1, only doubly resonant decays along with the simplest non-resonant component $(K^+K^-)_{S} \, (\pi^+ \pi^-)_{S}$ are considered. Figure \ref{fig:BIC}(b) shows a plot of the complexity factor $\lambda$, against the resulting BIC values. We found that the fit cannot distinguish between amplitudes with $K^*(1680)^+ \to K^*(892)^0\, \pip$ and $K^*(1410)^+ \to K^*(892)^0 \, \pip$, which both peak outside the kinematic range of the \D\ decay's phase space. We therefore only include $K^*(1680)^+ \to K^*(892)^0\, \pip$ in our nominal model. An alternative fit with the $K^*(1410)^{+}$, which has marginally worse fit quality is presented in Table~\ref{tab:alternativeModels2}.

In Stage 2, the LASSO procedure is again performed with the components selected by Stage 1 and all single-resonant components. It should be noted in the case of 
cascade decays that if LASSO picked an amplitude component but not its conjugate decay in the first stage, the conjugate is also considered again in this stage. Once more, the interplay between $D \rightarrow SS$ amplitudes leads to very large interference terms, and thus $f_0(980) \,(\pi^+ \pi^-)_{S}$ and $f_0(980) \, (K^+ K^-)_{S}$ components are considered as a replacement for the non-resonant $(K^+K^-)_{S} \, (\pi^+ \pi^-)_{S}$ component in an alternative model. 
The final fit merges the components chosen in Stage 1 and Stage 2 and includes the \CP-tagged data. Within this set of amplitudes, 6 are considered insignificant relative to their error and removed from the fit with no significant impact on fit quality.

 % end input /Users/pnaik/Documents/CLEO/Papers/latexpand/latex/model.tex
 %
% start input /Users/pnaik/Documents/CLEO/Papers/latexpand/latex/results.tex
%

\clearpage\section{\texorpdfstring{$\Dz \to \fourpi$ Amplitude Analysis Results}{D0 to pi+ pi- pi+ pi- Amplitude Analysis Results}}
\label{sec:results4pi}

\subsection{Amplitude Model Fit Results}
\label{ssec:4piComponents}
\begin{figure}[h]
\centering
	\includegraphics[width=0.49\textwidth, height = !]{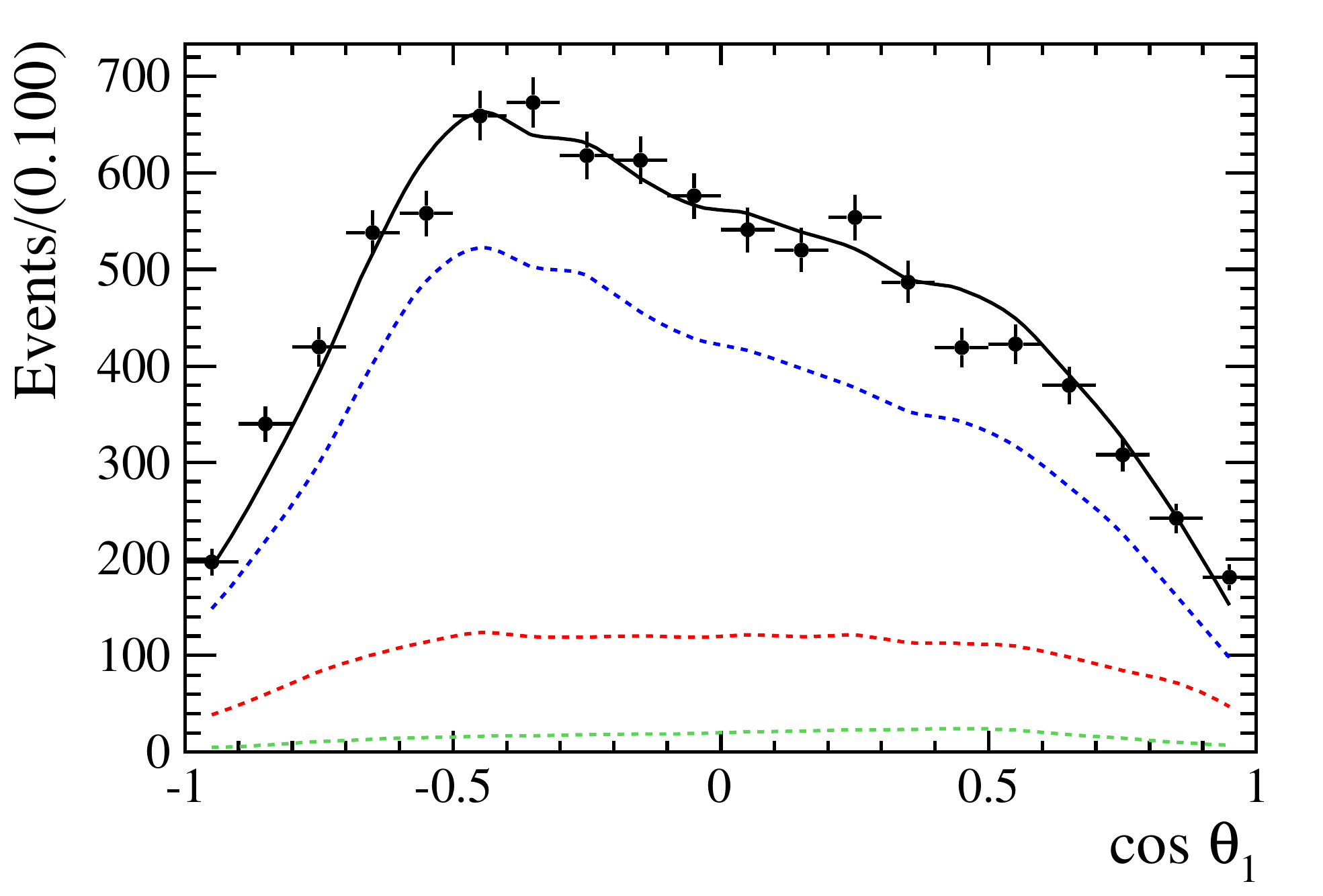} 
	\includegraphics[width=0.49\textwidth, height = !]{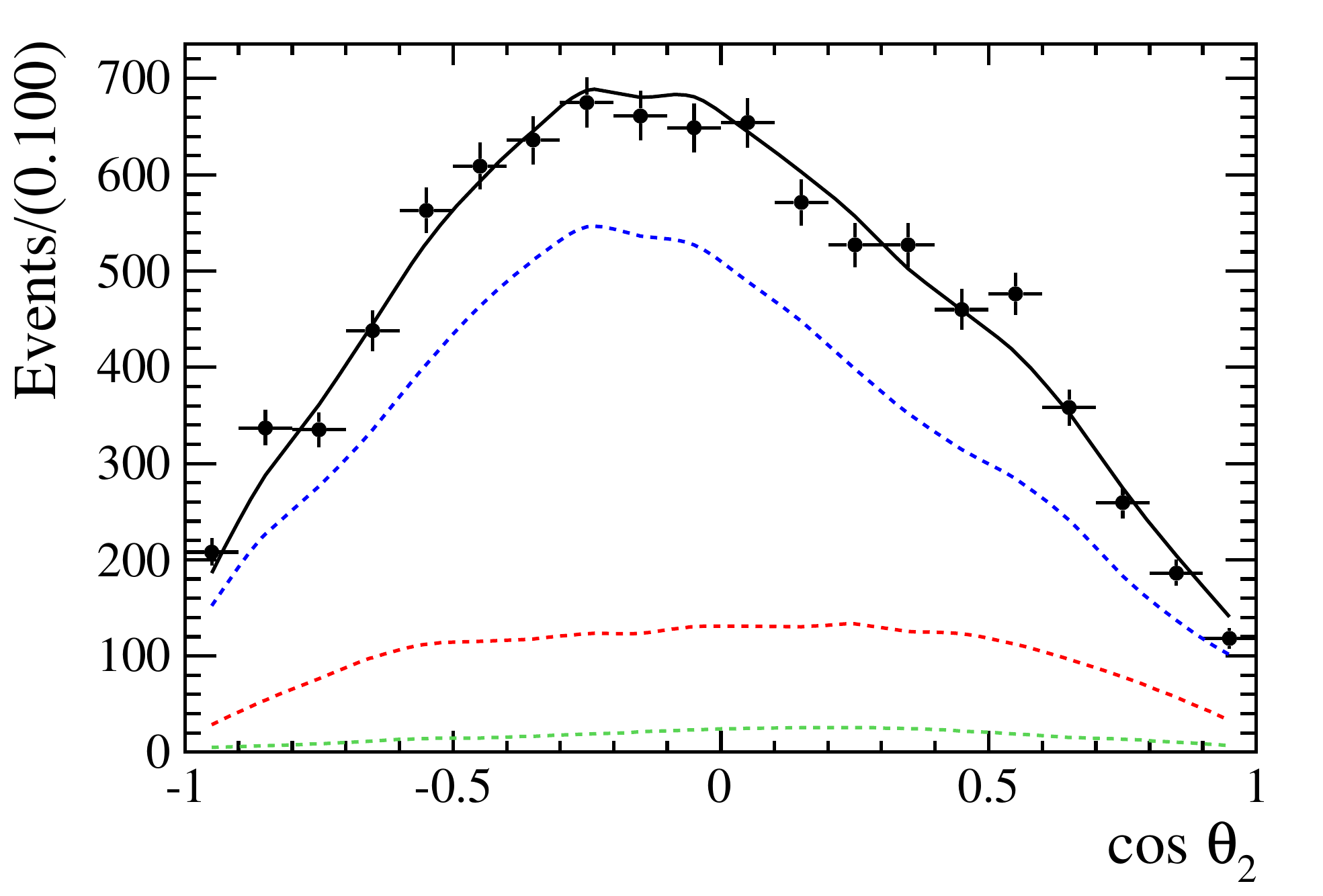} 
	\centering
	\includegraphics[width=0.49\textwidth, height = !]{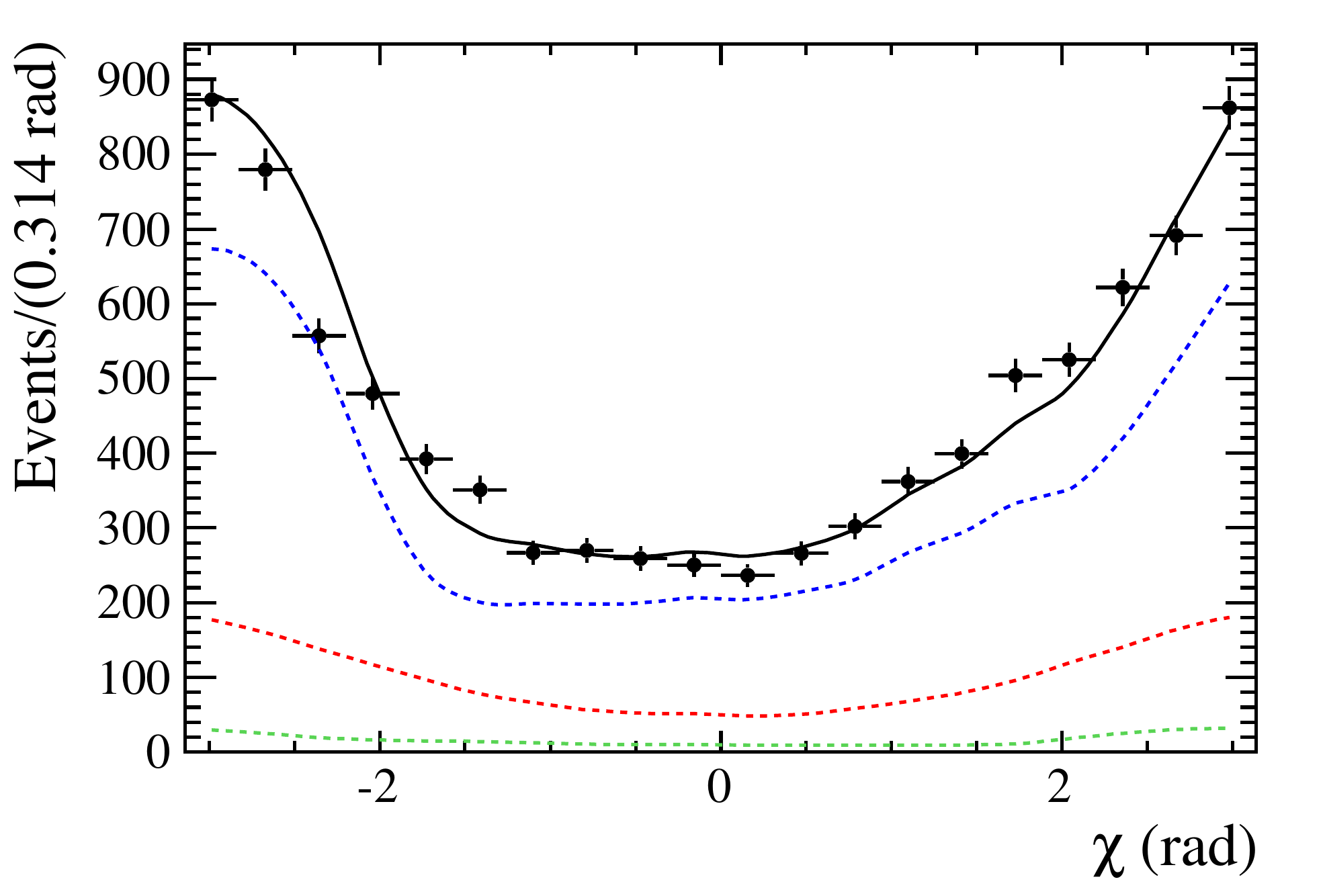} 

	\caption{Angular projections of the $\Dz \to \fourpi$ fit results (black solid line) in the transversity basis.
	The signal component is shown in blue (dashed), the background component in red (dashed) and the wrongly tagged contribution in green (dashed).
	}
	\label{fig:baselineFit2}
\end{figure}

Table \ref{tab:lassoModel} 
lists the real and imaginary part of the complex amplitude coefficients $a_{i}$, 
obtained by fitting the LASSO model to the data,
along with the corresponding fit fractions. 
The letters in square brackets refer to the relative orbital angular momentum of the decay products. 
If no angular momentum is specified, the lowest angular momentum state consistent with angular momentum conservation and, where appropriate, parity conservation is used.
The interference fractions are given in Appendix~\ref{a:interference}. 
Figure \ref{fig:baselineFit} shows the distributions of 
selected
phase space 
observables, which demonstrate 
reasonable agreement between data and the fit model. 
We also project into the transversity basis to demonstrate good description of the overall angular structure in
Fig.~\ref{fig:baselineFit2}: 
The acoplanarity angle 
${\chi}$, 
is the angle between the two decay planes formed by 
the $\pi^+\pi^-$ combination with minimum invariant mass, ${\rm min}[m(\pi^+\pi^-)]$,  
and the remaining $\pip \pim$ 
combination
in the $D$ rest frame; boosting into the rest frames of the two-body systems defining these decay planes,
the two helicity variables 
are defined as the cosine of the angle, ${\theta}$, 
of each \pip\ momentum with the $D$ flight direction.

In order to quantify the quality of the fit in the five-dimensional phase space,
a \chisq value is determined by binning the data;
\begin{equation}
	\chi^{2} = \sum_{b=1}^{N_{\rm bins}} \frac{(N_{b}-N_{b}^{\rm exp})^{2}}{N_{b}^{\rm exp}},
\end{equation}
where $N_{b}$ is the number of data events in a given bin, 
$N_{b}^{\rm exp}$ is the event count predicted by the fitted PDF
and $N_{\rm bins}$ is the number of bins.
An adaptive binning used in Ref.~\cite{KKpipi}
is used to ensure sufficient statistics in each bin for a robust $\chi^{2}$ calculation.
At least $25$ events per bin are required.
The number of degrees of freedom $\nu$, in an unbinned fit is bounded by $N_{\rm bins}-1$ and $(N_{\rm bins}- 1) - N_{\rm par}$, 
where $N_{\rm par}$ is the number of free fit parameters.
We use the \chisq value divided by $\nu = (N_{\rm bins}-1) - N_{\rm par}$ as a conservative estimate.
For the LASSO model, this 
amounts to $\chisq/\nu = 1.40$ with $\nu = 221$ and $N_{\rm par}=34$, 
indicating a decent fit quality.

\begin{sidewaystable}[p]
  \footnotesize
  \centering
  \caption{ \small Real and imaginary part of the complex amplitude coefficients 
    and fit fraction of each component of the $\Dz \to \fourpi$ LASSO model. 
    The complex fit parameter listed for $\Dz \to \pip \, a_{1}(1260)^{-}$ describes the relative magnitude and phase of $\Dz \to \pip \, a_{1}(1260)^{-}$ and $\Dz \to \pim \, a_{1}(1260)^{+}$ as described in \secref{sec:analysis}.
    For the fit coefficients, the first quoted uncertainty is statistical, while the second 
    arises from systematic sources. The third uncertainty in the fit fraction arises from the alternative models considered.}
  \begin{tabular}
     {@{\hspace{0.5cm}}l@{\hspace{0.25cm}}  @{\hspace{0.25cm}}c@{\hspace{0.25cm}}  @{\hspace{0.25cm}}c@{\hspace{0.25cm}}  @{\hspace{0.25cm}}c@{\hspace{0.5cm}}}
     \hline \hline
     Decay channel & $\Re (a_{i}) $ & $\Im (a_{i}) $ & $F_{i} \, (\%)$ \\ \hline
     $\Dz \to \pim \, \left[ a_{1}(1260)^{+}\to \pip \, \rho(770)^{0} \right] $ & 100.00 (fixed) & 0.00 (fixed) & $ 38.1 \pm 2.3 \pm 3.2 \pm 1.7$   \\
     $\Dz \to \pim \, \left[ a_{1}(1260)^{+} \to \pip \, \sigma \right] $ & $56.46 \pm 13.85 \pm 14.49$ &$ 167.87 \pm 14.51 \pm 19.38$ &   $ 10.2 \pm 1.4 \pm 2.1 \pm 2.5$\\

     $\Dz \to \pip \, a_{1}(1260)^{-}$  & $0.218 \pm 0.028 \pm 0.036$ & $0.180 \pm 0.024 \pm 0.017$& -   \\
     \hspace{10pt}$\Dz \to \pip \, \left[ a_{1}(1260)^{-}\to \pim \, \rho(770)^{0} \right] $  & - & - & $ 3.1 \pm 0.6 \pm 0.5 \pm 0.9$   \\
     \hspace{10pt}$\Dz \to \pip \, \left[ a_{1}(1260)^{-} \to \pim \, \sigma \right] $ & - & - &   $ 0.8 \pm 0.2 \pm 0.1 \pm 0.4$\\

      $\Dz \to \pim \, \left[ \pi(1300)^{+} \to \pip \, \sigma \right] $ &$  -15.11 \pm 3.08 \pm 9.44$&$ 19.80 \pm 3.54 \pm 5.90$ &  $ 6.8 \pm 0.9 \pm 1.5 \pm 3.1$\\
	$\Dz \to \pip \, \left[ \pi(1300)^{-} \to \pim \, \sigma \right] $ & $-6.48 \pm 2.39 \pm 6.08$ & $15.19 \pm 2.62 \pm 7.52$  &   $ 3.0 \pm 0.6 \pm 2.0 \pm 2.0$\\

	$\Dz \to \pim \, \left[ a_{1}(1640)^{+}[D] \to \pip \, \rho(770)^{0} \right] $ & $-125.40 \pm 20.59 \pm 28.50$ &$-10.89 \pm 15.07 \pm 13.75$&  $ 4.2 \pm 0.6 \pm 0.9 \pm 1.8 $\\
	$\Dz \to \pim \, \left[ a_{1}(1640)^{+}\to \pip \, \sigma \right] $ & $77.57 \pm 21.59 \pm 31.24$&$ -94.98 \pm 21.12 \pm 34.54$ &  $ 2.4 \pm 0.7 \pm 1.1 \pm 1.3$ \\

	$\Dz \to \pim \, \left[ \pi_{2}(1670)^{+}\to \pip \, f_{2}(1270) \right] $  &$ -49.93 \pm 42.23 \pm 77.44$& $348.39 \pm 40.95 \pm 42.87$ & $ 2.7 \pm 0.6 \pm 0.7 \pm 0.9$  \\
	$\Dz \to \pim \, \left[ \pi_{2}(1670)^{+} \to \pip \, \sigma \right] $  & $-51.35 \pm 22.21 \pm 15.18$ & $-209.98 \pm 22.21 \pm 41.58$ & $ 3.5 \pm 0.6 \pm 0.8 \pm 0.9$ \\

	$\Dz \to \sigma \, f_{0}(1370)  $ &$27.71 \pm 6.81 \pm 19.04$& $71.93 \pm 6.41 \pm 17.44$&  $ 21.2 \pm 1.8 \pm 4.2 \pm 5.2$ \\
	
	$\Dz \to \sigma \,  \rho(770)^{0}  $ &$  41.99 \pm 4.19 \pm 4.42$& $-25.42 \pm 3.62 \pm 6.53$ & $ 6.6 \pm 1.0 \pm 1.2 \pm 3.0$\\

	$\Dz[S] \to \rho(770)^{0} \, \rho(770)^{0}$  & $2.37 \pm 1.24 \pm 2.00$ & $8.89 \pm 1.35 \pm 1.83$&  $ 2.4 \pm 0.7 \pm 1.1 \pm 1.0$ \\
	$\Dz[P] \to \rho(770)^{0} \, \rho(770)^{0}$  & $ -2.51 \pm 1.33 \pm 1.46$ & $-20.80 \pm 1.48 \pm 3.67$ &    $ 7.0 \pm 0.5 \pm 1.6 \pm 0.3$\\
	$\Dz[D] \to \rho(770)^{0} \, \rho(770)^{0}$ &$-33.99 \pm 3.34 \pm 5.11$&$-7.64 \pm 2.62 \pm 4.77$&  $ 8.2 \pm 1.0 \pm 1.7 \pm 3.5$\\
	
	$\Dz \to f_{2}(1270) \,  f_{2}(1270) $  &$ -34.47 \pm 21.71 \pm 22.46$& $-172.87 \pm 21.71 \pm 27.01$&  $ 2.1 \pm 0.5 \pm 0.3 \pm 2.3$\\
	
	\hline
	Sum & & &  $122.0 \pm 4.0 \pm 6.4 \pm 7.6$ \\
	\hline\hline
	\end{tabular}
	\label{tab:lassoModel}
\end{sidewaystable}

\begin{figure}[p]
\centering
	\includegraphics[width=0.43\textwidth, height = !]{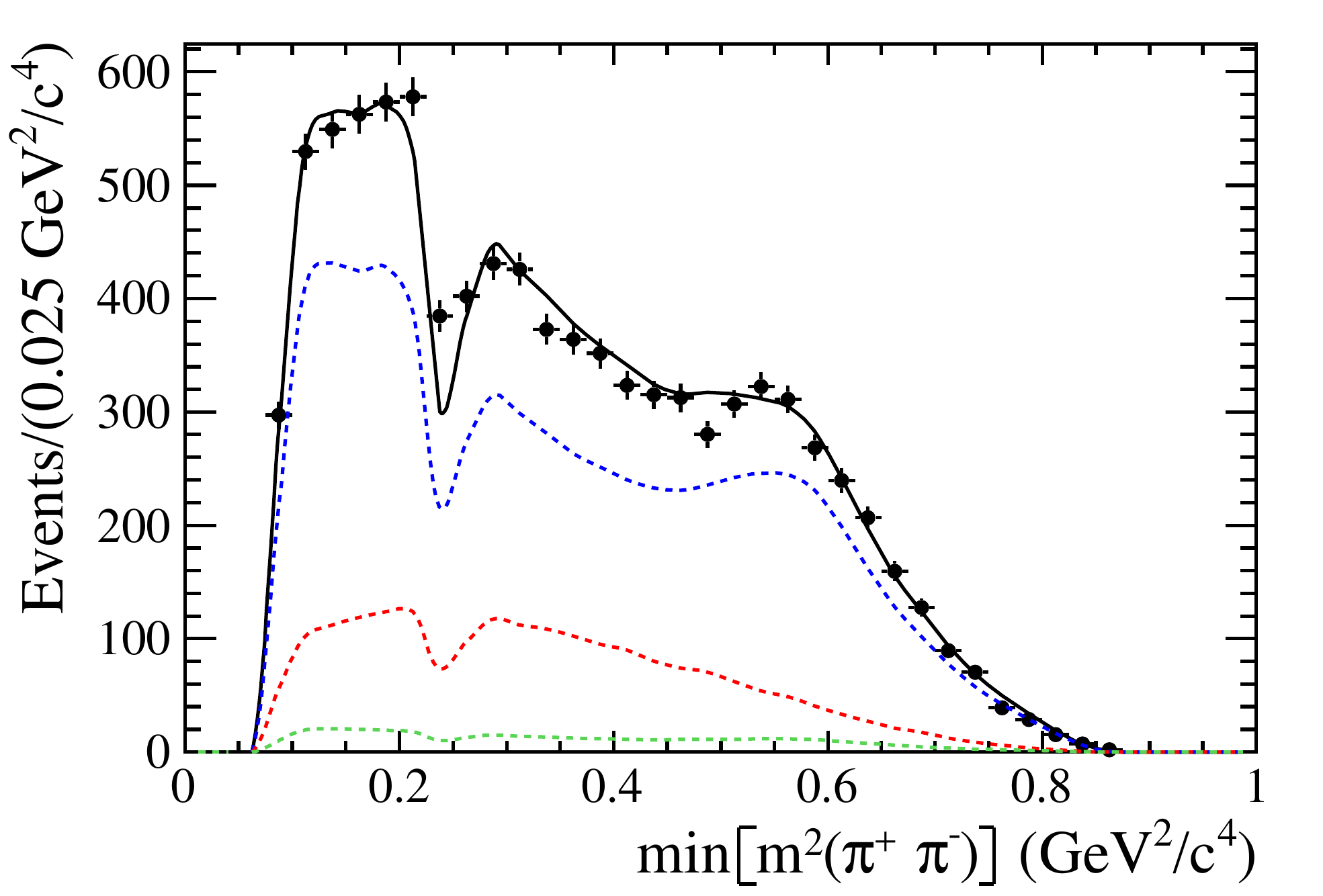} 
	\includegraphics[width=0.43\textwidth, height = !]{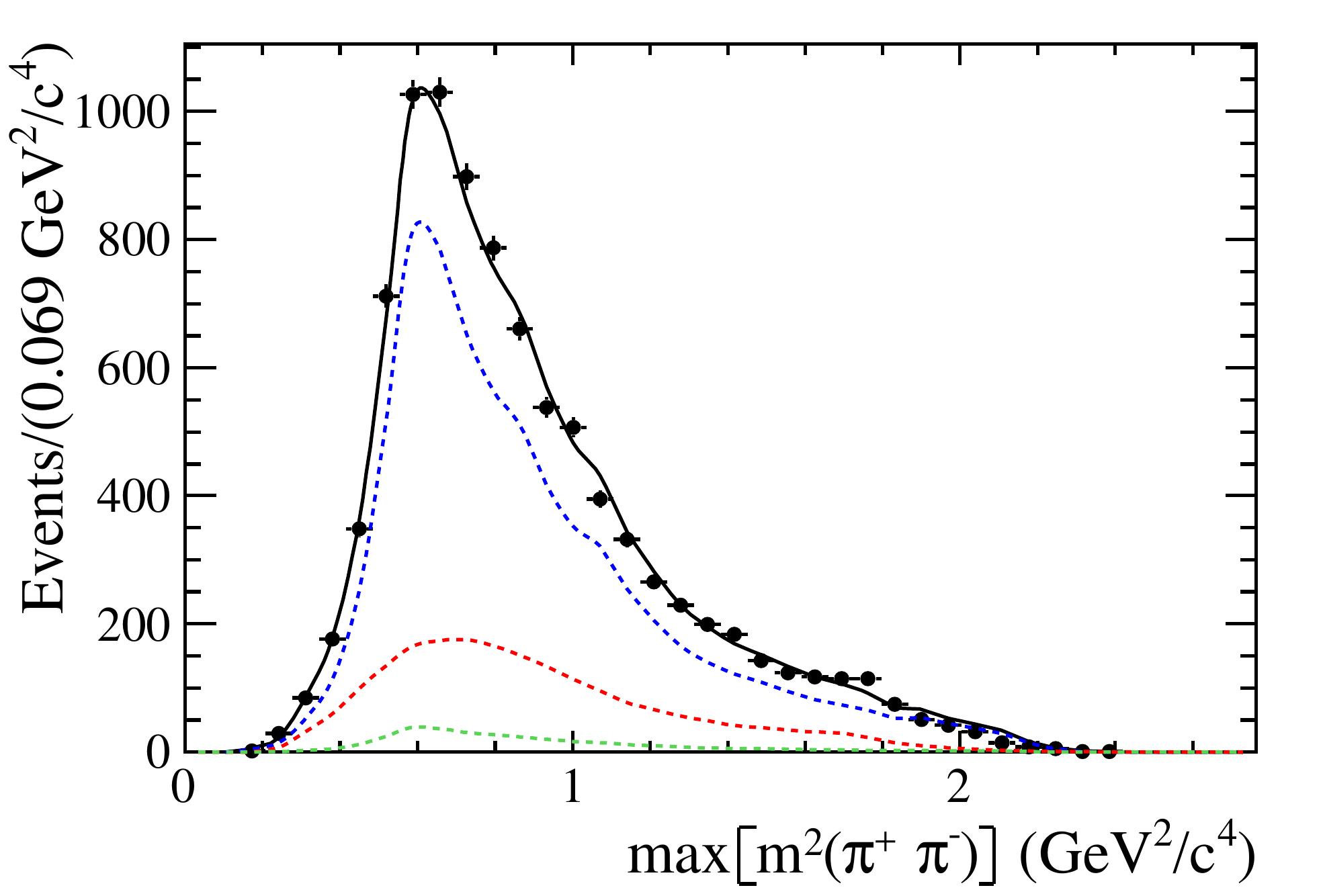} 

	\includegraphics[width=0.43\textwidth, height = !]{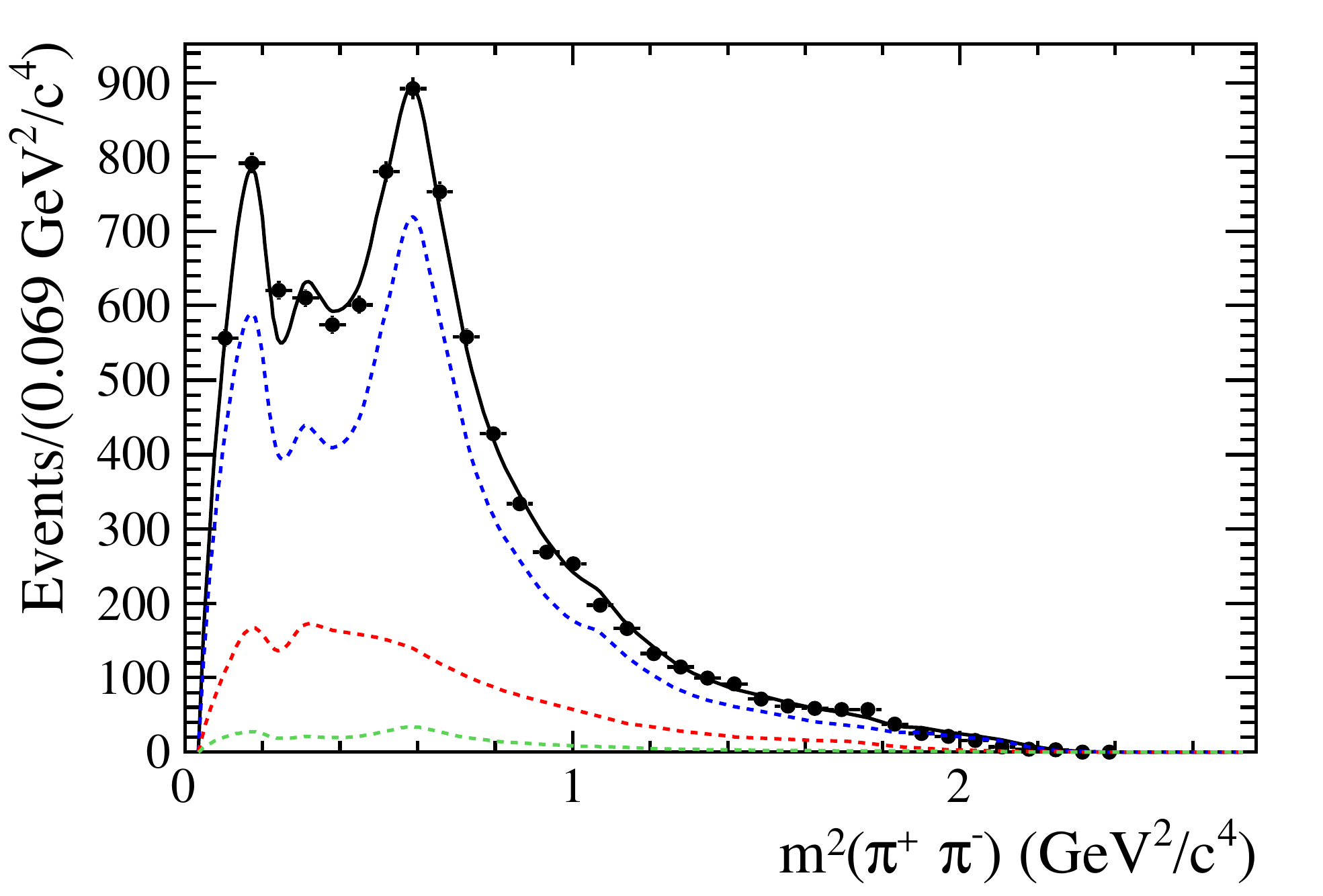} 
	\includegraphics[width=0.43\textwidth, height = !]{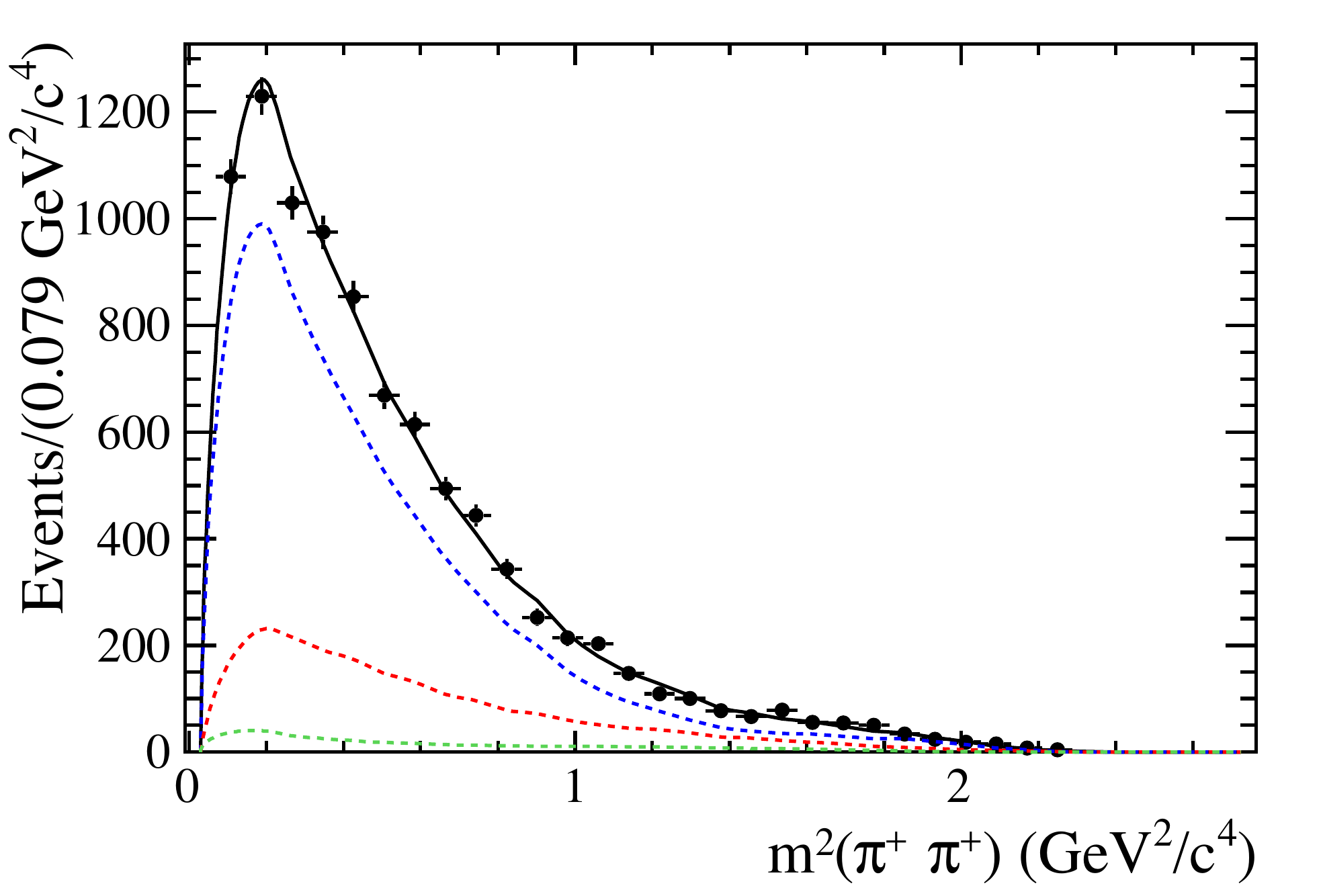} 

	\includegraphics[width=0.43\textwidth, height = !]{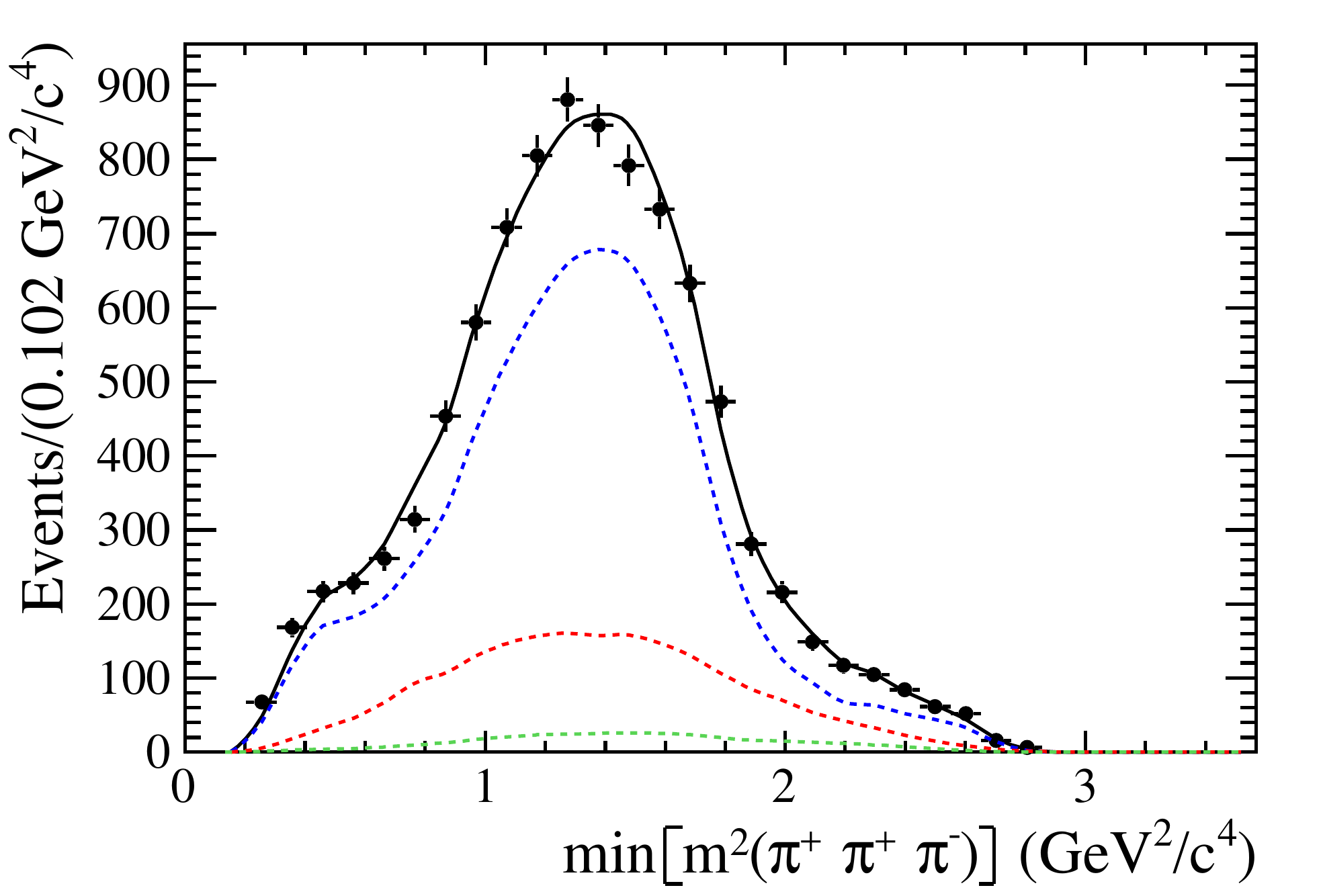} 
	\includegraphics[width=0.43\textwidth, height = !]{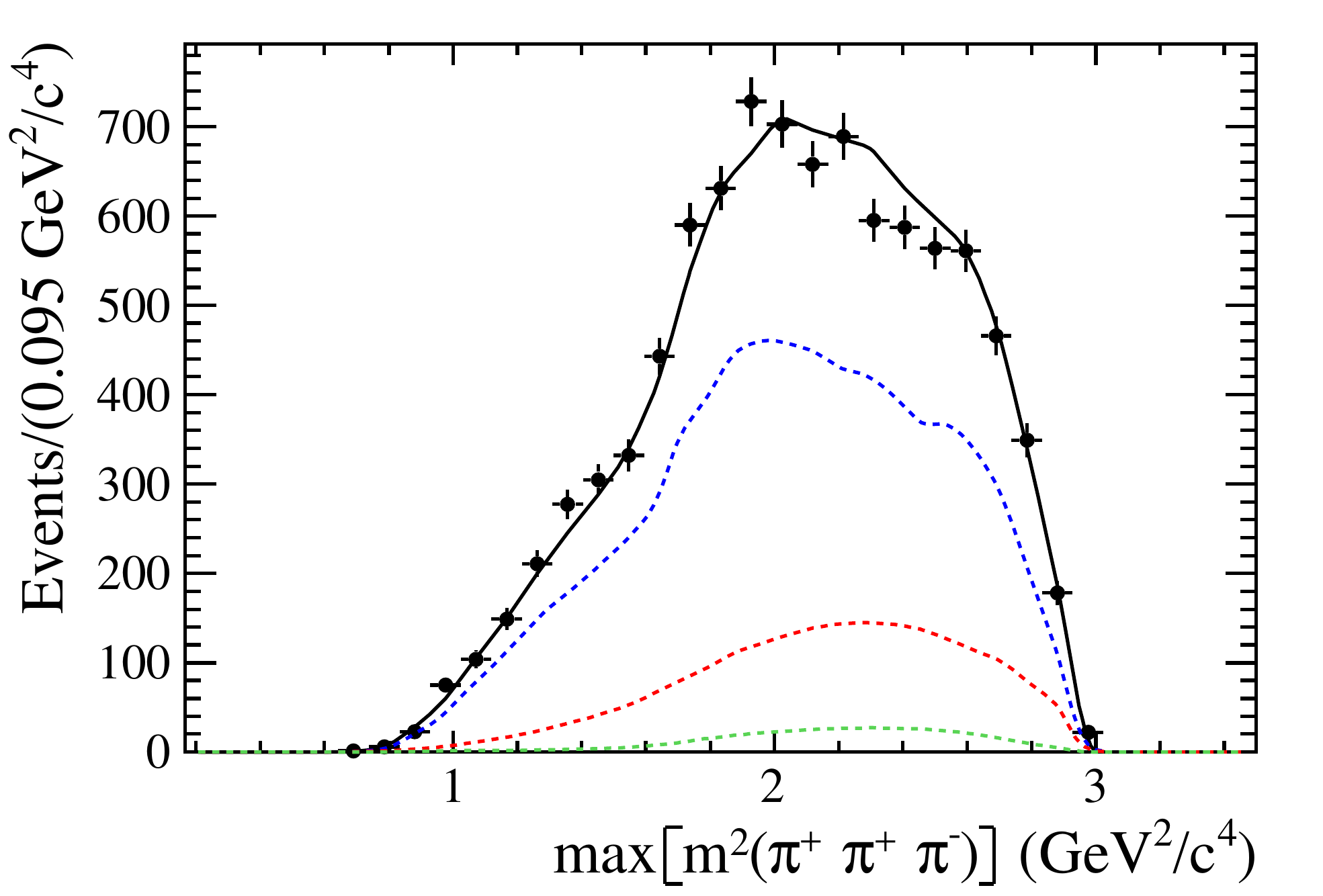} 
	
	\includegraphics[width=0.43\textwidth, height = !]{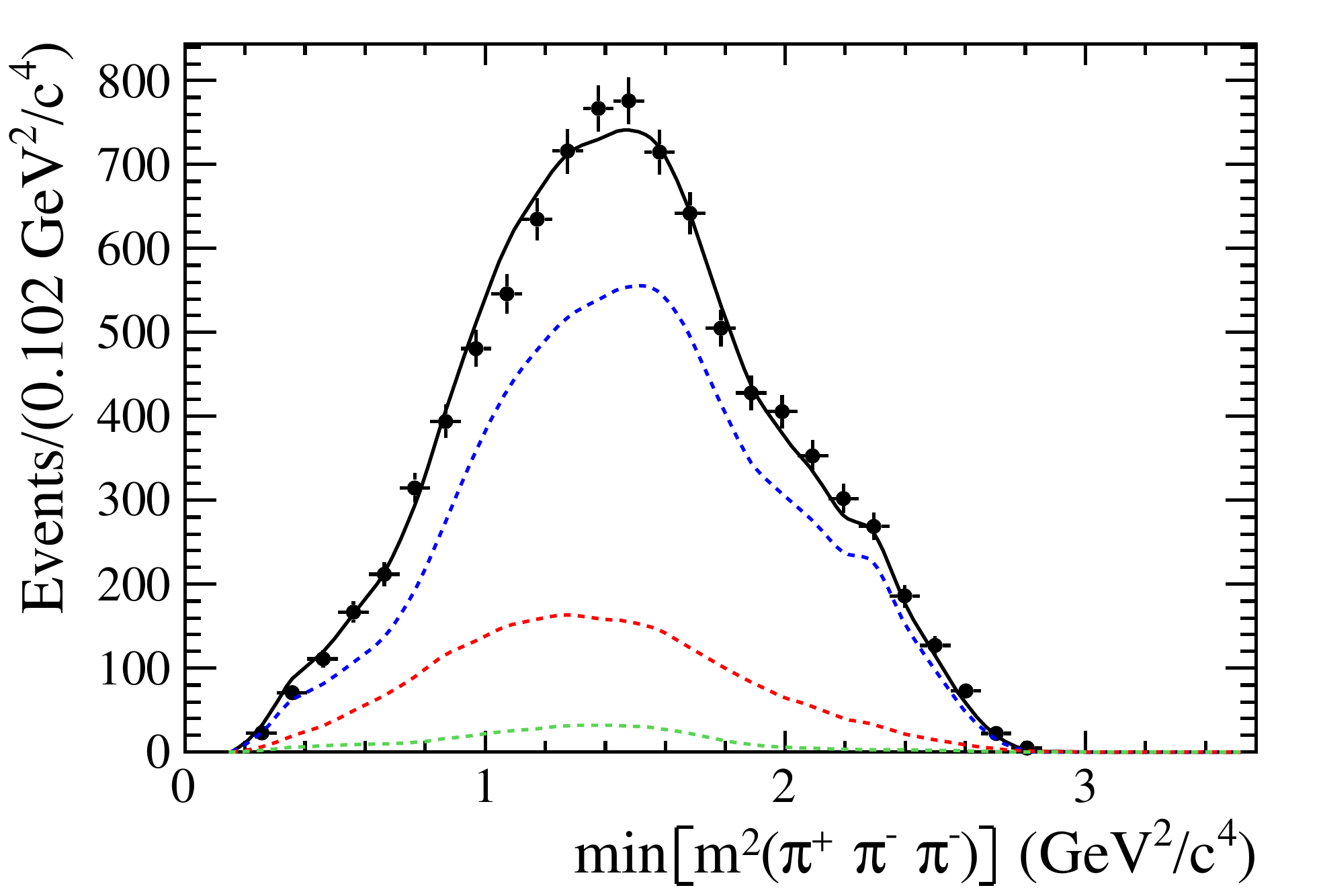} 
	\includegraphics[width=0.43\textwidth, height = !]{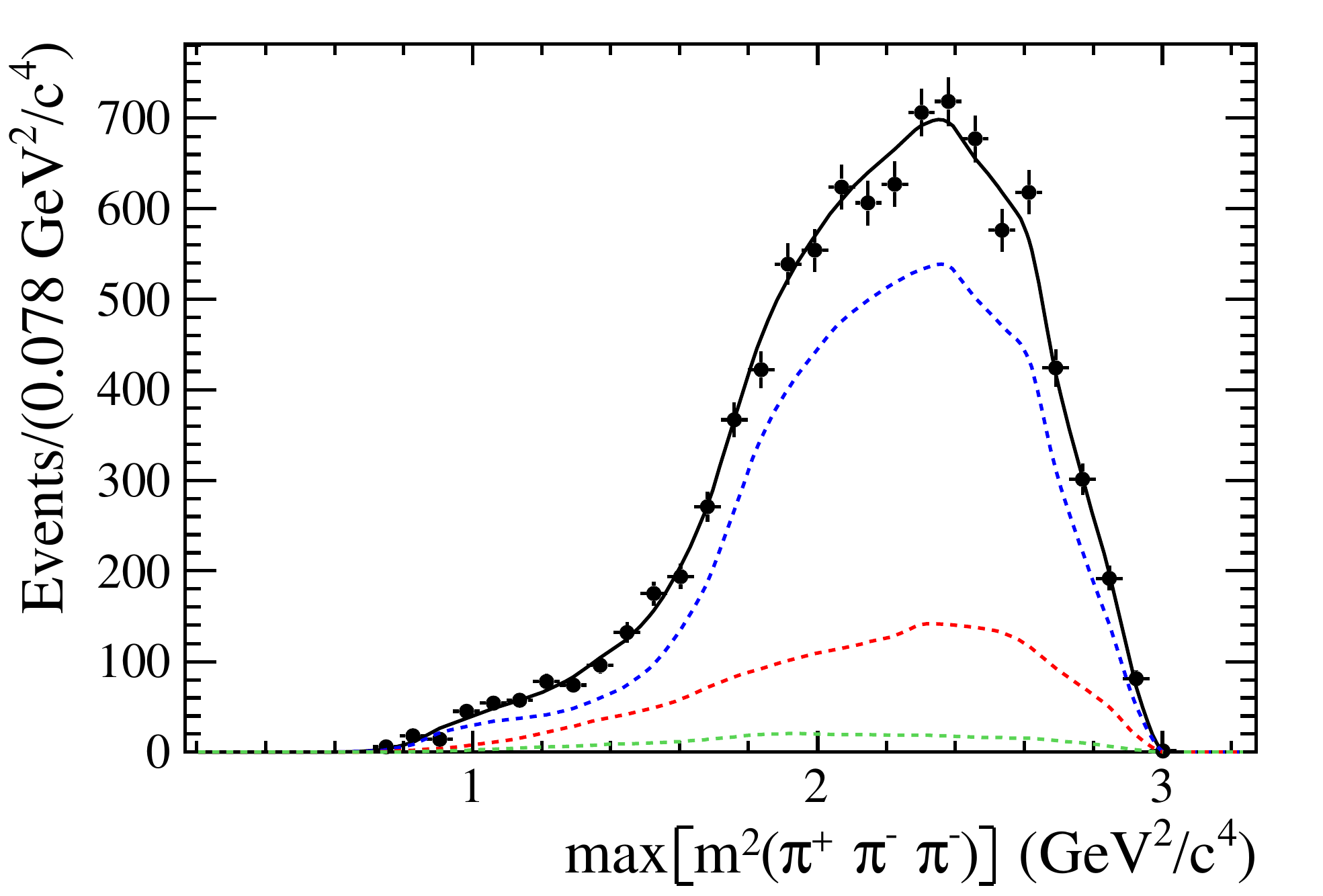} 

	\caption{Invariant mass distributions of $\Dz \to \fourpi$ signal candidates (points with error bars) and fit projections (black solid line).  
	The signal component is shown in blue (dashed), the background component in red (dashed) and the wrongly tagged contribution in green (dashed).
	While the $m^{2}(\pip \pim)$ includes all four possible $\pip \pim$ combinations, the 
	$\text{min}[m^{2}(\pip \pim)]$ ($\text{max}[m^{2}(\pip \pim)]$) distribution includes the two 
        $\pip \pim$ combinations with the lowest (highest) invariant mass. 
	The $\text{min}[m^{2}(\pip \pi^\pm \pim)]$ ($\text{max}[m^{2}(\pip \pi^\pm \pim)]$) distribution 
        includes the %
        $\pip \pi^\pm \pim$ combination %
        with the lowest (highest) invariant mass. 
        The effect of the $K^0_S$ veto can clearly be seen in the top left projection. }
	
	\label{fig:baselineFit}
\end{figure}

In addition to the best five models as determined by the LASSO procedure, a further four alternative models are studied and presented in Table~\ref{tab:alternativeModels1}. 
These comprise an ``Extended'' model whereby 
all conjugate partners of non-self-conjugate intermediate states 
chosen by the LASSO procedure are included. Two involving the removal of the $\pi(1300)$ and $a_1(1640)$ resonances are described in the next section, while another based on the FOCUS model~\cite{FOCUS4pi} is also considered.
From this sample of alternative models, except the one based on the 
FOCUS model due to its poor fit quality, a model-dependent error on the fit fractions 
and the resonance parameters is derived from the variance. If one of the nominal amplitudes 
is not included in an alternative model, the corresponding fraction is set to zero.

The dominant contribution is the $a_{1}(1260)$ resonance in the decay
modes $a_{1}(1260) \to \rho(770)^{0} \pi$ and $a_{1}(1260) \to \sigma
\pi$ followed by the quasi-two-body decays $D \to \sigma f_{0}(1370)$
and $D \to \rho(770)^{0} \rho(770)^{0}$. 
We find
that the decay $\Dz \to a_1(1260)^+ \pi^-$ dominates over $\Dz \to a_1(1260)^- \pi^+$, which is similar to the pattern observed in the $B$ sector,
where $B^0 \to a_1(1260)^+ \pi^-$ is preferred over $B^0 \to a_1(1260)^- \pi^+$ ~\cite{Aubert:2006gb,Dalseno:2012hp}.

\subsection{\texorpdfstring{Lineshapes of $a_1(1260)$, $\pi(1300)$, $a_1(1640)$}{Lineshapes of a1(1260), pi(1300), a1(1640)}}
Resonance properties that
were also determined from the fit to data are given in
Tables~\ref{tab:lassoModelmass} and \ref{tab:correlation}. 
\begin{table}[t]
  \footnotesize
  \centering
  \caption{Resonance parameters determined from the fit to $\Dz \to \fourpi$ decays. The uncertainties are statistical, systematic and model-dependent, respectively.}
  \begin{tabular}
     {@{\hspace{0.5cm}}c@{\hspace{0.25cm}}  @{\hspace{0.25cm}}c@{\hspace{0.5cm}}}
     \hline \hline
     Parameter & Value \\ \hline
	$m_{a_{1}(1260)}$ $(\mev/c^2)$& $1225 \pm 9 \pm 17 \pm 10$ \\
	$\Gamma_{a_{1}(1260)}$ $(\mev)$&  $430 \pm 24 \pm 25 \pm 18$\\	
	$m_{\pi(1300)}$ $(\mev/c^2)$& $1128 \pm 26 \pm 59 \pm 37$\\
	$\Gamma_{\pi(1300)}$ $(\mev)$& $314 \pm 39 \pm 61 \pm 26$\\
	$m_{a_{1}(1640)}$ $(\mev/c^2)$& $1691 \pm 18 \pm 16 \pm 25$\\
	$\Gamma_{a_{1}(1640)}$ $(\mev)$&  $171 \pm 33 \pm 20 \pm 35$\\
	\hline\hline
	\end{tabular}
	\label{tab:lassoModelmass}

  \caption{The statistical correlation coefficients between the resonance parameters determined from the $\Dz \to \fourpi$ fit.}
  \begin{tabular}
     {@{\hspace{0.5cm}}c@{\hspace{0.25cm}}  |@{\hspace{0.25cm}}c@{\hspace{0.25cm}} @{\hspace{0.25cm}}c@{\hspace{0.25cm}}  @{\hspace{0.25cm}}c@{\hspace{0.25cm}} @{\hspace{0.25cm}}c@{\hspace{0.25cm}} @{\hspace{0.25cm}}c@{\hspace{0.25cm}}   @{\hspace{0.25cm}}c@{\hspace{0.5cm}}}
     \hline \hline
	& $m_{a_{1}(1260)}$ & $\Gamma_{a_{1}(1260)}$ & $m_{a_{1}(1640)}$ & $\Gamma_{a_{1}(1640)}$ & $m_{\pi(1300)}$ & $\Gamma_{\pi(1300)}$ \\\hline
	$m_{a_{1}(1260)}$ & $+1.000$ & $+$0.689 & $-$0.065 & $-$0.282 & $+$0.116 & $-$0.258 \\
	$\Gamma_{a_{1}(1260)}$ & & $+1.000$ &$-$0.114 & $-$0.176 & $+$0.013 & $-$0.004 \\	
	$m_{a_{1}(1640)}$ & & &$+1.000$ &$-$0.335 &$-$0.136 &$-$0.119\\
	$\Gamma_{a_{1}(1640)}$ & &&&$+1.000$ &$-$0.258  &$+$0.370 \\
	$m_{\pi(1300)}$ &&&&&$+1.000$ &$-$0.425 \\
	$\Gamma_{\pi(1300)}$&&&&&&$+1.000$\\
	\hline\hline
	\end{tabular}
	\label{tab:correlation}
\end{table}
The mass and width of the $a_{1}(1260)$ meson are in good agreement with the PDG estimates,
$m_{a_{1}(1260)} = 1230 \pm 40 \mev/c^2$ and  $\Gamma_{a_{1}(1260)} = 250-600 \mev$; 
however they differ somewhat from 
one of the most precise single measurements to date, 
$m_{a_{1}(1260)} = 1255 \pm 6 \textrm{ (stat)} ^{+7}_{-17} \textrm{ (syst)} \mev/c^2$ and  $\Gamma_{a_{1}(1260)} = 367 \pm 9  \textrm{ (stat)} ^{+28}_{-25}  \textrm{ (syst)}\mev$,
performed by the COMPASS Collaboration~\cite{Alekseev:2009aa}.
It is, however, not straightforward to compare these values to our measurement 
since the COMPASS analysis was performed assuming a relativistic Breit-Wigner, $cf.$ \eqnPRDref{eq:gamma2}, for the lineshape of the $a_{1}(1260)$ resonance.
When fitting our data with a relativistic Breit-Wigner for the $a_{1}(1260)$ propagator we obtain the values
$m_{a_{1}(1260),{\rm RBW}} = 1221\, \pm\, 8 \textrm{ (stat)}\mev/c^2$ and  
$\Gamma_{a_{1}(1260),{\rm RBW}} = 387\, \pm\, 18 \textrm{ (stat)}\mev$. 
When fitting our data with a constant width for the $a_{1}(1260)$ propagator, we obtain the values
$m_{a_{1}(1260),{\rm SBW}} = 1134\, \pm\, 8 \textrm{ (stat)}\mev/c^2$ and  $\Gamma_{a_{1}(1260),{\rm SBW}} = 367\, \pm\, 15 \textrm{ (stat)}\mev$.
Our nominal lineshape model is preferred over the relativistic Breit-Wigner (constant width Breit-Wigner) with a significance of $10 \sigma$ ($7 \sigma$), 
determined from the log-likelihood difference $\sigma = \sqrt{\Delta ( -2 \, \log \mathcal L) }$.
The $a_1(1260)$ lineshape parameters have also been measured in the three-pion decay of the tau-lepton. The most recent measurement using this decay is by CLEO and finds $m_{a_1(1260)} = 1331 \pm 10 \pm 3 \mev/c^2$ and $\Gamma_{a_1(1260)} = 814 \pm 36 \pm 13 \mev$~\cite{PhysRevD.61.012002}. The unusually large value for the width might be related to the specific choice of lineshape parametrization. In Ref.~\cite{Vojik:2010ua}, the three-pion decay of the $\tau$ lepton was studied using a similar model for the $a_{1}(1260)$ propagator as used in the analysis presented here. From a simultaneous fit to ALEPH~\cite{Schael:2005am}, ARGUS~\cite{Albrecht1993}, OPAL~\cite{Akers:1995vy} and CLEO~\cite{PhysRevD.61.012002} data, the following results are obtained: $m_{a_{1}(1260)} = 1233\, \pm\, 18 \mev/c^2$ and $\Gamma_{a_{1}(1260)} = 431\, \pm\, 20 \mev$, which are in
very good agreement with our measurement.  
The results of the FOCUS
amplitude analysis~\cite{FOCUS4pi} are $m_{a_{1}(1260)} = 1240^{+30}_{-10} \mev/c^2$ and
$\Gamma_{a_{1}(1260)} = 560^{+120}_{-40} \mev$; a potentially relevant
difference between their model and ours is that the only intermediate
state decaying to three pions included is the $a_{1}(1260)$ resonance,
while our LASSO model also includes the $\pi(1300), a_{1}(1640)$ and
$\pi_{2}(1670)$ resonances. 

The $a_{1}(1640)$ resonance, the first radial excitation of the $a_{1}(1260)$ meson, 
was observed in Ref.~\cite{Baker1999114} decaying to $\sigma \pi$ and $f_{2}(1270) \pi$,
and in Ref.~\cite{PhysRevD.65.072001} decaying to $(\rho(770)^{0} \pi)_{D}$, though confirmation is still needed.
We find the decay modes $a_{1}(1640) \to (\rho(770)^{0} \pi)_{D}$ and $a_{1}(1640) \to (\sigma \pi)$
with a combined fit fraction of $6.6 \%$.
The mass and width obtained from the fit are 
compatible with the PDG average of $m_{a_{1}(1640)} = 1647 \pm 22 \mev/c^2$ 
and  $\Gamma_{a_{1}(1640)} = 254 \pm 27 \mev$.
The scalar $\pi(1300)^{+}$ resonance is seen decaying to $\sigma \pip$ and its mass and width are also measured to be 
in agreement with other experiments \cite{PDG2016}.

It is important to note that even though the $a_{1}(1640)$ and the $\pi(1300)$ resonances are selected by the model building, 
satisfactory fit results can also be obtained without them.
The LASSO models obtained when explicitly excluding the  $a_{1}(1640)$ and the $\pi(1300)$ resonance from the pool of amplitudes are 
given in Appendix \ref{a:alternative}.
These models are used to %
generate many pseudo-data sets according to the ``no-$a_{1}(1640)$'' or ``no-$\pi(1300)$'' hypotheses denoted as $H_{0}$.
The pseudo-data is then fitted with $H_{0}$
and the alternative hypotheses, \eg $a_{1}(1640)$ hypothesis $H_{1}$, 
in order to predict the distributions of the log-likelihood differences $\Delta ( -2 \, \log \mathcal L) = 2 \, \log (\mathcal L(H_{1})/\mathcal L(H_{0}) )$  
under the $H_{0}$ hypotheses.   
We use a Gaussian function to parameterize 
the $\Delta ( -2 \, \log \mathcal L) $ distributions.
By integrating the tails of the Gaussians above the $\Delta ( -2 \, \log \mathcal L) $ value observed on the real data,
the $H_{0}$ hypotheses can be excluded in favor of the $a_{1}(1640)$ and $\pi(1300)$ alternate hypotheses 
at the $2.4 \sigma$ and $6.1 \sigma$ levels, respectively. 

\begin{figure}[p]
	\centering
	\includegraphics[width=0.32\textwidth, height=!]{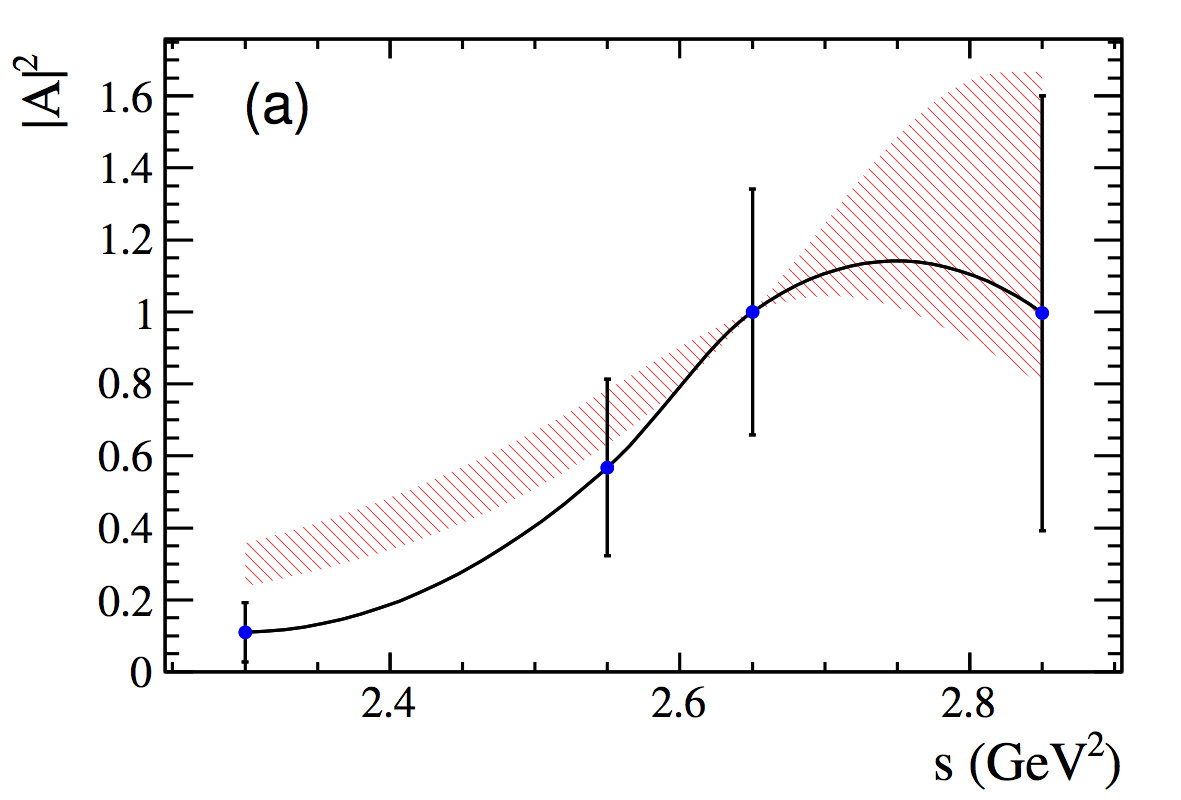} 
	\includegraphics[width=0.32\textwidth, height=!]{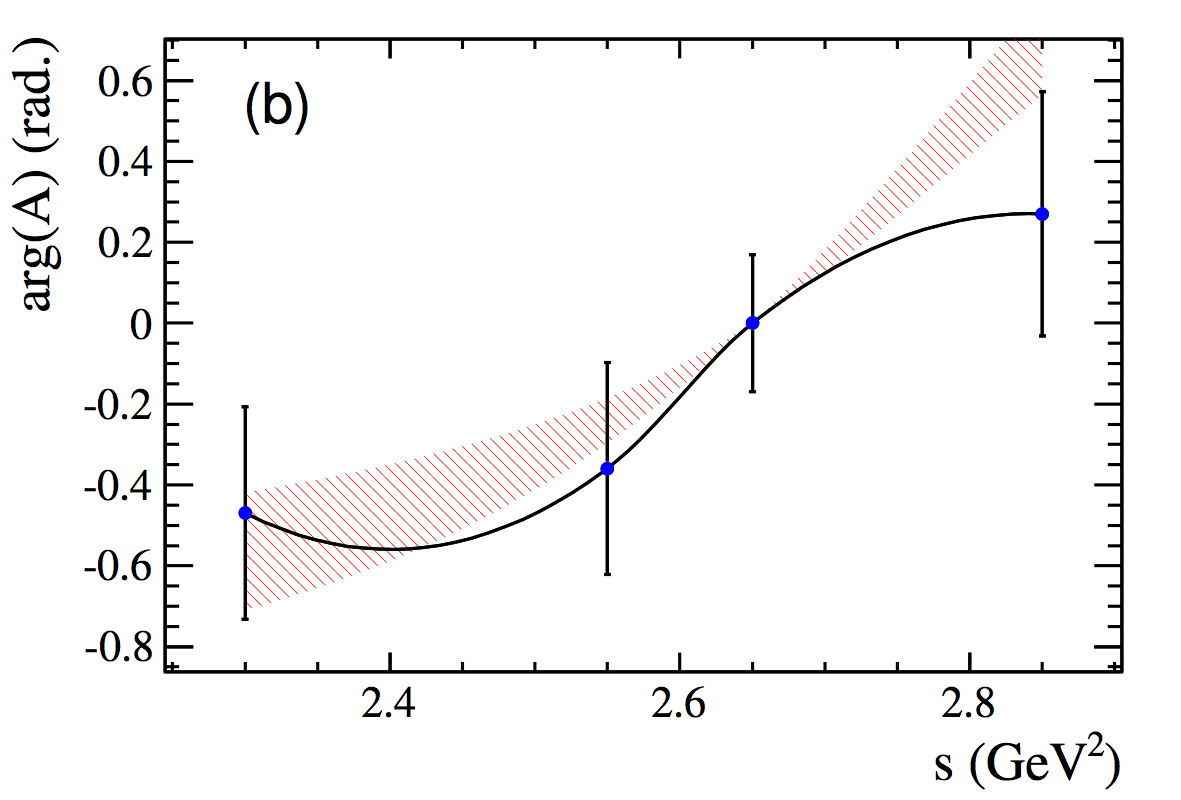} 
	\includegraphics[width=0.32\textwidth, height=!]{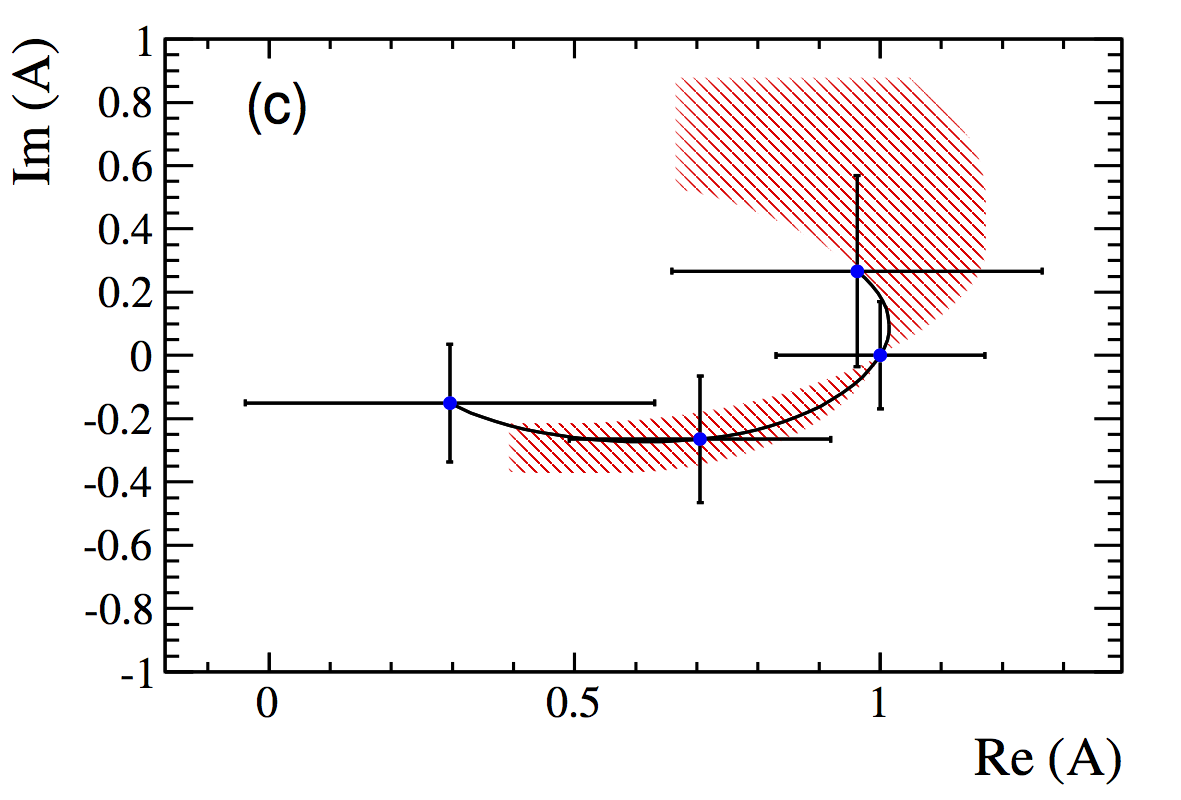} 
	\caption{Magnitude-squared (a), phase (b) and Argand diagram (c) of 
	the quasi-model-independent $a_{1}(1640)$ lineshape. 
	The fitted knots are displayed as points with error bars and the black line shows the interpolated spline.
	The Breit-Wigner lineshape with the mass and width from the nominal fit is superimposed (red area). 
	The latter is chosen to agree with the interpolated spline at the point $\Re(A) = 1$, $\Im(A) = 0$.}
	\label{fig:argand}
	
	\hspace{0.1cm}
	
	\includegraphics[width=0.32\textwidth, height=!]{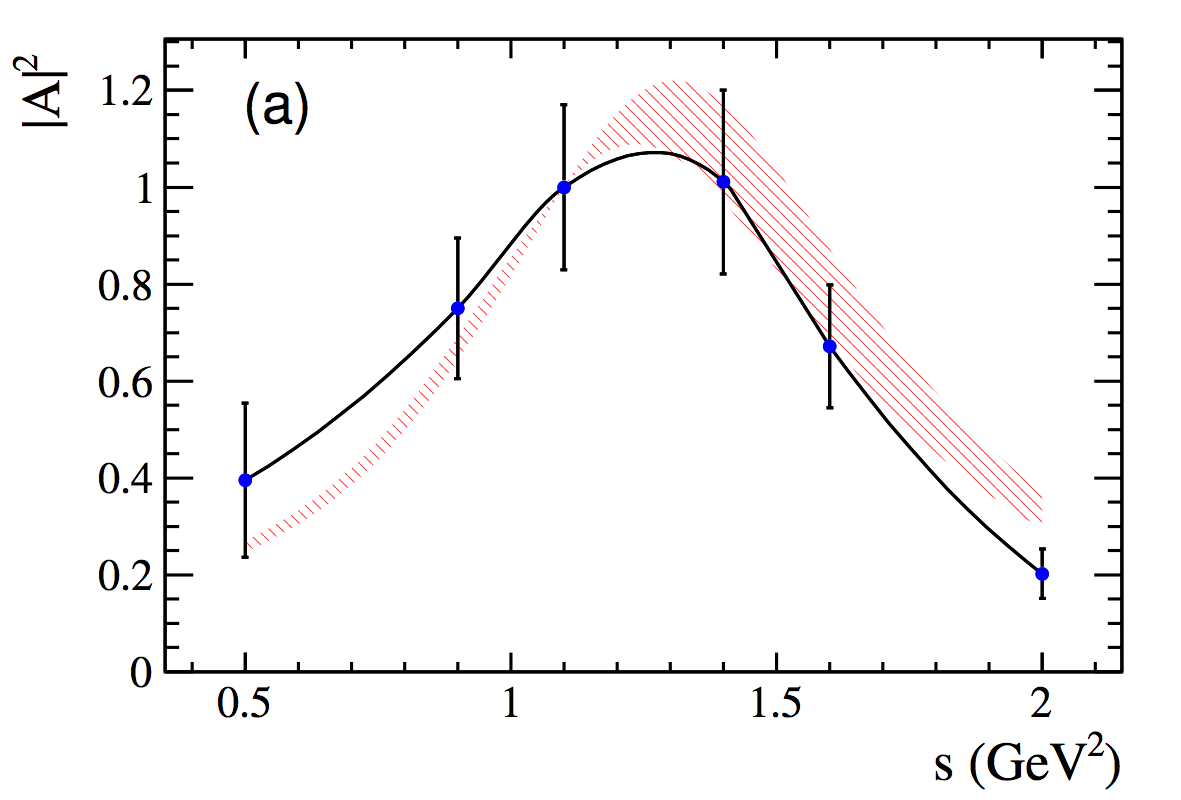} 
	\includegraphics[width=0.32\textwidth, height=!]{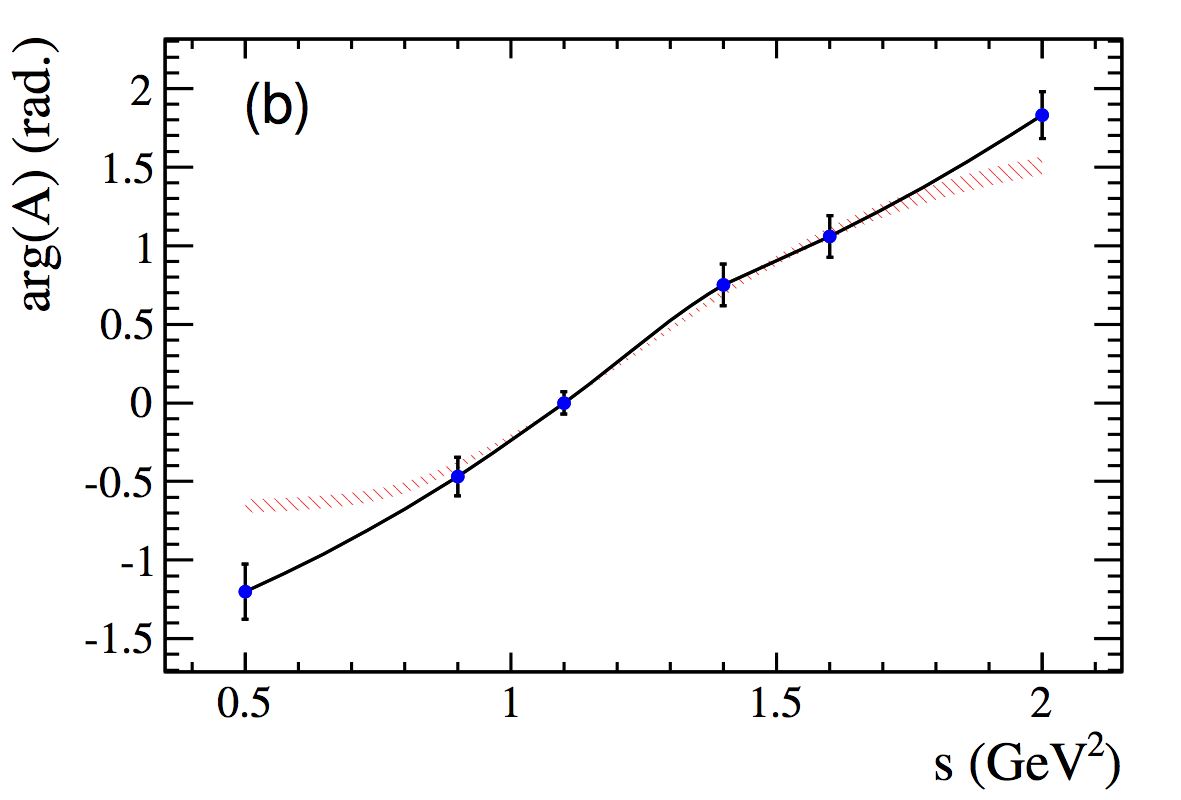} 
	\includegraphics[width=0.32\textwidth, height=!]{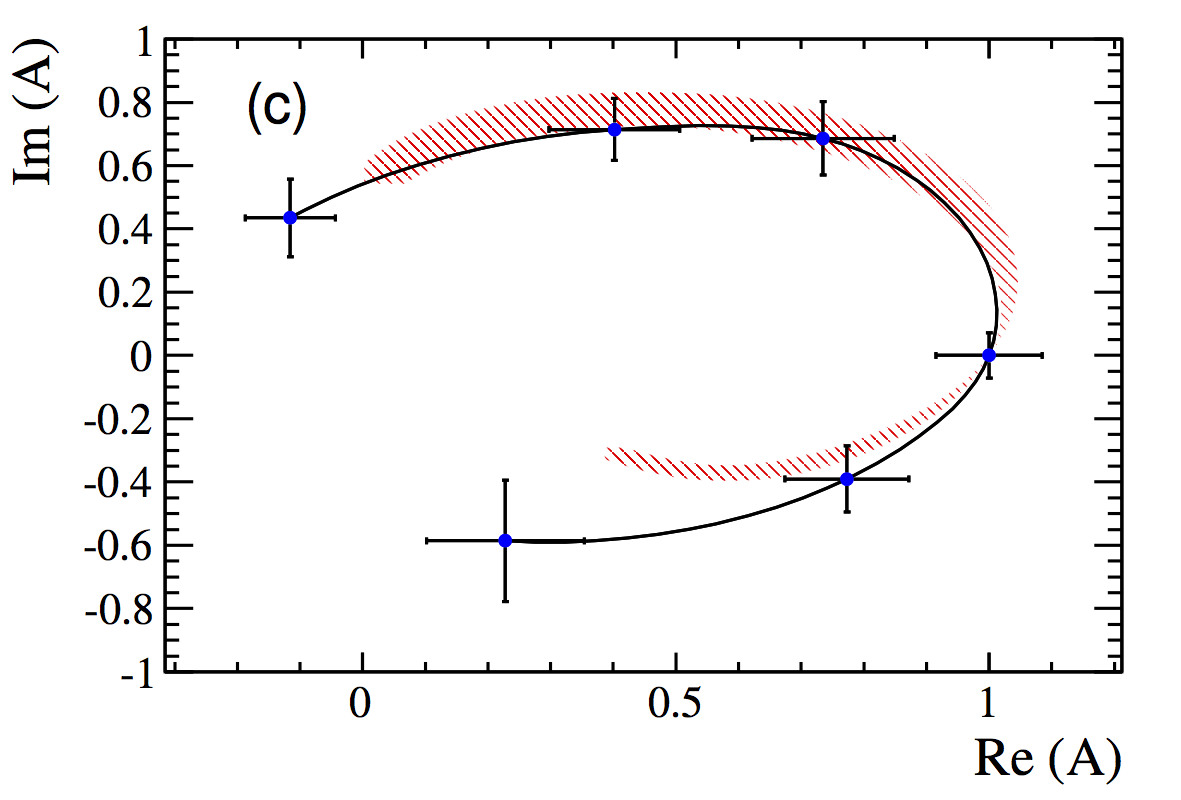} 
	\caption{Magnitude-squared (a), phase (b) and Argand diagram (c) of 
	the quasi-model-independent $a_{1}(1260)$ lineshape. 
	The fitted knots are displayed as points with error bars and the black line shows the interpolated spline.
	The Breit-Wigner lineshape with the mass and width from the nominal fit is superimposed (red area). 
	The latter is chosen to agree with the interpolated spline at the point $\Re(A) = 1$, $\Im(A) = 0$.}
	\label{fig:argand_a1}

	\hspace{0.1cm}

	\includegraphics[width=0.32\textwidth, height=!]{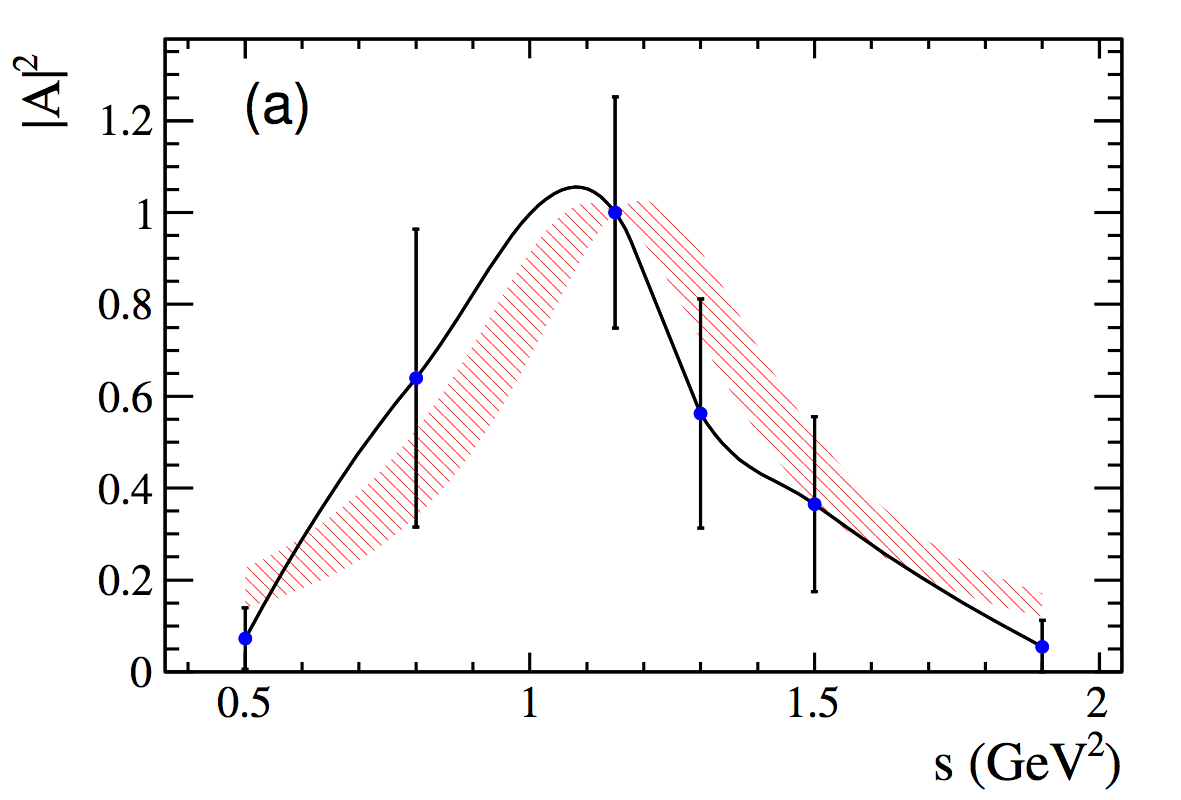} 
	\includegraphics[width=0.32\textwidth, height=!]{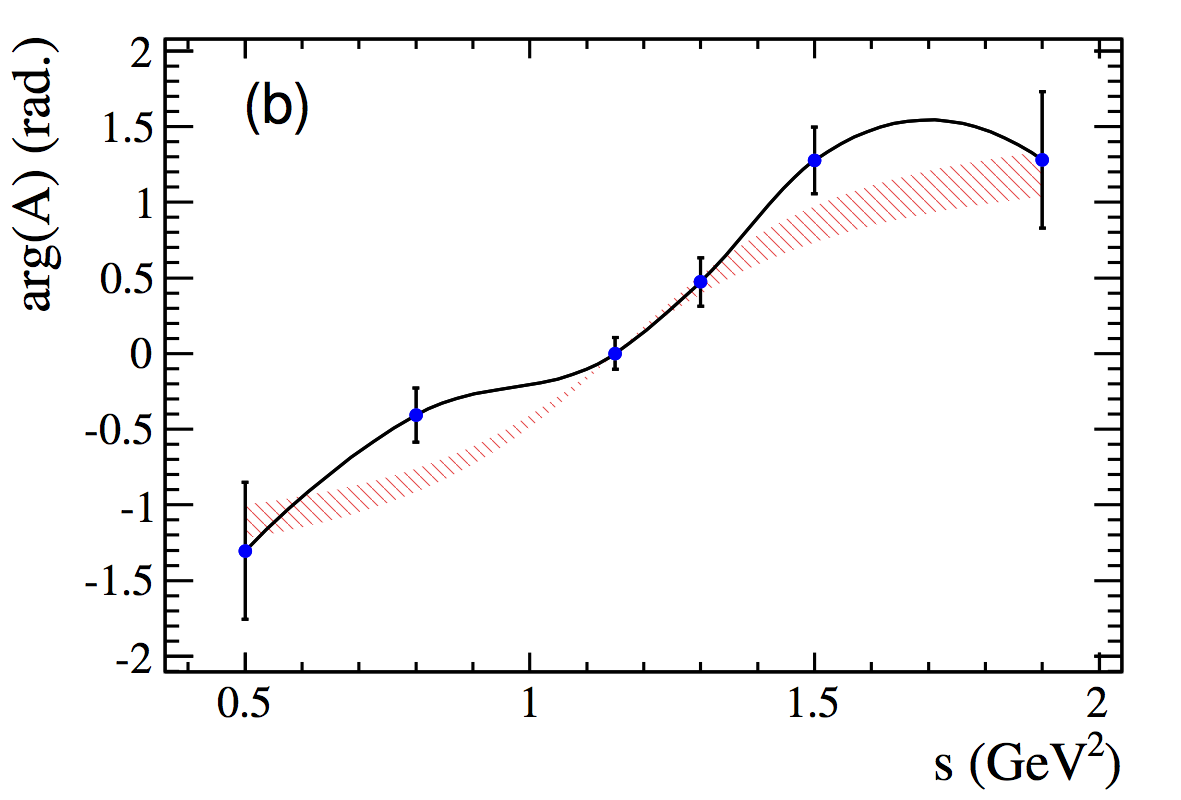} 
	\includegraphics[width=0.32\textwidth, height=!]{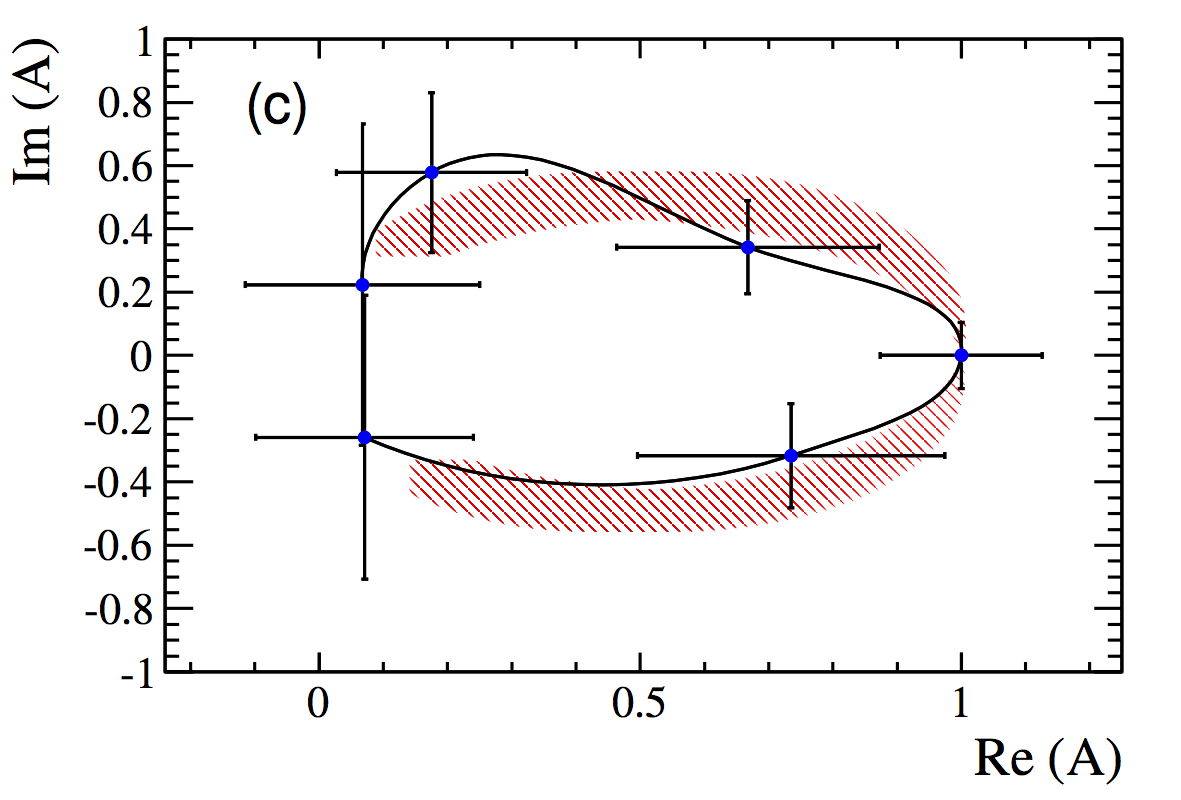} 
	\caption{Magnitude-squared (a), phase (b) and Argand diagram (c) of 
	the quasi-model-independent $\pi(1300)$ lineshape. 
	The fitted knots are displayed as points with error bars and the black line shows the interpolated spline.
	The Breit-Wigner lineshape with the mass and width from the nominal fit is superimposed (red area). 
	The latter is chosen to agree with the interpolated spline at the point $\Re(A) = 1$, $\Im(A) = 0$.}
	\label{fig:argand_pi}
	
\end{figure}

Since the $a_{1}(1640)^{+}$ resonance is not yet well established, we verify 
its resonant phase motion in a quasi-model-independent way as pioneered in Ref.~\cite{Aaij:2014jqa}.
For this purpose, the Breit-Wigner lineshape is replaced by a complex-valued cubic spline. 
The interpolated cubic spline has to pass through 
independent complex knots
spaced in the $m^{2}(\pip \pip \pim)$ region around the nominal mass.
The position of the knots is chosen ad-hoc. 
We verified on simulated experiments that with this choice a Breit-Wigner lineshape can be properly reproduced, given there is a real resonance.
The fitted magnitudes and phases of the knots are shown in Fig.~\ref{fig:argand},
where the expectations from a Breit-Wigner shape with 
 the mass and width from the nominal fit 
are superimposed 
taking only the statistical uncertainties on the 
mass and width into account. 
The interpolated spline generally reproduces the features of the Breit-Wigner parametrization.
In particular, the resulting Argand diagram %
shows a circular, counter-clockwise trajectory which is the expected behavior of a resonance. 
Note that the high-mass tail of the $a_{1}(1640)$ is outside of the phase space boundary 
such that it is not possible to investigate the full phase motion. 
Similar quasi-model-independent studies are performed for the $a_{1}(1260)$ and $\pi(1300)$ resonances 
as shown in Figs.~\ref{fig:argand_a1} and \ref{fig:argand_pi}, respectively.
Since the investigated resonances are all very broad, the quasi-model-independent lineshapes 
can absorb statistical fluctuations in the data, especially near the phase space boundaries. 
Therefore, the agreement with the Breit-Wigner expectation in all cases
indicates that it is qualitatively reasonable that these resonances are indeed real features of the data. %

\clearpage

\subsection{Global \CP\ Content Measurement}

The fractional \CP-even content, $F_{+}^{4\pi}$, is determined from the integral in \eqnPRDref{eqn:CPcont}, using the nominal model for $A^{4\pi}_{\Dz}$ and $A^{4\pi}_{\Dzb}$ assuming no direct \CP\ violation in the $D$ meson decay. 
The uncertainty on $F_{+}^{4\pi}$ is calculated from pseudo-experiments by randomly varying the free parameters of the amplitude fit within their measured statistical and systematic uncertainties. 
For each variation, $F_{+}^{4\pi}$ is redetermined, and the 
square root of the sample variance
of these values is taken as the uncertainty. 
An additional systematic uncertainty is assigned by computing $F_{+}^{4\pi}$ for each of the alternative amplitude models. %
The standard deviation of these values is taken as the additional model uncertainty. 
The obtained result,
\begin{equation}
F_{+}^{4\pi} (\text{flavor-tagged, model-dependent}) = \left[72.9 \pm 0.9 \stat \pm 1.5 \syst \pm 1.0 \; (\mathrm{model})\right] \%,
\end{equation}
is consistent with a previous model-independent analysis of \CP-tagged events \cite{Malde:2015mha},
\begin{equation}
F_{+}^{4\pi} (\text{\CP-tagged, model-independent}) = \left(73.7 \pm 2.8\right) \% .
\end{equation}

\subsection{Search for Direct \CP\ Violation}
\label{sec:CPV}

A search for \CP violation is performed by fitting the LASSO model
to the flavor-tagged $\Dz$ and $\Dzb$ samples.
In contrast to our default fit described in \secref{sec:likelihood},
we now allow the amplitude coefficients for 
$\Dz \to \fourpi$ and $\Dzb \to \pim \pip \pim \pip$ decays to differ, as described in \secref{subsec:MQ}.

The masses and widths of the resonances are fixed to the values obtained in the nominal fit.
Possible additional biases due to this assumption are included in the systematic uncertainties which
are otherwise determined as described in \secref{sec:systematics}.
Table \ref{tab:CPV} compares 
the resulting fit fractions for the $\Dz$ and $\Dzb$ decays.
The sensitivity to $\mathcal A_{CP}^{i}$ is at the level of $4 \%$ to $22 \%$ depending on the decay mode. 
No significant \CP violation is observed for any of the amplitudes. 
Also, the integrated \CP asymmetry over phase space is found to be
\begin{equation}
	\mathcal A^{4\pi}_{CP} = [+0.54 \pm 1.04 \stat \pm 0.51
	\syst] \%,
\end{equation}
which is consistent with $CP$ conservation. 
Due to the 
cancellation of systematic uncertainties in asymmetry-like quantities, the only remaining 
source considered for the global \CP asymmetry is the tagging efficiency ratio, which is set to unity for this purpose.
This nominal value of $A^{4\pi}_{CP}$ is consistent with that which can be found from the amplitude model via  \eqnPRDref{eq:AcpModel}, 
$\mathcal A^{4\pi}_{CP} = [+0.60 \pm 0.56 \textrm{ (stat)} ] \%$.

\begin{table}[h]
  \footnotesize
  \centering
  \renewcommand{\arraystretch}{1.3}
  \caption{Direct $CP$ asymmetry and significance for each component of the $\Dz \to \fourpi$ LASSO model. 
  The first uncertainty is statistical, the second systematic and the third due to alternative models.}
  \begin{tabular}
    {@{\hspace{0.5cm}}l@{\hspace{0.25cm}}  @{\hspace{0.25cm}}c@{\hspace{0.25cm}}  @{\hspace{0.25cm}}c@{\hspace{0.5cm}}}
    \hline \hline 
    Decay channel & ${\mathcal A}_{CP}^i$ $(\%)$ & Significance ($\sigma$) \\
    \hline
	$\Dz \to \pim \,  a_{1}(1260)^{+} $ & $ +4.7 \pm 2.6 \pm 4.3 \pm 2.4$ & $ 0.9 $    \\

	$\Dz \to \pip \, a_{1}(1260)^{-}$ & $ +13.7 \pm 13.8 \pm 9.8 \pm 5.8$ & $ 0.8 $\\

	$\Dz \to \pim \,  \pi(1300)^{+}$ & $ -1.6 \pm 12.9 \pm 5.0 \pm 4.4$ & $ 0.1 $\\

	$\Dz \to \pip \, \pi(1300)^{-} $ & $ -5.6 \pm 11.9 \pm 25.6 \pm 10.3$ & $ 0.2 $ \\

	$\Dz \to \pim \, a_{1}(1640)^{+} $ & $ +8.6 \pm 17.8 \pm 16.0 \pm 10.8$ & $ 0.3 $\\

	$\Dz \to \pim \, \pi_{2}(1670)^{+}$ & $ +7.3 \pm 15.1 \pm 8.0 \pm 6.6$ & $ 0.4 $ \\

	$\Dz \to \sigma \, f_{0}(1370)  $ &  $ -14.6 \pm 16.5 \pm 9.3 \pm 1.3$ & $ 0.8 $ \\
	
	$\Dz \to \sigma \,  \rho(770)^{0}  $ & $ +2.5 \pm 16.8 \pm 13.8 \pm 14.6$ & $ 0.1 $ \\

	$\Dz \to \rho(770)^{0} \, \rho(770)^{0}$ & $ -5.6 \pm 5.0 \pm 2.2 \pm 1.9$ & $ 1.0 $ \\

	$\Dz \to f_{2}(1270) \,  f_{2}(1270) $ & $ -28.3 \pm 12.3 \pm 18.5 \pm 9.7$ & $ 1.2 $ \\

    \hline \hline

	\end{tabular}
	\label{tab:CPV}
\end{table}

\clearpage\section{\texorpdfstring{$\Dz \to \KKpipi$ Amplitude Analysis Results}{D0 to K+ K- pi+ pi- Amplitude Analysis Results}}
\label{sec:resultsKKpi}\label{sec:resultsKKpipi}
\subsection{Amplitude Model Fit Results}
\label{ssec:KKpiComponents}
\begin{figure}[h]
\centering
  \includegraphics[width=0.49\textwidth, height=!]{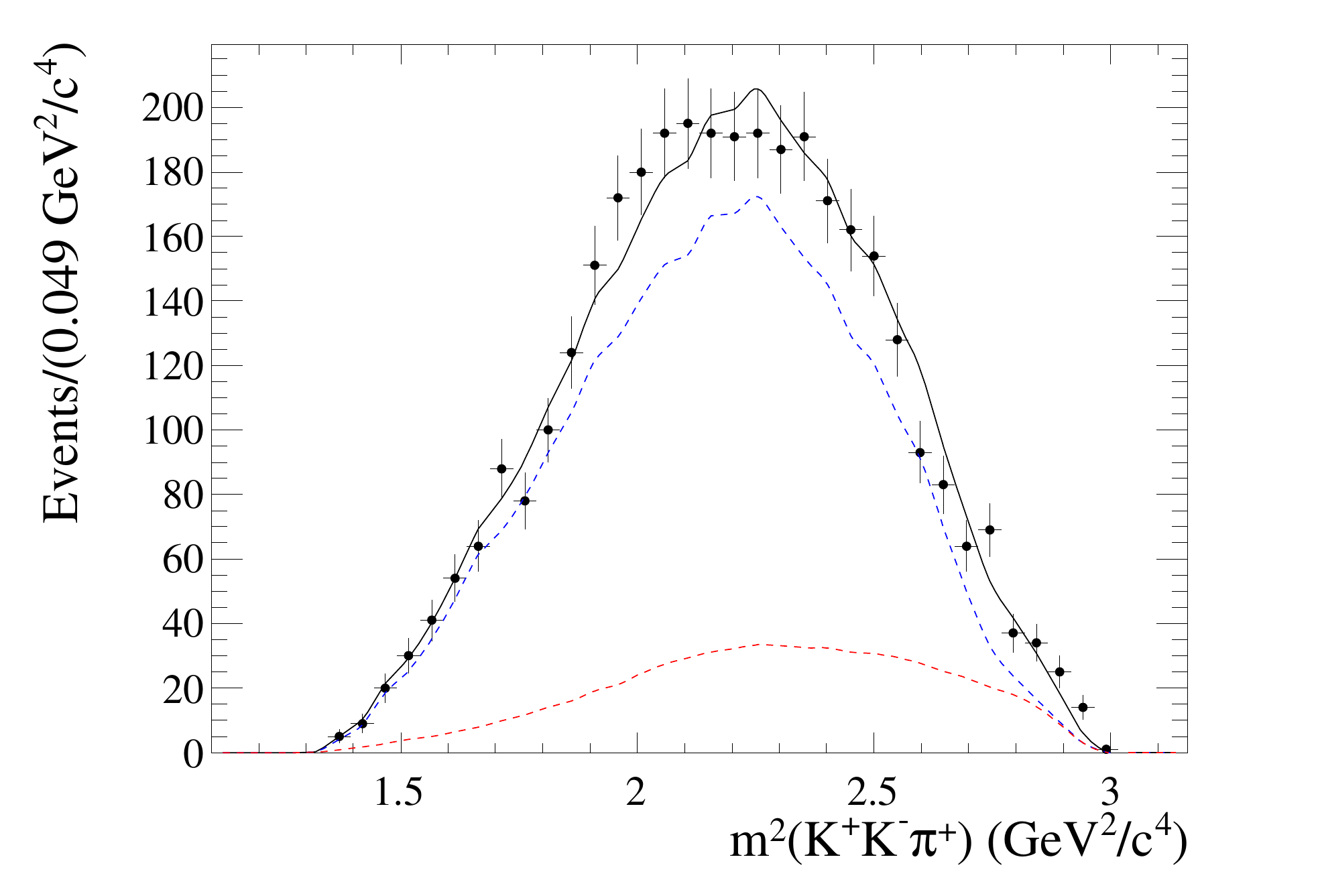}
  \includegraphics[width=0.49\textwidth, height=!]{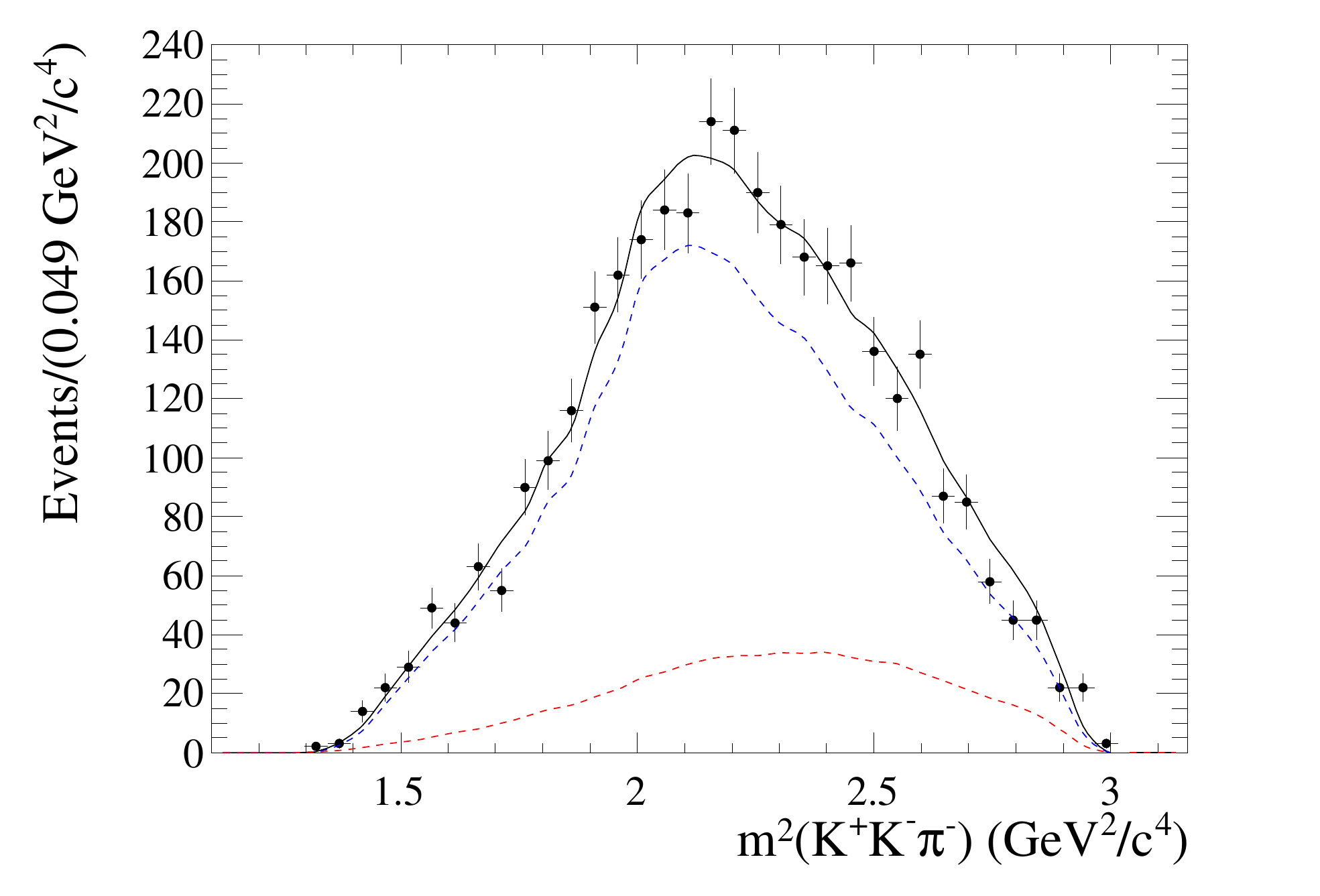}\\ 

  \includegraphics[width=0.49\textwidth, height=!]{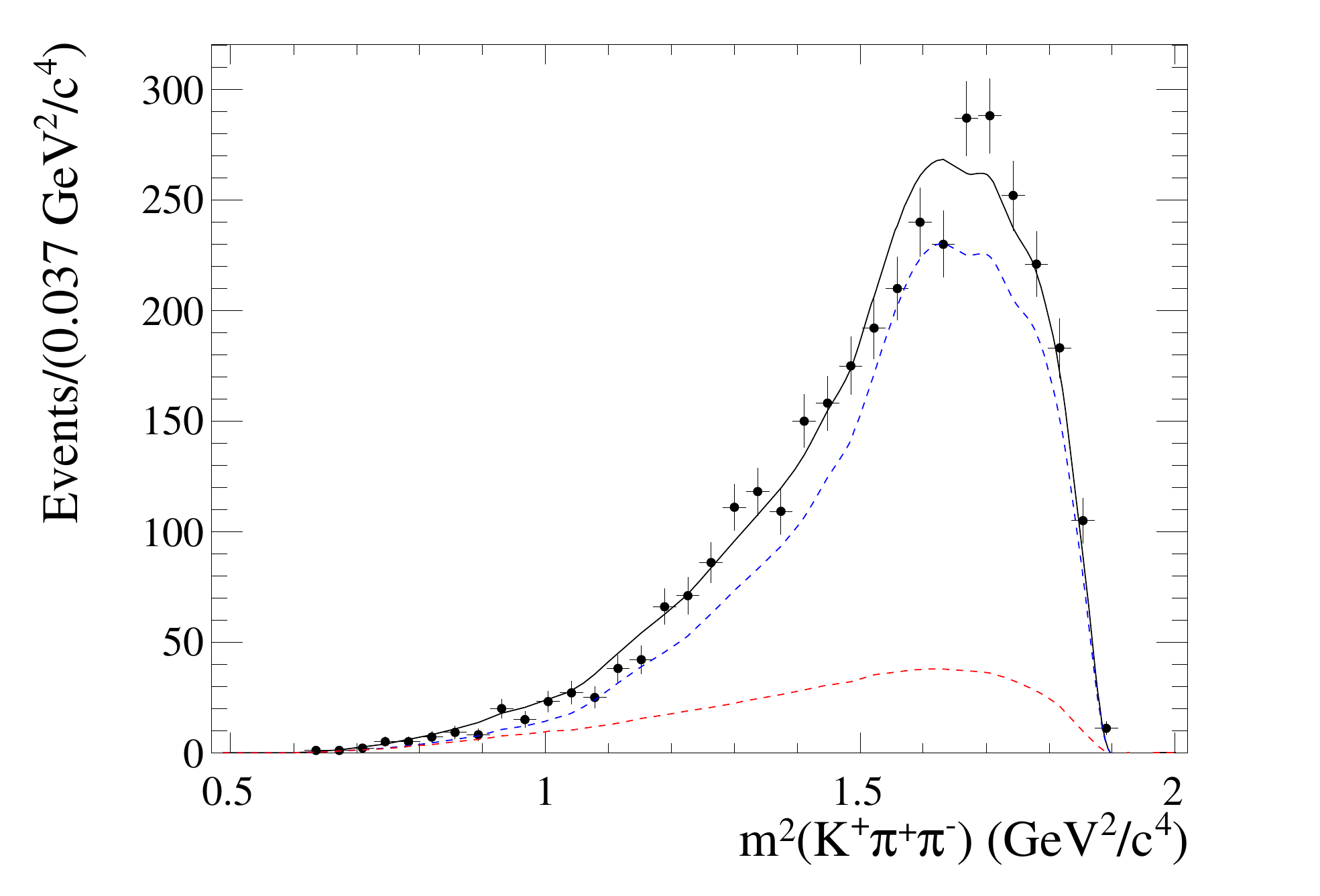}
  \includegraphics[width=0.49\textwidth, height=!]{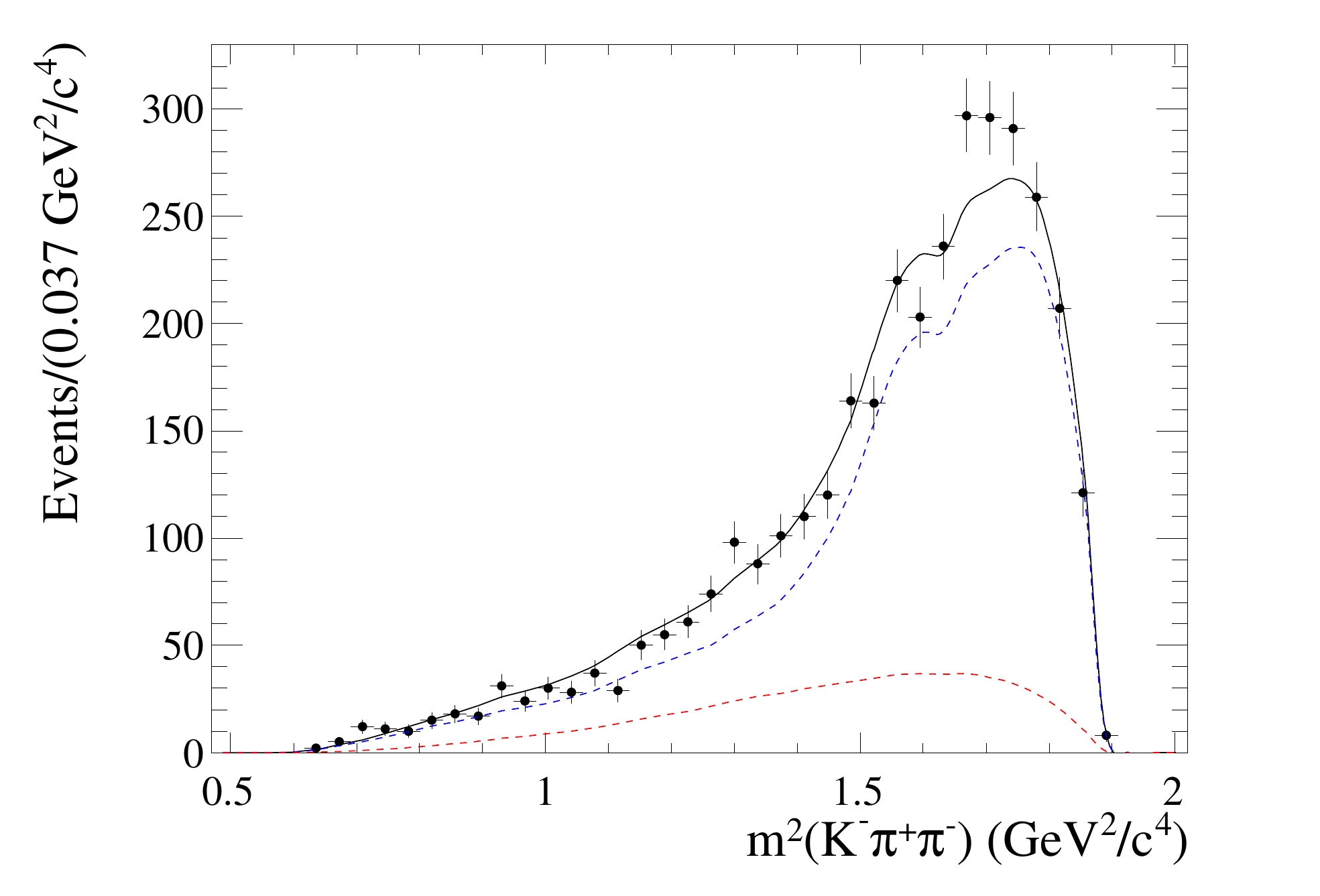} 
  \caption{Invariant 3-body mass distributions of $\Dz \to \KKpipi$ signal events shown as points with error bars. 
           The overall fit projection is shown in black, the signal in blue and the background in red.}
  \label{fig:baselineFitKKpipi3body}
\end{figure}

\begin{figure}[h]
\centering
	\includegraphics[width=0.49\textwidth, height = !]{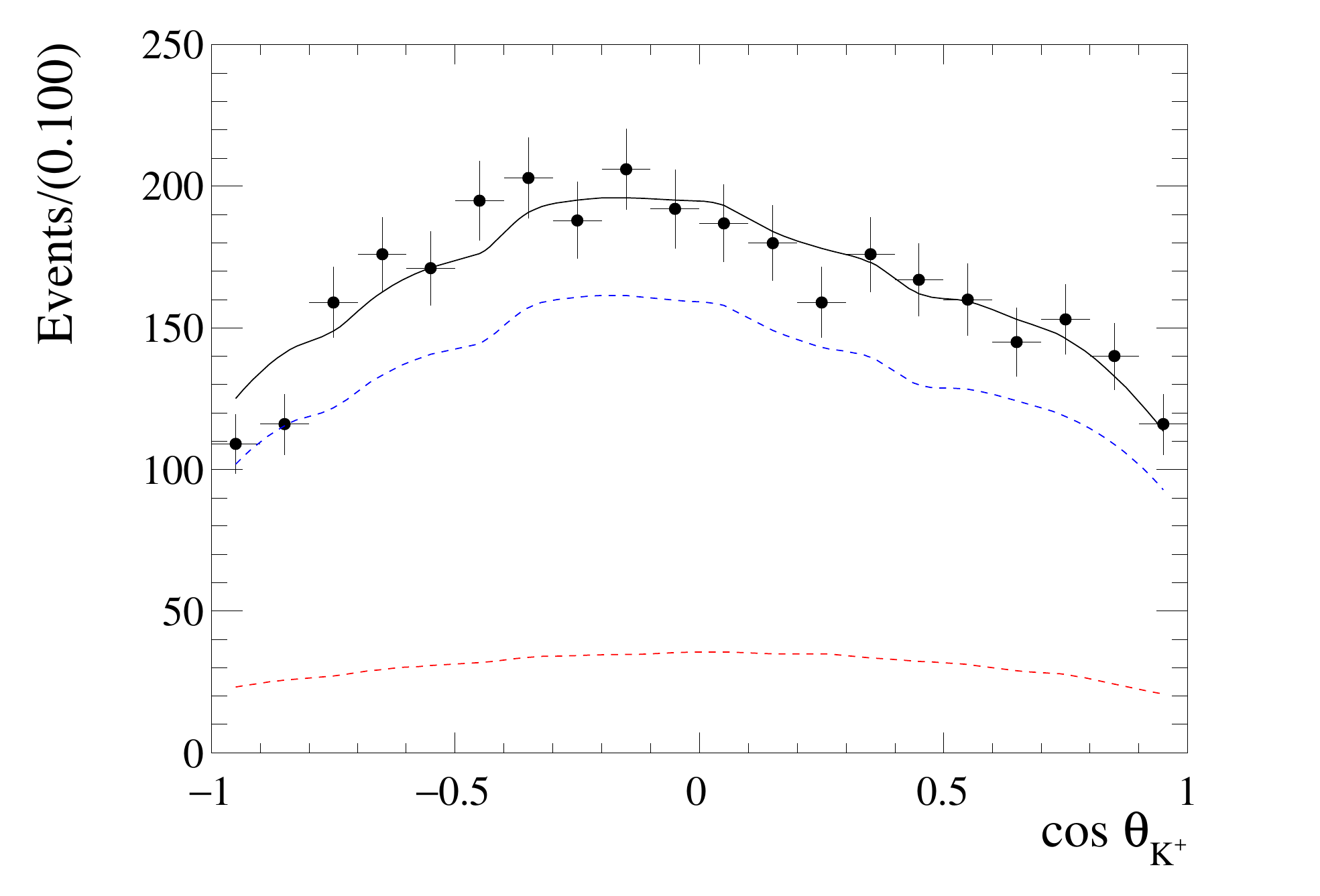} 
	\includegraphics[width=0.49\textwidth, height = !]{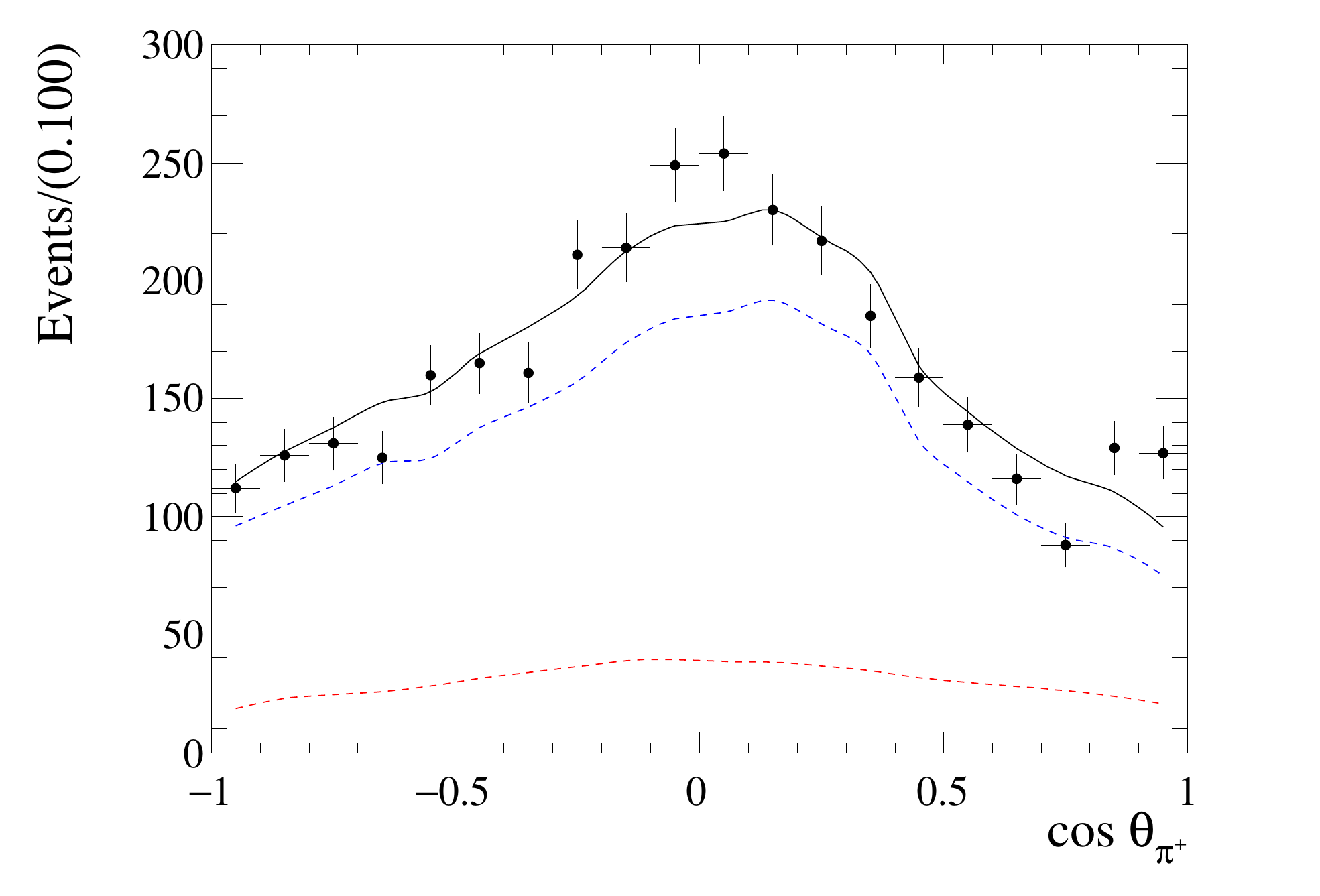} 
	\centering
	\includegraphics[width=0.49\textwidth, height = !]{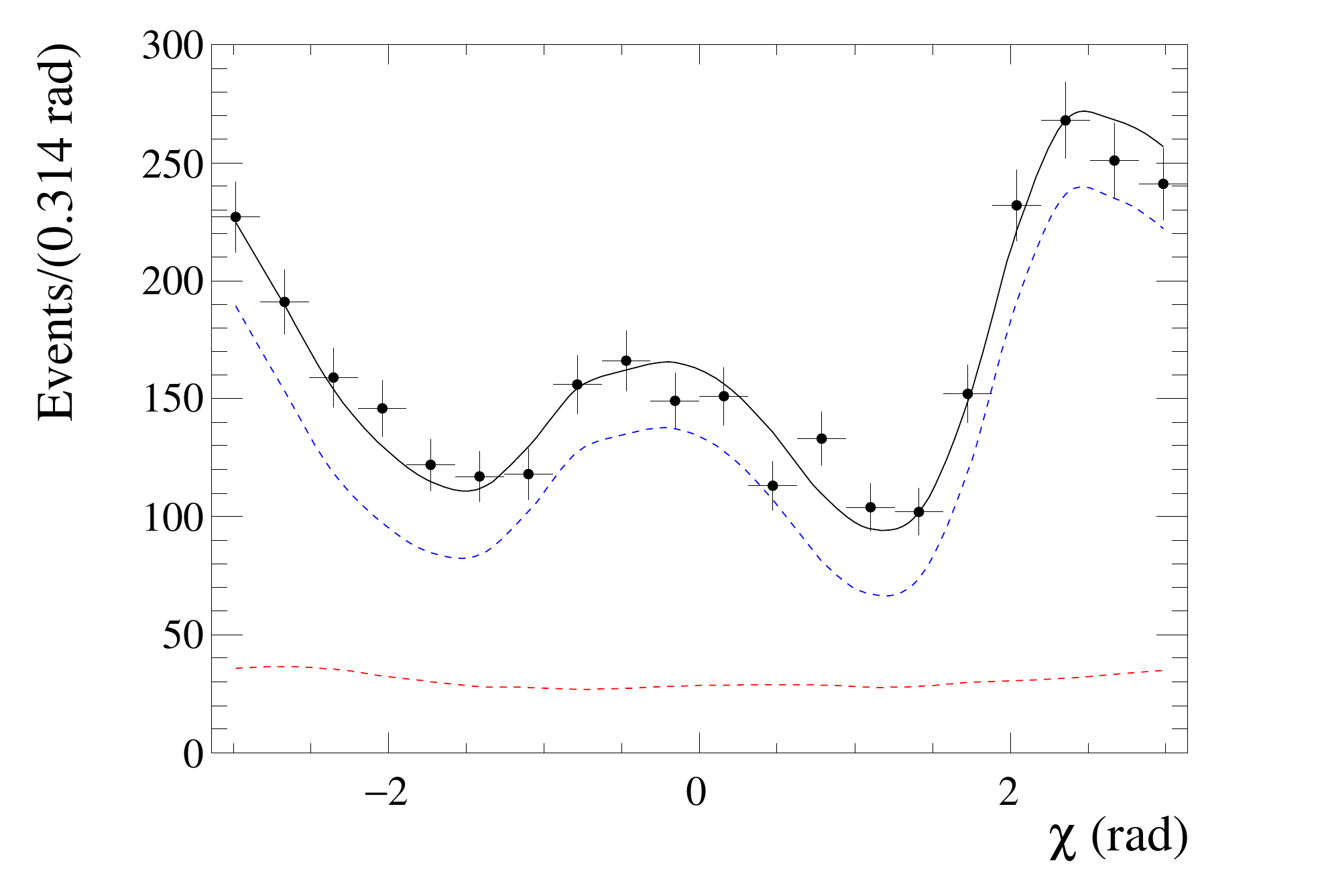} 

	\caption{Angular projections of the $\Dz \to \KKpipi$ fit results (black solid line) in the transversity basis, 
        to the flavor-tagged data sample only.
	The signal component is shown in blue (dashed) and the background component in red (dashed). 
	}
	\label{fig:baselineFitKKpipiAngular}
\end{figure}

Table \ref{tab:lassoModelKKpipi} lists the real and imaginary part of the complex amplitude coefficients $a_{i}$, along with the corresponding fit fractions. 
The interference fractions are given in Appendix~\ref{a:interference}. 
Figures~\ref{fig:baselineFitKKpipi2body}~and~\ref{fig:baselineFitKKpipi3body} show the distributions of 
selected
phase space 
observables, which demonstrate
reasonable agreement between data and the fit model.
For the flavor-tagged data only, we also project into the transversity basis to demonstrate good description of the overall angular structure in
Fig.~\ref{fig:baselineFitKKpipiAngular}:
The acoplanarity angle
${\chi}$,
is the angle between the two decay planes formed by
the $K^+ K^-$ 
combination
and the $\pip \pim$
combination
in the $D$ rest frame; 
boosting into the rest frames of the two-body systems defining these decay planes,
the two helicity variables
are defined as the cosine of the angle, ${\theta_{K^+}}$, 
of the $K^+$ 
momentum with the $D$ flight direction, 
and the cosine of the angle, ${\theta_{\pip}}$, 
of the $\pip$ 
momentum with the $D$ flight direction.
%
% start input /Users/pnaik/Documents/CLEO/Papers/latexpand/latex/KKpipiresultstable_new.tex
\begin{sidewaystable}[p]
  \footnotesize
  \centering
  \caption{\small Real and imaginary part of the complex amplitude coefficients 
    and fit fraction of each component of the 
    $\Dz \to \KKpipi$
    LASSO model. For the fit coefficients, the first quoted uncertainty is statistical, while the second 
    arises from systematic sources. The third uncertainty in the fit fraction arises from the alternative models considered.}
  \begin{tabular}
     {@{\hspace{0.5cm}}l@{\hspace{0.25cm}}  @{\hspace{0.25cm}}c@{\hspace{0.25cm}}  @{\hspace{0.25cm}}c@{\hspace{0.25cm}}  @{\hspace{0.25cm}}c@{\hspace{0.5cm}}}
     \hline \hline
     Decay channel & $\Re (a_{i})$ & $\Im (a_{i})$ & $F_{i} (\%)$ \\ \hline
           $\Dz \to K^- \, [K_{1}(1270)^{+} \to \pi^+ \, K^{*}(892)^{0}]$&  6.36 $\pm$ 1.24 $\pm$ 3.41&  -6.86 $\pm$ 1.37 $\pm$ 3.38 & 5.5 $\pm$ 1.4 $\pm$ 2.7 $\pm$ 2.0\\ 
           $\Dz \to K^- \, [K_{1}(1270)^{+} \to \pi^+ \, K^{*}(1430)^{0}]$& 34.93 $\pm$ 3.74 $\pm$ 8.39& 5.66 $\pm$ 4.50 $\pm$ 6.92 & 6.1 $\pm$ 1.2 $\pm$ 1.3 $\pm$ 1.3 \\ 
              $\Dz \to K^- \, [K_{1}(1270)^{+} \to K^+ \, \rho(770)^{0}]$& -17.63 $\pm$ 2.44 $\pm$ 5.26&-11.51 $\pm$ 2.27 $\pm$ 3.38 & 9.1 $\pm$ 1.5 $\pm$ 1.9 $\pm$ 0.1 \\ 
              $\Dz \to K^+ \, [\bar{K_{1}}(1270)^{-} \to K^- \, \rho(770)^{0}]$& 9.87 $\pm$ 1.75 $\pm$ 2.61&-12.91 $\pm$ 1.49 $\pm$ 3.82 & 5.4 $\pm$ 0.7  $\pm$ 1.1 $\pm$ 0.7 \\ 
              $\Dz \to K^- \, [K_{1}(1270)^{+} \to K^+ \, \omega(782)]$ & 3.78 $\pm$ 3.34 $\pm$ 3.81&-9.36 $\pm$ 2.35 $\pm$ 5.87 & 0.6 $\pm$ 0.3 $\pm$ 0.4 $\pm$ 0.2 \\ 
              $\Dz \to K^- \, [K_{1}(1400)^{+} \to \pi^+ \, K^{*}(892)^{0}]$& -8.08 $\pm$ 5.11 $\pm$ 8.93&-36.92 $\pm$ 4.35 $\pm$ 10.98 & 12.4 $\pm$ 2.6 $\pm$ 3.9 $\pm$ 5.0 \\ 
              $\Dz \to K^- \, [K^*(1680)^{+} \to \pi^+ \, K^{*}(892)^{0}]$& 132.69 $\pm$ 15.81 $\pm$ 27.13&-35.45 $\pm$ 19.13 $\pm$ 29.21 & 3.6 $\pm$ 0.8 $\pm$ 1.0 $\pm$ 0.3 \\ 
            
              $\Dz[S] \to K^{*}(892)^{0} \, \bar{K}^{*}(892)^{0}$ & 822.04 $\pm$ 81.84 $\pm$ 126.47& -350.52 $\pm$ 105.58 $\pm$ 234.41 & 4.5 $\pm$ 0.8 $\pm$ 1.1 $\pm$ 1.7 \\ 
              $\Dz[P] \to K^{*}(892)^{0} \, \bar{K}^{*}(892)^{0}$ & -975.58 $\pm$ 115.30 $\pm$ 241.90&-476.35 $\pm$ 148.15 $\pm$ 162.11 & 3.6 $\pm$ 0.7 $\pm$ 1.4 $\pm$ 0.5 \\ 
              $\Dz[D] \to K^{*}(892)^{0} \, \bar{K}^{*}(892)^{0}$ & -4786.72 $\pm$ 408.23 $\pm$ 503.72&985.18 $\pm$ 533.86 $\pm$ 941.34 & 4.0 $\pm$ 0.6 $\pm$ 0.7 $\pm$ 0.2 \\ 
              $\Dz[S] \to \phi(1020) \, \rho(770)^{0}$ & 5158.65 (fixed) &0.00 (fixed) &28.1 $\pm$ 1.3 $\pm$ 1.7 $\pm$ 0.3  \\
              $\Dz[P] \to \phi(1020) \, \rho(770)^{0}$ & -1552.19 $\pm$ 203.06 $\pm$ 271.29&-354.72 $\pm$ 432.81 $\pm$ 531.39 & 1.6 $\pm$ 0.3 $\pm$ 0.6 $\pm$ 0.3 \\ 
              $\Dz[D] \to \phi(1020) \, \rho(770)^{0}$ & 5208.02 $\pm$ 685.26 $\pm$ 840.57&791.86 $\pm$ 954.63 $\pm$ 689.75 & 1.7 $\pm$ 0.4 $\pm$ 0.4 $\pm$ 0.2 \\ 
             
              $\Dz \to K^{*}(892)^{0} \, (K^- \pi^+)_{S}$ & 43.52 $\pm$ 4.59 $\pm$ 9.73&-1.20 $\pm$ 5.20 $\pm$ 5.27 & 5.8 $\pm$ 1.2 $\pm$ 2.1 $\pm$ 0.0 \\ 
              $\Dz \to \phi(1020) \, (\pi^+ \pi^-)_{S}$& 29.62 $\pm$ 7.39 $\pm$ 12.19&-54.93 $\pm$ 5.65 $\pm$ 11.92 & 4.0 $\pm$ 0.6 $\pm$ 1.3 $\pm$ 1.7 \\ 
            
              $\Dz \to (K^+K^-)_{S} \, (\pi^+ \pi^-)_{S}$& 0.42 $\pm$ 0.05 $\pm$ 0.10&0.49 $\pm$ 0.05 $\pm$ 0.09 & 11.1 $\pm$ 1.2 $\pm$ 2.1 $\pm$ 0.7 \\ 
    \hline
    Sum & & & 106.9  $\pm$  4.5  $\pm$  6.9  $\pm$  6.1  \\
    \hline\hline
    \end{tabular}
    \label{tab:lassoModelKKpipi}
\end{sidewaystable}
 % end input /Users/pnaik/Documents/CLEO/Papers/latexpand/latex/KKpipiresultstable_new.tex
 In contrast to the treatment of the $a_1(1260)$ and $\pi(1300)$ substructure
in the $\Dz \to \fourpi$ analysis, we do not enforce the
same amplitude substructure for the $K(1270)^+$, $K(1400)^+$,
$K(1680)^+$ decays as for $K(1270)^-$, $K(1400)^-$, $K(1680)^-$; this
choice has historical reasons. It is re-assuring to see that the
results we obtain without these constraints are consistent with what
one would expect if such constraints had been applied (\cf model A in Table~\ref{tab:alternativeModels2}).
For the LASSO model, the $\chi^2/\nu$ is 1.5 with $\nu = 116$, where the effective number of degrees of freedom is determined with a pseudo-experiment technique.
Its value is chosen to be the one that best converts the distribution of $\chi^2$ values for each experiment into the standard uniform distribution. This method differs from that used in $\Dz \to \fourpi$ as the relatively small size of the data sample here would otherwise result in negative degrees of freedom.

Four alternate models are presented in Appendix \ref{a:alternative}:
\begin{enumerate}[label=(\Alph*)]
  \item a model that requires the use of conjugate amplitudes for all present non-self-conjugate decays
  \item replacing $K^*(1680)^+ \to K^*(892)^0 \, \pip$ with the $K^*(1410)^+ \to K^*(892)^0 \, \pip$ amplitude
  \item replacing the flat non-resonant term with the $f_0(980)\,(\pi^+ \pi^-)_{S}$ and $f_0(980)\,(K^+ K^-)_{S}$ amplitudes
  \item the model previously reported in Ref.~\cite{KKpipi}
\end{enumerate}
\begin{figure}[h]
\centering
  \includegraphics[width=0.49\textwidth, height=!]{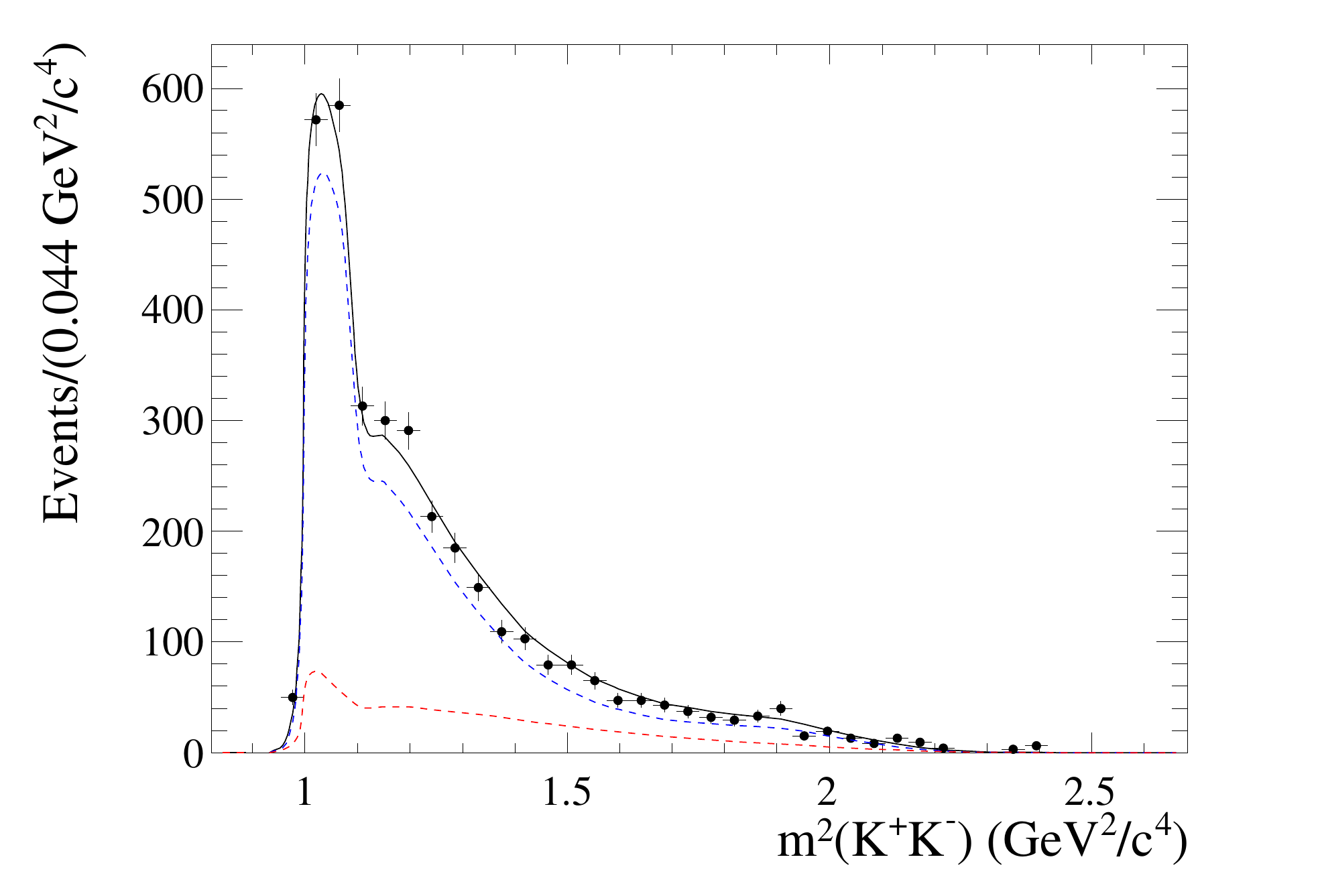}
  \includegraphics[width=0.49\textwidth, height=!]{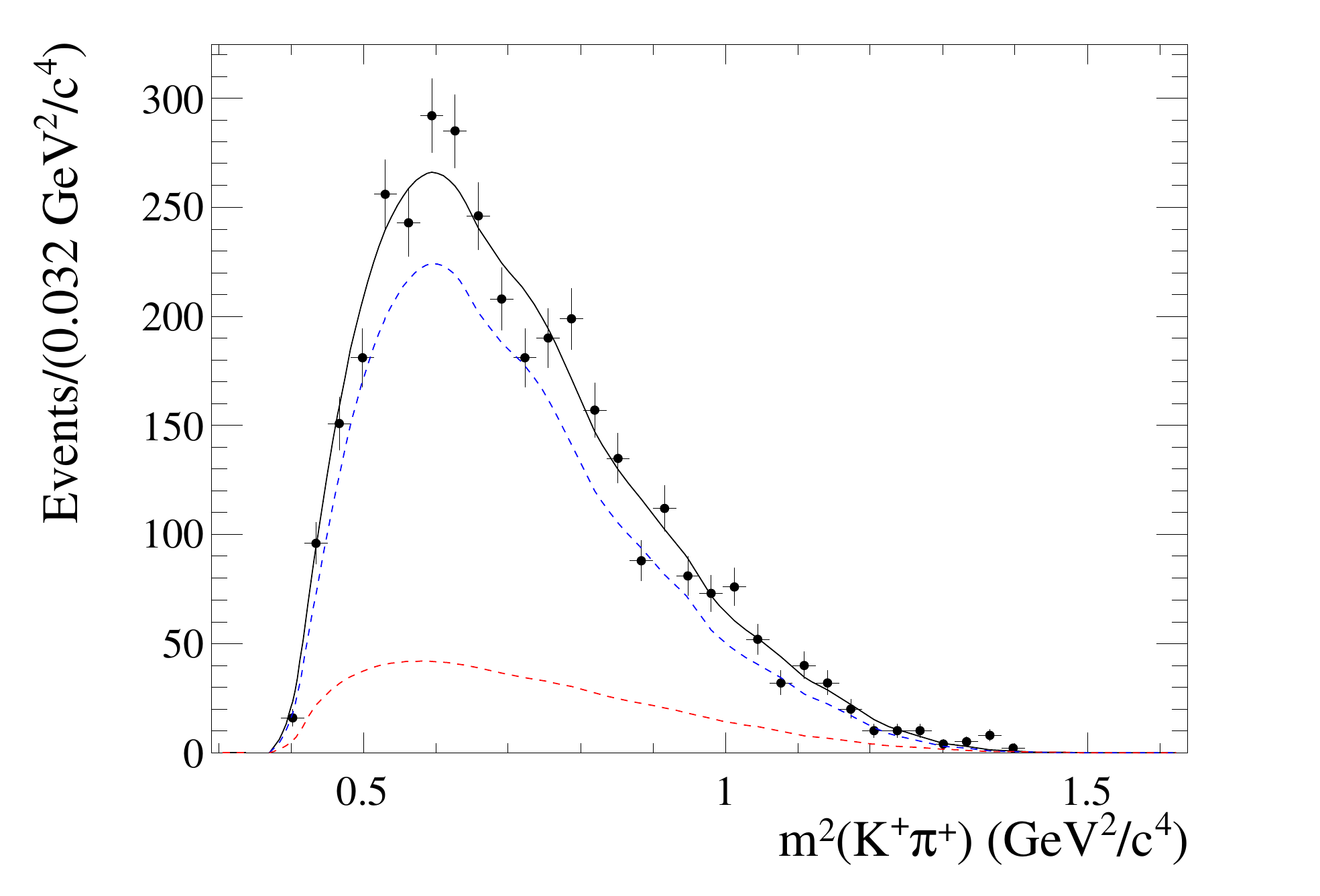}\\ 

  \includegraphics[width=0.49\textwidth, height=!]{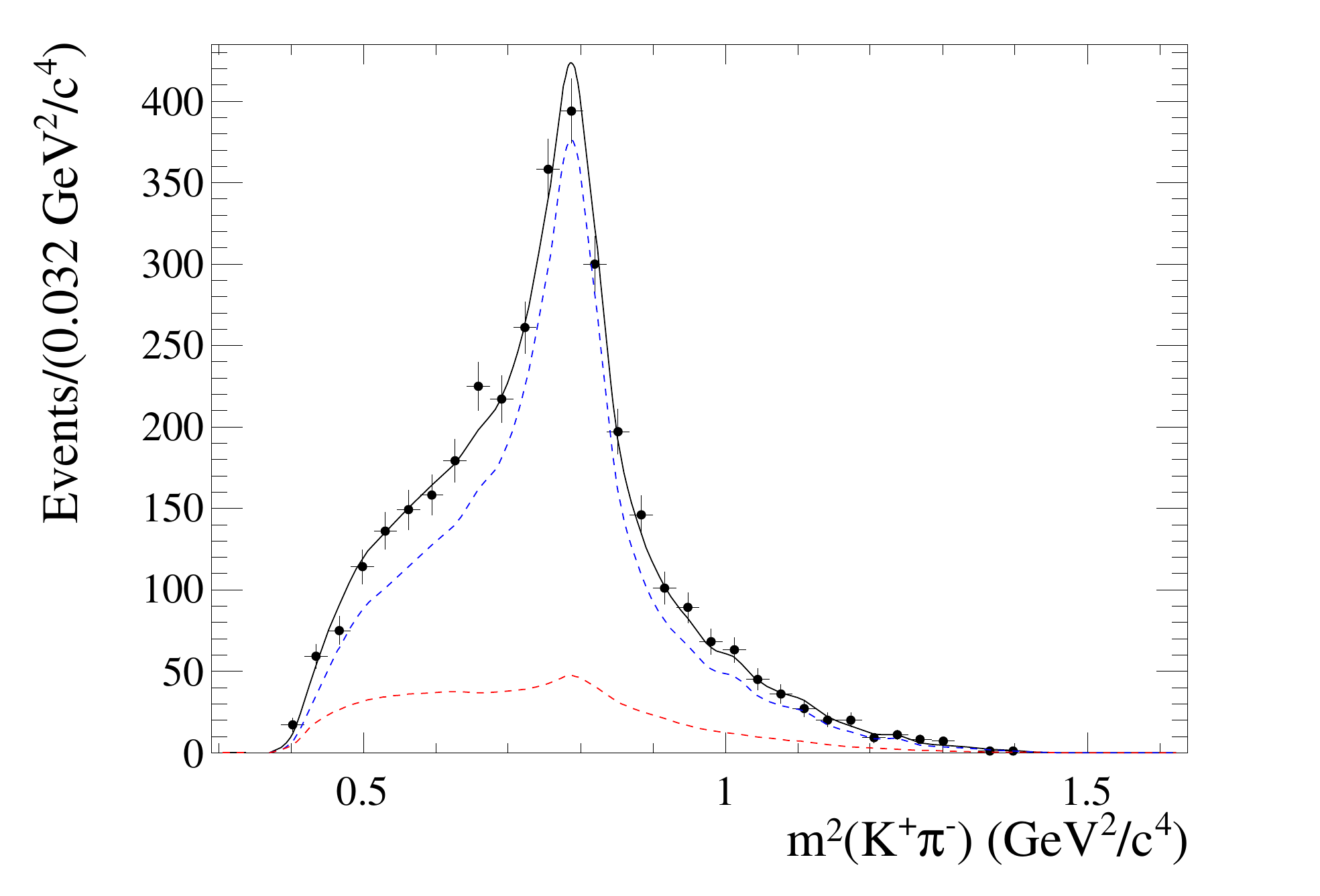}
  \includegraphics[width=0.49\textwidth, height=!]{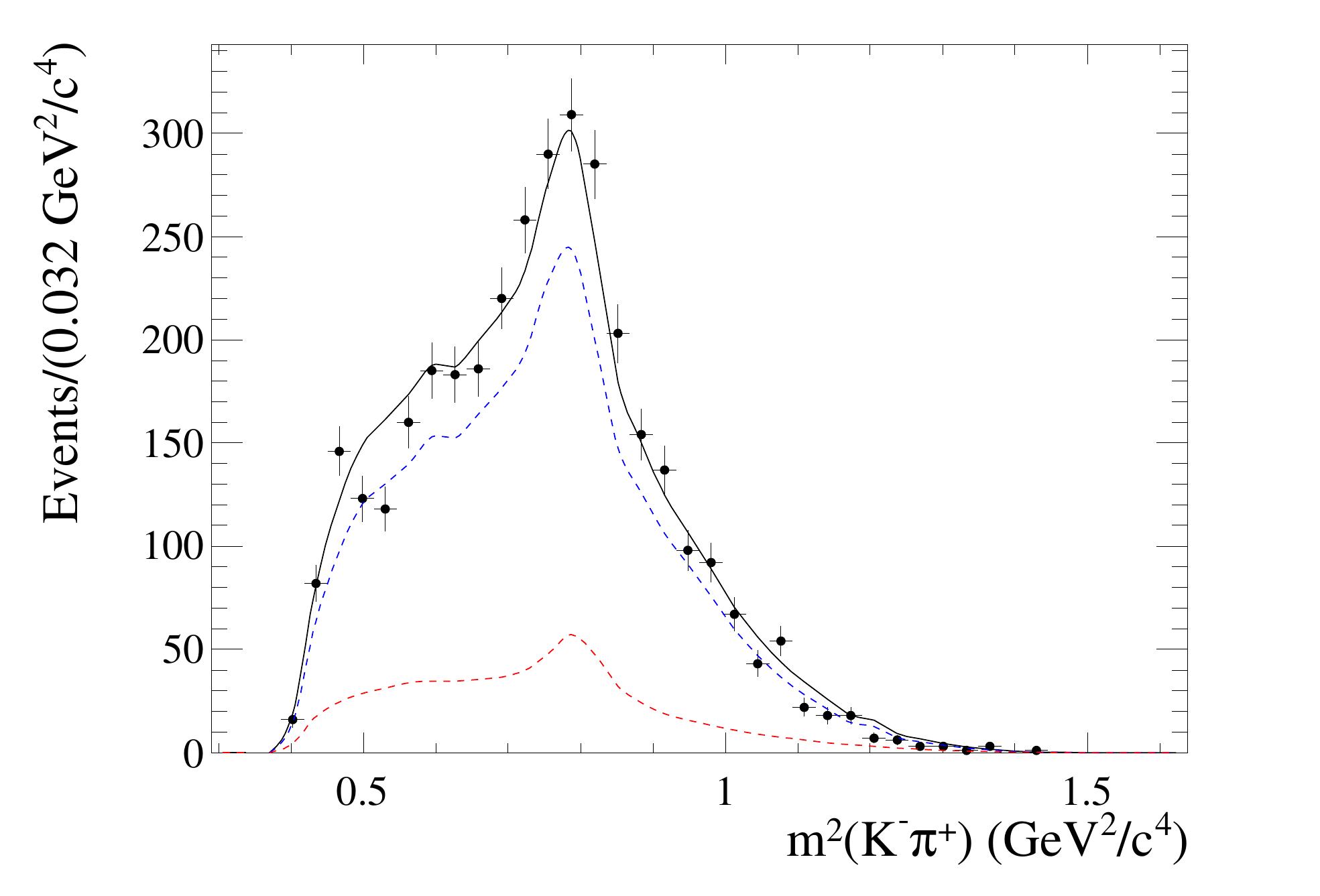}\\ 

  \includegraphics[width=0.49\textwidth, height=!]{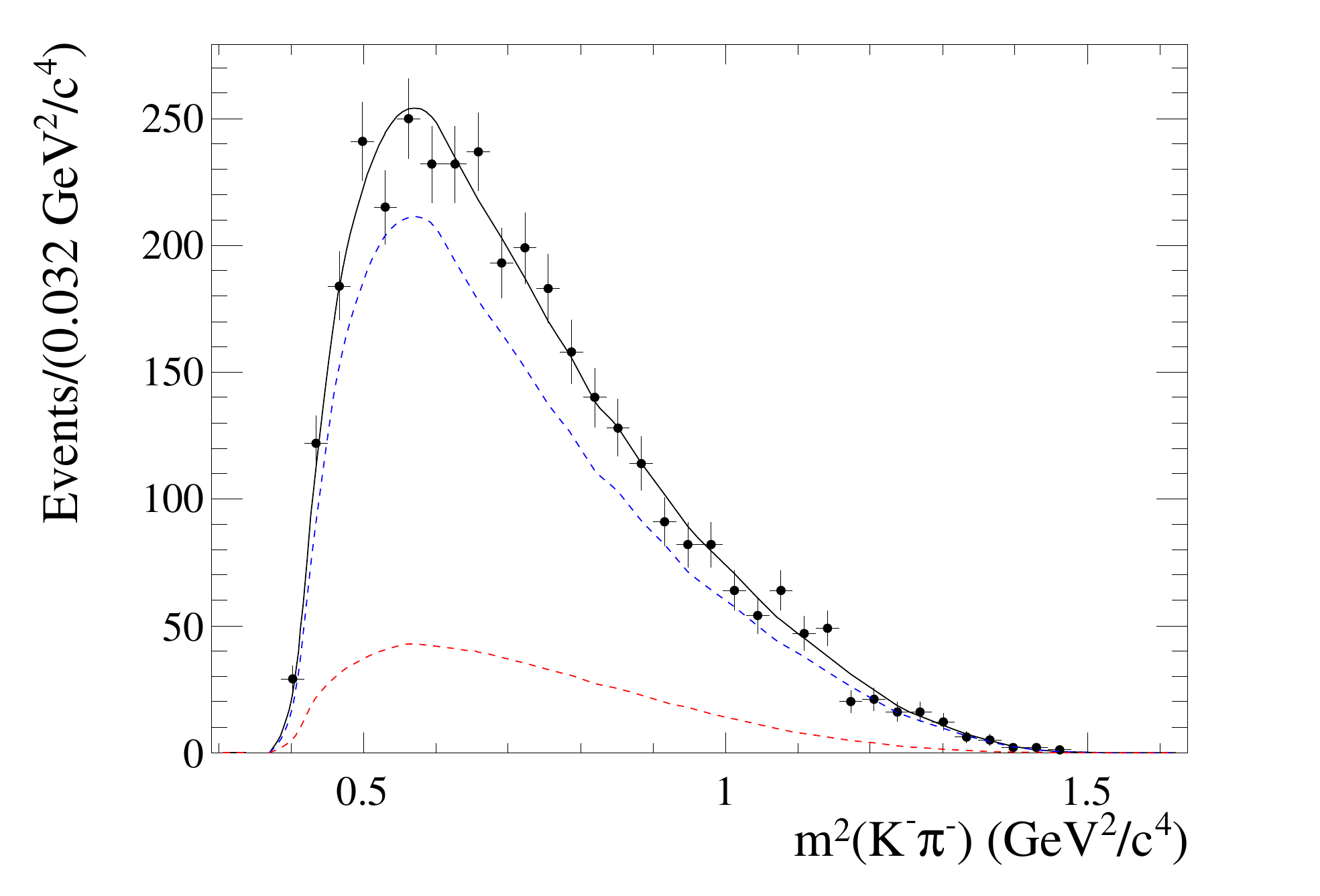}
  \includegraphics[width=0.49\textwidth, height=!]{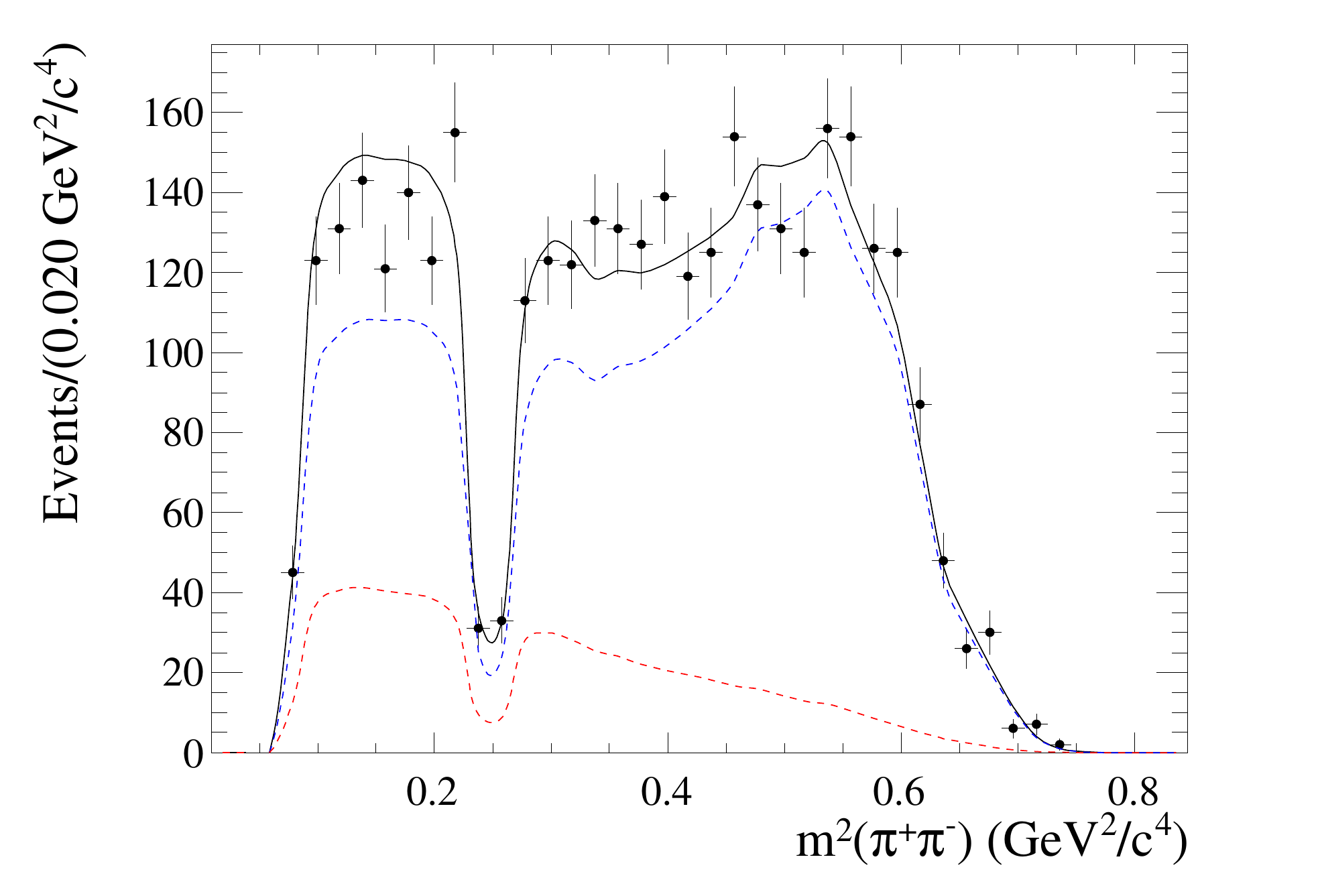}
  \caption{Invariant 2-body mass distributions of $\Dz \to \KKpipi$ signal candidates shown as points with error bars. 
  The overall fit projection is shown in black, the signal in blue and the background in red. 
  The effect of the $K^0_S$ veto can clearly be seen in the bottom right projection.}
  \label{fig:baselineFitKKpipi2body}
\end{figure}

The results between models are broadly consistent where the largest individual fit fraction corresponds to the $\Dz \to \phi(1020) \, \rho(770)^{0}$ amplitude. %
We found that we cannot distinguish 
between the $K^*(1680)$ meson in our default model 
and %
the $K^*(1410)$ meson trialled in alternative model B. Both of these components peak outside the kinematically allowed range.

Relative to the previous analysis of the same data set~\cite{KKpipi}, the most notable apparent difference in our default model is the fit fraction of the $\phi(1020)\,\rho(770)^{0}$ $S$-wave, which was 38.3\% in Ref.~\cite{KKpipi}, but only 28.1\% in our current analysis. This is because of our modified description of the $VV$ $D$-wave. In Ref.~\cite{KKpipi}, the component labeled as $D$-wave is a superposition of $D$ and $S$ waves, a choice which was motivated by the convention used in four-body amplitude analyses at the time. This led to a large 
interference between the components labeled as $S$ wave and $D$ wave of -15.7\%. In this analysis, as we parametrize a pure $D$-wave, we find an interference fraction between the $\phi(1020)\,\rho(770)^{0}$ $S$- and $D$-waves of -3.7\%. Taking these interference fractions into account, the combined $\phi(1020)\,\rho(770)^{0}$ $S$- and $D$-wave fraction of 26\% is therefore consistent between both analyses. In contrast to Ref.~\cite{KKpipi}, we also find a small, but significant $\phi(1020)\,\rho(770)^{0}$ $P$-wave component.
Another difference in the two-resonance topology is in the $K^*(892)^0  \, \bar{K}^*(892)^0$ mode, where our results indicate a significant $P$- and $D$-wave contribution, while in Ref.~\cite{KKpipi}, only an $S$-wave contribution was observed. Note though, that model 6 in Ref.~\cite{KKpipi} has a $P$-wave in the $K^*(892)\,(K\pi)_P$ decay of a similar size as our 
$K^*(892)^0 \, \bar{K}^*(892)^0$ $P$-wave.
The largest differences in our results are, as might be expected, in the cascade topology, because of the significant changes we implemented to improve the description of the lineshapes of resonance decays to three-body final states. We find that the process $\Dz \to K^{**+}  \, K^-$, where $K^{**}$ represents any excited kaon, dominates over $\Dz \to K^{**-} \, K^+$, analogous to the dominance of $\Dz \to a_1(1260)^+ \, \pi^-$ over $\Dz \to a_1(1260)^- \, \pi^+$ decays.  %
In Ref.~\cite{KKpipi}, this was only the case for the $K(1270) \to K^*(890) \, \pi$ amplitude. 
We also observe a significant $K(1270) \to K^*(1430) \, \pi$ component in agreement with Ref.~\cite{FOCUSKKpipi} but not with Ref.~\cite{KKpipi}.
The description of this type of decay chain, with a daughter whose mean mass is outside 
the kinematically allowed region, benefits particularly from our improved 
lineshapes. As in Ref.~\cite{FOCUSKKpipi}, but unlike in Ref.~\cite{KKpipi}, we also see a 
significant $K(1400) \to K^*(890) \, \pi$ contribution, albeit at a lower level.

\subsection{Global \CP\ Content Measurement}
Following the same approach as for $\Dz\ \to \fourpi$ decays, for the fractional \CP-even content we obtain
\begin{align}
F_{+}^{KK\pi\pi} = \left[ 75.3 \pm 1.8\,(\mathrm{stat}) \pm 3.3\,(\mathrm{syst}) \pm 3.5\,(\mathrm{model})\right] \%,
\end{align}
for the nominal $\Dz \to \KKpipi$ model, the first such measurement in this final state.

\subsection{Search for Direct \CP\ Violation}

Following the same approach as for $\Dz\ \to \fourpi$ decays, we measure the direct \CP\ violating parameters given in 
Table~\ref{tab:CPVKKpipi}. The \CP\ asymmetry over phase space is found to be
\begin{equation}
	\mathcal A^{KK\pi\pi}_{CP} = [+1.84 \pm 1.74 \stat \pm 0.30
	\syst] \%.
\end{equation}
All measurements are consistent with \CP\ conservation.

\begin{table}[h]
  \footnotesize
  \centering
  \renewcommand{\arraystretch}{1.3}
  \caption{Direct $CP$ asymmetry and significance for each component of the $\Dz \to \KKpipi$ LASSO model. 
  The first uncertainty is statistical, the second systematic and the third due to alternative models.}
  \begin{tabular}
    {@{\hspace{0.5cm}}l@{\hspace{0.25cm}}  @{\hspace{0.25cm}}c@{\hspace{0.25cm}}  @{\hspace{0.25cm}}c@{\hspace{0.5cm}}}
    \hline \hline 
    Decay channel & ${\cal A}_{CP}^i$ $(\%)$ & Significance ($\sigma$) \\
    \hline
%
% start input /Users/pnaik/Documents/CLEO/Papers/latexpand/latex/Acp_KKpipi_new.tex
$\Dz \to K^- \, K_{1}(1270)^{+}$          & $+25.3 \pm 9.7 \pm 9.2 \pm 8.8$&1.6\\    
$\Dz \to K^+ \, \bar{K_{1}}(1270)^{-}$    &$ -50.4\pm12.0\pm15.9 \pm2.4$& 2.5 \\
$\Dz \to K^- \, K_{1}(1400)^{+}$          &$ +9.2  \pm 15.1\pm	20.3 \pm 1.1$& 0.4\\    
$\Dz \to K^- \, K^*(1680)^{+}$            &$ -17.1 \pm 21.8\pm	18.0 \pm 4.2$& 0.6\\
$\Dz \to K^{*}(892)^{0}\bar{K}^{*}(892)^{0}$           &$ -4.6 \pm	9.0 \pm	9.8\pm5.7 $&	0.3\\
$\Dz \to \phi(1020) \, \rho(770)^{0}$                 &$ +1.5 \pm	4.6 \pm 8.0 \pm 0.5$&		0.1\\
$\Dz \to K^{*}(892)^{0} \, (K^- \pi^+)_{S}$                       &$ -13.1 \pm 17.9\pm	29.7 \pm 9.4$& 0.4\\
$\Dz \to \phi(1020) \, ( \pi^+ \pi^- )_{S}$              &$-4.0  \pm 18.0\pm	44.6 \pm 1.2$& 0.1\\
$\Dz \to (K^+K^-)_{S} \, (\pi^+ \pi^-)_{S}$             &$ +8.2  \pm	10.9\pm	16.9 \pm 2.7$& 0.4\\

 % end input /Users/pnaik/Documents/CLEO/Papers/latexpand/latex/Acp_KKpipi_new.tex
     \hline \hline
	\end{tabular}
	\label{tab:CPVKKpipi}
\end{table}
 % end input /Users/pnaik/Documents/CLEO/Papers/latexpand/latex/results.tex
 %
% start input /Users/pnaik/Documents/CLEO/Papers/latexpand/latex/systematics.tex
%

\clearpage\section{Systematic Uncertainties}
\label{sec:systematics}

There are three main sources of systematic uncertainties on the fit parameters to be considered;
an intrinsic fit bias, as well as experimental and model-dependent uncertainties. 
For each four-body decay, the fit bias itself is determined from a large ensemble of MC pseudo-experiments generated from the nominal 
LASSO model. The mean difference between the generated and fitted parameters are taken as a systematic uncertainty.

The experimental systematic uncertainties occur due to imperfect knowledge of
the yield of background events and their distribution in phase space, 
the wrong tag probability, and various effects on the efficiency variation over phase space.
To estimate the systematic uncertainty related
to the background shape that was fixed from sideband, the amplitude fit is repeated where the
background parameters are allowed to vary within their statistical uncertainties.  
In addition, several alternative background PDFs are tested whereby
each background contribution is replaced, one at a time, by a flat, non-resonant model.
The largest deviations from the nominal values are assigned as
systematic uncertainties.  
The uncertainty due to both the signal
fraction and the wrong tag probabilities in the flavor-tagged samples are estimated by repeating the fit
and allowing them to vary under Gaussian constraints. The signal fraction uncertainty for the \CP-tagged sample is determined by fixing the fraction to unity and repeating the fit. 
Various assumptions made on the acceptance in the fit model are also considered. As the acceptance comes from MC, we account for differences between data and MC arising from tracking and particle identification as a function of momentum of the daughter particles. Using correction factors obtained from independent internal CLEO studies, the MC is reweighted separately for each effect and the fit to data repeated.
While detector resolution can be safely ignored in $\Dz \to \fourpi$ decays, neglecting the effect of finite momentum resolution on the $\phi(1020)$ resonance in $\Dz \to \KKpipi$ decays may lead to a bias. To counter this, a large number of pseudo-experiments were generated by distributing MC events that have passed full selection and weighted by the LASSO model found from data. Each experiment is then fit with the signal model where the mean difference between the generated and fitted parameters are assigned as systematic uncertainties.
Finally, the integration error due to the limited size of the MC
sample is of the order of $0.5\%$, so it is neglected as a source of systematic uncertainty.

Model-dependent uncertainties arise from fixed lineshape parameters 
and the effects of interference from Cabbibo-suppressed decays on the tag-side in the \cleo-c flavor-tagged data samples.
The uncertainties due to fixed masses and widths of resonances 
are evaluated by varying them one-by-one within their quoted errors.
In our nominal fit, the Blatt-Weisskopf radial parameter is set to
$r_{\rm BW} = 1.5  \; (\textrm{GeV}/c)^{-1}$. 
As a systematic check, we set the radial parameter to zero.
For the calculation of the energy-dependent widths, the partial widths into the $\pi \pi \pi$ channel are obtained using an iterative procedure described in Sec.~\ref{ssec:lineshapes}. 
The systematic error of this approach is estimated by repeating the fit using the iteration previous to the final.
In some cases, the energy-dependent width relies on external measurements of intermediate branching fractions.
In $\Dz \to \fourpi$, their impact is studied by recalculating the width considering only decays into the $\pi \pi \pi$ ($\pi \pi$) final state for
three-body (two-body) resonances. 
For $\Dz \to \KKpipi$, the energy-dependent widths of the three-body resonances are recalculated assuming a flat phase space distribution.
Similarly, the energy-dependent mass of the $a_{1}(1260)$ resonance is approximated by a constant and the resulting shifts of the fit parameters are assigned as systematic errors.

The systematic uncertainty related to interference from the tag-side arising between the CKM-favored $c \to s$ and CKM-suppressed $c \to d$ amplitudes in the final states used for flavor-tagging is accounted for by using an alternative signal PDF at the cost of two additional fit parameters as described in Ref.~\cite{KKpipi}.

All systematic uncertainties are added in quadrature and summarized in Tables~\ref{tab:sys4pi} and \ref{tab:sysFractions4pi} for $\Dz \to \fourpi$
and in Tables~\ref{tab:syskkpipi} and \ref{tab:sysFractionskkpipi} for $\Dz \to \KKpipi$.

\begin{sidewaystable}[h]
  \scriptsize
  \centering
  \caption{\small Systematic uncertainties on the fit parameters of our nominal $\Dz \to \fourpi$ model in units of statistical standard deviations, $\sigma$. The different contributions are:
  1) Blatt-Weisskopf barrier factors, 2) Masses and widths of resonances, 3) Background model, 4) Signal fraction, 5) Wrong tag fractions, 6) Tag-side interference, 7) Efficiency, 8) Fit bias, 9) Energy-dependent masses and widths.
  }
  \begin{tabular}{@{\hspace{0.5cm}}l@{\hspace{0.25cm}}  @{\hspace{0.25cm}}c@{\hspace{0.25cm}}  @{\hspace{0.25cm}}c@{\hspace{0.25cm}}  @{\hspace{0.25cm}}c@{\hspace{0.25cm}}  @{\hspace{0.25cm}}c@{\hspace{0.25cm}}  @{\hspace{0.25cm}}c@{\hspace{0.25cm}}  @{\hspace{0.25cm}}c@{\hspace{0.25cm}}  @{\hspace{0.25cm}}c@{\hspace{0.25cm}}  @{\hspace{0.25cm}}c@{\hspace{0.25cm}}  @{\hspace{0.25cm}}c@{\hspace{0.25cm}}  @{\hspace{0.25cm}}c@{\hspace{0.5cm}}}
    \hline \hline 
    Fit parameter & 1 & 2 & 3 & 4 & 5 & 6 & 7 & 8 & 9 & Total \\
    \hline 
$\Dz \to \pim \, \left[ a_{1}(1260)^{+} \to \pip \, \sigma \right] \enspace\Re (a_{i})$ &0.55 & 0.11 & 0.16 & 0.09 & 0.20 & 0.78 & 0.17 & 0.22 & 0.14 & 1.05 \\ 
$\Dz \to \pim \, \left[ a_{1}(1260)^{+} \to \pip \, \sigma \right] \enspace\Im (a_{i})$ &1.20 & 0.19 & 0.28 & 0.16 & 0.19 & 0.28 & 0.11 & 0.14 & 0.18 & 1.33 \\ 
$\Dz \to \pip \,  a_{1}(1260)^{-} \enspace \Re (a_{i})$ & 0.43 & 1.13 & 0.24 & 0.11 & 0.22 & 0.16 & 0.18 & 0.09 & 0.11 & 1.29 \\ 
$\Dz \to \pip \,  a_{1}(1260)^{-} \enspace \Im (a_{i})$ &0.56 & 0.35 & 0.10 & 0.12 & 0.10 & 0.04 & 0.15 & 0.07 & 0.09 & 0.71 \\ 
$\Dz \to \pim \, \left[ \pi(1300)^{+} \to \pip \, \sigma \right] \enspace\Re (a_{i})$&1.32 & 1.72 & 0.06 & 0.19 & 0.29 & 0.22 & 0.54 & 0.28 & 2.07 & 3.08 \\ 
$\Dz \to \pim \, \left[ \pi(1300)^{+} \to \pip \, \sigma \right] \enspace\Im (a_{i})$&0.12 & 0.53 & 0.20 & 0.17 & 0.25 & 0.22 & 0.08 & 0.09 & 1.51 & 1.67 \\ 
$\Dz \to \pip \, \left[ \pi(1300)^{-} \to \pim \, \sigma \right]\enspace \Re (a_{i})$&0.74 & 1.83 & 0.17 & 0.03 & 0.10 & 0.45 & 0.33 & 0.03 & 1.39 & 2.51 \\ 
$\Dz \to \pip \, \left[ \pi(1300)^{-} \to \pim \, \sigma \right] \enspace\Im (a_{i})$&0.01 & 1.01 & 0.23 & 0.08 & 0.40 & 0.33 & 0.10 & 0.17 & 2.50 & 2.76 \\ 
$\Dz \to \pim \, \left[ a_{1}(1640)^{+}[D] \to \pip \, \rho(770)^{0} \right] \enspace\Re (a_{i})$&0.85 & 0.44 & 0.09 & 0.11 & 0.06 & 0.90 & 0.14 & 0.34 & 0.33 & 1.39 \\ 
$\Dz \to \pim \, \left[ a_{1}(1640)^{+}[D] \to \pip \, \rho(770)^{0} \right] \enspace\Im (a_{i})$&0.53 & 0.26 & 0.39 & 0.08 & 0.03 & 0.38 & 0.05 & 0.17 & 0.40 & 0.90 \\ 
$\Dz \to \pim \, \left[ a_{1}(1640)^{+}\to \pip \, \sigma \right] \enspace\Re (a_{i})$&0.97 & 0.44 & 0.37 & 0.15 & 0.03 & 0.69 & 0.08 & 0.17 & 0.42 & 1.42 \\ 
$\Dz \to \pim \, \left[ a_{1}(1640)^{+}\to \pip \, \sigma \right]\enspace \Im (a_{i})$&0.45 & 1.23 & 0.19 & 0.13 & 0.23 & 0.80 & 0.12 & 0.01 & 0.35 & 1.61 \\ 
$\Dz \to \pim \, \left[ \pi_{2}(1670)^{+}\to \pip \, f_{2}(1270) \right] \enspace\Re (a_{i})$&0.70 & 1.57 & 0.44 & 0.02 & 0.16 & 0.24 & 0.11 & 0.13 & 0.41 & 1.84 \\ 
$\Dz \to \pim \, \left[ \pi_{2}(1670)^{+}\to \pip \, f_{2}(1270) \right] \enspace\Im (a_{i})$&0.21 & 0.38 & 0.14 & 0.08 & 0.06 & 0.89 & 0.15 & 0.07 & 0.16 & 1.03 \\ 
$\Dz \to \pim \, \left[ \pi_{2}(1670)^{+} \to \pip \, \sigma \right] \enspace\Re (a_{i})$&0.47 & 0.05 & 0.03 & 0.08 & 0.15 & 0.26 & 0.02 & 0.19 & 0.32 & 0.68 \\ 
$\Dz \to \pim \, \left[ \pi_{2}(1670)^{+} \to \pip \, \sigma \right] \enspace\Im (a_{i})$&0.07 & 1.82 & 0.16 & 0.11 & 0.11 & 0.29 & 0.10 & 0.03 & 0.24 & 1.87 \\ 
$\Dz \to \sigma \, f_{0}(1370)  \enspace\Re (a_{i})$ &0.37 & 0.71 & 0.29 & 0.04 & 0.25 & 0.26 & 0.11 & 0.01 & 2.62 & 2.79 \\ 
$\Dz \to \sigma \, f_{0}(1370)  \enspace\Im (a_{i})$ &0.85 & 2.24 & 0.15 & 0.04 & 0.18 & 0.28 & 0.53 & 0.10 & 1.09 & 2.71 \\ 
$\Dz \to \sigma \,  \rho(770)^{0}  \enspace\Re (a_{i})$ &0.01 & 0.86 & 0.22 & 0.07 & 0.07 & 0.49 & 0.07 & 0.20 & 0.24 & 1.06 \\ 
$\Dz \to \sigma \,  \rho(770)^{0}  \enspace\Im (a_{i})$ &0.81 & 0.99 & 0.18 & 0.09 & 0.11 & 1.05 & 0.19 & 0.05 & 0.66 & 1.81 \\ 
$\Dz[S] \to \rho(770)^{0} \, \rho(770)^{0} \enspace\Re (a_{i})$&0.25 & 1.05 & 0.06 & 0.07 & 0.13 & 1.11 & 0.13 & 0.15 & 0.40 & 1.62 \\ 
$\Dz[S] \to \rho(770)^{0} \, \rho(770)^{0} \enspace\Im (a_{i})$&1.24 & 0.35 & 0.25 & 0.05 & 0.08 & 0.26 & 0.13 & 0.05 & 0.03 & 1.35 \\ 
$\Dz[P] \to \rho(770)^{0} \, \rho(770)^{0} \enspace\Re (a_{i})$ &0.17 & 0.97 & 0.16 & 0.08 & 0.18 & 0.31 & 0.02 & 0.11 & 0.32 & 1.11 \\ 
$\Dz[P] \to \rho(770)^{0} \, \rho(770)^{0} \enspace\Im (a_{i})$ &0.18 & 2.08 & 0.10 & 0.11 & 0.20 & 1.32 & 0.09 & 0.17 & 0.15 & 2.50 \\ 
$\Dz[D] \to \rho(770)^{0} \, \rho(770)^{0} \enspace\Re (a_{i})$&0.45 & 1.33 & 0.05 & 0.08 & 0.09 & 0.43 & 0.04 & 0.01 & 0.31 & 1.51 \\ 
$\Dz[D] \to \rho(770)^{0} \, \rho(770)^{0} \enspace\Im (a_{i})$&0.05 & 1.60 & 0.34 & 0.07 & 0.03 & 0.78 & 0.19 & 0.20 & 0.05 & 1.83 \\ 
$\Dz \to f_{2}(1270) \,  f_{2}(1270)\enspace \Re (a_{i})$ &0.67 & 0.52 & 0.23 & 0.10 & 0.22 & 0.37 & 0.10 & 0.24 & 0.17 & 1.03 \\ 
$\Dz \to f_{2}(1270) \,  f_{2}(1270) \enspace \Im (a_{i})$ &0.50 & 1.11 & 0.19 & 0.02 & 0.07 & 0.02 & 0.06 & 0.04 & 0.15 & 1.24 \\ 
$m_{a_1(1260)}$ &1.31 & 1.15 & 0.14 & 0.04 & 0.14 & 0.10 & 0.13 & 0.15 & 0.52 & 1.85 \\ 
$\Gamma_{a_1(1260)}$ &0.10 & 0.48 & 0.04 & 0.04 & 0.01 & 0.04 & 0.13 & 0.03 & 0.88 & 1.03 \\ 
$m_{\pi(1300)}$ &0.75 & 0.78 & 0.19 & 0.15 & 0.36 & 0.10 & 0.39 & 0.12 & 1.90 & 2.26 \\ 
$\Gamma_{\pi(1300)}$ &0.11 & 0.82 & 0.06 & 0.02 & 0.25 & 0.05 & 0.16 & 0.17 & 1.27 & 1.55 \\ 
$m_{a_1(1640)}$&0.64 & 0.45 & 0.06 & 0.05 & 0.13 & 0.03 & 0.07 & 0.22 & 0.37 & 0.89 \\ 
$\Gamma_{a_1(1640)}$ &0.54 & 0.21 & 0.12 & 0.09 & 0.05 & 0.04 & 0.05 & 0.38 & 0.07 & 0.61 \\
    \hline \hline
	\end{tabular}
	\label{tab:sys4pi}
\end{sidewaystable}

\begin{sidewaystable}[h]
  \scriptsize
  \centering
  \caption{\small Systematic uncertainties on the fit fractions from our nominal $\Dz \to \fourpi$ model in units of statistical standard deviations, $\sigma$. The different contributions are:
  1) Blatt-Weisskopf barrier factors, 2) Masses and widths of resonances, 3) Background model, 4) Signal fraction, 5) Wrong tag fractions, 6) Tag-side interference, 7) Efficiency, 8) Fit bias, 9) Energy-dependent masses and widths.
  }
  \begin{tabular}{@{\hspace{0.5cm}}l@{\hspace{0.25cm}}  @{\hspace{0.25cm}}c@{\hspace{0.25cm}}  @{\hspace{0.25cm}}c@{\hspace{0.25cm}}  @{\hspace{0.25cm}}c@{\hspace{0.25cm}}  @{\hspace{0.25cm}}c@{\hspace{0.25cm}}  @{\hspace{0.25cm}}c@{\hspace{0.25cm}}  @{\hspace{0.25cm}}c@{\hspace{0.25cm}}  @{\hspace{0.25cm}}c@{\hspace{0.25cm}}  @{\hspace{0.25cm}}c@{\hspace{0.25cm}}  @{\hspace{0.25cm}}c@{\hspace{0.25cm}}  @{\hspace{0.25cm}}c@{\hspace{0.5cm}}}
    \hline \hline 
    Decay channel & 1 & 2 & 3 & 4 & 5 & 6 & 7 & 8 & 9 & Total \\
    \hline 
$\Dz \to \pim \, \left[ a_{1}(1260)^{+}\to \pip \, \rho(770)^{0} \right] $&0.19 & 1.32 & 0.18 & 0.15 & 0.18 & 0.17 & 0.08 & 0.08   & 0.15 & 1.39 \\ 
$\Dz \to \pim \, \left[ a_{1}(1260)^{+} \to \pip \, \sigma \right] $ &1.09 & 0.76 & 0.22 & 0.13 & 0.17 & 0.54 & 0.18 & 0.03    & 0.25 & 1.50 \\ 
$\Dz \to \pip \, \left[ a_{1}(1260)^{-}\to \pim \, \rho(770)^{0} \right]$&0.55 & 0.56 & 0.19 & 0.06 & 0.11 & 0.10 & 0.13 &   0.09  & 0.06 & 0.83 \\ 
$\Dz \to \pip \, \left[ a_{1}(1260)^{-} \to \pim \, \sigma \right] $&0.14 & 0.24 & 0.24 & 0.11 & 0.20 & 0.21 & 0.13 &   0.06  & 0.06 & 0.50 \\ 
$\Dz \to \pim \, \left[ \pi(1300)^{+} \to \pip \, \sigma \right] $&0.55 & 1.25 & 0.26 & 0.05 & 0.20 & 0.51 & 0.40 & 0.07    & 0.62 & 1.67 \\ 
$\Dz \to \pip \, \left[ \pi(1300)^{-} \to \pim \, \sigma \right] $&0.08 & 1.34 & 0.26 & 0.09 & 0.32 & 1.01 & 0.11 &   0.06  & 2.85 & 3.33 \\ 
$\Dz \to \pim \, \left[ a_{1}(1640)^{+}[D] \to \pip \, \rho(770)^{0} \right] $&0.68 & 0.55 & 0.14 & 0.11 & 0.09 & 1.08 & 0.22 &   0.09  & 0.47 & 1.50 \\ 
$\Dz \to \pim \, \left[ a_{1}(1640)^{+}\to \pip \, \sigma \right] $&0.65 & 1.07 & 0.12 & 0.15 & 0.20 & 0.89 & 0.13 &  0.02   & 0.13 & 1.57 \\ 
$\Dz \to \pim \, \left[ \pi_{2}(1670)^{+}\to \pip \, f_{2}(1270) \right] $&0.18 & 0.79 & 0.19 & 0.01 & 0.06 & 0.79 & 0.09 & 0.06    & 0.18 & 1.17 \\ 
$\Dz \to \pim \, \left[ \pi_{2}(1670)^{+} \to \pip \, \sigma \right] $&0.35 & 1.23 & 0.12 & 0.07 & 0.08 & 0.31 & 0.06 & 0.02    & 0.14 & 1.33 \\ 
$\Dz \to \sigma \, f_{0}(1370)  $ &0.70 & 1.79 & 0.09 & 0.10 & 0.15 & 0.81 & 0.52 &  0.03   & 0.88 & 2.33 \\ 
$\Dz \to \sigma \,  \rho(770)^{0}  $ &0.58 & 0.22 & 0.14 & 0.06 & 0.04 & 0.83 & 0.07 &  0.05  & 0.58 & 1.20 \\ 
$\Dz[S] \to \rho(770)^{0} \, \rho(770)^{0}$ &1.12 & 0.93 & 0.27 & 0.05 & 0.15 & 0.46 & 0.12 &  0.07   & 0.17 & 1.57 \\ 
$\Dz[P] \to \rho(770)^{0} \, \rho(770)^{0}$ &0.09 & 1.84 & 0.24 & 0.03 & 0.02 & 2.60 & 0.13 &  0.26   & 0.13 & 3.20 \\ 
$\Dz[D] \to \rho(770)^{0} \, \rho(770)^{0}$&0.74 & 1.19 & 0.09 & 0.07 & 0.06 & 0.90 & 0.17 &   0.07  & 0.26 & 1.70 \\ 
$\Dz \to f_{2}(1270) \,  f_{2}(1270) $ &0.17 & 0.48 & 0.11 & 0.07 & 0.16 & 0.17 & 0.04 &   0.11  & 0.16 & 0.60 \\
    \hline \hline
	\end{tabular}
	\label{tab:sysFractions4pi}
\end{sidewaystable}

\begin{sidewaystable}[h]
  \scriptsize
  \centering
\caption{Systematic uncertainties on the fit parameters of our nominal $\Dz \to \KKpipi$ model in units of statistical standard deviations, $\sigma$. The different contributions are: 1) Blatt-Weisskopf barrier factors, 2) Masses and widths of resonances, 3) \CP-tagged signal fraction, 4) Flavor-tagged background model, 5) Flavor-tagged signal fractions, 6) Mistag rates, 7) Tag-side interference, 8) Efficiency, 9) Fit bias, 10) Detector resolution, 11) Energy-dependent masses and widths.}
           \begin{tabular}{@{\hspace{0.5cm}}l@{\hspace{0.25cm}}  @{\hspace{0.25cm}}c@{\hspace{0.25cm}}  @{\hspace{0.25cm}}c@{\hspace{0.25cm}}  @{\hspace{0.25cm}}c@{\hspace{0.25cm}}  @{\hspace{0.25cm}}c@{\hspace{0.25cm}}  @{\hspace{0.25cm}}c@{\hspace{0.25cm}}  @{\hspace{0.25cm}}c@{\hspace{0.25cm}}  @{\hspace{0.25cm}}c@{\hspace{0.25cm}}  @{\hspace{0.25cm}}c@{\hspace{0.25cm}}  @{\hspace{0.25cm}}c@{\hspace{0.25cm}}  @{\hspace{0.25cm}}c@{\hspace{0.25cm}}  @{\hspace{0.25cm}}c@{\hspace{0.25cm}}  @{\hspace{0.25cm}}c@{\hspace{0.5cm}}}
              \hline
              \hline
          Fit parameter & 1 & 2 & 3 & 4 & 5 & 6 & 7 & 8 & 9 & 10 & 11 & Total \\
              \hline           
              $\Dz \to K^- \, \left[ K_{1}(1270)^{+} \to \pi^+ \, K^{*}(892)^{0}\right] \enspace\Re (a_{i})$& 0.732&	2.126&	0.190&	0.199&	0.019&	0.001&	0.101&	0.376&	0.090&	0.211&	1.505&	2.757\\     
              $\Dz \to K^- \, \left[ K_{1}(1270)^{+} \to \pi^+ \, K^{*}(892)^{0}\right] \enspace \Im (a_{i})$ & 0.331&	2.396&	0.207&	0.147&	0.087&	0.001&	0.064&	0.300&	0.040&	0.233&	0.054&	2.465\\
              $\Dz \to K^- \, \left[ K_{1}(1270)^{+} \to \pi^+ \, K^{*}(1430)^{0}\right] \enspace\Re (a_{i})$& 0.493&	2.050&	0.244&	0.277&	0.005&	0.001&	0.212&	0.090&	0.274&	0.434&	0.344&	2.240\\
              $\Dz \to K^- \, \left[ K_{1}(1270)^{+} \to \pi^+ \, K^{*}(1430)^{0}\right] \enspace \Im (a_{i})$& 0.295&	1.125&	0.196&	0.147&	0.024&	0.002&	0.842&	0.060&	0.190&	0.085&	0.445&	1.538\\
              $\Dz \to K^- \, \left[ K_{1}(1270)^{+} \to K^+ \, \rho(770)^{0}\right] \enspace\Re (a_{i})$ & 0.470&	2.011&	0.199&	0.154&	0.009&	0.002&	0.207&	0.063&	0.292&	0.283&	0.332&	2.157\\
              $\Dz \to K^- \, \left[ K_{1}(1270)^{+} \to K^+ \, \rho(770)^{0}\right] \enspace \Im (a_{i})$ & 0.014&	1.127&	0.165&	0.181&	0.027&	0.001&	0.466&	0.215&	0.042&	0.021&	0.780&	1.485\\
              $\Dz \to K^+ \, \left[ K_{1}(1270)^{-} \to K^- \, \rho(770)^{0}\right] \enspace\Re (a_{i})$& 0.560&	0.972&	0.160&	0.169&	0.172&	0.001&	0.512&	0.139&	0.100&	0.226&	0.746&	1.497\\
              $\Dz \to K^+ \, \left[ K_{1}(1270)^{-} \to K^- \, \rho(770)^{0}\right] \enspace \Im (a_{i})$& 1.008&	2.125&	0.213&	0.298&	0.123&	0.000&	0.346&	0.216&	0.306&	0.178&	0.773&	2.564\\
              $\Dz \to K^- \, \left[ K_{1}(1270)^{+} \to K^+ \, \omega(782)\right] \enspace\Re (a_{i})$& 	0.549&	0.713&	0.149&	0.216&	0.024&	0.001&	0.165&	0.025&	0.157&	0.087&	0.605&	1.142\\
              $\Dz \to K^- \, \left[ K_{1}(1270)^{+} \to K^+ \, \omega(782)\right] \enspace \Im (a_{i})$& 	0.030&	1.506&	0.097&	0.033&	0.032&	0.000&	0.714&	0.274&	0.178&	0.387&	1.785&	2.497\\
              $\Dz \to K^- \, \left[ K_{1}(1400)^{+} \to \pi^+ \, K^{*}(892)^{0}\right] \enspace\Re (a_{i})$&	0.228&	1.587&	0.292&	0.176&	0.132&	0.000&	0.221&	0.180&	0.057&	0.169&	0.489&	1.748\\
              $\Dz \to K^- \, \left[ K_{1}(1400)^{+} \to \pi^+ \, K^{*}(892)^{0}\right] \enspace \Im (a_{i})$ & 0.090&	1.880&	0.101&	0.146&	0.047&	0.001&	0.821&	0.151&	0.312&	0.082&	1.409&	2.522\\
              $\Dz \to K^- \, \left[ K^*(1680)^{+} \to \pi^+ \, K^{*}(892)^{0}\right] \enspace\Re (a_{i})$ & 0.636&	0.732&	0.639&	0.095&	0.106&	0.000&	0.989&	0.246&	0.010&	0.567&	0.467&	1.716\\
              $\Dz \to K^- \, \left[ K^*(1680)^{+} \to \pi^+ \, K^{*}(892)^{0}\right] \enspace \Im (a_{i})$ & 0.423&	0.716&	0.254&	0.136&	0.066&	0.003&	0.051&	0.060&	0.043&	0.301&	1.206&	1.527\\
              $\Dz[S] \to K^{*}(892)^{0}\bar{K}^{*}(892)^{0} \enspace \Re (a_{i})$ & 0.044&	0.547&	0.218&	0.177&	0.054&	0.002&	1.006&	0.221&	0.357&	0.146&	0.892&	1.545\\
              $\Dz[S] \to K^{*}(892)^{0}\bar{K}^{*}(892)^{0} \enspace \Im (a_{i})$ & 0.536&	1.151&	0.169&	0.135&	0.026&	0.001&	0.237&	0.028&	0.091&	0.019&	1.790&	2.220\\
              $\Dz[P] \to K^{*}(892)^{0}\bar{K}^{*}(892)^{0} \enspace \Re (a_{i})$ & 0.575&	0.437&	0.171&	0.261&	0.016&	0.001&	1.909&	0.198&	0.015&	0.236&	0.213&	2.098\\
              $\Dz[P] \to K^{*}(892)^{0}\bar{K}^{*}(892)^{0} \enspace \Im (a_{i})$ & 0.518&	0.480&	0.412&	0.205&	0.011&	0.001&	0.401&	0.154&	0.226&	0.020&	0.501&	1.094\\
              $\Dz[D] \to K^{*}(892)^{0}\bar{K}^{*}(892)^{0} \enspace \Re (a_{i})$ & 0.863&	0.252&	0.636&	0.273&	0.068&	0.000&	0.055&	0.141&	0.069&	0.430&	0.132&	1.234\\
              $\Dz[D] \to K^{*}(892)^{0}\bar{K}^{*}(892)^{0} \enspace \Im (a_{i})$ & 0.445&	0.595&	0.162&	0.228&	0.085&	0.002&	0.453&	0.154&	0.013&	0.162&	1.489&	1.763\\         
              $\Dz[P] \to \phi(1020) \rho(770)^{0} \enspace \Re (a_{i})$ & 0.695	&       0.161&	0.145&	0.435&	0.009&	0.003&	1.006&	0.023&	0.128&	0.132&	0.139&	1.336\\
              $\Dz[P] \to \phi(1020) \rho(770)^{0} \enspace \Im (a_{i})$ & 0.243&	0.289&	0.209&	0.793&	0.014&	0.000&	0.085&	0.099&	0.112&	0.322&	0.747&	1.228\\
              $\Dz[D] \to \phi(1020) \rho(770)^{0} \enspace \Re (a_{i})$ & 0.709&	0.100&	0.750&	0.179&	0.013&	0.001&	0.040&	0.185&	0.009&	0.399&	0.449&	1.227\\
              $\Dz[D] \to \phi(1020) \rho(770)^{0} \enspace \Im  (a_{i})$ & 0.064&	0.227&	0.524&	0.349&	0.002&	0.001&	0.053&	0.023&	0.152&	0.107&	0.178&	0.723\\
              $\Dz \to K^{*}(892)^{0}(K^- \pi^+)_{S}\enspace\Re (a_{i})$ & 0.762& 1.161&	0.464&	0.114&	0.146&	0.002&	1.357&	0.134&	0.309&	0.233&	0.548&	2.119\\
              $\Dz \to K^{*}(892)^{0}(K^- \pi^+)_{S}\enspace \Im (a_{i})$ & 0.693& 0.358&	0.162&	0.145&	0.118&	0.001&	0.070&	0.105&	0.228&	0.194&	0.502&	1.013\\
              $\Dz \to \phi(1020) ( \pi^+ \pi^- )_{S}\enspace\Re (a_{i})$& 0.132&	0.209&	0.210&	0.247&	0.013&	0.001&	0.670&	0.107&	0.446&	0.657&	1.208&	1.648\\
              $\Dz \to \phi(1020) ( \pi^+ \pi^- )_{S}\enspace \Im (a_{i})$ & 0.685&	0.194&	0.302&	0.755&	0.087&	0.001&	1.720&	0.057&	0.292&	0.402&	0.266&	2.111\\
              $\Dz \to (K^+K^-)_{S}(\pi^+ \pi^-)_{S}\enspace\Re (a_{i})$ &1.163&	0.519&	0.201&	0.499&	0.049&	0.001&	0.255&	0.169&	0.044&	0.087&	1.301&	1.926\\
              $\Dz \to (K^+K^-)_{S}(\pi^+ \pi^-)_{S}\enspace \Im (a_{i})$ &0.999&	0.525&	0.202&	0.606&	0.319&	0.001&	1.223&	0.291&	0.020&	0.038&	0.531&	1.910\\
              \hline\hline
              \label{tab:syskkpipi}
           \end{tabular}
\end{sidewaystable}

\begin{sidewaystable}[h]
  \scriptsize
  \centering
\caption{Systematic uncertainties on the fit fractions from our nominal $\Dz \to \KKpipi$ model in units of statistical standard deviations, $\sigma$. The different contributions are: 1) Blatt-Weisskopf barrier factors, 2) Masses and widths of resonances, 3) \CP-tagged signal fraction, 4) Flavor-tagged background model, 5) Flavor-tagged signal fractions, 6) Mistag rates, 7) Tag-side interference, 8) Efficiency, 9) Fit bias, 10) Detector resolution, 11) Energy-dependent masses and widths.}
           \begin{tabular}{@{\hspace{0.5cm}}l@{\hspace{0.25cm}}  @{\hspace{0.25cm}}c@{\hspace{0.25cm}}  @{\hspace{0.25cm}}c@{\hspace{0.25cm}}  @{\hspace{0.25cm}}c@{\hspace{0.25cm}}  @{\hspace{0.25cm}}c@{\hspace{0.25cm}}  @{\hspace{0.25cm}}c@{\hspace{0.25cm}}  @{\hspace{0.25cm}}c@{\hspace{0.25cm}}  @{\hspace{0.25cm}}c@{\hspace{0.25cm}}  @{\hspace{0.25cm}}c@{\hspace{0.25cm}}  @{\hspace{0.25cm}}c@{\hspace{0.25cm}}  @{\hspace{0.25cm}}c@{\hspace{0.25cm}}  @{\hspace{0.25cm}}c@{\hspace{0.25cm}}  @{\hspace{0.25cm}}c@{\hspace{0.5cm}}}
              \hline
              \hline
          Decay channel & 1 & 2 & 3 & 4 & 5 & 6 & 7 & 8 & 9 & 10 & 11 & Total \\
              \hline           
              $\Dz \to K^- \, \left[ K_{1}(1270)^{+} \to \pi^+ \, K^{*}(892)^{0} \right]$ & 0.224&	0.960&	0.084&	0.071&	0.028&	0.000&	0.034&	0.140&	0.027&	0.094&	0.406&	1.087\\     
              
              $\Dz \to K^- \, \left[ K_{1}(1270)^{+} \to \pi^+ \, K^{*}(1430)^{0} \right]$ & 0.436&	1.812&	0.217&	0.245&	0.006&	0.001&	0.236&	0.080&	0.243&	0.382&	0.312&	1.987\\
              
              $\Dz \to K^- \, \left[ K_{1}(1270)^{+} \to K^+ \, \rho(770)^{0} \right]$  & 0.319&	1.443&	0.151&	0.129&	0.013&	0.001&	0.239&	0.099&	0.199&	0.192&	0.394&	1.588\\
              
              $\Dz \to K^+ \, \left[ {K_{1}}(1270)^{-} \to K^- \, \rho(770)^{0} \right]$ & 0.495&	1.009&	0.113&	0.147&	0.087&	0.000&	0.252&	0.110&	0.140&	0.118&	0.449&	1.271\\
              
              $\Dz \to K^- \, \left[ K_{1}(1270)^{+} \to K^+ \, \omega(782) \right]$ & 0.178&	0.879&	0.073&	0.072&	0.020&	0.000&	0.406&	0.155&	0.112&	0.220&	1.025&	1.454\\
              
              $\Dz \to K^- \, \left[ K_{1}(1400)^{+} \to \pi^+ \, K^{*}(892)^{0} \right]$ & 0.063&	1.125&	0.074&	0.089&	0.034&	0.001&	0.481&	0.092&	0.183&	0.054&	0.827&	1.498\\
              
              $\Dz \to K^- \, \left[ K^*(1680)^{+} \to \pi^+ \, K^{*}(892)^{0} \right]$  & 0.420&	0.496&	0.416&	0.068&	0.070&	0.001&	0.639&	0.159&	0.011&	0.371&	0.393&	1.153\\
              
              $\Dz[S] \to K^{*}(892)^{0}\bar{K}^{*}(892)^{0}$  & 0.224&	0.630&	0.178&	0.145&	0.042&	0.002&	0.764&	0.167&	0.271&	0.110&	1.000&	1.483\\
              
              $\Dz[P] \to K^{*}(892)^{0}\bar{K}^{*}(892)^{0}$   & 0.528&	0.424&	0.248&	0.233&	0.014&	0.001&	1.540&	0.176&	0.114&	0.189&	0.304&	1.766\\
              
              $\Dz[D] \to K^{*}(892)^{0}\bar{K}^{*}(892)^{0}$   & 1.182&	0.405&	0.865&	0.379&	0.097&	0.001&	0.181&	0.199&	0.094&	0.586&	0.572&	1.793\\
              
              $\Dz[S] \to \phi(1020) \rho(770)^{0}$  & 1.190&	0.265&	0.832&	0.622&	0.021&	0.003&	0.807&	0.211&	0.137&	0.524&	0.712&	2.015\\

              $\Dz[P] \to \phi(1020) \rho(770)^{0}$  & 0.635&	0.193&	0.159&	0.524&	0.010&	0.003&	0.906&	0.048&	0.125&	0.185&	0.351&	1.318\\
              
              $\Dz[D] \to \phi(1020) \rho(770)^{0}$   & 0.620&	0.097&	0.663&	0.169&	0.011&	0.001&	0.036&	0.162&	0.029&	0.349&	0.394&	1.080\\
              
              $\Dz \to K^{*}(892)^{0}(K^- \pi^+)_{S}$  & 0.328&	0.500&	0.200&	0.049&	0.063&	0.001&	0.584&	0.058&	0.133&	0.100&	0.236&	0.912\\
              
              $\Dz \to \phi(1020) ( \pi^+ \pi^- )_{S}$ & 0.606&	0.214&	0.295&	0.679&	0.077&	0.001&	1.564&	0.083&	0.376&	0.538&	0.783&	2.113\\
              
              $\Dz \to (K^+K^-)_{S}(\pi^+ \pi^-)_{S}$  & 0.946&	0.458&	0.177&	0.490&	0.206&	0.001&	0.797&	0.212&	0.029&	0.058&	0.851&	1.681\\
              
              \hline
              \label{tab:sysFractionskkpipi}
           \end{tabular}
\end{sidewaystable}

 % end input /Users/pnaik/Documents/CLEO/Papers/latexpand/latex/systematics.tex
 %
% start input /Users/pnaik/Documents/CLEO/Papers/latexpand/latex/conclusions.tex
%

\clearpage\section{Conclusion}
\label{sec:conclusions}

The first amplitude analysis of flavor-tagged $\Dz \to \fourpi$ decays 
has been presented based on CLEO-c data.
Due to the large amount of possible intermediate resonance components, a model-building procedure has been applied 
which balances the fit quality against the number of free fit parameters.
The prominent contribution is found to be the $a_{1}(1260)$ resonance in the decay modes
$a_{1}(1260) \to \rho(770)^{0} \, \pi$ and $a_{1}(1260) \to \sigma \, \pi$.
Along with the $a_{1}(1260)$, further cascade decays involving the resonances $\pi(1300)$ and $a_{1}(1640)$ are also seen.
The masses and widths of these resonances are determined using
an advanced lineshape parametrization taking into account the resonant three-pion substructure.
The resonant phase motion of 
these states has been verified by means of a quasi-model-independent study.
In addition to these cascade topologies, there is a significant contribution from the quasi-two-body decays 
$\Dz \to \rho(770)^{0} \, \rho(770)^{0}$ and $\Dz \to \sigma \, f_{0}(1370)$.
The \CP-even fraction of the decay $\Dz \to \fourpi$ as predicted by the amplitude model 
is consistent with a previous model-independent study.
The amplitude model has also been used to search for %
\CP\ violation in $\Dz \to \fourpi$ decays, where no \CP\ violation 
among the amplitudes is observed within the given precision of a few percent.

Moreover, the amplitude analysis of 
$\D \to \KKpipi$ decays performed by CLEO~\cite{KKpipi} has 
been revisited by applying the significantly improved formalism
presented in this paper,
using decays obtained from CLEO
II.V, CLEO III, and CLEO-c data.
The largest components are the processes $\Dz \to \phi(1020) \, \rho(770)^{0}$, $\Dz
\to K_1(1270)^+  \, K^-$ and $\Dz \to K(1400)^+ \, K^-$, which together account for
over half of the %
 $\Dz \to \KKpipi$ decay rate.
The
fractional \CP-even content is measured for the first time %
and a search for \CP asymmetries in the amplitude components 
yields no
evidence for \CP violation. 

In addition to shedding light on the dynamics of \prt{\Dz \to
  h^+ h^- \pi^+\pi^-} decays, these results are expected to provide
important input for a determination of the \CP-violating phase
$\gamma$~$(\phi_{3})$ in $B^{-} \to D \Km$ decays.

 % end input /Users/pnaik/Documents/CLEO/Papers/latexpand/latex/conclusions.tex
 %
%
% start input /Users/pnaik/Documents/CLEO/Papers/latexpand/latex/acknowledgements.tex
%

%
\section*{Acknowledgments}

This analysis was performed using CLEO II.V, CLEO III and CLEO-c data. 
The authors, some of whom were members of CLEO, are grateful to the 
collaboration for the privilege of using these data. 
We wish to thank Lauren Atkin and Andrew Powell who helped with their invaluable expertise; we also thank Jonathan Rosner and Roy Briere for their careful reading of the draft manuscript and their valuable suggestions.
We also acknowledge the support of the 
UK Science and Technology Facilities Council (STFC), the European Research Council 7 / ERC 
Grant Agreement number 307737, the German Federal Ministry of Education and Research (BMBF)
and the Particle Physics beyond the Standard Model research training group (GRK 1940).

 % end input /Users/pnaik/Documents/CLEO/Papers/latexpand/latex/acknowledgements.tex
 %
% start input /Users/pnaik/Documents/CLEO/Papers/latexpand/latex/appendix.tex
%

\appendix

\clearpage\section*{Appendices}

\section{Energy-Dependent Widths}
\label{a:rw}

\setcounter{figure}{0}
\setcounter{table}{0}

\renewcommand{\thefigure}{A.\arabic{figure}}

\begin{figure}[h]
\centering
  \includegraphics[width=0.49\textwidth, height = !]{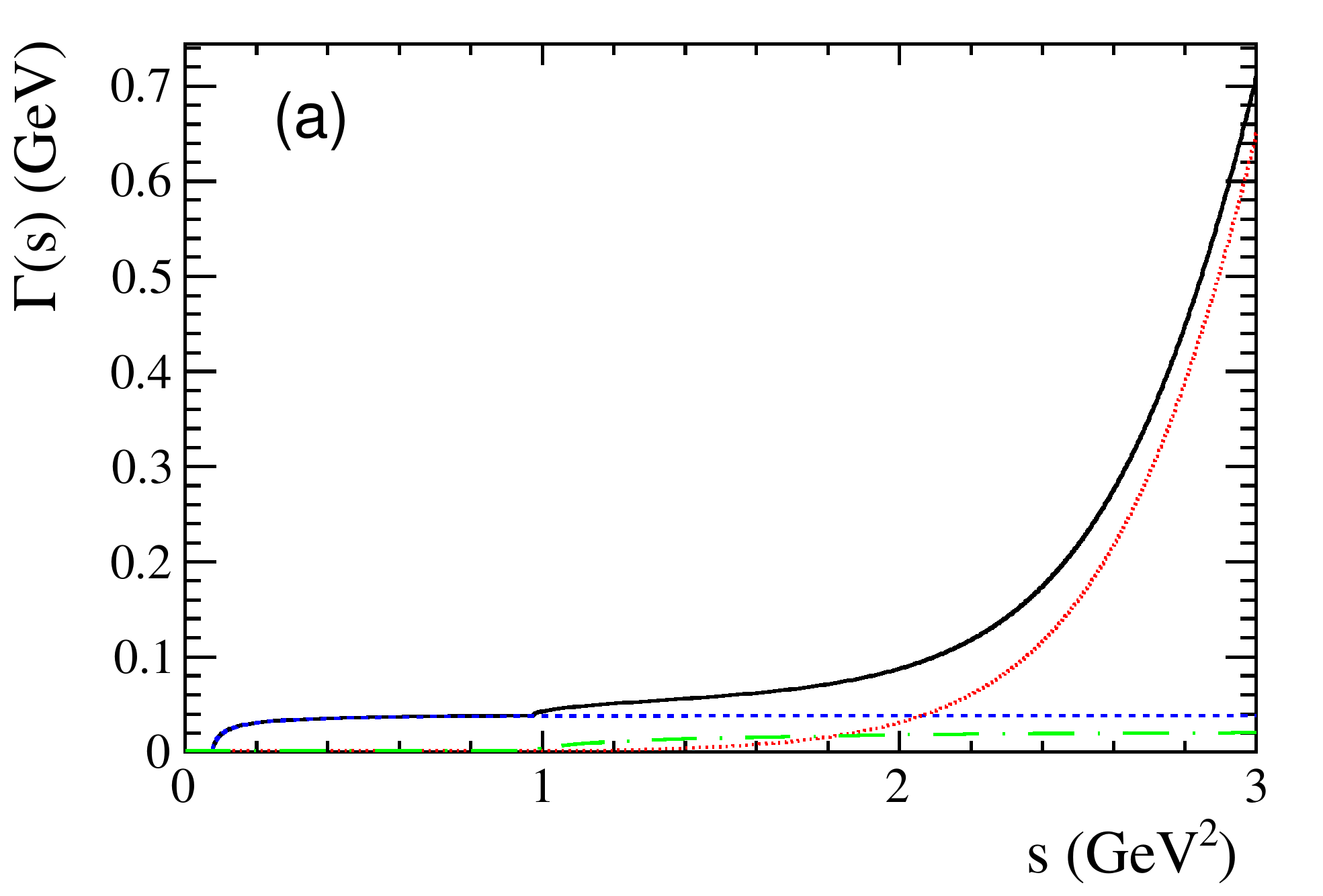} 
  \includegraphics[width=0.49\textwidth, height = !]{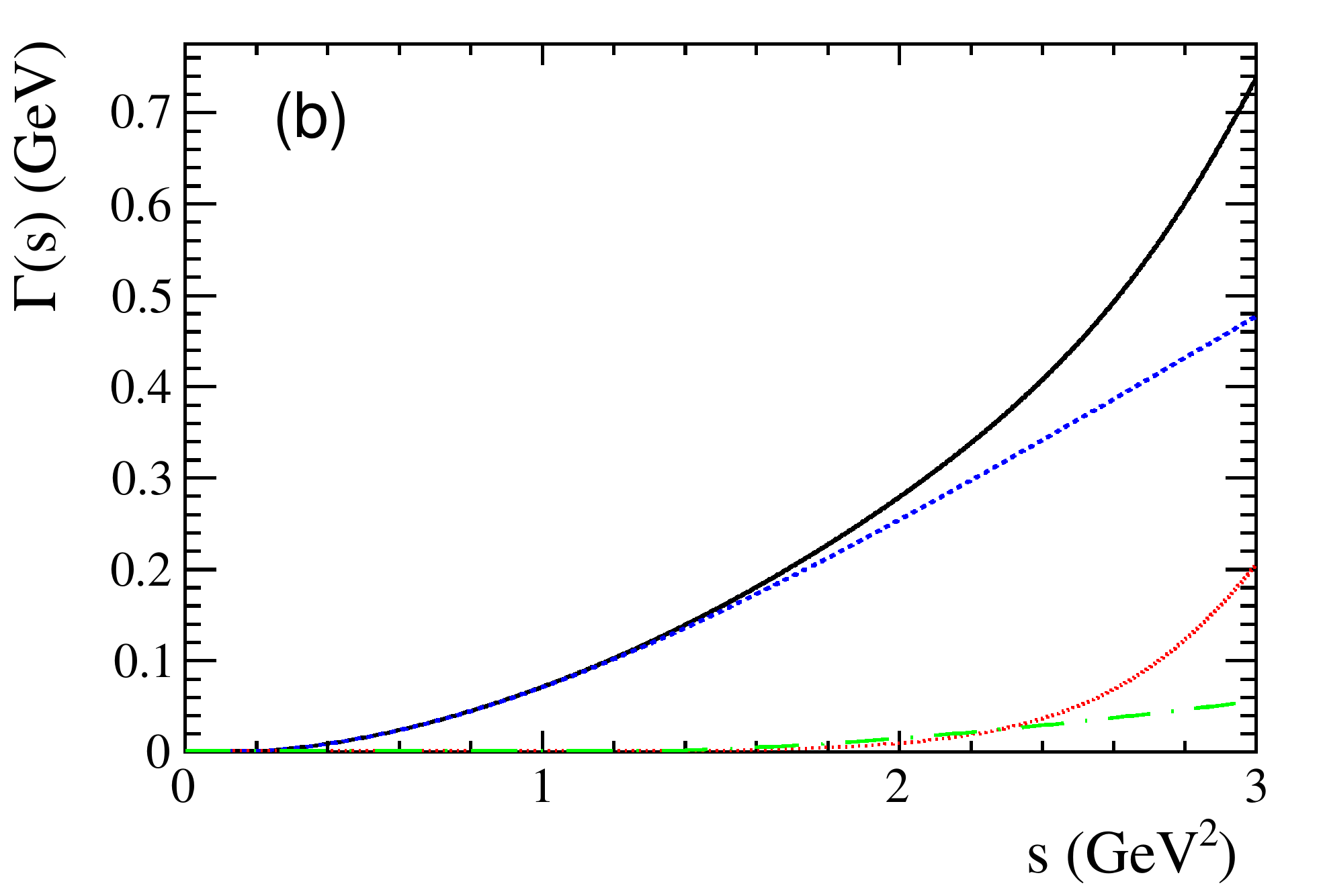} 
  \caption{Energy-dependent width for the $f_{0}(1370)$ (a) and $f_{2}(1270)$ (b) resonances. 
    The total width is shown in black (solid), while the partial widths into the channels 
    $\pi  \pi$, $\pi \pi \pi \pi$ and $K  K + \eta  \eta$ 
    are shown in blue (dashed), red (dotted) and green (dashed-dotted), respectively.\label{fig:gamma_f0}
  }
\end{figure}

\begin{figure}[h]
\centering
  \includegraphics[width=0.49\textwidth, height = !]{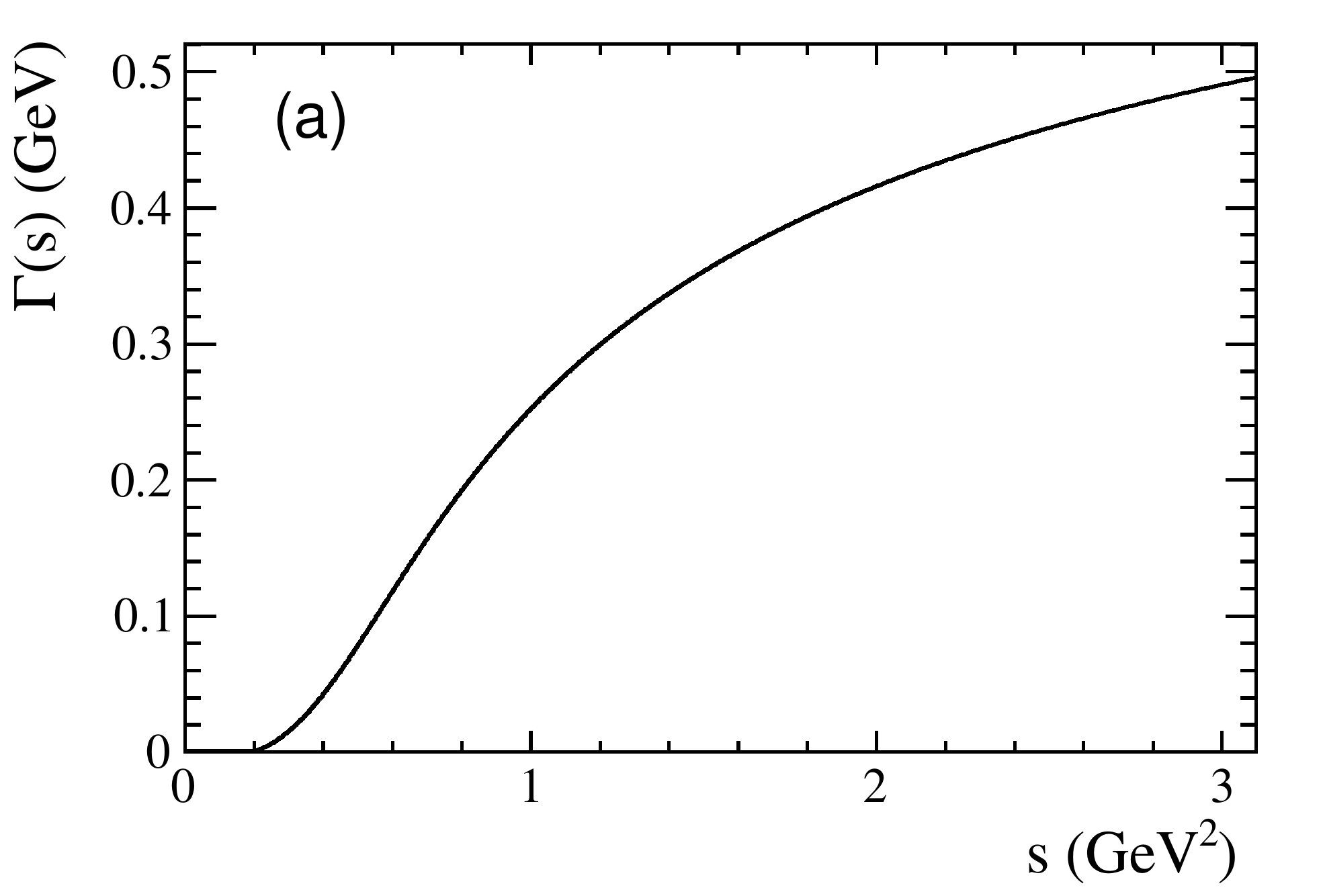} 
  \includegraphics[width=0.49\textwidth, height = !]{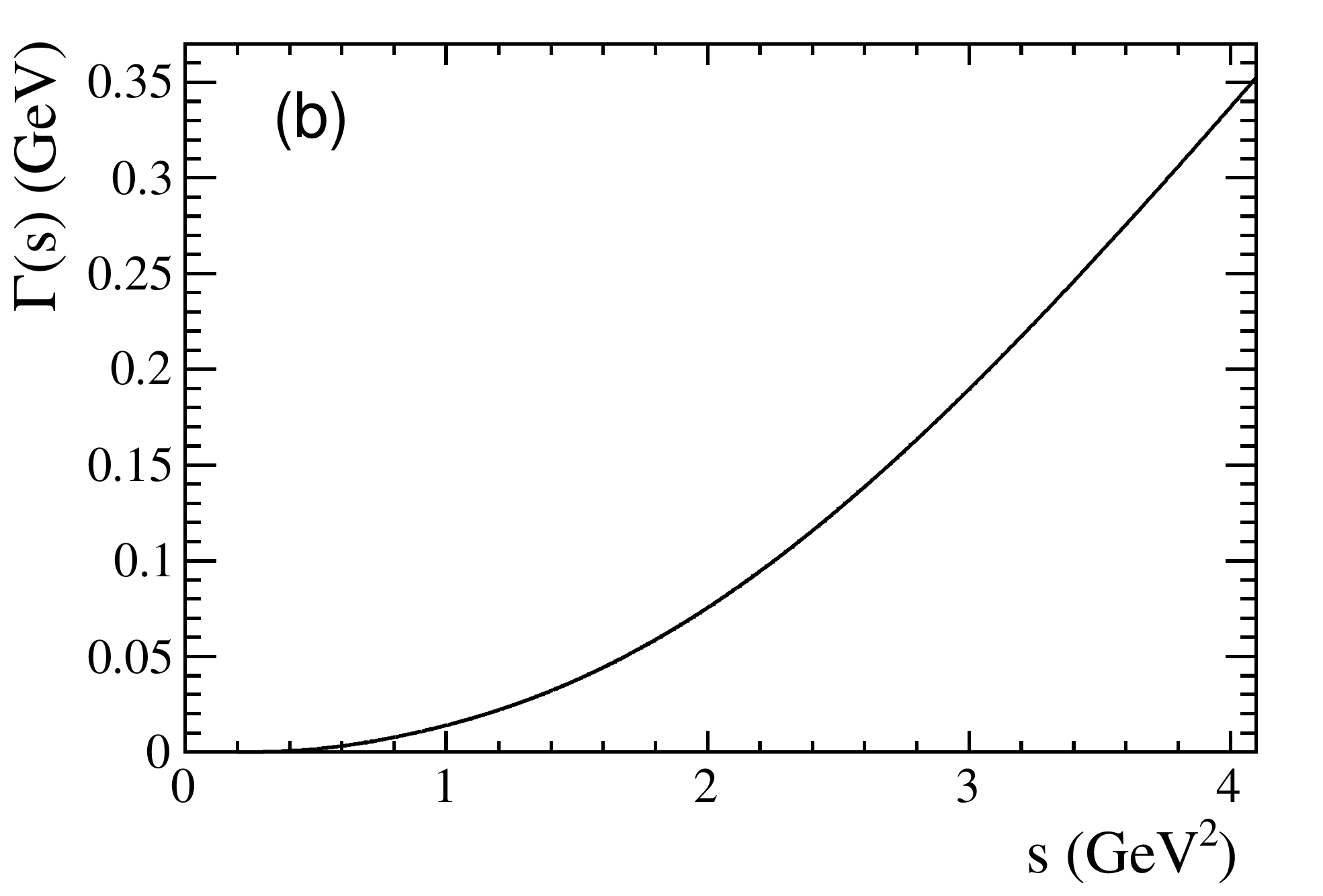} 
  \caption{Final iteration of the energy-dependent width for the $\pi(1300)$ (a) and $a_{1}(1640)$ (b) resonances.\label{fig:gamma_a1p}}
\end{figure}

\begin{figure}[h]
\centering
  \includegraphics[width=0.49\textwidth, height = !]{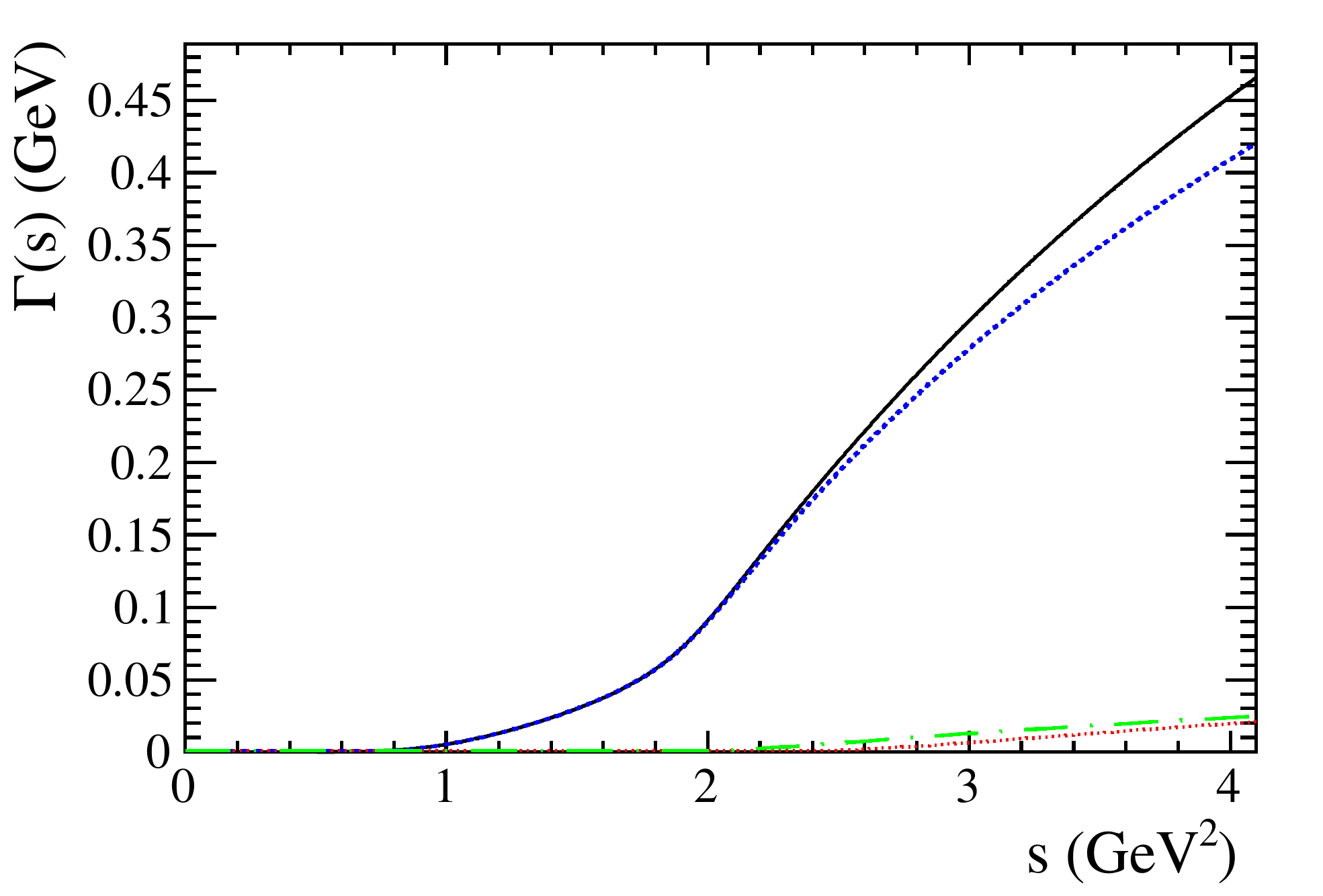} 
  \caption{Energy-dependent width for the $\pi_{2}(1670)$ resonance.
    The total width is shown in black (solid), while the partial widths into the channels 
    $\pi \pi \pi$,  $\omega \rho(770)$ and  $K \bar K \pi$
    are shown in blue (dashed), red (dotted) and green (dashed-dotted), respectively.\label{fig:gamma_pi2}
  }
  \includegraphics[width=0.49\textwidth, height = !]{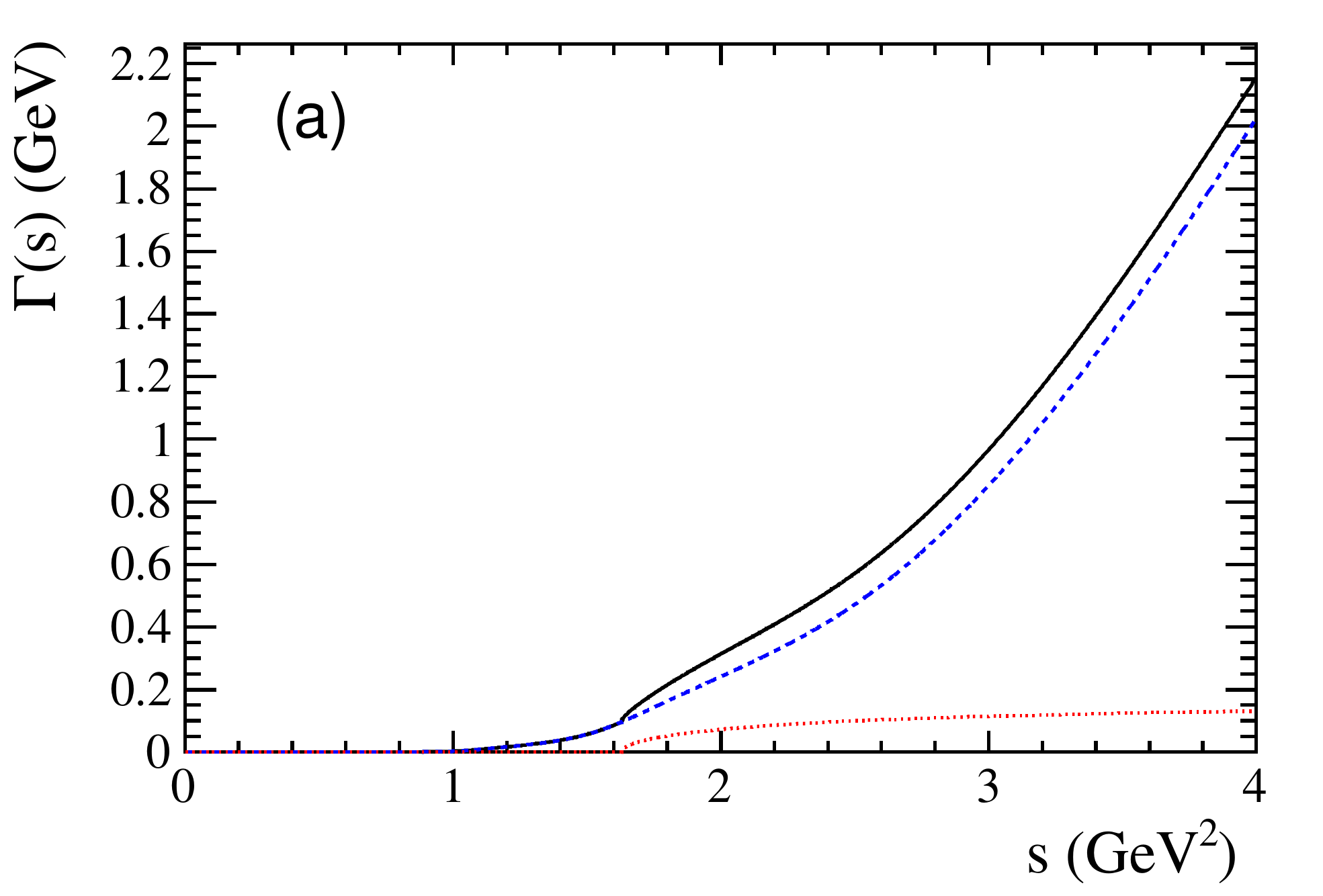} 
  \includegraphics[width=0.49\textwidth, height = !]{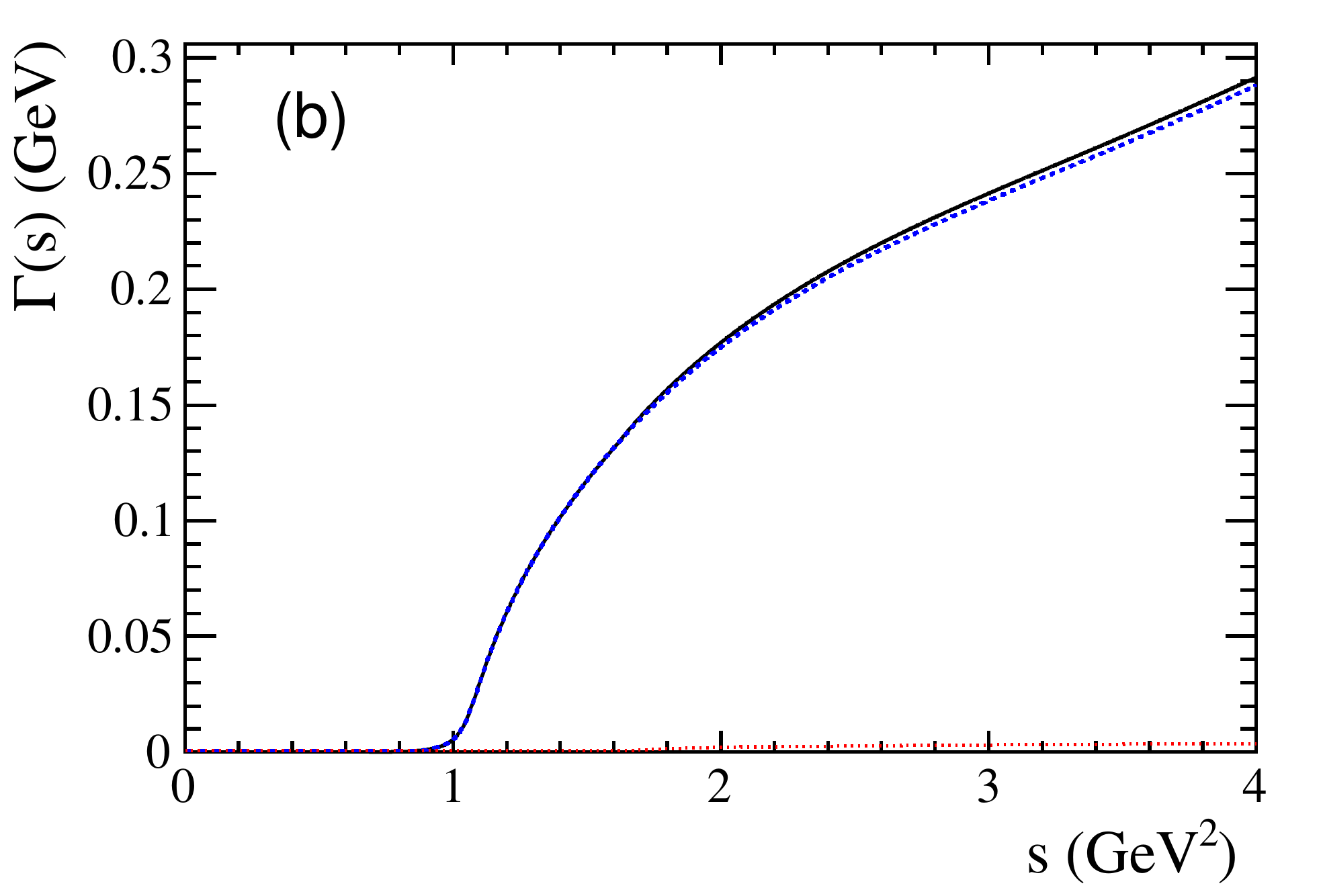} 
  \caption{Energy-dependent width for the $K_1(1270)$ (a) and $K_{1}(1400)$ (b) resonances. 
      The total width is shown in black (solid), while the partial widths into the channels 
    $K \pi \pi $ and $K \omega$ 
    are shown in blue (dashed) and red (dotted), respectively.\label{fig:gamma_K1} 
  }
  \includegraphics[width=0.49\textwidth, height = !]{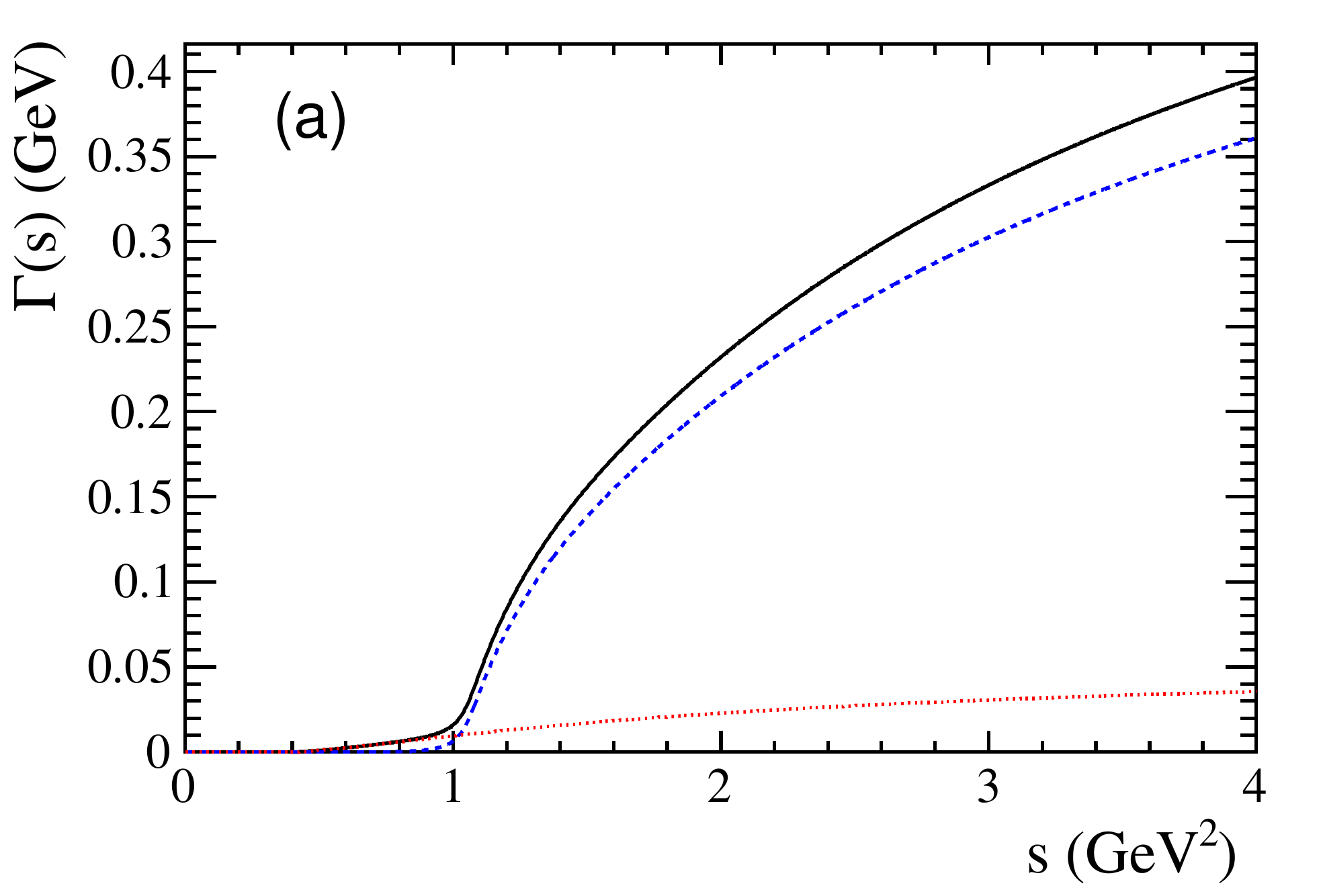} 
  \includegraphics[width=0.49\textwidth, height = !]{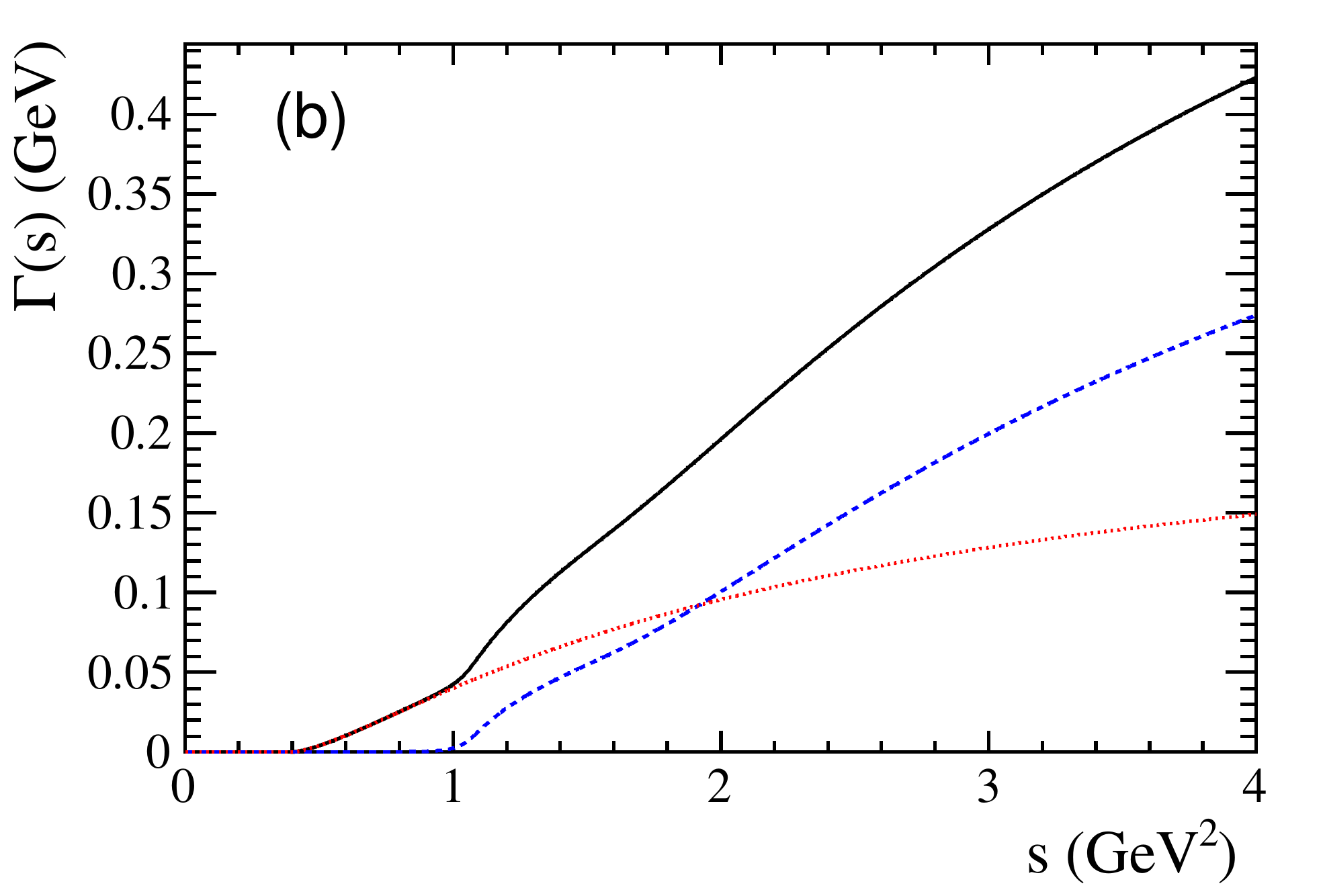} 
  \caption{Energy-dependent width for the $K^{*}(1410)$ (a) and $K^{*}(1680)$ (b) resonances.
        The total width is shown in black (solid), while the partial widths into the channels 
    $K \pi \pi $ and $K \pi$ 
    are shown in blue (dashed) and red (dotted), respectively.\label{fig:gamma_Ks}}
\end{figure}

\clearpage\section{Spin Amplitudes} 
\label{a:sf}

The spin factors used for $D \to P_1 \, P_2 \, P_3 \, P_4$ decays are given in Table \ref{tab:Sf}.
To fix our phase convention, we give the exact matching of the particles $P_1$, $P_2$, $P_3$, and $P_4$ in the spin factor definition to the final state particles in specific decay
chains in Tables \ref{tab:ordering1} and \ref{tab:ordering2}.

\renewcommand{\thetable}{B.\arabic{table}}

\begin{table}[h]
  \scriptsize
  \centering
  \caption{\small Spin factors for all topologies considered in this analysis. 
  In the decay chains, $S$, $P$, $V$, $A$, $T$ and $PT$ stand for scalar, pseudoscalar, vector, axial vector, tensor and pseudotensor, respectively. 
  If no angular momentum is specified, the lowest angular momentum state compatible with angular momentum conservation and, where appropriate, parity conservation, is used.}
  	\resizebox{\linewidth}{!}{
	\begin{tabular}{@{\hspace{0.5cm}}c@{\hspace{0.25cm}} @{\hspace{0.25cm}} l@{\hspace{0.25cm}}  @{\hspace{0.25cm}}l@{\hspace{0.5cm}}}
     \hline \hline
	Number & Decay chain & Spin amplitude  \\ \hline
	1 & $D \to (P \, P_{1})$, $P \to (S \, P_{2})$, $S \to (P_{3} \, P_{4})$ & $1$ \\
	 2 & $D \to (P \, P_{1})$, $P \to (V \, P_{2})$, $V \to (P_{3} \, P_{4})$ & $L_{(1)\alpha}(P) \,   \, L_{(1)}^{\alpha}(V) $ \\

	 3 & $D \to (A \, P_{1})$, $A \to (V \, P_{2})$, $V \to (P_{3} \, P_{4})$ & $L_{(1)\alpha}(D) \, P_{(1)}^{\alpha \beta}(A)  \, L_{(1)\beta}(V) $ \\
	 4 & $D \to (A \, P_{1})$, $A[D] \to (P_{2} \, V)$, $V \to (P_{3} \, P_{4})$ & $L_{(1)\alpha}(D) \, L_{(2)}^{\alpha \beta}(A)  \, L_{(1)\beta}(V) $   \\
	 5 & $D \to (A \, P_{1})$, $A \to (S \, P_{2})$, $S \to (P_{3} \, P_{4})$ & $L_{(1)\alpha}(D) \, L_{(1)}^{\alpha}(A) $    \\
	6 &  $D \to (A \, P_{1})$, $A \to (T \, P_{2})$, $T \to (P_{3} \, P_{4})$ &  $L_{(1)\alpha}(D) \, L_{(1)\beta}(A) \, L_{(2)}^{\alpha \beta}(T) $  \\
	
	7 &  $D \to (V_{1} \, P_{1})$, $V_{1} \to (V_{2} \, P_{2})$, $V_{2} \to (P_{3} \, P_{4})$ & 
	$ L_{(1)\mu}(D) \, P_{(1)}^{\mu \alpha }(V_{1})  \, \epsilon_{\alpha \beta \gamma \delta} \, L_{(1)}^{\beta}(V_{1}) \, p_{V_{1}}^{\gamma} \, L_{(1)}^{\delta}(V_{2}) $ \\
	
	 8 & $D \to (PT \, P_{1})$, $PT \to (V \, P_{2})$, $V \to (P_{3} \, P_{4})$ & 
	  $L_{(2)\alpha \beta}(D) \, P_{(2)}^{\alpha \beta \gamma \delta}(PT)  \, L_{(1)\gamma}(PT) \, L_{(1)\delta}(V) $ \\
	 9 &  $D \to (PT \, P_{1})$, $PT \to (S \, P_{2})$, $S \to (P_{3} \, P_{4})$ & 
	  $L_{(2)\alpha \beta}(D)  \, L_{(2)}(PT)^{\alpha \beta}  $ \\
	  10 & $D \to (PT \, P_{1})$, $PT \to (T \, P_{2})$, $T \to (P_{3} \, P_{4})$ & 
	  $L_{(2)\alpha \beta}(D) \, P_{(2)}^{\alpha \beta \gamma \delta}(PT)  \, L_{(2)\gamma \delta}(T)  $ \\
	  
	   11 & $D \to (T \, P_{1})$, $T \to (V \, P_{2})$, $V \to (P_{3} \, P_{4})$ & 
	  $L_{(2)\mu \nu}(D) \, P_{(2)}^{\mu \nu \rho \alpha}(T)  \, \epsilon_{\alpha \beta \gamma \delta}  
	  \, L_{(2)\rho}^{\beta}(T) \, p_{T}^{\gamma} \,  P_{(1)}^{\delta \sigma}(T) \, L_{(1)\sigma}(V) $ \\
	  
	   12 & $D \to (T_{1} \, P_{1})$, $T_{1} \to (T_{2} \, P_{2})$, $T_{2} \to (P_{3} \, P_{4})$ & 
	  $L_{(2)\mu \nu}(D) \, P_{(2)}^{\mu \nu \rho \alpha}(T_{1})  \, \epsilon_{\alpha \beta \gamma \delta}  
	  \, L_{(1)}^{\beta}(T_{1}) \, p_{T_{1}}^{\gamma} \,  L_{(2)\rho}^{\delta}(T_{2}) $ \\

	  13 & $D \to (S_{1} \, S_{2})$, $S_{1} \to (P_{1} \, P_{2})$, $S_{2} \to (P_{3} \, P_{4})$ & $ 1$ \\	  
	 
	  14 & $D \to (V \, S)$, $V \to (P_{1} \, P_{2})$, $S \to (P_{3} \, P_{4})$ & $L_{(1)\alpha}(D) \, L_{(1)}^{\alpha}(V) $ \\
	 
	  15 & $D \to (V_{1} \, V_{2})$, $V_{1} \to (P_{1} \, P_{2})$, $V_{2} \to (P_{3} \, P_{4})$ & $L_{(1)\alpha}(V_{1})  \, L_{(1)}^{\alpha}(V_{2}) $ \\
	  16 & $D[P] \to (V_{1} \, V_{2})$, $V_{1} \to (P_{1} \, P_{2})$, $V_{2} \to (P_{3} \, P_{4})$ & 
	 $\epsilon_{\alpha \beta \gamma \delta} \, L_{(1)}^{\alpha}(D) \, L_{(1)}^{\beta}(V_{1}) \, L_{(1)}^{\gamma}(V_{2}) \, p_{D}^{\delta}$ \\
	 17 &  $D[D] \to (V_{1} \, V_{2})$, $V_{1} \to (P_{1} \, P_{2})$, $V_{2} \to (P_{3} \, P_{4})$ &   $ L_{(2)\alpha \beta}(D)  \,L_{(1)}^{\alpha}(V_{1}) \, L_{(1)}^{\beta}(V_{2}) $ \\

	   18 & $D \to (T \, S)$, $T \to (P_{1} \, P_{2})$, $S \to (P_{3} \, P_{4})$ & $L_{(2)\alpha \beta}(D) \, L_{(2)}^{\alpha \beta}(T) $ \\
	   19 & $D \to (V \, T)$, $T \to (P_{1} \, P_{2})$, $V \to (P_{3} \, P_{4})$ & $L_{(1)\alpha}(D) \, L_{(2)}^{\alpha \beta}(T) \, L_{(1)\beta}(V) $ \\
	   20 & $D[D] \to (T \, V)$, $T \to (P_{1} \, P_{2})$, $V \to (P_{3} \, P_{4})$ &
	   $\epsilon_{\alpha \beta \delta \gamma} \, L_{2}^{\alpha \mu}(D) \, L_{2\mu}^{\beta} \, L_{(1)}^{\gamma}(V) \, p_{D}^{\delta} $ \\
	
	   21 & $D \to (T_{1} \, T_{2})$, $T_{1} \to (P_{1} \, P_{2})$, $T_{2} \to (P_{3} \, P_{4})$ & $L_{(2)\alpha \beta}(T_{1}) \, L_{(2)}^{\alpha \beta}(T_{2}) $ \\
	  22 &  $D[P] \to (T_{1} \, T_{2})$, $T_{1} \to (P_{1} \, P_{2})$, $T_{2} \to (P_{3} \, P_{4})$ & 
	  $\epsilon_{\alpha \beta \gamma \delta} \, L_{(1)}^{\alpha}(D) \, L_{(2)}^{\beta \mu}(T_{1}) \, L_{(2)\mu}^{\gamma}(T_{2}) \, p_{D}^{\delta} $ \\
	  23 &  $D[D] \to (T_{1} \, T_{2})$, $T_{1} \to (P_{1} \, P_{2})$, $T_{2} \to (P_{3} \, P_{4})$ & 
	  $L_{(2)\alpha \beta}(D) \, L_{(2)}^{\alpha \gamma}(T_{1}) \, L_{(2)\gamma}^{\beta}(T_{2}) $ \\

     \hline \hline
	\end{tabular}
	}
	\label{tab:Sf}
\end{table}

\begin{table}[h]
  \footnotesize
  \centering
  \caption{ \small Spin factors used for the decay chains included in the $\Dz \to \fourpi$ LASSO model, including the particle numbering scheme. 
The second column refers to the spin factors as numbered in Table \ref{tab:Sf}, and the particles $P_1$, $P_2$, $P_3$, and $P_4$ refer to those defined in Table \ref{tab:Sf}.}
  \begin{tabular}
     {@{\hspace{0.5cm}}l@{\hspace{0.25cm}}  @{\hspace{0.25cm}}c@{\hspace{0.25cm}}  @{\hspace{0.25cm}}c@{\hspace{0.25cm}}  @{\hspace{0.25cm}}c@{\hspace{0.25cm}} @{\hspace{0.25cm}}c@{\hspace{0.25cm}}  @{\hspace{0.25cm}}c@{\hspace{0.5cm}}}
     \hline \hline
     Decay channel & Spin factor number & $P_1$ & $P_2$ & $P_3$ & $P_4$ \\ \hline
     $\Dz \to \pim \, \left[ a_{1}(1260)^{+}\to \pip \, \rho(770)^{0} \right] $ & 3 & \pim & \pip & \pip & \pim  \\
     $\Dz \to \pim \, \left[ a_{1}(1260)^{+} \to \pip \, \sigma \right] $ & 5 & \pim & \pip & \pip & \pim \\
     $\Dz \to \pip \, \left[ a_{1}(1260)^{-}\to \pim \, \rho(770)^{0} \right] $ & 3 & \pip & \pim & \pim & \pip  \\
     $\Dz \to \pip \, \left[ a_{1}(1260)^{-} \to \pim \, \sigma \right] $ & 5 &  \pip & \pim & \pim & \pip \\
  
     $\Dz \to \pim \, \left[ \pi(1300)^{+} \to \pip \, \sigma \right] $ & 1 & \pim & \pip & \pip & \pim \\
     $\Dz \to \pip \, \left[ \pi(1300)^{-} \to \pim \, \sigma \right] $ & 1 & \pip & \pim & \pim & \pip \\

	$\Dz \to \pim \, \left[ a_{1}(1640)^{+}[D] \to \pip \, \rho(770)^{0} \right] $ & 4 & \pim & \pip & \pip & \pim \\
	$\Dz \to \pim \, \left[ a_{1}(1640)^{+}\to \pip \, \sigma \right] $ & 5 & \pim & \pip & \pip & \pim  \\

	$\Dz \to \pim \, \left[ \pi_{2}(1670)^{+}\to \pip \, f_{2}(1270) \right] $  & 10 & \pim & \pip & \pip & \pim  \\
	$\Dz \to \pim \, \left[ \pi_{2}(1670)^{+} \to \pip \, \sigma \right] $  & 9 & \pim & \pip & \pip & \pim  \\

	$\Dz \to \sigma \, f_{0}(1370)  $ & 13 & \pip & \pim & \pip & \pim  \\
	
	$\Dz \to \sigma \,  \rho(770)^{0}  $ & 14 & \pip & \pim & \pip & \pim \\

	$\Dz[S] \to \rho(770)^{0} \, \rho(770)^{0}$  & 15 & \pip & \pim & \pip & \pim  \\
	$\Dz[P] \to \rho(770)^{0} \, \rho(770)^{0}$  & 16 & \pip & \pim & \pip & \pim \\
	$\Dz[D] \to \rho(770)^{0} \, \rho(770)^{0}$ & 17 & \pip & \pim & \pip & \pim \\
	
	$\Dz \to f_{2}(1270) \,  f_{2}(1270) $  & 21 & \pip & \pim & \pip & \pim \\
	
	\hline\hline
	\end{tabular}
	\label{tab:ordering1}
\end{table}

\begin{table}[h]
  \footnotesize
  \centering
  \caption{  \small Spin factors used for the decay chains included in the $\Dz \to \KKpipi$ LASSO model, including the particle numbering scheme. 
  The second column refers to the spin factors as numbered in Table \ref{tab:Sf}, and the particles $P_1$, $P_2$, $P_3$, and $P_4$ refer to those defined in Table \ref{tab:Sf}.}
  \begin{tabular}
     {@{\hspace{0.5cm}}l@{\hspace{0.25cm}}  @{\hspace{0.25cm}}c@{\hspace{0.25cm}} @{\hspace{0.25cm}}c@{\hspace{0.25cm}}  @{\hspace{0.25cm}}c@{\hspace{0.25cm}} @{\hspace{0.25cm}}c@{\hspace{0.25cm}}  @{\hspace{0.25cm}}c@{\hspace{0.5cm}}}
     \hline \hline
     Decay channel & Spin factor number & $P_1$ & $P_2$ & $P_3$ & $P_4$ \\ \hline
           $\Dz \to K^- \, [K_{1}(1270)^{+} \to \pi^+ \, K^{*}(892)^{0}]$ & 3 & \Km & \pip & \Kp & \pim\\ 
           $\Dz \to K^- \, [K_{1}(1270)^{+} \to \pi^+ \, K^{*}(1430)^{0}]$& 5 & \Km & \pip & \Kp & \pim \\ 
              $\Dz \to K^- \, [K_{1}(1270)^{+} \to K^+ \, \rho(770)^{0}]$& 3 & \Km & \Kp & \pip & \pim\\ 
              $\Dz \to K^+ \, [\bar{K_{1}}(1270)^{-} \to K^- \, \rho(770)^{0}]$& 3 & \Kp & \Km & \pim & \pip \\ 
              $\Dz \to K^- \, [K_{1}(1270)^{+} \to K^+ \, \omega(782)]$  &  3 & \Km & \Kp & \pip & \pim\\ 
              $\Dz \to K^- \, [K_{1}(1400)^{+} \to \pi^+ \, K^{*}(892)^{0}]$& 3 & \Km & \pip & \Kp & \pim \\ 
              $\Dz \to K^- \, [K^*(1680)^{+} \to \pi^+ \, K^{*}(892)^{0}]$& 7 &\Km & \pip & \Kp & \pim \\ 
            
              $\Dz[S] \to K^{*}(892)^{0} \, \bar{K}^{*}(892)^{0}$ & 15 & \Kp & \pim & \Km & \pip \\ 
              $\Dz[P] \to K^{*}(892)^{0} \, \bar{K}^{*}(892)^{0}$ & 16 & \Kp & \pim & \Km & \pip \\ 
              $\Dz[D] \to K^{*}(892)^{0} \, \bar{K}^{*}(892)^{0}$ & 17 & \Kp & \pim & \Km & \pip \\ 
              $\Dz[S] \to \phi(1020) \, \rho(770)^{0}$ & 15 & \Kp & \Km & \pip & \pim \\
              $\Dz[P] \to \phi(1020) \, \rho(770)^{0}$ & 16 & \Kp & \Km & \pip & \pim \\ 
              $\Dz[D] \to \phi(1020) \, \rho(770)^{0}$ & 17 &\Kp & \Km & \pip & \pim \\ 
             
              $\Dz \to K^{*}(892)^{0} \, (K^- \pi^+)_{S}$ & 14 &  \Kp & \pim & \Km & \pip \\ 
              $\Dz \to \phi(1020) \, (\pi^+ \pi^-)_{S}$& 14 & \Kp & \Km & \pip & \pim \\ 
            
              $\Dz \to (K^+K^-)_{S} \, (\pi^+ \pi^-)_{S}$& 13 & \Kp & \Km & \pip & \pim \\ 

	\hline\hline
	\end{tabular}
	\label{tab:ordering2}
\end{table}

\clearpage\section{Considered Decay Chains}
\label{a:decays}

The various decay channels considered in the model building are listed in Tables \ref{tab:components4pi} and \ref{tab:componentsKKpipi}.

\setcounter{table}{0}
\renewcommand{\thetable}{C.\arabic{table}}

\begin{table}[h]
  \footnotesize
  \centering
  \caption{Decays considered in $\Dz \to \fourpi$ LASSO model building. For cascade non-self-conjugate channels, the conjugate partner is implied.}
	\begin{tabular}{@{\hspace{0.5cm}}l@{\hspace{0.25cm}}}
     \hline \hline
	 Decay channel  \\ \hline

	$\Dz \to \pim \, \left[ a_{1}(1260)^{+} \to \pip \, \sigma \right] $  \\
	$\Dz \to \pim \, \left[ a_{1}(1260)^{+}[S,D] \to \pip \, \rho(770)^{0} \right] $  \\
	$\Dz \to \pim \, \left[ a_{1}(1260)^{+} \to \pip \, f_{0}(980) \right] $  \\
	$\Dz \to \pim \, \left[ a_{1}(1260)^{+} \to \pip \, f_{2}(1270) \right] $  \\
	$\Dz \to \pim \, \left[ a_{1}(1260)^{+} \to \pip \, f_{0}(1370) \right] $  \\
	$\Dz \to \pim \, \left[ a_{1}(1260)^{+}[S,D] \to \pip \, \rho(1450)^{0} \right] $  \\

	$\Dz \to \pim \, \left[ \pi(1300)^{+} \to \pip \, \sigma \right] $  \\	
	$\Dz \to \pim \, \left[ \pi(1300)^{+} \to \pip \, \rho(770)^{0} \right] $  \\
	$\Dz \to \pim \, \left[ \pi(1300)^{+} \to \pip \,( \pip \pim )_{P} \right] $  \\
	
	$\Dz \to \pim \, \left[ a_{2}(1320)^{+} \to \pip \, \rho(770)^{0} \right] $  \\
	$\Dz \to \pim \, \left[ a_{2}(1320)^{+} \to \pip \, f_{2}(1270) \right] $  \\
	
	$\Dz \to \pim \, \left[ a_{1}(1420)^{+} \to \pip \, f_{0}(980) \right] $  \\
	
	$\Dz \to \pim \, \left[ \pi_{1}(1600)^{+} \to \pip \, \rho(770)^{0} \right] $  \\

	$\Dz \to \pim \, \left[ a_{1}(1640)^{+} \to \pip \, \sigma \right] $  \\		
	$\Dz \to \pim \, \left[ a_{1}(1640)^{+}[S,D] \to \pip \, \rho(770)^{0} \right] $  \\
	$\Dz \to \pim \, \left[ a_{1}(1640)^{+} \to \pip \, f_{2}(1270) \right] $  \\

	$\Dz \to \pim \, \left[ \pi_{2}(1670)^{+} \to \pip \, \sigma \right] $  \\
	$\Dz \to \pim \, \left[ \pi_{2}(1670)^{+} \to \pip \, \rho(770)^{0} \right] $  \\
	$\Dz \to \pim \, \left[ \pi_{2}(1670)^{+} \to \pip \, f_{2}(1270) \right] $  \\

	$\Dz \to (\pi \, \pi)_{S} \, (\pi \, \pi)_{S}$\\ %
	$\Dz \to  \sigma \, (\pi \, \pi)_{S}$\\
        $\Dz \to   \sigma \, \sigma$\\
        $\Dz \to   \sigma \, f_{0}(980)$\\
        $\Dz \to   \sigma \, f_{0}(1370)$\\
	$\Dz \to f_{0}(980) \, f_{0}(980)$  \\
        $\Dz \to   f_{0}(1370) \, f_{0}(1370)$   \\

	$\Dz \to \rho(770)^{0} \, \sigma$\\
	$\Dz \to \rho(770)^{0} \, f_{0}(980)$  \\
        $\Dz \to \rho(770)^{0} \, f_{0}(1370)$ \\
        
	$\Dz \to \rho(1450)^{0} \, \sigma$\\

	$\Dz[S,P,D] \to (\pi \, \pi)_{P} \, (\pi \, \pi)_{P}$\\	
	$\Dz[S,P,D] \to \rho(770)^{0} \, (\pi \, \pi)_{P}$  \\
	$\Dz[S,P,D] \to \rho(770)^{0} \, \rho(770)^{0}$  \\
	$\Dz[S,P,D] \to \rho(770)^{0} \, \omega(782)^{0}$  \\
	$\Dz[S,P,D] \to \omega(782)^{0} \, \omega(782)^{0}$  \\
        $\Dz[S,P,D] \to \rho(1450)^{0} \, (\pi \, \pi)_{P}$ \\
        $\Dz[S,P,D] \to \rho(1450)^{0} \, \rho(1450)^{0}$ \\

	$\Dz \to f_{2}(1270) \, \sigma$  \\
	$\Dz \to f_{2}(1270) \,  f_{0}(980)$  \\

	$\Dz[P,D] \to f_{2}(1270) \, \rho(770)^{0}$  \\
	$\Dz[S,P,D] \to f_{2}(1270) \, f_{2}(1270) $    \\

	\hline\hline
	\end{tabular}
	\label{tab:components4pi}
\end{table}

\begin{table}[h]
  \footnotesize
  \centering
  \caption{Decays considered in $\Dz \to \KKpipi$ LASSO model building. For cascade non-self-conjugate channels, the conjugate partner is implied.\label{tab:componentsKKpipi}}
           \begin{tabular} {@{\hspace{0.5cm}}l@{\hspace{0.25cm}}}
     \hline \hline
	 Decay channel  \\ \hline
              $\Dz \to K^- \, [K^{*}(1410)^+\to \pi^+\, K^{*}(892)^{0}]$ \\
              $\Dz \to K^- \, [K_{1}(1270)^{+}[S,D]\to \pi^+ \, K^{*}(892)^{0}]$ \\
              $\Dz \to K^- \, [K_{1}(1270)^{+}[S,D]\to \pi^+ \, K^{*}(1430)^{0}]$ \\
              $\Dz \to K^- \, [K_{1}(1270)^{+}[S,D]\to K^+ \, \rho(770)^{0}]$ \\
              $\Dz \to K^- \, [K_{1}(1270)^{+}[S,D]\to K^+ \, \omega(782)]$ \\
              $\Dz \to K^- \, [K_{1}(1400)^{+}[S,D]\to \pi^+ \, K^{*}(892)^{0}]$ \\
              $\Dz \to K^- \, [K_{2}^*(1430)^{+}\to \pi^+ \, K^{*}(892)^{0}]$ \\
              $\Dz \to K^- \, [K_{2}^*(1430)^{+}\to K^+ \, \rho(770)^{0}]$ \\
              $\Dz \to K^- \, [K^*(1680)^{+}\to \pi^+ \, K^{*}(892)^{0}]$ \\
              $\Dz \to K^- \, [K^*(1680)^{+}\to K^+ \, \rho(770)^{0}]$ \\
              $\Dz[S,P,D] \to K^{*}(892)^{0} \, \bar{K}^{*}(892)^{0}$\\
              $\Dz[S,P,D] \to \phi(1020) \, \rho(770)^{0}$ \\
              $\Dz \to \phi(1020) \, \omega(782)$ \\
              $\Dz[P,D] \to f_2(1270)^0 \, \phi(1020)$ \\
              $\Dz \to \rho(770)^{0} \, (K^+ K^-)_{S}$\\
              $\Dz[S,P,D] \to \rho(770)^{0} \, (K^+ K^-)_{P}$\\
              $\Dz \to K^{*}(892)^{0} \, (K^- \pi^+)_{S}$ \\
              $\Dz[S,P,D] \to K^{*}(892)^{0} \, (K^- \pi^+)_{P}$ \\
              $\Dz \to \phi(1020) \, ( \pi^+ \pi^- )_{S}$ \\
              $\Dz[S,P,D] \to \phi(1020) \, ( \pi^+ \pi^- )_{P}$ \\
              $\Dz \to f_0(980) \, (\pi^+ \pi^-)_{S}$ \\
              $\Dz \to f_0(980) \, (K^+ K^-)_{S}$ \\
              $\Dz \to (K^+K^-)_{S} \, (\pi^+ \pi^-)_{S}$ \\
              \hline \hline
           \end{tabular}
\end{table}

\clearpage\section{Alternative Fit Models}
\label{a:alternative}

The fit fractions and \chisq values of the baseline and several alternative models are summarized in Tables~\ref{tab:alternativeModels1}-\ref{tab:alternativeModels2}.

\renewcommand{\thetable}{D.\arabic{table}}

\begin{table}[h]
  \scriptsize
  \centering
       	\caption{\small Fit fractions in percent for each component of specific alternative models for \Dz \to \fourpi.		
	Resonance parameters,  $F_{+}^{4\pi} $  and  $\chi^{2}/\nu$  are also given.
        The uncertainties are statistical only.
        \label{tab:alternativeModels1}}
	\begin{tabular}
        {@{\hspace{0.5cm}}l@{\hspace{0.25cm}}  @{\hspace{0.25cm}}c@{\hspace{0.25cm}}  @{\hspace{0.25cm}}c@{\hspace{0.25cm}}  @{\hspace{0.25cm}}c@{\hspace{0.25cm}}  @{\hspace{0.25cm}}c@{\hspace{0.5cm}}}
     \hline \hline
	Decay mode & Extended & No $\pi(1300)$   & No $a_{1}(1640)$ & FOCUS \\ \hline
	$\Dz \to \pim \, \left[ a_{1}(1260)^{+}\to \pip \, \rho(770)^{0} \right] $ &$37.3 \pm 1.9$ &$41.0 \pm 2.7$ & $36.7 \pm 2.9$ &  $38.2 \pm 2.8$   \\
	$\Dz \to \pim \, \left[ a_{1}(1260)^{+}[D] \to \pip \, \rho(770)^{0} \right] $ & - & - &$2.6 \pm 0.5$ &   $7.0 \pm 1.2$ \\
	$\Dz \to \pim \, \left[ a_{1}(1260)^{+} \to \pip \, \sigma \right] $ &$8.1 \pm 1.2$ &$ 5.5 \pm 0.7$ &$5.1 \pm 0.8$ &   $6.6 \pm 0.9$\\
	$\Dz \to \pip \, \left[ a_{1}(1260)^{-}\to \pim \, \rho(770)^{0} \right]  $ & $2.1 \pm 0.4$ &$ 3.0 \pm 0.5$ & $1.0 \pm 0.2$ &   - \\
	$\Dz \to \pip \,  \left[ a_{1}(1260)^{-}[D] \to \pim \, \rho(770)^{0} \right]  $ & - & - & $0.07 \pm 0.04$ &   - \\
	$\Dz \to \pip \, \left[ a_{1}(1260)^{-} \to \pim \, \sigma \right]  $ & $0.5 \pm 0.2 $ &$ 0.4 \pm 0.2$ & $0.14 \pm 0.06$ &   - \\

	$\Dz \to \pim \, \left[ \pi(1300)^{+} \to \pip \, \sigma \right] $ & $8.6 \pm 0.9$ & - & $10.7 \pm 1.8$  & -\\
	$\Dz \to \pip \, \left[ \pi(1300)^{-} \to \pim \, \sigma \right] $ &  $5.0 \pm 0.7$ & - & $ 2.8 \pm 0.8$  & - \\

	$\Dz \to \pim \, \left[ a_{1}(1640)^{+}[D] \to \pip \, \rho(770)^{0} \right] $ & $2.9 \pm 0.4$ &$6.5 \pm 0.8$ & - &  - \\
	$\Dz \to \pim \, \left[ a_{1}(1640)^{+} \to \pip \, \sigma \right] $ & $3.0 \pm 0.7$ & - &- &  - \\	
	$\Dz \to \pim \, \left[ a_{1}(1640)^{+}\to \pip \, f_{2}(1270) \right] $ & - &$2.1 \pm 0.8$ &- &  -   \\
	$\Dz \to \pip \, \left[ a_{1}(1640)^{-}[D] \to \pim \, \rho(770)^{0} \right]   $ & $1.0 \pm 0.6$ & - & - &  - \\
	$\Dz \to \pip \, \left[ a_{1}(1640)^{-} \to \pim \, \sigma \right]  $ & $1.1 \pm 0.6$ & - & - &  - \\

	$\Dz \to \pim \, \left[ \pi_{2}(1670)^{+}\to \pip \, f_{2}(1270) \right] $ & $0.8 \pm 0.3$ &$2.6 \pm 0.7$ & $3.4 \pm 0.8$  & -   \\
	$\Dz \to \pim \, \left[ \pi_{2}(1670)^{+} \to \pip \, \sigma \right] $ & $3.3 \pm 0.5$ &$3.4 \pm 0.6$ & $1.0 \pm 0.3$ &   - \\
	$\Dz \to \pip \, \left[ \pi_{2}(1670)^{-}\to \pim \, f_{2}(1270) \right]  $ & $0.3 \pm 0.2$ & - & - &   - \\
	$\Dz \to \pip \, \left[ \pi_{2}(1670)^{-}\to \pim \, \sigma \right]  $ & $1.3 \pm 0.6$ & - & - &   - \\

	$\Dz \to \sigma \, (\pi \, \pi)_{S} $ &- &- &- & $24.7 \pm 2.7$ \\
	$\Dz \to \sigma \, f_{0}(1370)  $ & $26.1 \pm 1.8$ &$9.4 \pm 1.0$ & $28.4 \pm 2.8$ &  - \\

	$\Dz \to f_{0}(980) \, (\pi \, \pi)_{S} $  &- & -& -&  $4.6 \pm 1.1$\\

	$\Dz \to \sigma \,  \rho(770)^{0}  $ & $10.6 \pm 1.1$ &$6.3 \pm 0.9$ & $7.4 \pm 1.2$ &  - \\

	$\Dz[S] \to \rho(770)^{0} \, \rho(770)^{0}$  & $ 0.9 \pm 0.3$& $3.2 \pm 0.7$ & $0.8 \pm 0.4$ &  $5.0 \pm 1.4$ \\
	$\Dz[P] \to \rho(770)^{0} \, \rho(770)^{0}$  & $6.8 \pm 0.5$ &$6.5 \pm 0.6$ & $6.9 \pm 0.5$ &   $6.3 \pm 0.7$\\
	$\Dz[D] \to \rho(770)^{0} \, \rho(770)^{0}$ & $13.2 \pm 1.0$ &$3.7 \pm 0.8$ & $11.8 \pm 1.6$ & $3.2 \pm 0.8$\\
	
	$\Dz \to f_{2}(1270) \, (\pi \, \pi)_{S} $  & -& - & - &  $2.4 \pm 0.6$\\
	$\Dz \to f_{2}(1270) \,  \sigma $  &- &$1.1 \pm 0.7$ & $1.4 \pm 0.4$ &  -\\
	$\Dz \to f_{2}(1270) \,  f_{0}(980) $  &- &$4.6 \pm 1.0$ & - &  -\\

	$\Dz \to f_{2}(1270) \,  f_{2}(1270) $  & $2.1 \pm 0.4$ &$7.9 \pm 1.7$ & $4.0 \pm 0.8$&  -\\

	\hline
	Sum & $135 \pm 4$ &$107 \pm 4$& $124 \pm 5$ &  $98 \pm 4$ \\ \hline
	$m_{a_{1}(1260)}  \, (\mev/c^{2})$ & $1225 \pm 10$ &$1225 \pm 9$ & $1230 \pm 9$ & $1304 \pm 14$\\
	$\Gamma_{a_{1}(1260)} \, (\mev)$ &$  442 \pm 26$ &$460 \pm 30$ & $421 \pm 26$ &  $529 \pm 38$\\	
	$m_{\pi(1300)} \, (\mev/c^{2})$& $1093 \pm 21$ & - & $1135 \pm 22$ & - \\
	$\Gamma_{\pi(1300)} \, (\mev)$ & $314 \pm 36$& - & $308 \pm 36$ & - \\	
	$m_{a_{1}(1640)} \, (\mev/c^{2})$ & $1710 \pm 20$ &$1727 \pm 20$ &- & -\\
	$\Gamma_{a_{1}(1640)} \, (\mev)$ & $201 \pm 38$ &$141 \pm 45$ &- & - \\	
	\hline
	$\chi^{2}/\nu$ & 1.52  & 1.79 & 1.55 & 2.36\\
	$\nu$ & 217 & 223 & 223 & 237\\
	$F_{+}^{4\pi} \, (\%)$ & $70.8 \pm 0.9$ & $70.8 \pm 0.9$ & $72.6 \pm 0.9$ &  $61.7 \pm 0.8$ \\  
	\hline\hline
	\end{tabular}
\end{table}

\begin{table}[h]
  \scriptsize
  \centering
	\caption{ \small Fit fractions in percent for each component of various alternative models for $\Dz \to \fourpi$ based on fit quality.	
	Resonance parameters,  $F_{+}^{4\pi} $  and  $\chi^{2}/\nu$  are also given.
        The uncertainties are statistical only.
}
	\begin{tabular}
	{@{\hspace{0.5cm}}l@{\hspace{0.25cm}}  @{\hspace{0.25cm}}c@{\hspace{0.25cm}}  @{\hspace{0.25cm}}c@{\hspace{0.25cm}}  @{\hspace{0.25cm}}c@{\hspace{0.25cm}}  @{\hspace{0.25cm}}c@{\hspace{0.25cm}}  @{\hspace{0.25cm}}c@{\hspace{0.5cm}}}
     \hline \hline
	Decay mode & Alt. 1 & Alt. 2    &  Alt. 3   & Alt. 4  & Alt. 5  \\ \hline
	$\Dz \to \pim \, \left[ a_{1}(1260)^{+}\to \pip \, \rho(770)^{0} \right] $ & $37.1 \pm 2.3$ &${38.3 \pm 2.4}$ & $35.2 \pm 2.6$ & $38.4 \pm 2.5$ & $35.7 \pm 2.7$ \\
	$\Dz \to \pim \, \left[ a_{1}(1260)^{+} \to \pip \, \sigma \right] $ & $11.3 \pm 1.0 $&$9.8 \pm 1.2$ &  $9.4 \pm 1.2$&$11.6 \pm 1.4$ & $11.4 \pm 1.7$\\
	$\Dz \to \pip \, \left[ a_{1}(1260)^{-}\to \pim \, \rho(770)^{0} \right]$   & $2.1 \pm 0.5$ &$ 3.3 \pm 0.6$ & $3.7 \pm 0.7$ &   $3.1 \pm 0.6$ & $4.1 \pm 0.7$ \\
	$\Dz \to \pip \, \left[ a_{1}(1260)^{-} \to \pim \, \sigma \right]  $    & $0.6 \pm 0.2$ &$ 0.9 \pm 0.2$ & $1.0 \pm 0.3$ &   $0.9 \pm 0.2$  & $1.3 \pm 0.3$\\

	$\Dz \to \pim \, \left[ \pi(1300)^{+}\to \pip \, (\pip \, \pim)_{P} \right] $ &- &- &- &- & $6.4 \pm 1.3$ \\
	$\Dz \to \pim \, \left[ \pi(1300)^{+} \to \pip \, \sigma \right] $  & $8.1 \pm 1.0$  & $8.6 \pm 1.4$ & $6.0 \pm 1.0$& $7.7 \pm 1.6$ & $4.3 \pm 1.1$\\
	$\Dz \to \pip \, \left[ \pi(1300)^{-}\to \pim \, (\pip \, \pim)_{P} \right] $ &- &- &- &- & $2.5 \pm 0.5$ \\
	$\Dz \to \pip \, \left[ \pi(1300)^{-} \to \pim \, \sigma \right]  $ & $4.3 \pm 0.9$ & $4.0 \pm 1.5$& $6.8 \pm 1.6$ & $4.9 \pm 1.6$& $1.7 \pm 0.4$\\

	$\Dz \to \pim \, \left[ a_{1}(1640)^{+}[D] \to \pip \, \rho(770)^{0} \right] $ & $2.7 \pm 0.9$ &$4.5 \pm 1.5$ &$3.9 \pm 1.6$ & $ 5.2 \pm 1.1$& $3.7 \pm 1.8$\\
	$\Dz \to \pim \, \left[ a_{1}(1640)^{+} \to \pip \, \sigma \right] $ &$ 3.2 \pm 1.3$ &$1.4 \pm 0.5$ & $2.4 \pm 1.0$& $3.0 \pm 0.9$& $1.2 \pm 0.7$\\	

	$\Dz \to \pim \, \left[ \pi_{2}(1670)^{+}\to \pip \, f_{2}(1270) \right] $ &$1.8 \pm 0.5$ &$0.6 \pm 0.2$ &$1.2 \pm 0.4$ & $1.7 \pm 0.5$ & $1.6 \pm 0.4$ \\
	$\Dz \to \pim \, \left[ \pi_{2}(1670)^{+}\to \pip \, \rho(770)^{0} \right] $ &$ 2.7 \pm 0.5$ & - & - &  - & - \\
	$\Dz \to \pim \, \left[ \pi_{2}(1670)^{+} \to \pip \, \sigma \right] $ & $2.1 \pm 0.4 $& $3.9 \pm 0.6$& $3.3 \pm 0.6$ & $3.8 \pm 0.6$& $3.5 \pm 0.6$\\
	$\Dz \to \sigma \, f_{0}(1370)  $ &$20.7 \pm 2.2$ &$19.3 \pm 2.4$ & $21.3 \pm 2.4$& $21.8 \pm 2.5$&$20.4 \pm 2.1$\\

	$\Dz \to \sigma \,  \rho(770)^{0}  $ &$5.5 \pm 1.0$ & $8.7 \pm 1.2$& $8.7 \pm 1.4$ &- &$ 4.8 \pm 1.2$ \\
	$\Dz \to f_{0}(980) \,  \rho(770)^{0}  $ & - & - & $3.6 \pm 0.8$& - & -\\
	$\Dz \to f_{0}(1370) \,  \rho(770)^{0}  $ & - & - & -& $5.8 \pm 1.0$ & -\\

	$\Dz[S] \to \rho(770)^{0} \, \rho(770)^{0}$  & - & $1.5 \pm 0.4$ &$0.8 \pm 0.4$ &$1.2 \pm 0.4$ & $0.9 \pm 0.4$\\
	$\Dz[P] \to \rho(770)^{0} \, \rho(770)^{0}$  & $7.3 \pm 0.5 $&$6.8 \pm 0.5$ & $6.9 \pm 0.5$ &$6.8 \pm 0.5$ &$6.4 \pm 0.5$ \\
	$\Dz[D] \to \rho(770)^{0} \, \rho(770)^{0}$ & $10.4 \pm 0.9$& $8.3 \pm 1.0$& $11.4 \pm 1.4$ &$10.9 \pm 1.2$ & $16.0 \pm 2.1$\\

	$\Dz \to f_{2}(1270) \,  f_{2}(1270) $  & $2.5 \pm 0.5$ & - & $1.2 \pm 0.3$ & $1.4 \pm 0.4$ & $1.1 \pm 0.3$\\

	\hline
	Sum & $122 \pm 4$ & $120 \pm 3$& $127 \pm 4$ & $128 \pm 4$ &$127 \pm 6$ \\\hline
	$m_{a_{1}(1260)} \, (\mev/c^{2})$ & $1198 \pm 8$ & $1220 \pm 8$ &$1213 \pm 9$ & $1215 \pm 8$ &  $1231 \pm 9 $\\
	$\Gamma_{a_{1}(1260)} \, (\mev)$ & $429 \pm 24$ & $408\pm 23$ & $434 \pm 24$ & $420 \pm 24$&$459 \pm 25 $ \\	
	$m_{\pi(1300)} \, (\mev/c^{2})$& $1110 \pm 17 $& $1079 \pm 25$ & $1075 \pm 22$& $1077 \pm 36$& $1180 \pm 15 $\\
	$\Gamma_{\pi(1300)} \, (\mev)$ & $314 \pm 39$ & $347 \pm 40$ & $330 \pm 39$& $377 \pm 41$ & $297 \pm 36 $ \\	
	$m_{a_{1}(1640)} \, (\mev/c^{2})$ & $1694 \pm 19$ & $1681 \pm 18$& $1672 \pm 22$& $1686 \pm 18$&  $1644 \pm 16 $\\
	$\Gamma_{a_{1}(1640)} \, (\mev)$ & $177 \pm 45$ & $ 171 \pm 36  $ & $250 \pm 59$& $209 \pm 28$& $222 \pm 56 $\\	
	\hline
	$\chi^{2}/\nu$ & $1.50$ & $1.42$& $1.43$ & $1.50$ & $1.33$ \\
	$\nu$ & 221 & 223 & 219 & 221 & 219\\
	$F_{+}^{4\pi} \, (\%)$ &$71.7 \pm 0.9 $ & $72.9 \pm 0.9$ & $73.0 \pm 0.9$ & $73.3 \pm 0.9$ & $73.5 \pm 0.9 $\\  
	\hline\hline
	\end{tabular}
\end{table}

\begin{table}[h]
  \scriptsize
  \centering
  \caption{ \small Fit fractions in percent for each component of various alternative models for $\Dz \to \KKpipi$ based on fit quality. 
  	The $F_{+}^{KK\pi\pi}$  and  $\chi^{2}/\nu$ values are also given.
        The uncertainties are statistical only. \label{tab:alternativeModels2}
}
        \begin{tabular}
	{@{\hspace{0.5cm}}l@{\hspace{0.25cm}}  @{\hspace{0.25cm}}c@{\hspace{0.25cm}}  @{\hspace{0.25cm}}c@{\hspace{0.25cm}}  @{\hspace{0.25cm}}c@{\hspace{0.25cm}}  @{\hspace{0.25cm}}c@{\hspace{0.5cm}}}
        \hline \hline
        Decay Mode& Model A & Model B & Model C & Model D\\ \hline
        \hline
$\Dz \to K^- \, [K_{1}(1270)^{+} \to \pi^+ \, K^{*}(892)^{0}]$&	5.76 $\pm$	1.65 &	6.06 $\pm$	1.45 &	8.23 $\pm$	1.29 &	9.38 $\pm$	0.98 \\
$\Dz \to K^+ \, [{K}_{1}(1270)^{-} \to \pi^- \, \bar{K}^{*}(892)^{0}]$&	1.12 $\pm$	0.76 &	-	&	-	&	0.50 $\pm$	0.28 \\
$\Dz \to K^- \, [K_{1}(1270)^{+} \to \pi^+ \, K^{*}(1430)^{0}]$&	5.78 $\pm$	1.63 &	6.31 $\pm$	1.20 &	9.51 $\pm$	1.64 &	-	\\
$\Dz \to K^+ \, [{K}_{1}(1270)^{-} \to \pi^- \, \bar{K}^{*}(1430)^{0}]$&	0.69 $\pm$	0.60 &	-	&	-	&	-	\\
$\Dz \to K^- \, [K_{1}(1270)^{+} \to K^+ \, \omega(782)]$&	0.78 $\pm$	0.41 &	0.58 $\pm$	0.26 &	0.94 $\pm$	0.34& 	-	\\
$\Dz \to K^+ \, [{K}_{1}(1270)^{-} \to K^- \, \omega(782)]$&	0.39 $\pm$	0.37 &	-	&	-	&	-	\\
$\Dz \to K^- \, [K_{1}(1270)^{+} \to K^+ \, \rho(770)^{0}]$&	9.06 $\pm$	1.85 &	9.43 $\pm$	1.56 &	10.45 $\pm$	1.79 &	7.58 $\pm$	0.95 \\
$\Dz \to K^+ \, [{K}_{1}(1270)^{-} \to K^- \, \rho(770)^{0}]$&	1.42 $\pm$	0.76 &	4.84 $\pm$	0.73 &	5.05 $\pm$	0.83 &	6.10 $\pm$	0.83 \\
$\Dz \to K^- \, [K_{1}(1400)^{+} \to \pi^+ \, K^{*}(892)^{0}]$&	14.05 $\pm$	3.13 &	14.51 $\pm$	2.82 &	22.28 $\pm$	3.52 &	-	\\
$\Dz \to K^+ \, [{K}_{1}(1400)^{-} \to \pi^- \, \bar{K}^{*}(892)^{0}]$&	1.17 $\pm$	1.00 &	-	&	-	&	-	\\
$\Dz \to K^- \, [K^*(1680)^{+} \to \pi^+ \, K^{*}(892)^{0}]$&	2.97 $\pm$	0.95 &	-	&	4.60 $\pm$	0.92 &	-	\\
$\Dz \to K^+ \, [\bar{K}^*(1680)^{+} \to \pi^- \, \bar{K}^{*}(892)^{0}]$&	0.68 $\pm$	0.43 &	-	&	-	&	-	\\

$\Dz[S] \to K^{*}(892)^{0} \, \bar{K}^{*}(892)^{0}$ &	4.60 $\pm$	1.19 &	4.54 $\pm$	0.77 &	4.84 $\pm$ 0.81 &	9.14 $\pm$	1.29 \\
$\Dz[P] \to K^{*}(892)^{0} \, \bar{K}^{*}(892)^{0}$&	3.06 $\pm$	1.10 &	3.91 $\pm$	0.70 &	5.14 $\pm$	0.78 &	-	\\
$\Dz[D] \to K^{*}(892)^{0} \, \bar{K}^{*}(892)^{0}$&	3.55 $\pm$	0.75 &	3.83 $\pm$	0.63 &	5.08 $\pm$	0.76& 	-\\
$\Dz[S] \to \phi(1020) \, \rho(770)^{0}$&	27.13 $\pm$	1.59 &	27.47 $\pm$	1.32 &	27.66 $\pm$ 	1.35 &	31.08 $\pm$	1.38 \\
$\Dz[P] \to \phi(1020) \, \rho(770)^{0}$&	1.91 $\pm$	0.47 &	1.80 $\pm$	0.39 &	1.70 $\pm$	0.37 &	-	\\
$\Dz[D] \to \phi(1020) \, \rho(770)^{0}$&	1.58 $\pm$	0.46 &	1.47 $\pm$	0.42 &	1.70 $\pm$	0.45 &	2.60 $\pm$	0.61 \\

$\Dz \to K^{*}(892)^{0} \, (K^- \pi^+)_{S}$&	5.33 $\pm$ 1.40 &	5.75 $\pm$	1.21 &	6.20 $\pm$	1.34 &	-	\\
$\Dz \to \bar{K}^{*}(892)^{0} \, (K^+ \pi^-)_{S}$&	1.26 $\pm$	0.83 &	-	&	-	&	-	\\
$\Dz \to \phi(1020) \, (\pi^+ \pi^-)_{S}$&	4.35 $\pm$	0.85 &	4.47 $\pm$	0.69& 	5.40 $\pm$	0.76 &	7.86 $\pm$	0.88 \\

$\Dz \to (K^+K^-)_{S} \, (\pi^+ \pi^-)_{S}$&	10.14 $\pm$	1.41 &	10.82 $\pm$	1.22 &	-	&	-	\\

$\Dz \to K^- \, [K_{1}(1410)^{+} \to \pi^+ \, K^{*}(892)^{0}]$&	-	&	3.35 $\pm$	0.78 &	-	&	3.23 $\pm$	0.69 \\
$\Dz \to K^+ \, [{K}_{1}(1410)^{-} \to \pi^- \, \bar{K}^{*}(892)^{0}]$&-	&	-	&	-	&	5.55 $\pm$	0.77 \\

$\Dz \to f_0(980) \, (\pi^+ \pi^-)_{S}$&	-	&	-	&	1.32 $\pm$	0.76 &	-	\\
$\Dz \to f_0(980) \, (K^+ K^-)_{S}$&	-	&	-	&	1.01 $\pm$	0.64 &	-	\\
$\Dz \to (K^-\pi^+)_{P}\, (K^+\pi^-)_{S}$&	-	&	-	&	-	&	10.69 $\pm$	1.10 \\
\hline
Sum&	106.76 $\pm$	5.83&	109.13 $\pm$	4.70&	121.11 $\pm$ 5.38	&	93.72 $\pm$ 3.10	\\	\hline
$\chi^2/\nu$ &1.490 &1.503 &1.707&1.754 \\
$\nu$ & 116 & 116 & 116 & 116 \\
	$F_{+}^{KK\pi\pi} \, (\%)$ &$77.5 \pm 3.0 $ & $74.2 \pm 1.9$ & $68.1 \pm 2.0$ & $73.8 \pm 2.0$ \\  
        \hline\hline
        \end{tabular}
\end{table}

\clearpage\section{Interference Fractions}
\label{a:interference}

\renewcommand{\thetable}{E.\arabic{table}}

Tables \ref{tab:interferenceFractions4pi}-\ref{tab:interferenceFractionsKKpipi2} list the interference fractions,
ordered by magnitude, for the nominal models of \prt{\Dz \to \fourpi}
and \prt{\Dz \to \KKpipi}.

\begin{table}[h]
  \footnotesize
  \centering
  \caption{\small Interference fractions $|I_{ij}| > 0.5 \%$, as defined in
    \eqnPRDref{eq:DefineInterferenceFractions}, ordered by magnitude, for
    the nominal $D\to \fourpi$ amplitude fit. Only the statistical uncertainties are given.
\label{tab:interferenceFractions4pi}
}
  \tiny
  \begin{tabular}
     {@{\hspace{0.5cm}}l@{\hspace{0.25cm}}  @{\hspace{0.25cm}}l@{\hspace{0.25cm}}  @{\hspace{0.25cm}}c@{\hspace{0.25cm}}}
     \hline \hline
Channel $i$ & Channel $j$ & $I_{ij}~(\%)$\\
\hline
%
% start input /Users/pnaik/Documents/CLEO/Papers/latexpand/latex/interferenceFractions1.tex
(1) $D^0 \to \pi^- [a_1(1260)^+ \to \sigma \pi^+]$ & $D^0 \to \pi^- [a_1(1260)^+ \to \rho(770)^0 \pi^+ ]$ & 20.010 $\pm$ 1.186 \\
(2) $D^0 \to \pi^- [\pi(1300)^+ \to \sigma \pi^+]$ & $D^0 \to f_0(1370) \, \sigma$ & -10.766 $\pm$ 0.835 \\
(3) $D^0 \to \rho(770)^0 \, \sigma$ & $D^0 \to \pi^- [a_1(1260)^+ \to \rho(770)^0 \pi^+ ]$ & -6.942 $\pm$ 0.752 \\
(4) $D^0 \to \pi^- [a_1(1260)^+ \to \sigma \pi^+]$ & $D^0 \to \pi^- [a_1(1640)^+ \to \sigma \pi^+]$ & -6.150 $\pm$ 1.186 \\
(5) $D^0 \to \pi^- [a_1(1260)^+ \to \rho(770)^0 \pi^+ ]$ & $D^0[D] \to \rho(770)^0 \, \rho(770)^0 $ & -5.244 $\pm$ 0.331 \\
(6) $D^0 \to \pi^- [a_1(1640)^+ \to \sigma \pi^+]$ & $D^0 \to \pi^- [a_1(1260)^+ \to \rho(770)^0 \pi^+ ]$ & -5.072 $\pm$ 0.686 \\
(7) $D^0 \to \pi^+ [\pi(1300)^- \to \sigma \pi^-]$ & $D^0 \to f_0(1370) \, \sigma$ & -4.495 $\pm$ 0.872 \\
(8) $D^0 \to \pi^- [a_1(1260)^+ \to \sigma \pi^+]$ & $D^0[D] \to \rho(770)^0 \, \rho(770)^0 $ & -4.301 $\pm$ 0.335 \\
(9) $D^0 \to \pi^- [\pi_2(1670)^+ \to \sigma \pi^+]$ & $D^0 \to \pi^- [\pi_2(1670)^+ \to f_2(1270)\pi^+]$ & -3.058 $\pm$ 0.429 \\
(10) $D^0 \to \pi^- [\pi(1300)^+ \to \sigma \pi^+]$ & $D^0 \to \pi^+ [\pi(1300)^- \to \sigma \pi^-]$ & 2.897 $\pm$ 0.338 \\
(11) $D^0 \to \pi^- [a_1(1260)^+ \to \rho(770)^0 \pi^+ ]$ & $D^0 \to \pi^+ [a_1(1260)^- \to \rho(770)^0 \pi^-]$ & 2.757 $\pm$ 0.128 \\
(12) $D^0 \to \pi^- [a_1(1260)^+ \to \sigma \pi^+]$ & $D^0 \to f_0(1370) \, \sigma$ & 2.653 $\pm$ 0.186 \\
(13) $D^0 \to f_2(1270) \, f_2(1270)$ & $D^0 \to \pi^- [\pi_2(1670)^+ \to f_2(1270)\pi^+]$ & -2.604 $\pm$ 0.531 \\
(14) $D^0 \to f_0(1370) \, \sigma$ & $D^0 \to \pi^- [a_1(1260)^+ \to \rho(770)^0 \pi^+ ]$ & 2.418 $\pm$ 0.135 \\
(15) $D^0 \to \pi^- [\pi_2(1670)^+ \to \sigma \pi^+]$ & $D^0 \to f_2(1270) \, f_2(1270)$ & 2.189 $\pm$ 0.273 \\
(16) $D^0[S] \to \rho(770)^0 \, \rho(770)^0 $ & $D^0[D] \to \rho(770)^0 \, \rho(770)^0 $ & 2.046 $\pm$ 0.438 \\
(17) $D^0 \to \pi^- [a_1(1640)^+[D] \to \rho(770)^0 \pi^+] $ & $D^0[D] \to \rho(770)^0 \, \rho(770)^0 $ & 1.995 $\pm$ 0.323 \\
(18) $D^0 \to \pi^- [a_1(1260)^+ \to \sigma \pi^+]$ & $D^0[S] \to \rho(770)^0 \, \rho(770)^0 $ & -1.805 $\pm$ 0.388 \\
(19) $D^0 \to \pi^+ [a_1(1260)^- \to \rho(770)^0 \pi^-]$ & $D^0[D] \to \rho(770)^0 \, \rho(770)^0 $ & -1.753 $\pm$ 0.089 \\
(20) $D^0 \to \pi^+ [a_1(1260)^- \to \rho(770)^0 \pi^-]$ & $D^0[S] \to \rho(770)^0 \, \rho(770)^0 $ & -1.747 $\pm$ 0.294 \\
(21) $D^0 \to \pi^+ [a _1(1260)^- \to \sigma \pi^-]$ & $D^0 \to \pi^+ [a_1(1260)^- \to \rho(770)^0 \pi^-]$ & 1.612 $\pm$ 0.095 \\
(22) $D^0 \to \pi^- [a_1(1260)^+ \to \sigma \pi^+]$ & $D^0 \to \pi^+ [a_1(1260)^- \to \rho(770)^0 \pi^-]$ & 1.600 $\pm$ 0.070 \\
(23) $D^0 \to \pi^- [a_1(1260)^+ \to \sigma \pi^+]$ & $D^0 \to \pi^+ [a _1(1260)^- \to \sigma \pi^-]$ & 1.511 $\pm$ 0.172 \\
(24) $D^0 \to f_0(1370) \, \sigma$ & $D^0[D] \to \rho(770)^0 \, \rho(770)^0 $ & -1.403 $\pm$ 0.096 \\
(25) $D^0 \to \pi^- [a_1(1260)^+ \to \sigma \pi^+]$ & $D^0 \to \pi^- [\pi_2(1670)^+ \to \sigma \pi^+]$ & 1.333 $\pm$ 0.120 \\
(26) $D^0 \to \pi^- [a_1(1640)^+[D] \to \rho(770)^0 \pi^+] $ & $D^0 \to f_2(1270) \, f_2(1270)$ & 1.286 $\pm$ 0.146 \\
(27) $D^0 \to \pi^+ [a_1(1260)^- \to \rho(770)^0 \pi^-]$ & $D^0 \to \pi^- [a_1(1640)^+[D] \to \rho(770)^0 \pi^+] $ & -1.219 $\pm$ 0.088 \\
(28) $D^0 \to \pi^- [a_1(1260)^+ \to \sigma \pi^+]$ & $D^0 \to \pi^+ [\pi(1300)^- \to \sigma \pi^-]$ & 1.192 $\pm$ 0.159 \\
(29) $D^0 \to \pi^- [a_1(1260)^+ \to \sigma \pi^+]$ & $D^0 \to \pi^- [\pi_2(1670)^+ \to f_2(1270)\pi^+]$ & -1.188 $\pm$ 0.161 \\
(30) $D^0 \to \pi^- [a_1(1260)^+ \to \sigma \pi^+]$ & $D^0 \to \pi^- [\pi(1300)^+ \to \sigma \pi^+]$ & -1.149 $\pm$ 0.097 \\
(31) $D^0 \to \pi^+ [a _1(1260)^- \to \sigma \pi^-]$ & $D^0[S] \to \rho(770)^0 \, \rho(770)^0 $ & -1.072 $\pm$ 0.124 \\
(32) $D^0 \to \pi^+ [a _1(1260)^- \to \sigma \pi^-]$ & $D^0 \to \rho(770)^0 \, \sigma$ & -1.029 $\pm$ 0.116 \\
(33) $D^0 \to \pi^- [a_1(1640)^+[D] \to \rho(770)^0 \pi^+] $ & $D^0[S] \to \rho(770)^0 \, \rho(770)^0 $ & -1.011 $\pm$ 0.129 \\
(34) $D^0 \to \pi^- [a_1(1640)^+ \to \sigma \pi^+]$ & $D^0 \to f_0(1370) \, \sigma$ & -1.000 $\pm$ 0.162 \\
(35) $D^0 \to \pi^- [a_1(1640)^+ \to \sigma \pi^+]$ & $D^0[D] \to \rho(770)^0 \, \rho(770)^0 $ & 0.966 $\pm$ 0.148 \\
(36) $D^0 \to \pi^- [\pi_2(1670)^+ \to \sigma \pi^+]$ & $D^0 \to f_0(1370) \, \sigma$ & -0.959 $\pm$ 0.081 \\
(37) $D^0 \to \pi^+ [a _1(1260)^- \to \sigma \pi^-]$ & $D^0[D] \to \rho(770)^0 \, \rho(770)^0 $ & -0.907 $\pm$ 0.098 \\
(38) $D^0 \to \pi^- [a_1(1640)^+ \to \sigma \pi^+]$ & $D^0 \to \pi^+ [a_1(1260)^- \to \rho(770)^0 \pi^-]$ & -0.892 $\pm$ 0.119 \\
(39) $D^0 \to \pi^+ [\pi(1300)^- \to \sigma \pi^-]$ & $D^0 \to \pi^- [\pi_2(1670)^+ \to \sigma \pi^+]$ & -0.865 $\pm$ 0.123 \\
(40) $D^0 \to \pi^- [\pi(1300)^+ \to \sigma \pi^+]$ & $D^0 \to \pi^- [a_1(1260)^+ \to \rho(770)^0 \pi^+ ]$ & -0.837 $\pm$ 0.096 \\
(41) $D^0 \to \pi^+ [a _1(1260)^- \to \sigma \pi^-]$ & $D^0 \to \pi^- [a_1(1640)^+ \to \sigma \pi^+]$ & -0.815 $\pm$ 0.184 \\
(42) $D^0 \to f_0(1370) \, \sigma$ & $D^0 \to \pi^+ [a_1(1260)^- \to \rho(770)^0 \pi^-]$ & 0.801 $\pm$ 0.033 \\
(43) $D^0 \to \rho(770)^0 \, \sigma$ & $D^0 \to \pi^- [a_1(1640)^+[D] \to \rho(770)^0 \pi^+] $ & 0.780 $\pm$ 0.115 \\
(44) $D^0 \to \pi^- [a_1(1640)^+ \to \sigma \pi^+]$ & $D^0 \to \pi^- [\pi(1300)^+ \to \sigma \pi^+]$ & 0.752 $\pm$ 0.104 \\
(45) $D^0 \to \pi^- [\pi_2(1670)^+ \to \sigma \pi^+]$ & $D^0 \to \pi^+ [a_1(1260)^- \to \rho(770)^0 \pi^-]$ & -0.689 $\pm$ 0.054 \\
(46) $D^0 \to \pi^- [a_1(1260)^+ \to \rho(770)^0 \pi^+ ]$ & $D^0 \to f_2(1270) \, f_2(1270)$ & 0.673 $\pm$ 0.073 \\
(47) $D^0 \to \pi^- [a_1(1640)^+ \to \sigma \pi^+]$ & $D^0 \to \pi^+ [\pi(1300)^- \to \sigma \pi^-]$ & -0.672 $\pm$ 0.155 \\
(48) $D^0 \to f_0(1370) \, \sigma$ & $D^0[S] \to \rho(770)^0 \, \rho(770)^0 $ & -0.665 $\pm$ 0.111 \\
(49) $D^0 \to \rho(770)^0 \, \sigma$ & $D^0 \to \pi^+ [a_1(1260)^- \to \rho(770)^0 \pi^-]$ & -0.649 $\pm$ 0.194 \\
(50) $D^0 \to \pi^- [a_1(1260)^+ \to \rho(770)^0 \pi^+ ]$ & $D^0 \to \pi^- [\pi_2(1670)^+ \to f_2(1270)\pi^+]$ & -0.634 $\pm$ 0.154 \\
(51) $D^0 \to \pi^- [\pi_2(1670)^+ \to \sigma \pi^+]$ & $D^0[S] \to \rho(770)^0 \, \rho(770)^0 $ & 0.627 $\pm$ 0.082 \\
(52) $D^0 \to \pi^- [a_1(1640)^+ \to \sigma \pi^+]$ & $D^0 \to f_2(1270) \, f_2(1270)$ & -0.623 $\pm$ 0.144 \\
(53) $D^0 \to \pi^- [a_1(1640)^+ \to \sigma \pi^+]$ & $D^0[S] \to \rho(770)^0 \, \rho(770)^0 $ & 0.616 $\pm$ 0.169 \\
(54) $D^0 \to \pi^- [\pi_2(1670)^+ \to \sigma \pi^+]$ & $D^0 \to \pi^- [a_1(1640)^+[D] \to \rho(770)^0 \pi^+] $ & -0.613 $\pm$ 0.063 \\
(55) $D^0 \to \pi^- [\pi(1300)^+ \to \sigma \pi^+]$ & $D^0[D] \to \rho(770)^0 \, \rho(770)^0 $ & -0.609 $\pm$ 0.067 \\
(56) $D^0 \to \pi^- [a_1(1260)^+ \to \sigma \pi^+]$ & $D^0 \to f_2(1270) \, f_2(1270)$ & 0.592 $\pm$ 0.130 \\
(57) $D^0 \to \pi^- [a_1(1640)^+ \to \sigma \pi^+]$ & $D^0 \to \pi^- [\pi_2(1670)^+ \to \sigma \pi^+]$ & -0.574 $\pm$ 0.094 \\
(58) $D^0 \to \pi^- [\pi_2(1670)^+ \to \sigma \pi^+]$ & $D^0 \to \rho(770)^0 \, \sigma$ & 0.522 $\pm$ 0.103 \\
(59) $D^0 \to \rho(770)^0 \, \sigma$ & $D^0 \to \pi^- [\pi_2(1670)^+ \to f_2(1270)\pi^+]$ & -0.521 $\pm$ 0.088 \\
(60) $D^0 \to \pi^+ [\pi(1300)^- \to \sigma \pi^-]$ & $D^0 \to \pi^- [a_1(1260)^+ \to \rho(770)^0 \pi^+ ]$ & 0.515 $\pm$ 0.054 \\
(61) $D^0 \to \pi^- [a_1(1640)^+ \to \sigma \pi^+]$ & $D^0 \to \pi^- [a_1(1640)^+[D] \to \rho(770)^0 \pi^+] $ & 0.513 $\pm$ 0.129 \\
(62) $D^0 \to \pi^- [a_1(1260)^+ \to \rho(770)^0 \pi^+ ]$ & $D^0 \to \pi^- [a_1(1640)^+[D] \to \rho(770)^0 \pi^+] $ & 0.507 $\pm$ 0.074 \\
 % end input /Users/pnaik/Documents/CLEO/Papers/latexpand/latex/interferenceFractions1.tex
 	\hline\hline
	\end{tabular}
\end{table}

\begin{table}[h]
  \footnotesize
  \centering
  \caption{\small Interference fractions $|I_{ij}| \leq 0.5 \%$, as defined in
    \eqnPRDref{eq:DefineInterferenceFractions}, ordered by magnitude, for
    the $D\to \fourpi$ amplitude fit using the LASSO model. Only the statistical uncertainties are given.
\label{tab:interferenceFractions4pi2}
}
  \tiny
  \begin{tabular}
     {@{\hspace{0.5cm}}l@{\hspace{0.25cm}}  @{\hspace{0.25cm}}l@{\hspace{0.25cm}}  @{\hspace{0.25cm}}c@{\hspace{0.25cm}}}
     \hline \hline
Channel $i$ & Channel $j$ & $I_{ij}~(\%)$\\
\hline
%
% start input /Users/pnaik/Documents/CLEO/Papers/latexpand/latex/interferenceFractions2.tex
(63) $D^0 \to \pi^- [a_1(1260)^+ \to \sigma \pi^+]$ & $D^0 \to \rho(770)^0 \, \sigma$ & -0.497 $\pm$ 0.373 \\
(64) $D^0[S] \to \rho(770)^0 \, \rho(770)^0 $ & $D^0 \to \pi^- [\pi_2(1670)^+ \to f_2(1270)\pi^+]$ & 0.496 $\pm$ 0.088 \\
(65) $D^0 \to \pi^- [\pi(1300)^+ \to \sigma \pi^+]$ & $D^0 \to \pi^- [\pi_2(1670)^+ \to f_2(1270)\pi^+]$ & 0.492 $\pm$ 0.054 \\
(66) $D^0 \to \pi^+ [a _1(1260)^- \to \sigma \pi^-]$ & $D^0 \to f_0(1370) \, \sigma$ & 0.452 $\pm$ 0.064 \\
(67) $D^0 \to \pi^- [a_1(1260)^+ \to \rho(770)^0 \pi^+ ]$ & $D^0[S] \to \rho(770)^0 \, \rho(770)^0 $ & 0.420 $\pm$ 0.873 \\
(68) $D^0[S] \to \rho(770)^0 \, \rho(770)^0 $ & $D^0 \to f_2(1270) \, f_2(1270)$ & -0.402 $\pm$ 0.103 \\
(69) $D^0 \to \pi^+ [\pi(1300)^- \to \sigma \pi^-]$ & $D^0[D] \to \rho(770)^0 \, \rho(770)^0 $ & -0.399 $\pm$ 0.046 \\
(70) $D^0 \to \pi^+ [a _1(1260)^- \to \sigma \pi^-]$ & $D^0 \to \pi^- [\pi(1300)^+ \to \sigma \pi^+]$ & 0.399 $\pm$ 0.057 \\
(71) $D^0 \to \pi^+ [a _1(1260)^- \to \sigma \pi^-]$ & $D^0 \to \pi^- [a_1(1640)^+[D] \to \rho(770)^0 \pi^+] $ & -0.393 $\pm$ 0.035 \\
(72) $D^0 \to \pi^- [\pi(1300)^+ \to \sigma \pi^+]$ & $D^0 \to \rho(770)^0 \, \sigma$ & -0.388 $\pm$ 0.238 \\
(73) $D^0 \to \pi^- [a_1(1260)^+ \to \sigma \pi^+]$ & $D^0 \to \pi^- [a_1(1640)^+[D] \to \rho(770)^0 \pi^+] $ & -0.333 $\pm$ 0.138 \\
(74) $D^0 \to \pi^- [a_1(1640)^+ \to \sigma \pi^+]$ & $D^0 \to \rho(770)^0 \, \sigma$ & 0.333 $\pm$ 0.241 \\
(75) $D^0 \to \pi^- [\pi_2(1670)^+ \to \sigma\pi^+]$ & $D^0[D] \to \rho(770)^0 \, \rho(770)^0 $ & 0.327 $\pm$ 0.051 \\
(76) $D^0 \to f_0(1370) \, \sigma$ & $D^0 \to \pi^- [\pi_2(1670)^+ \to f_2(1270)\pi^+]$ & 0.318 $\pm$ 0.033 \\
(77) $D^0 \to \pi^- [\pi_2(1670)^+ \to \sigma\pi^+]$ & $D^0 \to \pi^- [a_1(1260)^+ \to \rho(770)^0 \pi^+ ]$ & 0.314 $\pm$ 0.054 \\
(78) $D^0 \to \pi^+ [\pi(1300)^- \to \sigma \pi^-]$ & $D^0 \to \rho(770)^0 \, \sigma$ & -0.313 $\pm$ 0.207 \\
(79) $D^0 \to \pi^+ [a _1(1260)^- \to \sigma \pi^-]$ & $D^0 \to \pi^- [\pi_2(1670)^+ \to \sigma\pi^+]$ & -0.283 $\pm$ 0.032 \\
(80) $D^0 \to \pi^+ [\pi(1300)^- \to \sigma \pi^-]$ & $D^0 \to \pi^+ [a_1(1260)^- \to \rho(770)^0 \pi^-]$ & -0.245 $\pm$ 0.026 \\
(81) $D^0 \to \pi^- [\pi(1300)^+ \to \sigma \pi^+]$ & $D^0 \to \pi^- [\pi_2(1670)^+ \to \sigma\pi^+]$ & -0.243 $\pm$ 0.037 \\
(82) $D^0 \to \pi^- [\pi(1300)^+ \to \sigma \pi^+]$ & $D^0 \to \pi^+ [a_1(1260)^- \to \rho(770)^0 \pi^-]$ & 0.236 $\pm$ 0.014 \\
(83) $D^0 \to \pi^+ [\pi(1300)^- \to \sigma \pi^-]$ & $D^0 \to \pi^- [a_1(1640)^+[D] \to \rho(770)^0 \pi^+] $ & -0.233 $\pm$ 0.031 \\
(84) $D^0 \to \pi^- [a_1(1640)^+ \to \sigma \pi^+]$ & $D^0 \to \pi^- [\pi_2(1670)^+ \to f_2(1270)\pi^+]$ & 0.229 $\pm$ 0.061 \\
(85) $D^0 \to \pi^- [\pi(1300)^+ \to \sigma \pi^+]$ & $D^0 \to \pi^- [a_1(1640)^+[D] \to \rho(770)^0 \pi^+] $ & 0.226 $\pm$ 0.029 \\
(86) $D^0 \to \pi^+ [a _1(1260)^- \to \sigma \pi^-]$ & $D^0 \to \pi^- [\pi_2(1670)^+ \to f_2(1270)\pi^+]$ & 0.187 $\pm$ 0.022 \\
(87) $D^0 \to \pi^+ [\pi(1300)^- \to \sigma \pi^-]$ & $D^0 \to \pi^- [\pi_2(1670)^+ \to f_2(1270)\pi^+]$ & 0.180 $\pm$ 0.030 \\
(88) $D^0 \to \pi^+ [a _1(1260)^- \to \sigma \pi^-]$ & $D^0 \to \pi^+ [\pi(1300)^- \to \sigma \pi^-]$ & -0.173 $\pm$ 0.024 \\
(89) $D^0 \to \rho(770)^0 \, \sigma$ & $D^0[D] \to \rho(770)^0 \, \rho(770)^0 $ & 0.171 $\pm$ 0.012 \\
(90) $D^0[D] \to \rho(770)^0 \, \rho(770)^0 $ & $D^0 \to f_2(1270) \, f_2(1270)$ & 0.152 $\pm$ 0.115 \\
(91) $D^0 \to \pi^+ [a _1(1260)^- \to \sigma \pi^-]$ & $D^0 \to \pi^- [a_1(1260)^+ \to \rho(770)^0 \pi^+ ]$ & 0.146 $\pm$ 0.085 \\
(92) $D^0 \to \pi^+ [a_1(1260)^- \to \rho(770)^0 \pi^-]$ & $D^0 \to \pi^- [\pi_2(1670)^+ \to f_2(1270)\pi^+]$ & 0.143 $\pm$ 0.021 \\
(93) $D^0 \to \pi^- [a_1(1640)^+[D] \to \rho(770)^0 \pi^+] $ & $D^0 \to \pi^- [\pi_2(1670)^+ \to f_2(1270)\pi^+]$ & -0.128 $\pm$ 0.017 \\
(94) $D^0 \to f_0(1370) \, \sigma$ & $D^0 \to \pi^- [a_1(1640)^+[D] \to \rho(770)^0 \pi^+] $ & -0.110 $\pm$ 0.009 \\
(95) $D^0 \to \pi^+ [a_1(1260)^- \to \rho(770)^0 \pi^-]$ & $D^0 \to f_2(1270) \, f_2(1270)$ & 0.098 $\pm$ 0.022 \\
(96) $D^0 \to \pi^- [a_1(1260)^+ \to \rho(770)^0 \pi^+ ]$ & $D^0[P] \to \rho(770)^0 \, \rho(770)^0  $ & -0.096 $\pm$ 0.007 \\
(97) $D^0 \to f_0(1370) \, \sigma$ & $D^0 \to f_2(1270) \, f_2(1270)$ & -0.071 $\pm$ 0.042 \\
(98) $D^0 \to \pi^+ [a _1(1260)^- \to \sigma \pi^-]$ & $D^0 \to f_2(1270) \, f_2(1270)$ & -0.060 $\pm$ 0.032 \\
(99) $D^0 \to \pi^- [\pi(1300)^+ \to \sigma \pi^+]$ & $D^0[P] \to \rho(770)^0 \, \rho(770)^0  $ & 0.050 $\pm$ 0.003 \\
(100) $D^0 \to \pi^+ [a_1(1260)^- \to \rho(770)^0 \pi^-]$ & $D^0[P] \to \rho(770)^0 \, \rho(770)^0  $ & -0.043 $\pm$ 0.002 \\
(101) $D^0 \to \pi^- [a_1(1640)^+[D] \to \rho(770)^0 \pi^+] $ & $D^0[P] \to \rho(770)^0 \, \rho(770)^0  $ & 0.041 $\pm$ 0.003 \\
(102) $D^0 \to \rho(770)^0 \, \sigma$ & $D^0 \to f_0(1370) \, \sigma$ & 0.038 $\pm$ 0.006 \\
(103) $D^0 \to \pi^- [a_1(1260)^+ \to \sigma \pi^+]$ & $D^0[P] \to \rho(770)^0 \, \rho(770)^0  $ & -0.037 $\pm$ 0.002 \\
(104) $D^0[D] \to \rho(770)^0 \, \rho(770)^0 $ & $D^0 \to \pi^- [\pi_2(1670)^+ \to f_2(1270)\pi^+]$ & -0.035 $\pm$ 0.041 \\
(105) $D^0[S] \to \rho(770)^0 \, \rho(770)^0 $ & $D^0[P] \to \rho(770)^0 \, \rho(770)^0  $ & 0.033 $\pm$ 0.004 \\
(106) $D^0 \to f_0(1370) \, \sigma$ & $D^0[P] \to \rho(770)^0 \, \rho(770)^0  $ & -0.029 $\pm$ 0.003 \\
(107) $D^0 \to \pi^- [\pi(1300)^+ \to \sigma \pi^+]$ & $D^0[S] \to \rho(770)^0 \, \rho(770)^0 $ & -0.027 $\pm$ 0.003 \\
(108) $D^0[P] \to \rho(770)^0 \, \rho(770)^0  $ & $D^0 \to f_2(1270) \, f_2(1270)$ & 0.026 $\pm$ 0.003 \\
(109) $D^0 \to \rho(770)^0 \, \sigma$ & $D^0[S] \to \rho(770)^0 \, \rho(770)^0 $ & 0.024 $\pm$ 0.007 \\
(110) $D^0 \to \pi^+ [\pi(1300)^- \to \sigma \pi^-]$ & $D^0 \to f_2(1270) \, f_2(1270)$ & 0.019 $\pm$ 0.003 \\
(111) $D^0 \to \pi^- [\pi(1300)^+ \to \sigma \pi^+]$ & $D^0 \to f_2(1270) \, f_2(1270)$ & 0.014 $\pm$ 0.001 \\
(112) $D^0 \to \pi^+ [\pi(1300)^- \to \sigma \pi^-]$ & $D^0[S] \to \rho(770)^0 \, \rho(770)^0 $ & -0.012 $\pm$ 0.003 \\
(113) $D^0 \to \rho(770)^0 \, \sigma$ & $D^0[P] \to \rho(770)^0 \, \rho(770)^0  $ & 0.011 $\pm$ 0.001 \\
(114) $D^0 \to \pi^- [\pi_2(1670)^+ \to \sigma\pi^+]$ & $D^0[P] \to \rho(770)^0 \, \rho(770)^0  $ & -0.010 $\pm$ 0.001 \\
(115) $D^0 \to \pi^+ [a _1(1260)^- \to \sigma \pi^-]$ & $D^0[P] \to \rho(770)^0 \, \rho(770)^0  $ & -0.009 $\pm$ 0.001 \\
(116) $D^0 \to \rho(770)^0 \, \sigma$ & $D^0 \to f_2(1270) \, f_2(1270)$ & 0.009 $\pm$ 0.003 \\
(117) $D^0 \to \pi^+ [\pi(1300)^- \to \sigma \pi^-]$ & $D^0[P] \to \rho(770)^0 \, \rho(770)^0  $ & 0.006 $\pm$ 0.002 \\
(118) $D^0[D] \to \rho(770)^0 \, \rho(770)^0 $ & $D^0[P] \to \rho(770)^0 \, \rho(770)^0  $ & -0.005 $\pm$ 0.006 \\
(119) $D^0[P] \to \rho(770)^0 \, \rho(770)^0  $ & $D^0 \to \pi^- [\pi_2(1670)^+ \to f_2(1270)\pi^+]$ & 0.005 $\pm$ 0.002 \\
(120) $D^0 \to  \pi^- [a_1(1640)^+ \to \sigma \pi^+]$ & $D^0[P] \to \rho(770)^0 \, \rho(770)^0 $ & 0.003  $\pm$ 0.001 \\
 % end input /Users/pnaik/Documents/CLEO/Papers/latexpand/latex/interferenceFractions2.tex
 	\hline\hline
	\end{tabular}
\end{table}

%
%
% start input /Users/pnaik/Documents/CLEO/Papers/latexpand/latex/IntFractionsKKpipi.tex
\begin{table}[h]
  \footnotesize
  \centering
  \caption{\small Interference fractions $|I_{ij}| > 0.02$\%, as defined in
    \eqnPRDref{eq:DefineInterferenceFractions}, ordered by magnitude, for
    the nominal $D\to \KKpipi$ amplitude fit. Only the statistical uncertainties are given.}
\label{tab:interferenceFractionsKKpipi1}
\tiny
  \begin{tabular}
     {@{\hspace{0.5cm}}l@{\hspace{0.25cm}}  @{\hspace{0.25cm}}l@{\hspace{0.25cm}}  @{\hspace{0.25cm}}c@{\hspace{0.25cm}}}
     \hline \hline
Channel $i$ & Channel $j$ & $I_{ij}~(\%)$\\
\hline
(1)$\Dz \to K^- \, [K_{1}(1270)^{+} \to \pi^+ \, K^{*}(1430)^{0}]$&$\Dz \to K^- \, [K_{1}(1270)^{+} \to K^+ \, \rho(770)^{0}]$&	-8.145 	$\pm$	1.542 	 	\\
(2)$\Dz \to K^- \, [K_{1}(1270)^{+} \to \pi^+ \, K^{*}(892)^{0}]$&$\Dz \to K^{*}(892)^{0} \, (K^- \pi^+)_{S}$&	-5.650 	$\pm$	0.917 	\\
(3)$\Dz \to K^- \, [K_{1}(1400)^{+} \to \pi^+ \, K^{*}(892)^{0}]$&$\Dz[S] \to K^{*}(892)^{0} \, \bar{K}^{*}(892)^{0}$&	-3.686 	$\pm$	0.838 		\\
(4)$\Dz[S] \to \phi(1020) \, \rho(770)^{0}$&$ \Dz[D] \to \phi(1020) \, \rho(770)^{0}$&	-3.673 	$\pm$	0.490 	 	\\
(5)$\Dz \to K^- \, [K_{1}(1270)^{+} \to K^+ \, \rho(770)^{0}]$&$\Dz[S] \to \phi(1020) \, \rho(770)^{0}$&	3.338 	$\pm$	0.480 		\\
(6)$\Dz \to K^- \, [K_{1}(1270)^{+} \to \pi^+ \, K^{*}(892)^{0}]$&$\Dz \to K^- \, [K_{1}(1400)^{+} \to \pi^+ \, K^{*}(892)^{0}]$&	2.621 	$\pm$	1.832 	\\
(7)$\Dz \to K^- \, [K_{1}(1270)^{+} \to \pi^+ \, K^{*}(892)^{0}]$&$\Dz \to K^- \, [K_{1}(1270)^{+} \to K^+ \, \rho(770)^{0}]$&	-2.615 	$\pm$	0.462  	\\
(8)$\Dz \to K^+ \, [{K_{1}}(1270)^{-} \to K^- \, \rho(770)^{0}]$&$\Dz[S] \to \phi(1020) \, \rho(770)^{0}$&	2.321 	$\pm$	0.335 	 	\\
(9)$\Dz \to K^- \, [K_{1}(1400)^{+} \to \pi^+ \, K^{*}(892)^{0}]$&$\Dz[S] \to \phi(1020) \, \rho(770)^{0}$&	2.211 	$\pm$	0.253 	\\
(10)$\Dz \to K^- \, [K_{1}(1270)^{+} \to K^+ \, \omega(782)]$&$\Dz \to K^- \, [K_{1}(1270)^{+} \to K^+ \, \rho(770)^{0}]$&	1.941 	$\pm$	0.740 		\\
(11)$\Dz \to K^- \, [K_{1}(1270)^{+} \to \pi^+ \, K^{*}(892)^{0}]$&$\Dz[S] \to K^{*}(892)^{0} \, \bar{K}^{*}(892)^{0}$&	1.614 	$\pm$	0.426 	 	\\
(12)$\Dz \to K^- \, [K_{1}(1270)^{+} \to K^+ \, \rho(770)^{0}]$&$\Dz \to K^{*}(892)^{0} \, (K^- \pi^+)_{S}$&	1.565 	$\pm$	0.206  	\\
(13)$\Dz[S] \to K^{*}(892)^{0} \, \bar{K}^{*}(892)^{0}$&$\Dz[D] \to K^{*}(892)^{0} \, \bar{K}^{*}(892)^{0}$&	1.417 	$\pm$	0.145  	\\
(14)$\Dz \to K^- \, [K_{1}(1270)^{+} \to \pi^+ \, K^{*}(1430)^{0}]$&$\Dz \to K^+ \, [{K_{1}}(1270)^{-} \to K^- \, \rho(770)^{0}]$&	1.244 	$\pm$	0.260  	\\
(15)$\Dz \to K^- \, [K_{1}(1270)^{+} \to K^+ \, \rho(770)^{0}]$&$\Dz[S] \to K^{*}(892)^{0} \, \bar{K}^{*}(892)^{0}$&	-1.182 	$\pm$	0.166 	 	\\
(16)$\Dz \to K^- \, [K_{1}(1270)^{+} \to \pi^+ \, K^{*}(892)^{0}]$&$\Dz[D] \to K^{*}(892)^{0} \, \bar{K}^{*}(892)^{0}$&	-1.144 	$\pm$	0.212 	 	\\
(17)$\Dz \to K^- \, [K_{1}(1270)^{+} \to K^+ \, \rho(770)^{0}]$&$\Dz \to K^- \, [K_{1}(1400)^{+} \to \pi^+ \, K^{*}(892)^{0}]$&	1.119 	$\pm$	0.516  	\\
(18)$\Dz \to K^- \, [K_{1}(1400)^{+} \to \pi^+ \, K^{*}(892)^{0}]$&$\Dz \to K^{*}(892)^{0} \, (K^- \pi^+)_{S}$&	-1.052 	$\pm$	1.575 	\\
(19)$\Dz \to K^- \, [K_{1}(1400)^{+} \to \pi^+ \, K^{*}(892)^{0}]$&$\Dz[D] \to K^{*}(892)^{0} \, \bar{K}^{*}(892)^{0}$&	-0.966 	$\pm$	0.222 	\\
(20)$\Dz \to K^- \, [K_{1}(1270)^{+} \to K^+ \, \omega(782)]$&$\Dz[S] \to \phi(1020) \, \rho(770)^{0}$&	0.849 	$\pm$	0.201 	\\
(21)$\Dz[S] \to K^{*}(892)^{0} \, \bar{K}^{*}(892)^{0}$&$\Dz[S] \to \phi(1020) \, \rho(770)^{0}$&	-0.729 	$\pm$	0.164 	 	\\
(22)$\Dz \to K^{*}(892)^{0} \, (K^- \pi^+)_{S}$&$\Dz \to \phi(1020) \, (\pi^+ \pi^-)_{S}$&	0.691 	$\pm$	0.098 	\\
(23)$\Dz \to K^- \, [K^*(1680)^{+} \to \pi^+ \, K^{*}(892)^{0}]$&$\Dz[P] \to K^{*}(892)^{0} \, \bar{K}^{*}(892)^{0}$	& -0.689 	$\pm$	0.620 	 	\\
(24)$\Dz[S] \to \phi(1020) \, \rho(770)^{0}$&$\Dz[D] \to K^{*}(892)^{0} \, \bar{K}^{*}(892)^{0}$&	-0.687 	$\pm$	0.055 	 	\\
(25)$\Dz \to K^- \, [K_{1}(1270)^{+} \to \pi^+ \, K^{*}(1430)^{0}]$&$\Dz[S] \to \phi(1020) \, \rho(770)^{0}$&	0.647 	$\pm$	0.405 	\\
(26)$\Dz \to K^- \, [K_{1}(1270)^{+} \to K^+ \, \omega(782)]$&$\Dz \to K^- \, [K_{1}(1400)^{+} \to \pi^+ \, K^{*}(892)^{0}]$&	0.526 	$\pm$	0.136  	\\
(27)$\Dz \to K^- \, [K_{1}(1270)^{+} \to \pi^+ \, K^{*}(1430)^{0}]$&$\Dz \to \phi(1020) \, (\pi^+ \pi^-)_{S}$&	0.485 	$\pm$	0.085 	\\
(28)$\Dz \to K^- \, [K_{1}(1270)^{+} \to \pi^+ \, K^{*}(892)^{0}]$&$\Dz \to K^+ \, [{K_{1}}(1270)^{-} \to K^- \, \rho(770)^{0}]$&	0.424 	$\pm$	0.061\\
(29)$\Dz \to K^+ \, [{K_{1}}(1270)^{-} \to K^- \, \rho(770)^{0}]$&$  \Dz[S] \to K^{*}(892)^{0} \, \bar{K}^{*}(892)^{0}$&	-0.398 	$\pm$	0.123 	\\
(30)$\Dz \to K^- \, [K_{1}(1270)^{+} \to \pi^+ \, K^{*}(892)^{0}]$&$\Dz \to \phi(1020) \, (\pi^+ \pi^-)_{S}$&	-0.354 	$\pm$	0.055  	\\
(31)$\Dz \to K^- \, [K_{1}(1270)^{+} \to K^+ \, \omega(782)]$&$\Dz \to K^+ \, [{K_{1}}(1270)^{-} \to K^- \, \rho(770)^{0}]$&	0.346 	$\pm$	0.162 	\\
(32)$\Dz \to K^{*}(892)^{0} \, (K^- \pi^+)_{S}$&$\Dz[S] \to \phi(1020) \, \rho(770)^{0}$&	-0.341 	$\pm$	0.052 	 	\\
(33)$\Dz[P] \to K^{*}(892)^{0} \, \bar{K}^{*}(892)^{0}$&$\Dz[P] \to \phi(1020) \, \rho(770)^{0}$&	0.330 	$\pm$	0.079 	 	\\
(34)$\Dz \to K^- \, [K^*(1680)^{+} \to \pi^+ \, K^{*}(892)^{0}]$&$\Dz[P] \to \phi(1020) \, \rho(770)^{0}$&	0.303 	$\pm$	0.126 	\\
(35)$\Dz \to K^- \, [K_{1}(1400)^{+} \to \pi^+ \, K^{*}(892)^{0}]$&$\Dz \to \phi(1020) \, (\pi^+ \pi^-)_{S}$&	0.302 	$\pm$	0.125  	\\
(36)$\Dz \to K^+ \, [{K_{1}}(1270)^{-} \to K^- \, \rho(770)^{0}]$&$\Dz \to K^- \, [K_{1}(1400)^{+} \to \pi^+ \, K^{*}(892)^{0}]$&	0.280 	$\pm$	0.110 	\\
(37)$\Dz \to K^- \, [K_{1}(1270)^{+} \to \pi^+ \, K^{*}(1430)^{0}]$&$\Dz \to K^- \, [K_{1}(1270)^{+} \to K^+ \, \omega(782)]$&	0.225 	$\pm$	0.533  	\\
(38)$\Dz \to K^- \, [K_{1}(1270)^{+} \to K^+ \, \rho(770)^{0}]$&$\Dz \to K^+ \, [{K_{1}}(1270)^{-} \to K^- \, \rho(770)^{0}]$&	-0.220 	$\pm$	0.452  	\\
(39)$\Dz \to K^- \, [K_{1}(1270)^{+} \to \pi^+ \, K^{*}(1430)^{0}]$&$\Dz \to (K^+K^-)_{S} \, (\pi^+ \pi^-)_{S}$&	0.218 	$\pm$	0.022  	\\
(40)$\Dz \to K^+ \, [{K_{1}}(1270)^{-} \to K^- \, \rho(770)^{0}]$&$\Dz \to (K^+K^-)_{S} \, (\pi^+ \pi^-)_{S}$&	-0.207 	$\pm$	0.020  	\\
(41)$\Dz \to K^+ \, [{K_{1}}(1270)^{-} \to K^- \, \rho(770)^{0}]$&$ \Dz[D] \to \phi(1020) \, \rho(770)^{0}$&	-0.204 	$\pm$	0.031  	\\
(42)$\Dz \to K^- \, [K_{1}(1270)^{+} \to K^+ \, \rho(770)^{0}]$&$\Dz[D] \to K^{*}(892)^{0} \, \bar{K}^{*}(892)^{0}$&	0.197 	$\pm$	0.049 	 	\\
(43)$\Dz \to K^- \, [K_{1}(1270)^{+} \to \pi^+ \, K^{*}(1430)^{0}]$&$ \Dz[D] \to \phi(1020) \, \rho(770)^{0}$&	-0.196 	$\pm$	0.040  	\\
(44)$\Dz \to K^- \, [K_{1}(1270)^{+} \to \pi^+ \, K^{*}(892)^{0}]$&$\Dz[S] \to \phi(1020) \, \rho(770)^{0}$&	0.195 	$\pm$	0.149  	\\
(45)$\Dz \to K^+ \, [{K_{1}}(1270)^{-} \to K^- \, \rho(770)^{0}]$&$\Dz[D] \to K^{*}(892)^{0} \, \bar{K}^{*}(892)^{0}$&	-0.190 	$\pm$	0.025 	\\
(46)$\Dz \to K^- \, [K_{1}(1270)^{+} \to K^+ \, \rho(770)^{0}]$&$\Dz \to (K^+K^-)_{S} \, (\pi^+ \pi^-)_{S}$&	0.144 	$\pm$	0.015 	\\
(47)$\Dz \to K^- \, [K_{1}(1270)^{+} \to K^+ \, \omega(782)]$&$  \Dz[S] \to K^{*}(892)^{0} \, \bar{K}^{*}(892)^{0}$&	-0.142 	$\pm$	0.054  	\\
(48)$\Dz \to K^{*}(892)^{0} \, (K^- \pi^+)_{S}$&$\Dz \to (K^+K^-)_{S} \, (\pi^+ \pi^-)_{S}$&	0.127 	$\pm$	0.015 		\\
(49)$\Dz \to K^- \, [K_{1}(1270)^{+} \to \pi^+ \, K^{*}(1430)^{0}]$&$\Dz \to K^{*}(892)^{0} \, (K^- \pi^+)_{S}$&	-0.103 	$\pm$	0.015  	\\
(50)$\Dz \to K^{*}(892)^{0} \, (K^- \pi^+)_{S}$&$ \Dz[D] \to \phi(1020) \, \rho(770)^{0}$&	-0.095 	$\pm$	0.035 		\\
(51)$\Dz \to K^- \, [K_{1}(1270)^{+} \to \pi^+ \, K^{*}(892)^{0}]$&$ \Dz[D] \to \phi(1020) \, \rho(770)^{0}$&	0.080 	$\pm$	0.015 	 	\\
(52)$\Dz \to K^- \, [K_{1}(1270)^{+} \to \pi^+ \, K^{*}(892)^{0}]$&$\Dz \to (K^+K^-)_{S} \, (\pi^+ \pi^-)_{S}$&	-0.075 	$\pm$	0.010  	\\
(53)$\Dz \to K^- \, [K_{1}(1400)^{+} \to \pi^+ \, K^{*}(892)^{0}]$&$ \Dz[D] \to \phi(1020) \, \rho(770)^{0}$&	-0.075 	$\pm$	0.042 	\\
(54)$\Dz \to (K^+K^-)_{S} \, (\pi^+ \pi^-)_{S}$&$\Dz[P] \to K^{*}(892)^{0} \, \bar{K}^{*}(892)^{0}$&	-0.066 	$\pm$	0.007 	\\
(55)$\Dz \to K^+ \, [{K_{1}}(1270)^{-} \to K^- \, \rho(770)^{0}]$&$\Dz \to K^{*}(892)^{0} \, (K^- \pi^+)_{S}$&	0.064 	$\pm$	0.097  	\\
(56) $\Dz \to K^- \, [K^*(1680)^{+} \to \pi^+ \, K^{*}(892)^{0}]$&$\Dz[D] \to K^{*}(892)^{0} \, \bar{K}^{*}(892)^{0}$&	0.061 	$\pm$	0.009 	 	\\
(57)$\Dz \to K^{*}(892)^{0} \, (K^- \pi^+)_{S}$&$\Dz[D] \to K^{*}(892)^{0} \, \bar{K}^{*}(892)^{0}$&	0.057 	$\pm$	0.008 	\\
(58)$\Dz \to K^- \, [K_{1}(1400)^{+} \to \pi^+ \, K^{*}(892)^{0}]$&$\Dz \to (K^+K^-)_{S} \, (\pi^+ \pi^-)_{S}$&	0.048 	$\pm$	0.019  	\\
(59)$\Dz \to K^- \, [K_{1}(1400)^{+} \to \pi^+ \, K^{*}(892)^{0}]$&$\Dz \to K^- \, [K^*(1680)^{+} \to \pi^+ \, K^{*}(892)^{0}]$&	-0.048 	$\pm$	0.016 	\\
(60)$\Dz \to K^- \, [K_{1}(1270)^{+} \to \pi^+ \, K^{*}(892)^{0}]$&$\Dz \to K^- \, [K_{1}(1270)^{+} \to K^+ \, \omega(782)]$&	0.044 	$\pm$	0.173  	\\
(61)$\Dz \to K^- \, [K_{1}(1270)^{+} \to \pi^+ \, K^{*}(1430)^{0}]$&$\Dz \to K^- \, [K_{1}(1270)^{+} \to \pi^+ \, K^{*}(892)^{0}]$&	0.044 	$\pm$	0.007 		\\
(62)$\Dz \to K^- \, [K_{1}(1270)^{+} \to \pi^+ \, K^{*}(892)^{0}]$&$\Dz[P] \to K^{*}(892)^{0} \, \bar{K}^{*}(892)^{0}$&	0.044 	$\pm$	0.008 		\\
(63)$\Dz \to K^- \, [K_{1}(1270)^{+} \to K^+ \, \omega(782)]$&$ \Dz[D] \to \phi(1020) \, \rho(770)^{0}$&	-0.042 	$\pm$	0.015 	 	\\
(64)$\Dz \to K^- \, [K_{1}(1270)^{+} \to K^+ \, \rho(770)^{0}]$&$\Dz[P] \to K^{*}(892)^{0} \, \bar{K}^{*}(892)^{0}$&	-0.036 	$\pm$	0.004  	\\
(65)$\Dz \to K^- \, [K^*(1680)^{+} \to \pi^+ \, K^{*}(892)^{0}]$&$\Dz[S] \to K^{*}(892)^{0} \, \bar{K}^{*}(892)^{0}$&	-0.034 	$\pm$	0.007  	\\
(66)$\Dz[S] \to K^{*}(892)^{0} \, \bar{K}^{*}(892)^{0}$&$\Dz[P] \to K^{*}(892)^{0} \, \bar{K}^{*}(892)^{0}$&	-0.033 	$\pm$	0.007 		\\
(67)$\Dz[S] \to \phi(1020) \, \rho(770)^{0}$&$\Dz[P] \to \phi(1020) \, \rho(770)^{0}$&	0.033 	$\pm$	0.004 	 	\\
(68)$\Dz \to K^- \, [K_{1}(1270)^{+} \to \pi^+ \, K^{*}(892)^{0}]$&$\Dz \to K^- \, [K^*(1680)^{+} \to \pi^+ \, K^{*}(892)^{0}]$&	-0.033 	$\pm$	0.008  	\\
(69)$\Dz \to K^- \, [K_{1}(1270)^{+} \to K^+ \, \omega(782)]$&$\Dz \to K^{*}(892)^{0} \, (K^- \pi^+)_{S}$&	0.027 	$\pm$	0.069  	\\
(70)$\Dz \to K^- \, [K_{1}(1270)^{+} \to \pi^+ \, K^{*}(1430)^{0}]$&$\Dz[P] \to K^{*}(892)^{0} \, \bar{K}^{*}(892)^{0}$&	0.024 	$\pm$	0.003  	\\
(71)$\Dz \to K^- \, [K_{1}(1270)^{+} \to K^+ \, \omega(782)]$&$\Dz[D] \to K^{*}(892)^{0} \, \bar{K}^{*}(892)^{0}$&	-0.023 	$\pm$	0.008 		\\
	\hline\hline
	\end{tabular}
\end{table}

\begin{table}[h]
  \footnotesize
  \centering
  \caption{\small Interference fractions $|I_{ij}| < 0.02$\%, as defined in
    \eqnPRDref{eq:DefineInterferenceFractions}, ordered by magnitude, for
    the nominal $D\to \KKpipi$ amplitude fit. Only the statistical uncertainties are given.}
\label{tab:interferenceFractionsKKpipi2}
\tiny
  \begin{tabular}
     {@{\hspace{0.5cm}}l@{\hspace{0.25cm}}  @{\hspace{0.25cm}}l@{\hspace{0.25cm}}  @{\hspace{0.25cm}}c@{\hspace{0.25cm}}}
     \hline \hline
Channel $i$ & Channel $j$ & $I_{ij}~(\%)$\\
\hline
(72)$\Dz \to \phi(1020) \, (\pi^+ \pi^-)_{S}$&$\Dz[S] \to \phi(1020) \, \rho(770)^{0}$&	-0.019 	$\pm$	0.021  	\\
(73)$\Dz \to (K^+K^-)_{S} \, (\pi^+ \pi^-)_{S}$&$\Dz[P] \to \phi(1020) \, \rho(770)^{0}$&	-0.019 	$\pm$	0.003 	\\
(74)$\Dz \to (K^+K^-)_{S} \, (\pi^+ \pi^-)_{S}$&$\Dz[S] \to \phi(1020) \, \rho(770)^{0}$&	-0.018 	$\pm$	0.004 	\\
(75)$\Dz \to K^+ \, [{K_{1}}(1270)^{-} \to K^- \, \rho(770)^{0}]$&$\Dz[P] \to \phi(1020) \, \rho(770)^{0}$&	0.017 	$\pm$	0.003  	\\
(76)$\Dz \to K^- \, [K_{1}(1400)^{+} \to \pi^+ \, K^{*}(892)^{0}]$&$\Dz[P] \to K^{*}(892)^{0} \, \bar{K}^{*}(892)^{0}$&	0.017 	$\pm$	0.014  	\\
(77)$\Dz \to K^- \, [K_{1}(1270)^{+} \to K^+ \, \rho(770)^{0}]$&$ \Dz[D] \to \phi(1020) \, \rho(770)^{0}$&	0.017 	$\pm$	0.064 	\\
(78)$\Dz[S] \to K^{*}(892)^{0} \, \bar{K}^{*}(892)^{0}$&$\Dz \to \phi(1020) \, (\pi^+ \pi^-)_{S}$&	0.016 	$\pm$	0.004  	\\
(79)$\Dz \to K^- \, [K_{1}(1270)^{+} \to \pi^+ \, K^{*}(1430)^{0}]$&$\Dz \to K^- \, [K_{1}(1400)^{+} \to \pi^+ \, K^{*}(892)^{0}]$&	-0.015 	$\pm$	0.005  	\\
(80)$\Dz \to K^- \, [K_{1}(1270)^{+} \to K^+ \, \omega(782)]$&$\Dz \to (K^+K^-)_{S} \, (\pi^+ \pi^-)_{S}$&	0.013 	$\pm$	0.008  	\\
(81)$\Dz \to (K^+K^-)_{S} \, (\pi^+ \pi^-)_{S}$&$\Dz \to \phi(1020) \, (\pi^+ \pi^-)_{S}$&	-0.013 	$\pm$	0.007  	\\
(82)$\Dz \to K^- \, [K_{1}(1270)^{+} \to \pi^+ \, K^{*}(1430)^{0}]$&$\Dz[S] \to K^{*}(892)^{0} \, \bar{K}^{*}(892)^{0}$&	0.012 	$\pm$	0.007 	\\
(83)$\Dz[D] \to K^{*}(892)^{0} \, \bar{K}^{*}(892)^{0}$&$ \Dz[D] \to \phi(1020) \, \rho(770)^{0}$&	0.012 	$\pm$	0.033  	\\
(84)$\Dz \to K^+ \, [{K_{1}}(1270)^{-} \to K^- \, \rho(770)^{0}]$&$\Dz[P] \to K^{*}(892)^{0} \, \bar{K}^{*}(892)^{0}$&	0.011 	$\pm$	0.003 	\\
(85)$\Dz \to K^{*}(892)^{0} \, (K^- \pi^+)_{S}$&$\Dz[P] \to K^{*}(892)^{0} \, \bar{K}^{*}(892)^{0}$&	0.011 	$\pm$	0.002 	\\
(86)$\Dz \to K^- \, [K_{1}(1270)^{+} \to K^+ \, \rho(770)^{0}]$&$\Dz \to \phi(1020) \, (\pi^+ \pi^-)_{S}$&	-0.010 	$\pm$	0.002  	\\
(87)$\Dz[S] \to K^{*}(892)^{0} \, \bar{K}^{*}(892)^{0}$&$\Dz \to K^{*}(892)^{0} \, (K^- \pi^+)_{S}$&	-0.008 	$\pm$	0.001 	\\
(88)$ \Dz[D] \to \phi(1020) \, \rho(770)^{0}$&$\Dz[P] \to \phi(1020) \, \rho(770)^{0}$&	0.008 	$\pm$	0.001  	\\
(89)$\Dz[S] \to K^{*}(892)^{0} \, \bar{K}^{*}(892)^{0}$&$\Dz[P] \to \phi(1020) \, \rho(770)^{0}$&	-0.008 	$\pm$	0.001\\
(90)$\Dz \to K^- \, [K_{1}(1270)^{+} \to \pi^+ \, K^{*}(1430)^{0}]$&$\Dz \to K^- \, [K^*(1680)^{+} \to \pi^+ \, K^{*}(892)^{0}]$&	0.007 	$\pm$	0.018  	\\
(91)$\Dz \to \phi(1020) \, (\pi^+ \pi^-)_{S}$&$\Dz[D] \to K^{*}(892)^{0} \, \bar{K}^{*}(892)^{0}$&	0.007 	$\pm$	0.003  	\\
(92)$\Dz \to K^- \, [K_{1}(1270)^{+} \to K^+ \, \omega(782)]$&$\Dz \to K^- \, [K^*(1680)^{+} \to \pi^+ \, K^{*}(892)^{0}]$&	-0.006 	$\pm$	0.002  	\\
(93)$\Dz \to K^+ \, [{K_{1}}(1270)^{-} \to K^- \, \rho(770)^{0}]$&$\Dz \to K^- \, [K^*(1680)^{+} \to \pi^+ \, K^{*}(892)^{0}]$&	0.006 	$\pm$	0.006  	\\
(94)$\Dz \to K^- \, [K^*(1680)^{+} \to \pi^+ \, K^{*}(892)^{0}]$&$\Dz \to (K^+K^-)_{S} \, (\pi^+ \pi^-)_{S}$&	0.006 	$\pm$	0.020  	\\
(95)$\Dz \to K^- \, [K^*(1680)^{+} \to \pi^+ \, K^{*}(892)^{0}]$&$\Dz \to K^{*}(892)^{0} \, (K^- \pi^+)_{S}$&	-0.006 	$\pm$	0.001 	\\
(96)$\Dz \to K^- \, [K_{1}(1400)^{+} \to \pi^+ \, K^{*}(892)^{0}]$&$\Dz[P] \to \phi(1020) \, \rho(770)^{0}$&	-0.006 	$\pm$	0.002  	\\
(97)$\Dz \to (K^+K^-)_{S} \, (\pi^+ \pi^-)_{S}$&$\Dz[D] \to K^{*}(892)^{0} \, \bar{K}^{*}(892)^{0}$&	-0.005 	$\pm$	0.008  	\\
(98)$\Dz \to K^- \, [K^*(1680)^{+} \to \pi^+ \, K^{*}(892)^{0}]$&$\Dz[S] \to \phi(1020) \, \rho(770)^{0}$&	-0.005 	$\pm$	0.001  	\\
(99)$\Dz \to K^- \, [K_{1}(1270)^{+} \to \pi^+ \, K^{*}(892)^{0}]$&$\Dz[P] \to \phi(1020) \, \rho(770)^{0}$&	-0.005 	$\pm$	0.002 	\\
(100)$\Dz \to \phi(1020) \, (\pi^+ \pi^-)_{S}$&$\Dz[P] \to K^{*}(892)^{0} \, \bar{K}^{*}(892)^{0}$&	0.004 	$\pm$	0.001  	\\
(101)$\Dz \to K^- \, [K_{1}(1270)^{+} \to \pi^+ \, K^{*}(1430)^{0}]$&$\Dz[D] \to K^{*}(892)^{0} \, \bar{K}^{*}(892)^{0}$&	-0.004 	$\pm$	0.004  	\\
(102)$\Dz \to K^- \, [K_{1}(1270)^{+} \to K^+ \, \rho(770)^{0}]$&$\Dz[P] \to \phi(1020) \, \rho(770)^{0}$&	-0.004 	$\pm$	0.007 	\\
(103)$\Dz \to K^- \, [K^*(1680)^{+} \to \pi^+ \, K^{*}(892)^{0}]$&$ \Dz[D] \to \phi(1020) \, \rho(770)^{0}$&	-0.004 	$\pm$	0.001 		\\
(104)$\Dz \to \phi(1020) \, (\pi^+ \pi^-)_{S}$&$ \Dz[D] \to \phi(1020) \, \rho(770)^{0}$&	0.003 	$\pm$	0.014  	\\
(105)$\Dz[D] \to K^{*}(892)^{0} \, \bar{K}^{*}(892)^{0}$&$\Dz[P] \to K^{*}(892)^{0} \, \bar{K}^{*}(892)^{0}$&	0.003 	$\pm$	0.001  	\\
(106)$\Dz[D] \to K^{*}(892)^{0} \, \bar{K}^{*}(892)^{0}$&$\Dz[P] \to \phi(1020) \, \rho(770)^{0}$&	0.003 	$\pm$	0.001  	\\
(107)$\Dz \to K^{*}(892)^{0} \, (K^- \pi^+)_{S}$&$\Dz[P] \to \phi(1020) \, \rho(770)^{0}$&	0.002 	$\pm$	0.001 	\\
(108)$\Dz \to (K^+K^-)_{S} \, (\pi^+ \pi^-)_{S}$&$ \Dz[D] \to \phi(1020) \, \rho(770)^{0}$&	-0.002 	$\pm$	0.001  	\\
(109)$\Dz \to K^- \, [K^*(1680)^{+} \to \pi^+ \, K^{*}(892)^{0}]$&$\Dz \to \phi(1020) \, (\pi^+ \pi^-)_{S}$&	0.002 	$\pm$	0.002  	\\
(110)$\Dz \to K^- \, [K_{1}(1270)^{+} \to K^+ \, \omega(782)]$&$\Dz[S] \to K^{*}(892)^{0} \, \bar{K}^{*}(892)^{0}$&	-0.002 	$\pm$	0.002 	\\
(111)$\Dz \to K^- \, [K_{1}(1270)^{+} \to K^+ \, \rho(770)^{0}]$&$\Dz \to K^- \, [K^*(1680)^{+} \to \pi^+ \, K^{*}(892)^{0}]$&	0.002 	$\pm$	0.007  	\\
(112)$\Dz[S] \to K^{*}(892)^{0} \, \bar{K}^{*}(892)^{0}$&$\Dz \to (K^+K^-)_{S} \, (\pi^+ \pi^-)_{S}$&	-0.001 	$\pm$	0.001 	\\
(113)$\Dz[S] \to \phi(1020) \, \rho(770)^{0}$&$\Dz[P] \to K^{*}(892)^{0} \, \bar{K}^{*}(892)^{0}$&	0.001 	$\pm$	0.003 	\\
(114)$\Dz \to K^- \, [K_{1}(1270)^{+} \to K^+ \, \omega(782)]$&$\Dz[P] \to \phi(1020) \, \rho(770)^{0}$&	0.001 	$\pm$	0.001  	\\
(115)$ \Dz[D] \to \phi(1020) \, \rho(770)^{0}$&$\Dz[P] \to K^{*}(892)^{0} \, \bar{K}^{*}(892)^{0}$&	-0.001 	$\pm$	0.001 	\\
(116)$\Dz \to K^- \, [K_{1}(1270)^{+} \to \pi^+ \, K^{*}(1430)^{0}]$&$\Dz[P] \to \phi(1020) \, \rho(770)^{0}$&	0.001 	$\pm$	0.002  	\\
(117)$\Dz \to K^- \, [K_{1}(1270)^{+} \to K^+ \, \omega(782)]$&$\Dz \to \phi(1020) \, (\pi^+ \pi^-)_{S}$&	0.001 	$\pm$	0.001  	\\
(118)$\Dz \to K^+ \, [{K_{1}}(1270)^{-} \to K^- \, \rho(770)^{0}]$&$\Dz \to \phi(1020) \, (\pi^+ \pi^-)_{S}$&	0.000 	$\pm$	0.003 	\\
(119)$\Dz[S] \to K^{*}(892)^{0} \, \bar{K}^{*}(892)^{0}$&$ \Dz[D] \to \phi(1020) \, \rho(770)^{0}$&	0.000 	$\pm$	0.010  	\\
(120)$\Dz \to \phi(1020) \, (\pi^+ \pi^-)_{S}$&$\Dz[P] \to \phi(1020) \, \rho(770)^{0}$&	0.000 	$\pm$	0.004  	\\
	\hline\hline
	\end{tabular}
\end{table}
 % end input /Users/pnaik/Documents/CLEO/Papers/latexpand/latex/IntFractionsKKpipi.tex
  % end input /Users/pnaik/Documents/CLEO/Papers/latexpand/latex/appendix.tex
 %
% start input /Users/pnaik/Documents/CLEO/Papers/latexpand/latex/supplemental.tex
\clearpage
\setcounter{section}{0}
\section*{Supplemental Material}

We provide a collection of \verb|C| macros to reproduce all energy-dependent masses and widths described in Sec.~\ref{ssec:lineshapes}. These are intended to be parsed by the \verb|ROOT| software and have names indicating which energy-dependent quantity and resonance they correspond to.

Two additional text files containing the statistical correlation matrices of the nominal results for $\Dz \to \fourpi$ and $\Dz \to \KKpipi$ are provided. Their filenames are \verb|Correlations4pi.txt| and \verb|CorrelationsKKpipi.txt|, respectively. The format of each file is as follows. Firstly, each free parameter is assigned a numerical identifier. Following this, the lower diagonal correlation matrix is given for these indices.
 % end input /Users/pnaik/Documents/CLEO/Papers/latexpand/latex/supplemental.tex
 
\clearpage
\ifx\mcitethebibliography\mciteundefinedmacro
\PackageError{LHCb.bst}{mciteplus.sty has not been loaded}
{This bibstyle requires the use of the mciteplus package.}\fi
\providecommand{\href}[2]{#2}

\bibliographystyle{LHCb} 

\end{document}